\documentclass[
11pt, 
english, 
singlespacing, 
headsepline, 
]{MastersDoctoralThesis} 

\usepackage{mystyle}

\begin{document}

\pagestyle{plain} 

\pagenumbering{gobble}




\frontmatter 

\begin{titlepage}
\begin{center}

{\scshape\LARGE \univname\par}\vspace{1.5cm} 
\textsc{\Large Doctoral Thesis}\\[0.5cm] 

\HRule \\[0.4cm] 
{\huge \bfseries \ttitle\par}\vspace{0.4cm} 
\HRule \\[1.5cm] 
 
\begin{minipage}[t]{0.4\textwidth}
\begin{flushleft} \large
\emph{Author:}\\
\href{https://www.icfo.eu/people/people_details?people_id=924}{\authorname} 
\end{flushleft}
\end{minipage}
\begin{minipage}[t]{0.4\textwidth}
\begin{flushright} \large
\emph{Supervisor:} \\
\href{http://www.southampton.ac.uk/maths/about/staff/cl2u07.page}{\supname} 

\href{https://www.icfo.eu/research/84-group-member-details.html?gid=23&people_id=25}{Prof.~Maciej \textsc{Lewenstein}}
\end{flushright}
\end{minipage}\\[3cm]
 
\large \textit{A thesis submitted in fulfillment of the requirements\\ for the degree of \degreename}\\[0.3cm] 
\textit{in the}\\[0.4cm]
\facname\\[2cm]
 
{\large \today}\\[4cm] 

\vfill
\end{center}
\end{titlepage}


\vspace*{0.2\textheight}
\noindent\enquote{\itshape If you haven't found something strange during the day, it hasn't been much of a day.}\bigbreak
\hfill John Archibald Wheeler


\begin{abstract}
\addchaptertocentry{\abstractname} 

The central goal of this thesis is to develop methods to experimentally study topological phases. We do so by applying the powerful toolbox of quantum simulation techniques with cold atoms in optical lattices. To this day, a complete classification of topological phases remains elusive. In this context, experimental studies are key, both for studying the interplay between topology and complex effects and for identifying new forms of topological order. It is therefore crucial to find complementary means to measure topological properties in order to reach a fundamental understanding of topological phases. In one dimensional chiral systems, we suggest a new way to construct and identify topologically protected bound states, which are the smoking gun of these materials. In two dimensional Hofstadter strips (i.e: systems which are very short along one dimension), we suggest a new way to measure the topological invariant directly from the atomic dynamics.

In one dimensional optical lattices, topological bound states are difficult to generate due to the absence of sharp boundaries, and harder still to identify unambiguously. By periodically driving a one dimensional dilute gas of atoms with a pair of Raman lasers, we find that a system analogous to the two-step quantum walk can be realised. This system can host two flavours of topologically protected bound states, meaning that it escapes the standard classification of topological phases. This study details the considerations and many of the relevant experimental tools to design a topologically non-trivial system. In particular, we show that we can build a topological boundary by using the lasers' finite beam width, and that the topologically protected states which live at this boundary can be identified, and differentiated, by studying their spin distribution.

The bulk-boundary correspondence states that a system's bulk and edge properties are indissociable. It is unclear, however, how this principle extends to systems with vanishingly small bulks, as for instance the Hofstadter strip, which was recently realised using a one dimensional gas of spinful atoms. We define a topological invariant for this system which accurately counts the number of topological bound states. This suggests that, even in such an extreme situation, the bulk-boundary correspondence applies. We suggest a method for experimentally measuring this invariant from the atomic dynamics which relies on three main ingredients: the adiabatic loading of a well localised wavepacket in the ground state of the lattice, the application of a weak force along the axis of the strip, and the measurement of the centre of mass position after a Bloch oscillation.

\end{abstract}


\tableofcontents 

\listoffigures 

\listoftables 


\begin{declaration}
\addchaptertocentry{\authorshipname}

\noindent I, \authorname, declare that the thesis entitled \emph{\ttitle} and the work presented in the thesis are both my own, and have been generated by me as the result of my own original research. I confirm that:

\begin{itemize} 
\item this work was done wholly or mainly while in candidature for a research degree
at this University;
\item where any part of this thesis has previously been submitted for a degree or any
other qualification at this University or any other institution, this has been clearly
stated;
\item where I have consulted the published work of others, this is always clearly
attributed;
\item where I have quoted from the work of others, the source is always given. With
the exception of such quotations, this thesis is entirely my own work;
\item I have acknowledged all main sources of help;
\item where the thesis is based on work done by myself jointly with others, I have made clear exactly what was done by others and what I have contributed myself;
\item parts of this work have been published \citep{Mugel} and submitted for publication \parencite{Mugel2017}\\
\end{itemize}
 
\noindent Signed:\\
\rule[0.5em]{25em}{0.5pt} 
 
\noindent Date:\\
\rule[0.5em]{25em}{0.5pt} 
\end{declaration}

\cleardoublepage


\begin{acknowledgements}
\addchaptertocentry{\acknowledgementname}

First and foremost, I would like to thank my supervisors, Carlos Lobo and Maciej Lewenstein. Amongst his many qualities, what I admire the most in Carlos is his skill as a teacher, and the formidable patience he has when faced with a confused student (usually me). Maciej, I admire for his inexorable drive, his indisputable fairness, and his admirable self discipline. I could not have wished for two better forces to drive my work.

I would also like to thank warmly my collaborators, Pietro Massignan, Alessio Celi, Alexandre Dauphin, Leticia Tarruell and J\'anosh Asb\'oth. I cannot say how many times they helped me solve problems that I did not know how to tackle and quite frankly I think I owe them my sanity.

I would like to thank my friend and fellow PhD student Johnathan Lau for our stimulating discussions and the help he gave me when I first arrived in Southampton. I was shocked to hear about his tragic demise, and regret I could not be with our friends at the time.

I can confidently say that I have always been greatly attracted to research. I don't have words to express how proud I am to be able to acknowledge Charles de Llamby, Pauline Ascher, Bernard Buigues and Pierre Baudy. There is no doubt in my mind that it was our discussions that led me down the road of physics. I look forward to sharing more discussions like these in the future. An equally great source of inspiration was my grandfather, David Shrewsbury. He would have been so proud to be able to read my thesis and I'm heartbroken that I can't give him this pleasure.

Both my parents have this amazing, unquenchable thirst of knowledge, and I think that they've succeeded in passing some of that on to my little sister and I. I've been lucky to grow up in an exceptional family gifted with very special qualities, who have always encouraged me to follow my passions, and never questioned my decisions. I'm eager to enjoy many more merry travels with the three of you in the future.

Lena, you have succeeded in filling me with joy for this past year. Joy and delicious food. I'm so absolutely, deeply happy to have you in my life, I don't know what I've done to deserve you but it can't have been enough.

Jake, you are a really special friend. You've made me laugh to tears more than once. Sharing a flat with you has been and continues to be an absolute pleasure. Matt, every single time we've seen each other has been thoroughly enjoyable. I have never seen a passion for science as absolute as yours, and I can honestly say that without it to inspire me, I would probably not be where I am today.

I've made some close friends during my PhD, amongst which Anna, Vee, Abi, Poppie, Simon, Janeck, Giles, Martha, Konstanze, Thor, Olivia, Clara and Clara. I don't know where we'll see each other next but I sure hope it is soon. I'd particularly like to thank my fellow PhD students, Antoine, Nello, Angello, Alessandro, Cesar, Stan, Paco, Marta, Vanessa, Yafet, Tristan and Will. You have impressed me, inspired me, and made me proud time and time again to be able to count you among my friends.

Finally, I'd like to acknowledge my oldest friends, that are sadly also those that I miss the most cruelly: Jeanne, Aladin, Thomas, Damien, Mathias, Rose, Greg, Guy, Gab, Thibaut, Olivier, Naya, Monika, Antoine, Alasdair, Marie, Kat, Raphael, Nikolai, Cyril, Thibaut, Emm, Totor, Kailas and Hippie Tim. No matter how far apart we are and how long we go without meeting, I continue to unconditionally treasure our friendship. Jack and Jojo, every moment we had together was exceptional; I keep, now and forever, a loving memory of you.

With the oversight of Carlos Lobo, Maciej Lewenstein, Pietro Massignan and Alexandre Dauphin, editorial advice has been sought. No changes of intellectual content were made as a result of this advice.

\end{acknowledgements}


\begin{abbreviations}{ll} 

$\uparrow$ & Spin up (state)\\
$\downarrow$ & Spin down (state)\\
1D, 2D, 3D & One, two, three dimensions\\
\textbf{H}.\textbf{c}. & \textbf{H}ermitian \textbf{C}onjugate\\
\textbf{BZ} & \textbf{B}rillouin \textbf{Z}one \\
\textbf{CS} & \textbf{C}hiral \textbf{S}ymmetry\\
\textbf{TRS} & \textbf{T}ime \textbf{R}eversal \textbf{S}ymmetry\\
\textbf{PHS} & \textbf{P}article-\textbf{H}ole \textbf{S}ymmetry\\
\textbf{SSH} & \textbf{S}u-\textbf{S}chrieffer–\textbf{H}eeger (model) \\
\textbf{FHS} & \textbf{F}ukui-\textbf{H}atsugai-\textbf{S}uzuki (algorithm)\\
\textbf{TKNN} & \textbf{T}houless-\textbf{K}ohmoto-\textbf{N}ightingale-den \textbf{N}ijs (formula)

\end{abbreviations}


%
%
%


\begin{symbols}{lll} 


\addlinespace
\multicolumn{3}{c}{ \emph{Variables} }\\
\addlinespace

$n,m$ & Site index & \\
$x, y, z$ & Position & $m$ \\
$k$ & Quasimomentum & $m^{-1}$\\
$E$ & Energy & $J$ \\
$\varepsilon$ & Quasienergy & $J$ \\
$j$ & Band index & \\
$d$ & Lattice spacing & $m$ \\
$J$ & Tunnelling amplitude & $J$ \\
$\delta$ & Tunnelling amplitude & $J$ \\
$\theta$ & Spin mixing angle & \\
$\nu$ & Winding number & \\
$\mathcal{C}$ & Chern number & \\

\addlinespace
\multicolumn{3}{c}{ \emph{Tensors and operators} }\\
\addlinespace

$\bm{e}_i$ & Unit vector, with $i \in \{x,y,z\}$ &  \\
$\ket{u_j}$ & $j^{\rm th}$ Bloch state & \\
$\sigma_i$ & $i^{\rm th}$ Pauli matrix acting in spin space & \\
$\tau_i$ & $i^{\rm th}$ Pauli matrix acting in sublattice space & \\
$\hat{\mathcal{K}}$ & Complex conjugation operator & \\
$\hat{\Gamma}$ & Chiral symmetry operator & \\
$\hat{\mathcal{T}}$ & Time reversal symmetry operator & \\
$\hat{\mathcal{P}}$ & Particle-hole symmetry operator & \\

\addlinespace
\multicolumn{3}{c}{ \emph{Operators in the basis with four internal states} }\\
\addlinespace

$\ccc_n^\dagger, \ccc_n$ & Four vector of creation/annihilation operators & \\
$\h_S$ & Optical lattice Hamiltonian & \\
$\h_\theta$ & Raman field Hamiltonian & \\
\^U & Time evolution operator & \\
$\h_F$ & Floquet Hamiltonian & \\

\addlinespace
\multicolumn{3}{c}{ \emph{Operators in the two state (transformed) basis} }\\
\addlinespace

$\cc_n^\dagger, \cc_n$ & Vector of creation/annihilation operators & \\
$\h_S'$ & Optical lattice Hamiltonian & \\
$\h_\theta'$ & Raman field Hamiltonian & \\
\^U$'$ & Time evolution operator & \\
$\h_F'$ & Floquet Hamiltonian & \\

\end{symbols}


\dedicatory{For Jack Amiel and David Shrewsbury} 


\mainmatter 

\pagestyle{thesis} 



\chapter{General introduction}
\label{Chapter1}

\section{The integer quantum Hall effect}


At the heart of many-body physics lies the desire to understand the collective behaviour of particles. With the formulation of Bloch's theorem \parencite{Ashcroft1976} and density functional theory \parencite{Kohn1965,Engel2011}, it became possible to understand systems composed of many quantum mechanical bodies, thereby giving us the tools to describe a vast array of materials. It is from these theorems, for instance, that scientists can recognise at a glance
an electrical insulator.

Many years later, Klaus von Klitzing et al. made an extremely intriguing discovery: when exposing two dimensional silicon samples to a strong magnetic field, these systems could be made to conduct electricity, despite having all the characteristics of an electrical insulator. What is more, they found that the electrical current was perpendicular to the applied voltage, and that the sample's conductivity was exactly quantised \parencite{Klitzing1980}.

The phenomenon, which became famously known as the integer quantum Hall effect, was only explained a few years later by Laughlin and Thouless.
They showed that the integer quantum Hall effect occurs in new states of matter,
and that the transverse conductance in quantum Hall systems is quantised because the system can present different topological phases \parencite{Laughlin1981,Thouless1983}. By way of definition, two Hamiltonians correspond to distinct topological phases if one cannot be adiabatically deformed into the other. Of particular interest, the system's phase was found to be robust to small perturbations, such that all samples show the exact same conductivity for a given magnetic field, despite imperfections.

This discovery marked the birth of the first known topologically non-trivial material. 
These materials attracted a lot of attention because their phase transitions could not be characterised fully by specifying only the system’s local symmetries \parencite{Landau1936} but required additional topological information.

Indeed, many different topologically distinct phases can have the same local symmetry groups \parencite{Wen1989}.
In this sense, topological phases form an entirely new paradigm. The system's topological order is determined by an integer parameter, known as the topological invariant. What is more, this quantity is a global parameter: it cannot be determined by a local measurement \parencite{Nakahara2003}. It is this property which gives the conductance its robust character: local perturbations to the Hamiltonian cannot modify the system's phase.

It was not long before topologically non-trivial, finite systems were found to display unique surface properties. Indeed, these materials host robust excitations which are localised at the sample's edges. What is more, the number of these edge states is determined by the bulk topological invariant. This is the so-called bulk-boundary correspondence, which states that the properties of the system's boundary are determined by the state of the bulk \parencite{Wen2004QFT}.

The extremely precise quantisation of the Hall conductivity constituted the primary motivation to study topologically non-trivial systems, as it provides a mean to measure fundamental constants to a high accuracy \parencite{Klitzing1980}. Additionally, topologically non-trivial materials present surface Hamiltonians which could not exist without the bulk \parencite{Hasan2010}, which constitute an interesting resource to engineer new Hamiltonians. More recently, researchers suggested using the system's robust edge states to protect entangled states \parencite{Moulieras2013,Mittal2016} or to transfer quantum information \parencite{Rechtsman2016,Dlaska2017}. In the presence of interactions, these edge states can display fractional statistics \parencite{Yoshioka2002}. It is this discovery which has attracted the most attention, both for engineering exotic states of matter \parencite{Tsui1982,Laughlin1983}, and for the implementation of powerful quantum algorithms \parencite{Kitaev2003,Sarma2015}.


\section{Review of systems in which topology appears}


Almost ten years after the discovery of the integer quantum Hall effect, Haldane showed, with the discovery of the now famous Haldane model, that the dynamics particular to the integer quantum Hall effect can be obtained without an external magnetic field \parencite{Haldane1988}. It was this discovery which inspired researchers to look for new ways to induce momentum dependent forces, for instance by using spin-orbit coupling. This lead to a new generation of materials which do not have robust mass currents, which are characteristic of the integer quantum Hall effect. Their edge does, however, present a non-zero spin current \parencite{Murakami2004}, which was later shown to be robust and characterised by a new topological invariant \parencite{Kane2005a}. These systems became known as $\mathbb{Z}_2$ topological insulators.

One dimensional systems which accommodate topologically protected edge states were also discovered, such as the Kitaev chain \parencite{Kitaev2001}. This particular model attracted a lot of attention because its topologically protected excitations were found to be Majorana particles, states with anyonic statistics which have many applications for quantum algorithms.

New physical platforms outside of condensed matter physics were soon found which could also host topological phases. An important example is the Jackiw-Rebbi model, which shows that the one dimensional Dirac equation can host a zero energy mode which is topologically protected \parencite{Jackiw1976}. Another example which is relevant to quantum information is known as the Toric code, which uses topological operations to eliminate errors in stored data \parencite{Kitaev1997}.

Finally, three dimensional topological materials were also discovered, which were found to come in two flavours: weak topological insulators, which are not robust to disorder, and are obtained by layering $\mathbb{Z}_2$ topological insulators \parencite{Fu2007,Moore2007,Roy2009}, and strong topological insulators  \parencite{Moore2007}, which present topologically protected metallic surfaces.

\subsection{Table of topological phases}
\label{sec:IntroToTopology}

As we will come to understand, a topologically non-trivial system implies that its Hamiltonian is constrained. From this simple argument, we can see that it is crucial to know a system's symmetries to understand its topological properties. In this context, we define a symmetry as any constraint associated to an operator which either commutes or anti-commutes with the Hamiltonian.

\begin{table}[t]
\centering
\caption[Topological classification of Hamiltonians]{Classification of $d$ dimensional systems, with $d=\{1, 2, 3\}$, in the ten topological symmetry classes. The columns corresponding to time-reversal symmetry (TRS), particle-hole symmetry (PHS), and chiral symmetry (CS) show the symmetries of each class. The presence of a symmetry is marked by a $+1$ ($-1$) if the corresponding symmetry operator squares to the identity $\sigma_0$ ($-\sigma_0$). The symbol $\mathbb{Z}$ ($\mathbb{Z}_2$) indicates that the system's topological invariant can assume integer values (two values). This box is left blank if the class is topologically trivial.}
\label{tab:topo}
\begin{tabular}{ |c|c|c|c|c|c|c|c| }
\hline
category & symmetry class & TRS & PHS & CS & $d=1$ & $d=2$ & $d=3$ \\
 
\hline\hline
\multirow{3}{*}{Standard}
& A &   &   &   &   & $\mathbb{Z}$ &   \\
& AI & $+1$ &   &   &   &   &   \\
& AII & $-1$ &   &   &   & $\mathbb{Z}_2$ & $\mathbb{Z}_2$ \\

\hline
\multirow{3}{*}{Chiral}
& AIII &   &   & $+1$ & $\mathbb{Z}$ &   & $\mathbb{Z}$ \\
& BDI & $+1$ & $+1$ & $+1$ & $\mathbb{Z}$ &   &   \\
& CII & $-1$ & $-1$ &$+1$& $\mathbb{Z}$ &   & $\mathbb{Z}_2$\\

\hline
\multirow{4}{*}{BdG}
& D &   & $+1$ &   & $\mathbb{Z}_2$ & $\mathbb{Z}$ &  \\
& C &   & $-1$ &   &   & $\mathbb{Z}$ &   \\
& DIII & $-1$ & $+1$ &$+1$& $\mathbb{Z}_2$ & $\mathbb{Z}_2$ & $\mathbb{Z}$ \\
& CI & $+1$ & $-1$ &$+1$&   &   & $\mathbb{Z}$ \\

\hline
\end{tabular}
\end{table}


In fact, it was shown in Refs.\ \parencite{Schnyder2008,Kitaev2009a} that it is possible to infer which topological invariants can be non-zero simply from the system's dimensionality, and from the presence or absence of chiral symmetry, time-reversal symmetry, and particle-hole symmetry. Based on this, Schnyder et al.~built the topological classification of non-interacting Hamiltonians, which is presented in the Table \ref{tab:topo}. For illustration purposes, let us consider the example of the Kane-Mele model. This two dimensional model features time-reversal symmetry, implemented by an operator which squares to $-1$ \parencite{Kane2005a}. We can directly read off from the table that this system can have up to two distinct topological phases.

\subsection{Beyond the topological table}

This classification is not exhaustive, however, as some known systems cannot be adiabatically connected to any of the Hamiltonians in Table \ref{tab:topo}. A famous example is known as the fractional quantum Hall effect, which has topological properties as a result of electronic interactions \parencite{Laughlin1983}. Another altogether different example comes to us from Floquet systems, in which a time-dependent, cyclic Hamiltonian 
is approximated by a static one, which can have topological properties which do not exist in the standard classification \parencite{Rudner2013}. Finally, topological crystalline insulators are characterised by topological invariants which do not exist in any system from Table \ref{tab:topo} \parencite{Fu2011}. A more complete review of systems which lie outside the standard classification can be found in Ref.\ \parencite{Chiu2016a}.

\section{Experimental study of topology}


As we have seen in the above study, topologically non-trivial systems have a large number of real world outlets which underline the necessity of realising these systems experimentally. Amongst these, we mentioned potential applications to metrology, quantum information storage and communication, as well as the implementation of quantum algorithms.
Although it is likely that some of these applications represent long term research goals, there also exists immediate scientific interest to exploring topological phases experimentally. Indeed, experimental realisations allow us to study systems which go beyond our theoretical understanding and computational power. Many-body interacting systems, for instance, or strongly disordered systems, are extremely challenging to approach theoretically. In this context, experimental data is crucial to understand the interplay between complex effects and topological behaviour.

\subsection{Review of realisations}

Since the discovery of the integer quantum Hall effect, there have been many important experimental realisations of topologically non-trivial materials which have contributed to further our understanding of these systems. The Haldane model, for instance, was realised in solid state systems \parencite{Konig2007} following a proposal by \parencite{Bernevig2006}. This same system was later implemented using cold atoms in optical lattices \parencite{Jotzu2014} and photonic crystals \parencite{Rechtsman2016}. Other implementations of topologically non-trivial systems with cold atoms in optical lattices include the Hofstadter model \parencite{Aidelsburger2013}, the Thouless pump \parencite{Lohse2015,Nakajima2016},
and the SSH model \parencite{Atala2013}.
Concerning three dimensional strong topological insulators, most realisations to date have been solid state systems, as for instance Ref.\ \parencite{Hsieh2009}. Finally, the two-step quantum walk, a Floquet system which cannot be associated to any of the classes of Table \ref{tab:topo}, was realised using single photons \parencite{Kitagawa2012a, Cardano2015}.

\subsection{Quantum simulators applied to topology}

Amongst the experimental realisations listed above, we can immediately see that many rely on artificially engineered materials. This method is known as quantum simulation, and was originally suggested by Richard Feynman \parencite{Feynman1982}. Usually in material science, a Hamiltonian is found which describes the main attributes of a physical system. With quantum simulators, this methodology is reversed: a physical system is designed to implement a Hamiltonian, which is the subject of the study. Theoretically, this is extremely interesting because it allows us to identify which terms, or combination of terms lead to a given physical behaviour \parencite{Lewenstein2012}.

Beyond this, quantum simulators are extremely attractive due to their versatility \parencite{Bloch2012,Inguscio2013}. Indeed, optical lattices allow us to engineer flawless lattices, the parameters of which can be controlled dynamically by changing the intensity of the applied lasers. What is more, the dimensionality of these lattices can be controlled, either using real \parencite{Bloch2005} or synthetic dimensions \parencite{Kraus2012,Celi2013}, which is an important resource for realising all phases of Table \ref{tab:topo}. Raman coupling can be used to engineer transitions between an atom's magnetic sublevels, allowing for powerful cooling techniques \parencite{Kasevich1992}. Finally, it is also possible to engineer artificial gauge fields in synthetic materials, either by using optical fields \parencite{Goldman2014a}, lattice shaking techniques \parencite{Goldman2014}, or Raman transitions in synthetic dimensions \parencite{Celi2013}.

\subsection{Detection of topological phases}

Thus, quantum simulators provide numerous, powerful techniques for the realisation of artificial materials. The problem which we have not discussed, however, is \emph{once a topologically non-trivial system has been realised, how can we observe its topological properties?} This is an important open question, 
first and foremost
for physical applications: we must show that the system's interesting properties survive perturbations, which are potentially of unknown origin. The only way to do this it to have an unambiguous method for observing topological properties. What is more, the study of topological phases is a very new field, meaning that it is important to correlate topological properties to verify that they cannot exist independently. This requires the simultaneous application of different measurement methods.

A general property of topologically non-trivial systems is the presence of a robust edge state at its boundaries.
Much of the experimental evidence we have for topological properties relies on the detection of this bound state, particularly in solid state systems which are naturally finite.
Amongst these, Ref.\ \parencite{Mourik2012} considered a one dimensional junction, partially coupled to an s-wave superconductor. The experiment revealed the presence of an isolated state in the spectrum at $E=0$ which suggested that their system harboured a Majorana mode. There also exists two dimensional detection of edge modes, such as Ref.\
\parencite{Hart2014}, which measured the spatially resolved conductance of the sample to observe its edge states.

An alternative route relies on imaging the system's dispersion. This method is particularly relevant for three dimensional topological insulators, where the band structure is the smoking gun of the system's topological properties. Ref.\ \parencite{Hsieh2009}, for instance, showed that the surface dispersion of their system was that of a three dimensional topological insulator by using ARPES. This powerful method uses a high energy photon to eject an electron from the material's surface or bulk, and record its momentum. The band structure can be reconstructed by varying the photon's energy and incoming angle.

Finally, it is also possible to directly measure the system's topological invariant. This was done experimentally in the one dimensional SSH model by inducing Bloch oscillations and using Ramsey interferometry \parencite{Atala2013}. The topological invariants of the two-step quantum walk were recently also measured \parencite{Cardano2016} by measuring chiral currents in the system's bulk. Finally, in two dimensions, the Chern number was measured by recording the centre of mass position of an atomic cloud \parencite{Aidelsburger2014}, following the proposal in Ref.\ \parencite{Dauphin}.



\section{Topics addressed in this thesis}

There are three main topics addressed in this thesis:

\begin{itemize}
 \item We will study in detail how Floquet theory can be used to \emph{design a new, topologically non-trivial Hamiltonian} by applying the powerful toolbox of cold atoms in optical lattices. We further want to build a topological boundary so that we can generate topologically protected edge states in this system. There is no set protocol to do this, such that there is a need to detail the steps that must be taken to build the complete phase space of a system, and identify the relevant experimental tools to construct a topological boundary.
 
 \item The central theme of this thesis is the \emph{detection of topological properties}. As we have already emphasised, this is relevant for the study and application of topological systems. One way to identify a system's topological properties is to study its edge states. Unambiguously identifying a topologically protected state is a challenging task, particularly in one dimension,
 such that realistic methods 
 need to be developed to do this. We will also consider the possibility of measuring the topological invariant directly in single particle systems. Because this invariant is a global quantity, we will need to develop a method to address the entire system at once to measure it.
 
 \item The last subject which we consider in this thesis is the \emph{definition and determination of the topological invariants of small systems}. This is problematic because topological invariants are defined in periodic systems, or when the bulk is so large that it can be considered infinite. Whether or not topological invariants are well defined in systems with extremely small bulks is, however, an interesting and open problem.
\end{itemize}


In {\bf Chapter \ref{Chapter2}}, we discuss in detail how topological properties can arise in one and two dimensional systems.
For each system, we will focus on how its topological properties manifest themselves, which we will later use to identify their topological phase.

In {\bf Chapter \ref{Chapter3}}, we will briefly review the relevant experimental tools offered by cold atoms in optical lattices. Specifically, we will see how optical potentials can be designed for atoms, how atomic transitions can be induced through Raman transitions, and how artificial gauge fields can be constructed, allowing us to simulate, for instance, magnetic fields for neutral atoms.

In {\bf Chapter \ref{Chapter4}}, we suggest a new method to build a new topologically non-trivial system with cold atoms in a one dimensional optical lattice. By constructing the phase space of this system, we show its phases can go beyond the standard classification. We develop a method to generate a topological boundary in this system, such that it presents edge states. Finally, we describe a new protocol for identifying the topological bound states of this system, which is applicable to all chiral systems symmetric systems.

In {\bf Chapter \ref{Chapter5}}, we study the Hofstadter model on a thin strip, such that its bulk is vanishingly small. We suggest a definition for the system's topological invariant, and show that it is consistent and accurately counts the number of edge states. This invariant can be measured by preparing a wavepacket in the ground state of the lattice, then inducing a period of Bloch oscillations such that the atoms are made to explore the system, and finally measuring the atoms' centre of mass displacement.

In {\bf Chapter \ref{Chapter6}}, we conclude our work and describe interesting ways in which it could be extended.

Chapters \ref{Chapter4} and \ref{Chapter5} contain original work, which have been, respectively, published \parencite{Mugel} and submitted for publication \parencite{Mugel2017}.

\chapter{Review of the manifestation of topological properties}
\label{Chapter2}

Our goal in this chapter is to understand the constraints which can lead a Hamiltonian to display topologically non-trivial phases. Our main motivation, however, is to see how topology can manifest itself physically in these systems, as these will give us the experimental means to identify topological materials later in this thesis. Importantly, systems which belong to the same symmetry class display the same topological properties \parencite{Schnyder2008}. Thanks to this, the conclusions which we draw from the toy models encountered in this chapter will be applicable to the systems we study in Chapters \ref{Chapter4} and \ref{Chapter5}.

We will begin by introducing the Berry phase, which is defined as the geometric phase acquired by a particle when carried around a loop in phase space \parencite{Berry1984,Wilczek1989}. This quantity has been observed in a vast number of physical systems \parencite{Tomita1986,Delacretaz1986,Bitter1987,Tycko1987}. In particular, the Berry phase has proved essential to understanding the electron dynamics in crystals \parencite{Xiao}. To emphasise this point, we will consider in detail two well known, topologically non-trivial models, the one dimensional SSH model \parencite{Su1979} and the two dimensional Hofstadter model \parencite{Hofstadter1976}. We will show that their topological properties are in fact a direct consequence of the quantisation of the Berry phase. This study will give us the keys to understanding the physical consequences of topology in these systems. Finally, we study the two-step quantum walk \parencite{Kitagawa2010a}, which is an example of a system which is topologically inequivalent to all classes in Table \ref{tab:topo} due to its time dependent nature \parencite{Asboth2013}. We show that this system can be understood through Floquet theory, and state the formulas with which its topological invariants can be computed.

\begin{framed}
In this chapter:
\begin{enumerate}
 \item We derive the phase acquired by a state when it is carried around a loop in phase space, known as the Berry phase.
 \item We show that the Berry phase can be related to the number of edge states of the one dimensional SSH model.
 \item The Berry curvature is derived, which determines the conduction properties of the two dimensional Hofstadter model.
 \item We describe the two-step quantum walk, identify its topological invariants and the structure of its bound states.
\end{enumerate}
\end{framed} 

\section{The Berry phase}
\label{sec:BerryPhase}

In this section, we will introduce the Berry curvature by following the derivation
of \parencite{Xiao}. We find it useful to present this derivation to underline the assumptions that are made to reach our final result.

Let $\h(\R)$ be a general, $D$ dimensional Hamiltonian which depends on the set of parameters $\R$. We also define $\ket{u_j(\R)}$, the $j^{\rm th}$ eigenstate of $\h(\R)$:
\begin{equation}
 \h(\R) \ket{u_j(\R)} = E_j(\R) \ket{u_j(\R)}.
\end{equation}
Let us now vary the parameters $\R$ adiabatically in time, $\R=\R(t)$, along a path $\zeta$ in phase space. In other words, we are implicitly assuming that $\h(\R)$ is gapped for all values of $\R$ in $\zeta$, and that the rate with which we vary $\R(t)$ is much smaller than the band gap. When this is the case, we know from the quantum adiabatic theorem that, if the system is initially an eigenstate of the Hamiltonian, it will remain in an instantaneous eigenstate of $\h(\R(t))$ throughout the system's evolution \parencite{Kato1950,messiah1965}. Thus, if a particle in this system is initially in the eigenstate $\ket{u_j(\R(0))}$, then at time $t$ it is in the state:
\begin{equation}
\label{eq:StateTimeT}
 \ket{\Psi_j(t)} = e^{i \gamma_j (t)} \exp \left[ 
 -\frac{i}{\hbar} \int_0^t {\rm d} t' E_j(\R(t'))
 \right] \ket{u_j(\R(t))}.
\end{equation}
For simplicity we will omit the dependence of $\R$ on $t$ where possible. Note that we have included explicitly $\gamma_j (t)$, which represents a phase factor acquired by the particle as $\R$ is varied along $\zeta$. We can find an explicit expression for this quantity by substituting Eq.~\eqref{eq:StateTimeT} into the time-dependent Schr\"odinger equation:
\begin{equation}
 \label{eq:BerryPhase}
 \gamma_j= i \int_\zeta {\rm d} \R \cdot \bra{u_j(\R)} \partial_{\R} \ket{u_j(\R)}.
\end{equation}
Note that the $\ket{u_j(\R)}$ are only defined up to a phase, such that we have the freedom to perform the gauge transformation:
\begin{equation}
 \label{eq:GeneralGaugeTransformation}
 \ket{u_j(\R)} \xrightarrow{} e^{i g(\R)} \ket{u_j(\R)}
\end{equation}
for any smooth function $g(\R)$ without changing the physical properties of the system. When $\zeta$ is an open trajectory, Eq.~\eqref{eq:BerryPhase} changes under this transformation by:
\begin{equation}
 \gamma_j(t) \xrightarrow{} \gamma_j(t) + g(\R(t)) - g(\R(0)),
\end{equation}
where we have assumed that the phase of $\ket{u_j(\R)}$ is smooth along $\zeta$. If $\R(t)=\R(0)$, however, such that $\zeta$ is a closed trajectory, the single-valuedness of $\ket{u_j(\R)}$ implies that $g(\R(t)) - g(\R(0))$ can at most be a multiple of $2\pi$ \parencite{Berry1984}. When this is the case, $\gamma_j$ can no longer be gauged out, and it becomes an observable, physical quantity. This quantity, which is known as the Berry phase, can become quantised under certain conditions. This quantisation is a process through which topological invariants can appear in the system.

\section{One-dimensional example: the SSH model}
\label{sec:1dTopoIntro}

Although we have introduced the Berry phase, Eq.~\eqref{eq:BerryPhase}, as an abstract mathematical quantity, it is relevant in a vast number of real world systems. Amongst these, and perhaps most notably, the Berry phase arises in translationally invariant systems, which naturally present a periodic phase space as a result of Bloch's theorem \parencite{Zak1989}.
In the following, we illustrate this point by considering the one dimensional Su-Schrieffer-Heeger, or SSH model \parencite{Su1979}. This model is interesting to us because it constitutes a simple system in which the Berry phase becomes quantised, leading to highly non-trivial topological properties. Beyond this however, we will see that a detailed understanding of the SSH model will help us to understand how topological properties emerge in the two-step quantum walk, which we consider in Sec.~\ref{sec:TwoStepQW}.

\subsection{The model}
\label{sec:SSHmodel}

\begin{figure}
\centering
\includegraphics[width=\textwidth]{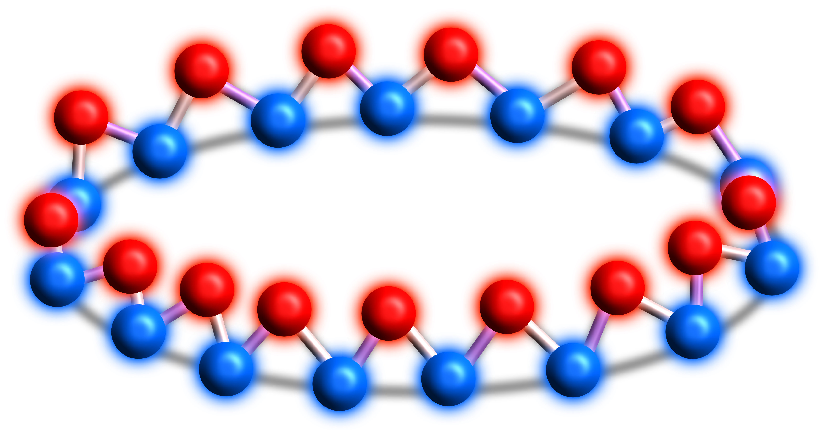}
\caption[Sketch of the SSH model]{Sketch of the SSH model, Eq. \eqref{eq:SSH}, with periodic boundary conditions. Sublattice sites $A$ ($B$) are represented in red (blue). Figure reproduced with the authorisation of Alexandre Dauphin.}
\label{fig:SSH}
\end{figure}

In its simplest form, the SSH model \parencite{Su1979} describes a single particle in a one dimensional chain of dimers:
\begin{equation}
 \label{eq:SSH}
  \h_{\rm SSH} = \sum_n J_1 \ket{n,B} \bra{n,A} + J_2 \ket{n+1,A} \bra{n,B} + \text{H.c.}
\end{equation}
where $J_1$ ($J_2$) is the intra-cell (inter-cell) hopping amplitude. We further define $\ket{n,A/B}=\ket{n}\otimes\ket{A/B}$ as the state which belongs to the $n^{\rm th}$ unit cell, with sublattice degree of freedom $A/B$. We will assume that the cell index $n$ can take any integer value in the interval $[1,N]$, and that the system is periodic, such that $\ket{N+1}=\ket{1}$. This system is schematically represented in the Fig.~\ref{fig:SSH}.


For $J_{1,2}=$ constant, this system is translationally invariant. When this is the case, we know from Bloch's theorem that we can block diagonalise $\h_{\rm SSH}$, Eq.~\eqref{eq:SSH}, by going to the basis of quasimomentum:
\begin{equation}
 \label{eq:SSH_mom}
 \h_{\rm SSH} = \int_{-\pi/d}^{\pi/d} {\rm d}k \ket{k}\bra{k} \otimes \h_{\rm SSH}(k),
\end{equation}
where $d$ is the lattice spacing and $\ket{k}$ is the state with well defined quasimomentum:
\begin{equation}
 \label{eq:FourierTransform}
 \ket{k} = \frac{1}{\sqrt{N}}\sum_n e^{i k n d} \ket{n}.
\end{equation}
The Hamiltonian $\h_{\rm SSH}$ can readily be calculated by substituting Eq.~\eqref{eq:FourierTransform} into Eq.~\eqref{eq:SSH_mom}:
\begin{equation}
 \label{eq:HSSH_k}
 \begin{split}
 & \h_{\rm SSH}(k) = J_1 \bm{h}(k) \cdot \bm{\sigma},\\
 & \bm{h}(k) = \left( 1+\frac{J_2}{J_1} \cos(k d), -\frac{J_2}{J_1} \sin(k d), 0 \right).
 \end{split}
\end{equation}
We have defined $\bm{\sigma}=(\sigma_1,\sigma_2,\sigma_3)$, the vector of Pauli matrices which act in sublattice space. We work in the basis where states which have amplitude only on sublattice $A$ ($B$) are eigenstates of $\sigma_3$ with eigenvalue $+1$ ($-1$).

The quasimomentum variable $k$ belongs to the interval $[-\pi/d,$ $\pi/d]$, which is known as the first Brillouin zone. As a consequence of Bloch's theorem, and of crucial importance to us, the point $k=-\pi/d$ is identified with $k=\pi/d$. The Brillouin zone therefore constitutes a closed path in phase space. Following Eq.~\eqref{eq:BerryPhase}, a particle which is carried around the Brillouin zone will acquire a Berry phase with value \parencite{Zak1989}:
\begin{equation}
 \label{eq:ZakPhase}
 \gamma_j=i \oint {\rm d}k \braket{u_j(k)}{\partial_k u_j(k)},
\end{equation}
where $\ket{u_j(k)}$ is the $j^{\rm th}$ eigenstate of $\h_{\rm SSH}(k)$. Because $\h_{\rm SSH}(k)$ is a $2\times 2$ matrix, $\ket{u_j(k)}$ is a two-component spinor and $j\in \{1,2\}$. By explicitly calculating $\ket{u_j(k)}$ from Eq.~\eqref{eq:HSSH_k}, it can be shown that the Berry phase measures half the solid angle enclosed by the vector $\bm{h}(k)$ as $k$ goes from $-\pi/d$ to $\pi/d$ \parencite{Delplace2011a}. While in general it can assume any value, we will show in the next subsection that a symmetry is present in the SSH model which restricts $\gamma_j$ to integer multiples of $\pi$.

\subsection{Chiral symmetry}
\label{sec:ChiralSymmetry}

As can be seen by inspecting $\h_{\rm SSH}(k)$, Eq.~\eqref{eq:HSSH_k}, the vector $\bm{h}(k)$ has no $z$ component, such that it is constrained to the $(x,y)$ plane. This constraint is a result of $\h_{\rm SSH}(k)$ being chiral symmetric. A general Hamiltonian \^H is said to be chiral symmetric if it anti-commutes with a unitary, Hermitian matrix $\hat{\Gamma}$:
\begin{equation}
\label{eq:ChiralSymmetry}
\hat{\Gamma} \h \hat{\Gamma} = -\h.
\end{equation}
In the basis of Eq.~\eqref{eq:HSSH_k}, for instance, the chiral symmetry (CS) operator is $\hat{\Gamma}=\sigma_3$. We will only be interested in CS operators which act within single unit cells. This way, translational invariance can be broken without breaking CS, making CS robust to local perturbations.

A useful property of Hamiltonians with CS is that they can always be put into the form \parencite{Chiu2016a}:
\begin{equation}
\label{eq:ChiralSymmetricForm}
 \h=
 \begin{pmatrix}
     0 & z\\
     z^\dagger & 0
    \end{pmatrix},
\end{equation}
where $z$ is an $N\times N$ matrix. In this basis, the CS operator takes the form: $\hat{\Gamma}=\sigma_3 \otimes \sigma_0^{N\times N}$, where $\sigma_0^{N\times N}$ is the $N\times N$ identity matrix.

\subsection{Winding number}
\label{sec:WindingNumber}

\begin{figure}
\centering
\includegraphics[width=\textwidth]{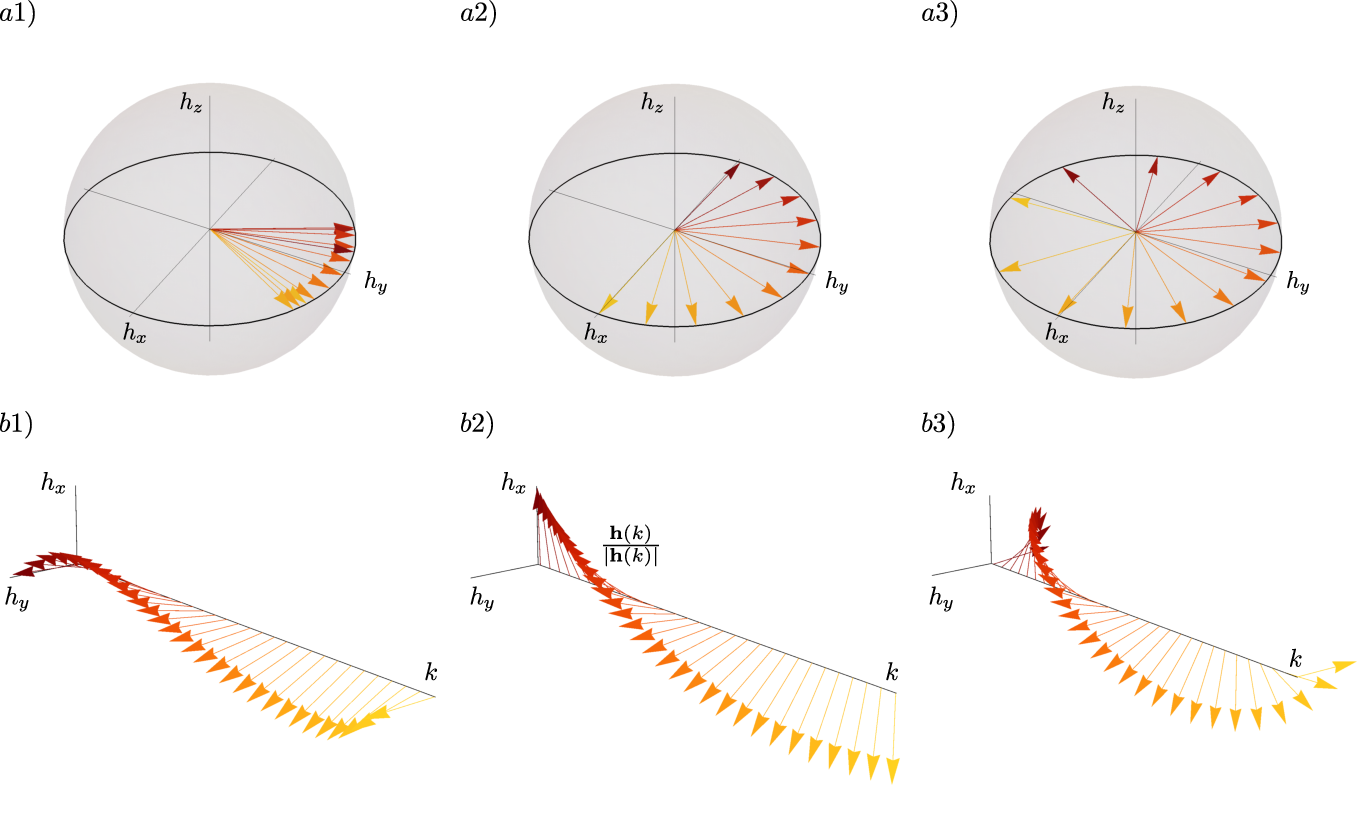}
\caption[Winding of the vector $\bm{h}(k)$]{\label{fig:BelenWinding} Normalised vector $\bm{h}(k)$, Eq.~\eqref{eq:HSSH_k}, plotted in the complex plane for $k\in[-\pi/d,\pi/d]$. Figs.~(a1)-(a3): $\bm{h}(k)$ plotted on the Bloch sphere; note that the area enclosed by the curve is determined by the number of times $\bm{h}(k)$ circles the origin. Figs.~(b1)-(b3): $\bm{h}(k)$ is plotted in the $(x,y)$ plane for each $k$ value in the Brillouin zone. The colour coding of the $\bm{h}(k)$ vectors indicates the value of $k$. Figs.~(a1), (b1): $J_2=J_1/2$; $\bm{h}(k)$ does not circle the origin. Figs.~(a2), (b2): $J_1=J_2$; we are at the phase transition and the winding number of $\bm{h}(k)$ is ill defined. Figs.~(a3), (b3): $J_2= 2 J_1$; $\bm{h}(k)$ circles the origin once anti-clockwise.}
\end{figure}

A direct consequence of CS is the quantisation of the Berry phase, given by Eq.~\eqref{eq:BerryPhase}. As was first realised by Ryu and Hatsugai, this is easiest to visualise by plotting the vector $\bm{h}(k)$ in the complex plane \parencite{Ryu2002}. In Figs.~\ref{fig:BelenWinding}(a1)-(a3), we plot $\bm{h}(k)$ on the Bloch sphere for all $k$ values, and for increasing values of $J_2/J_1$. Alongside these, we plot $\bm{h}(k)$ in the $(x,y)$ plane, presented in Figs.~\ref{fig:BelenWinding}(b1)-(b3), as $k$ is progressed around the Brillouin zone. In the case of Fig.~\ref{fig:BelenWinding}(a1), we have $J_2=J_1/2$, such that $\bm{h}(k)$ does not wind around the origin and encloses zero area. Following the discussion from Sec.~\ref{sec:SSHmodel}, the Berry phase, which measures half the solid angle enclosed by $\bm{h}(k)$, is zero in this case. In Fig.~\ref{fig:BelenWinding}(a2), $J_2=J_1$ and $\bm{h}(k)$ is discontinuous, such that the area which it encloses is ill defined. This is best seen by considering the Fig.~\ref{fig:BelenWinding}(b2). For Fig.~\ref{fig:BelenWinding}(a3), $J_2=2J_1$, such that $\bm{h}(k)$ winds once around the origin. In this case, the $\bm{h}(k)$ curve encloses an area of $2\pi$ (the Bloch sphere has unit radius), such that the Berry phase is $\pi$.

Visibly, the Berry phase is simply $\pi$ multiplied by the number of times that $\bm{h}(k)$ circles the origin, which we will refer to as the winding number $\nu$. Thus, we can appreciate how the single valuedness of the $\bm{h}(k)$ vectors leads to the appearance of an integer which determines the phase of the system. Note that, because the winding number is obtained by varying $k$ over the entire Brillouin zone, it is a characteristic feature of the entire energy band.

\begin{figure}
\centering
\includegraphics[width=\textwidth]{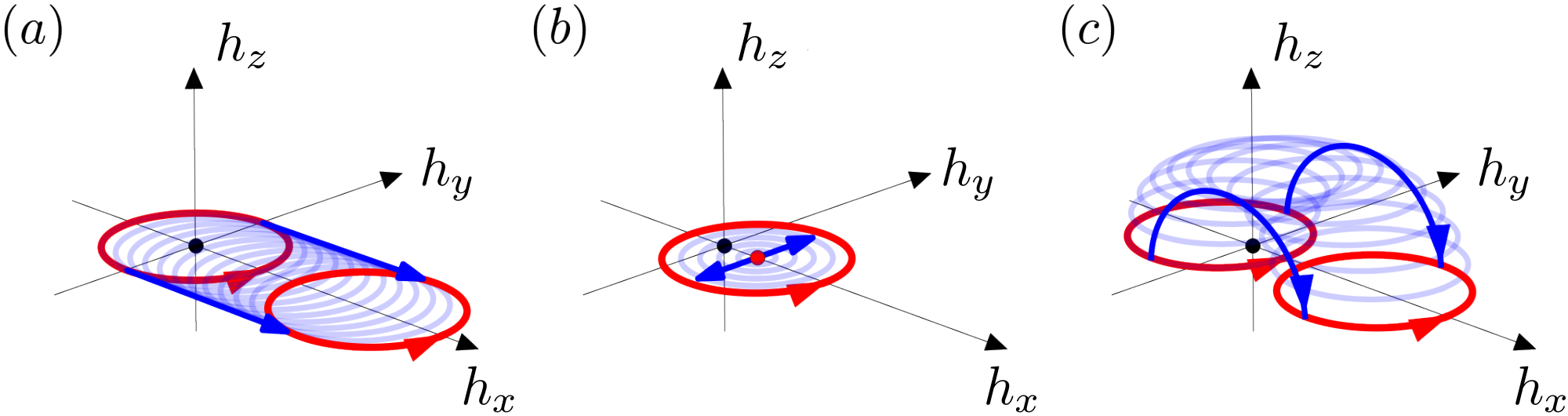}
\caption[Changing the winding number]{$\bm{h}(k)$ is plotted in the complex plane for $k\in[-\pi/d, \pi/d]$. Note that $\bm{h}(k)$ is not normalised, unlike in the Figs.~\ref{fig:BelenWinding}. To go from a situation where $\bm{h}(k)$ initially circles the origin to one where the origin lies outside of the loop, (a): the loop can be translated in plane, (b): deformed or (c): lifted out of the plane to avoid the origin. In cases (a) and (b), the circle must intercept the origin at some point, while case (c) requires a chiral symmetry breaking perturbation. Adapted from \parencite{Asboth2015}.}
\label{fig:AsbothWinding}
\end{figure}

As is visible from the figure \ref{fig:AsbothWinding}, the winding number cannot change continuously: it must go through at least one point where it is ill defined. This can happen either when the loop intercepts the origin, as in Figs.~\ref{fig:AsbothWinding}(a) and (b), or when it is not in the same plane as the origin, as in Fig.~\ref{fig:AsbothWinding}(c). Note however that chiral symmetric perturbations keep the $\bm{h}(k)$ constrained to the $(x,y)$ plane. Thus, for a CS conserving perturbation to change the winding number, the loop  must intercept the origin. This means that there must exist a quasimomentum for which the vector $\bm{h}(k)$ vanishes. As a consequence, 
the band gap must close for the winding number to change.

Formally, the winding number is what is known as a topological invariant; this quantity characterises a topological phase, in the same way as an order parameter characterises a phase of the system. Thus the SSH model admits two distinct topological phases, characterised by $\nu=0$ for $J_2<J_1$ and $\nu=1$ for $J_2>J_1$. The topological phase transition occurs when $J_2=J_1$, where $\nu$ is ill defined.

It is important to understand that, in one dimensional systems, topological properties cannot exist without restrictive symmetries \parencite{Verstraete,Chen}. In particular, it was shown by Zak that all one dimensional, chiral symmetric systems admit a winding number which is restricted to integer values \parencite{Zak1989}. A general formula for the winding number can be derived from Eqs.~\eqref{eq:ZakPhase} and \eqref{eq:ChiralSymmetricForm} \parencite{Asboth2014}:
\begin{equation}
\label{eq:evaluate_winding}
\nu[z]=\frac{1}{2\pi i}\oint{\rm d}k \frac{{\rm d}}{{\rm d}k}\log \det[z(k)].
\end{equation}
By inspecting Eq.~\eqref{eq:evaluate_winding}, we can understand the quantisation of the winding number from a complex analysis point of view. Indeed, if we extend the integral to the complex plane, the residue theorem then that tells us that Eq.~\eqref{eq:evaluate_winding} is simply counting the number poles of the integrand enclosed in the loop that is the Brillouin zone.

\subsection{Topological bound states}
\label{sec:TopoBS}

Interesting things start happening when two (gapped) systems with different winding numbers are brought into contact. We saw in Sec.~\ref{sec:WindingNumber} that, due to the Hamiltonian's constraints, the band gap must close at the boundary between the two systems for the winding number to change. Intuitively, we can imagine that states may occur at the interface which are not energetically allowed to exist elsewhere in the system.

For a general chiral symmetric Hamiltonian $\h$, the existence of CS implies that each of its eigenstates has a chiral partner with equal and opposite energy. This is readily derived from Eq.~\eqref{eq:ChiralSymmetry}:
\begin{equation}
\h \ket{\psi}=E \ket{\psi} \Rightarrow \h \left(\hat{\Gamma}\ket{\psi} \right)=-E \left(\hat{\Gamma}\ket{\psi} \right).
\end{equation}
States with energies such that $E=-E$ are special because they can transform into themselves under CS. This means that these states can exist without a chiral partner. When this is the case, they cannot be moved away from the energy satisfying $E=-E$ without breaking CS. These states can therefore not be coupled to other states in the system by chiral symmetric perturbations.

\begin{figure}[t]
\centering
\includegraphics[width=0.7\textwidth]{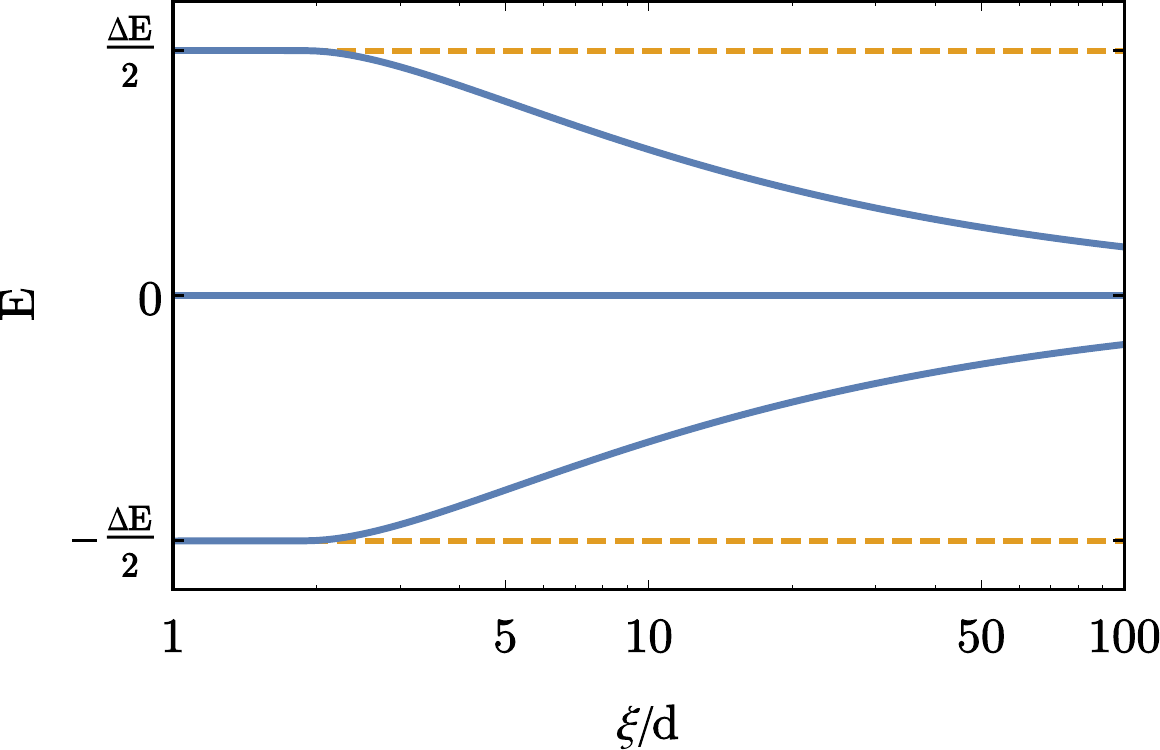}
\caption[Low energy states in the SSH model]{Energy of the three states closest to $E=0$ versus the topological boundary sharpness. The boundary width between the two regions $R_1$ and $R_2$ is $\xi$. The system is chiral symmetric and regions $R_1,R_2$ have a spectral gap $[-\Delta E/2,\Delta E/2]$; these energies are indicated by dashed yellow lines. As $\xi$ is reduced, the interface between $R_1$ and $R_2$ can accommodate fewer states, such that, when $\xi\lesssim 1$, only topological bound states can exist in the spectral gap of $R_1,R_2$. }
\label{fig:SSH_first_excited_state}
\end{figure}

Consider connecting two regions $R_1, R_2$ which have different winding numbers and present spectral gaps. The topological invariant cannot change without closing the spectral gap in $E=0$. We assume that
the parameters of the Hamiltonian vary smoothly at the boundary over a length scale $\xi$, where $\xi$ is large relative to the other length scales in the system. Under these conditions, the spectral gaps of $R_1, R_2$ are densely populated in this region. If we now make the interface between $R_1$ and $R_2$ sharper, fewer states can live in the boundary region, meaning that fewer states can populate the spectral gap of $R_1, R_2$. To illustrate this, we plot the energy of the three states closest in energy to $E=0$ versus the topological boundary sharpness in Fig.~\ref{fig:SSH_first_excited_state}. In the infinitely sharp boundary limit, all states have been lifted out of the spectral gap except for those which do not have a chiral partner. Thus, if a system presents an interface between regions $R_1, R_2$ with different winding numbers $\nu$, there must exist an $E=0$ state at this interface. Because $R_1$ and $R_2$ must be gapped to have well defined winding numbers, the $E=0$ state can only exist at the topological interface.

In fact, if $\Delta\nu$ is the difference in winding numbers between regions $R_1$ and $R_2$, then the system's Hamiltonian must have at least $\Delta\nu$ eigenstates which have energy $E=0$ and are bound to the interface between $R_1$ and $R_2$. This is in fact a result of the bulk boundary correspondence, which states that the properties of the bulk are indissociable from those of the boundary. Thus the robustness of a system's topological properties to local, symmetry preserving perturbations, and the robustness of the bound state, which occurs at its edges, are one and the same thing.

\subsection{Alternative view: SSH model in the continuum}
\label{sec:JRIntroduction}

We finish this section by pointing out that we made extensive use of Bloch's theorem, in Secs.~\ref{sec:WindingNumber} and \ref{sec:TopoBS}, to derive the quantisation of the Berry phase. This might seem inconsistent with the study presented in this section due to the absence of translational invariance. Periodic boundary conditions are not, however, required for the bulk's winding number to be well defined.
In this section, we present an alternative way to understand the appearance of bound states at a topological interface by taking the continuum limit of the SSH model. This will allow us to understand the system's topological properties without relying on a local density approximation, and will teach us some of the fundamental properties of the system's bound state.

We will begin by obtaining an ansatz for low energy eigenstates of the SSH model at the topological boundary. When $J_2=J_1$, we know that the Hamiltonian, Eq.~\eqref{eq:SSH}, has two $E=0$ eigenstates with quasimomentum $k=\pi/d$. If we take $J_2/J_1$ to be slowly varying, using the envelope function approximation, we can say express the $kd \approx \pi$ eigenstates of $\h_{\rm SSH}$ as \parencite{Asboth2015}:
\begin{equation}
\label{eq:PsiAnsatz}
\ket{\psi} = \sum_n e^{i \pi n}\bm{\varphi}(n)\otimes \ket{n},
\end{equation}
where $\bm{\varphi}(n)$ is the envelope function, which has two components, one for each of the particle's internal degrees of freedom. Let us now take $J_2$ constant and $J_1$ to be continuously spatially varying. Substituting Eq.~\eqref{eq:PsiAnsatz} into the Schr\"odinger equation and taking the continuum limit, we find that this state's envelope satisfies the one dimensional Dirac equation:
\begin{eqnarray}
\label{eq:SSHtoJR}
\left[-i v \sigma_2 \partial_x +m(x) \sigma_1 \right] \bm{\varphi}(x)= E \bm{\varphi}(n),
\end{eqnarray}
with $v = -J_2 d$ and $m(x) = J_1(x)- J_2$. This result was first derived using a local density approximation \parencite{Takayama1980}. Note that Eq.~\eqref{eq:SSHtoJR} is chiral symmetric, and its CS operator is $\hat{\Gamma}=\sigma_3$.

It was shown by Jackiw and Rebbi that there always exists a zero energy solution to Eq.~\eqref{eq:SSHtoJR} when the sign of $m(x)$ is different at $x\rightarrow -\infty$ and $x\rightarrow +\infty$ \parencite{Jackiw1976}. If we were to choose, for instance, $J_1<J_2$ for $x<0$ and $J_1>J_2$ for $x>0$, then $\bm{\varphi}(x)$ would takes the form:
\begin{equation}
\label{eq:JR_state}
\bm{\varphi}(x)=\psi_0 \exp\left[- \frac{1}{v}\int_0^x m(x')  {\rm d}x'\right]
\begin{pmatrix}
 1 \\ 0
\end{pmatrix},
\end{equation}
where the constant $\psi_0$ is the state's normalisation.

Thus, we see that there always exists a state with energy exactly zero when $m(x)$ changes sign. When this is the only such state, no local perturbation can lift this state away from $E=0$ without breaking CS. This is interesting because, following the study presented in Sec.~\ref{sec:WindingNumber}, the negative $x$ half lattice has a winding number $\nu=0$ while the positive $x$ half lattice has $\nu=1$. In Sec.~\ref{sec:TopoBS}, we used this difference in the winding number to show that a zero energy state which is robust to perturbations had to exist at $x=0$. Here, however, we were able to find the continuum wavefunction of this state without making a local density approximation. What is more, note that the state given in Eq.~\eqref{eq:JR_state} decays exponentially away from $x=0$. This is expected, as this state is energetically forbidden from existing in the bulk.

Once again in this example, we can see that the properties at a topological boundary are determined by the properties of the bulk. As a result, zero energy bound states can be said to be the real smoking gun of one dimensional topology. Unfortunately, unambiguously identifying a topological bound state is a difficult task, as this would require measuring both its energy and its robustness to perturbations.


\section{Two-dimensional example: the Hofstadter model}
\label{sec:2dTopoIntro}

In the Sec.~\ref{sec:1dTopoIntro}, we showed that in one dimensional systems, symmetry constraints can lead to the quantisation of the Berry phase. This, in turn, leads to the appearance of a global, integer order parameter, which determines the topological phase of the system. When two systems in different topological phases are brought into contact, a robust bound state is found in the system's spectral gap. In this section, we will show that many of these properties also exist in higher dimensional systems with one interesting exception: symmetry constraints are not required for topological properties to appear in two dimensions.

\subsection{The model}
\label{sec:Hofstadter}

In its simplest form, the Hofstadter model describes non-interacting spinless particles in a two-dimensional square lattice in presence of a uniform external magnetic field \parencite{Hofstadter1976}. In the Landau gauge, the Hofstadter Hamiltonian takes the form:
\begin{equation}
\label{eq:H0}
\text{\^H}_{\rm Hof}=\sum_{m,n} -J_x \hat{c}_{m+1,n}^\dagger \hat{c}_{m,n} 
-J_y e^{i m \Phi} \hat{c}_{m,n+1}^\dagger \hat{c}_{m,n} +
\text{H.c.}
\end{equation}
The operator $\hat{c}_{m,n}^\dagger$ ($\hat{c}_{m,n}$) creates (destroys) a particle at site $(m,n)$, where $m$ and $n$ are the site indices for the $x$ and $y$ directions respectively. We will assume that the system is finite, with $N_x$, $N_y$ sites in the $x$ and $y$ directions respectively, such that the site indices belong to the intervals $m \in [1, N_x]$ and $n \in [-(N_y-1)/2, (N_y-1)/2]$. Without loss of generality, we can take the hopping amplitudes in the $x$ and $y$ directions, $J_x$ and $J_y$ respectively, to be positive constants.

\begin{figure}[t]
\centering
\includegraphics[width=\textwidth]{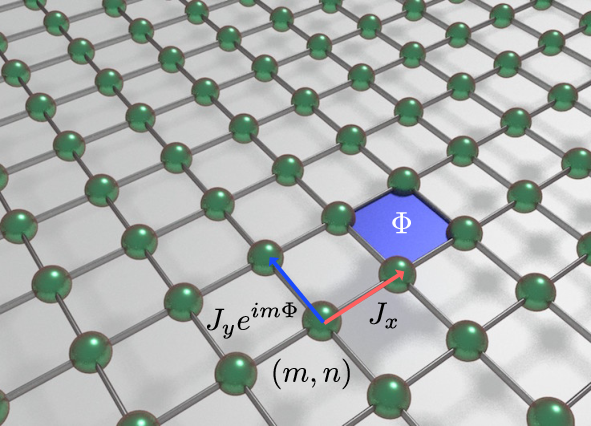}
\caption[Hoppings of the Hofstadter Hamiltonian]{\label{fig:Hofstadter} Sketch of the hoppings of the Hofstadter Hamiltonian, Eq.~\eqref{eq:H0}. The site indices in the $x,y$ directions are $m,n$ respectively. The total flux through each plaquette is $\Phi=2\pi p/q$ with $p, q \in \mathbb{Z}$.}
\end{figure}

Particles hopping in the $y$ direction acquire a Peierls phase $m \Phi$. This implies that the flux acquired when going around a plaquette is $\Phi$, as sketched in the Fig.~\ref{fig:Hofstadter}. Let the magnetic flux be of the the form $\Phi=2\pi p/q$, with $p$ and $q$ coprime integers. As a result, the phase factors acquired when hopping in the $x$ direction have a periodicity of $q$.
We say that the system
has magnetic unit cell dimensions $(q,1)$.


The Hofstadter model is, arguably, the simplest two dimensional model which presents topologically non-trivial properties. This makes it ideal to study the interplay between topology and inter-atomic interactions, as considered in Refs.\,\parencite{Powell2010,Powell2011}.

\subsection{Hofstadter model on a torus}
\label{sec:HofstadterTorus}

\begin{figure}[t]
\centering
\includegraphics[width=0.8\textwidth]{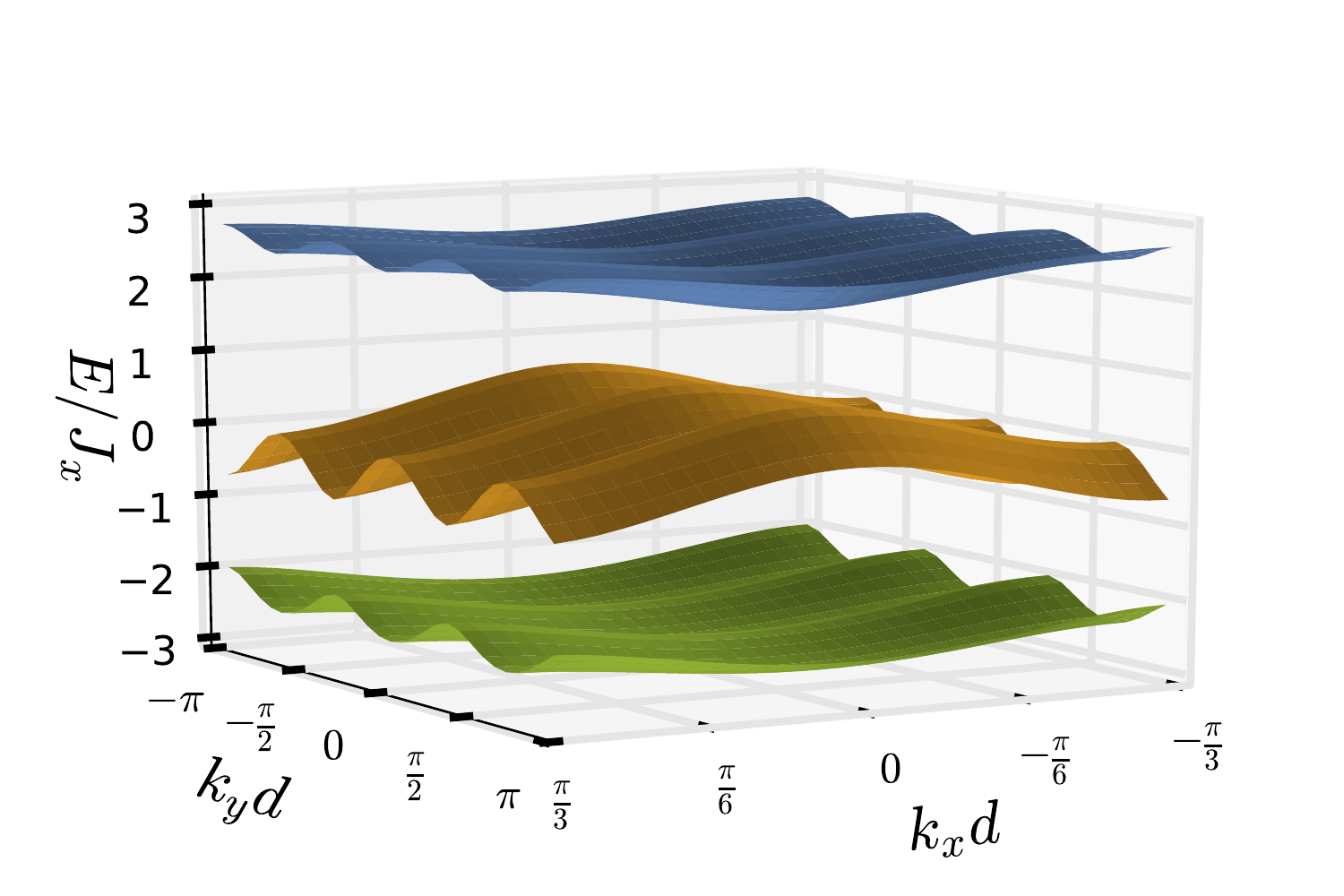}
\caption[Dispersion of the Hofstadter Hamiltonian]{\label{fig:dispersion}
Dispersion of the Hofstadter Hamiltonian Eq.~\eqref{eq:H0} with $J_x = J_y = 1$ and $\Phi = 2\pi/3$. The system presents two energy gaps.}
\end{figure}

In this section, we consider the single particle, translationally invariant Hofstadter model with periodic boundary conditions, such that the site $m=1$ and $m=N_x+1$ are identified, as are the sites $n=-(N_y-1)/2$ and $n=(N_y+1)/2$. Because the magnetic unit cell in the gauge of Eq.~\eqref{eq:H0} has dimensions $(q,1)$, we further require that $N_x$ is a multiple of $q$. In the single particle subspace, the quasimomentum variable $\kk=(k_x,k_y)$ is a good quantum number, such that we can block-diagonalise the Hamiltonian using Bloch's theorem:
\begin{equation}
\label{eq:HofMomBasis}
 \h_{\rm Hof}= \int {\rm d} \kk \ket{\kk}\bra{\kk} \otimes \h_{\rm Hof}(\kk),
\end{equation}
with:
\begin{equation}
\label{eq:HofFullMom}
 \h_{\rm Hof}(\kk)=
 \begin{pmatrix}
  \E_{k_y, 1} & -J_x & \ldots & -J_x e^{-i k_x q d} \\
  -J_x & \E_{k_y, 2} & & \\
  \ldots & & \ldots & \\
  & & & -J_x \\
  -J_x e^{i k_x q d} & & -J_x & \E_{k_y, (q-1)}
 \end{pmatrix},
\end{equation}
and $\E_{k_y, m}=-2 J_y \cos(k_y d + m \Phi)$. Within this gauge, the (magnetic) Brillouin zone has an extension of $(2\pi/(qd))\times(2\pi/d)$ along $k_x$ and $k_y$. For any given quasimomentum $\kk$, the Hamiltonian Eq.~\eqref{eq:HofFullMom} has $q$
eigenvalues $E_j(\bm{k})$. We compute these energies numerically for $\Phi=2\pi/3$ and $J_y=J_x$, and present the resulting dispersion in the Fig.~\ref{fig:dispersion}. Notice that for these parameters, the dispersion has three bands separated by band gaps.

\begin{figure}[t]
\centering
\includegraphics[width=\textwidth]{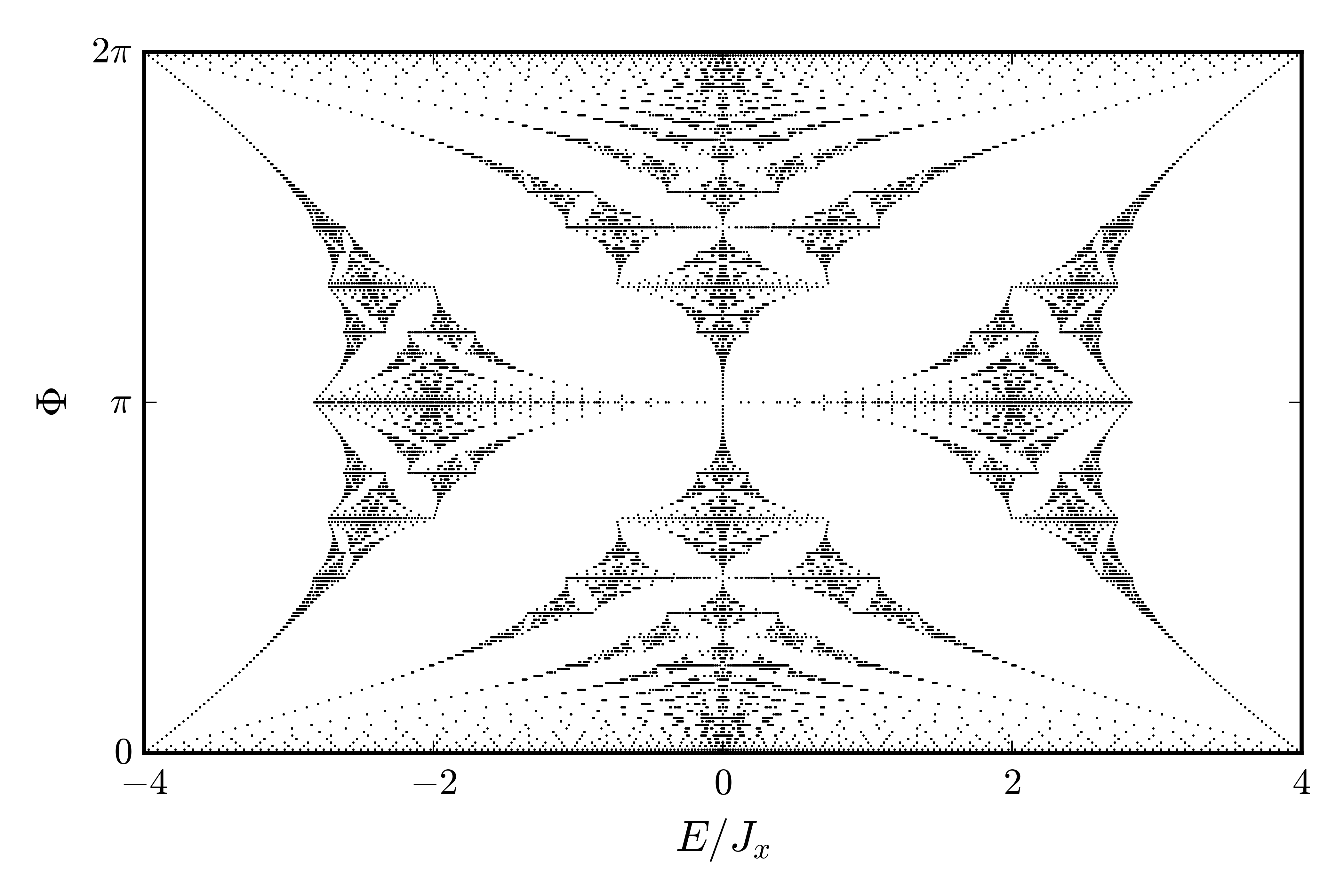}
\caption[Hofstadter butterfly]{\label{fig:Butterfly}
Integrated energy spectrum of the Hofstadter model with magnetic flux $\Phi$ and relative hopping amplitude $J_x=J_y$. This figure is known as the Hofstadter butterfly.}
\end{figure}

For a general external magnetic flux of $\Phi=2\pi p/q$, with $p$ and $q$ coprime integers and $J_x, J_y \neq 0$, the spectrum presents $q$ bands which are separated by $q-1$ band gaps. Systems with $q$ even are exceptions to this behaviour, as the spectrum presents a band touching in $E=0$, such that it presents $q-2$ band gaps. The integrated spectrum is plotted for varying values of $\Phi$ in Fig.~\ref{fig:Butterfly}. This is the famous Hofstadter's butterfly, one of the few known examples of a non-random fractal appearing in physics.

\subsubsection{The Chern number}
\label{sec:Chern}

We know from Bloch's theorem that the Brillouin zone has the topology of a torus, such that it forms a  closed surface in two dimensions. Following the discussion of Sec.~ \ref{sec:BerryPhase}, a particle which follows a closed loop on this surface acquires a Berry phase given by Eq.~\eqref{eq:BerryPhase}. We saw in Sec.~\ref{sec:SSHmodel} that the Berry phase was the key to understanding the topological properties of the SSH model. As we shall see shortly, the Berry phase in the Hofstadter model is also quantised when integrated over the entire system. This leads to topological properties appearing in the Hofstadter model in much the same way as they did in the SSH model.

For definiteness, we will start by showing that the Berry phase is quantised in the finite Hofstadter model, and then take the continuum limit. We do this to convince the reader that topological properties appear in two dimensional systems regardless of system size. In our case, the spacing between quasimomentum variables is:
\begin{equation}
\delta k_x=\frac{2\pi}{N_x q d}, \delta k_y=\frac{2\pi}{N_y d}.
\end{equation}
To simplify the notation, we introduce the vectors $\bm{e}_{k_x}$ and $\bm{e}_{k_y}$, defined by:
\begin{equation}
 (k_x, k_y) + \bm{e}_{k_x} = (k_x +\delta k_x, k_y), (k_x, k_y) + \bm{e}_{k_y} = (k_x, k_y +\delta k_y)
\end{equation}
We saw in the SSH model that the Berry phase, which a particle acquired when going around the entire Brillouin zone, was quantised. We would now like to derive a similar result for the Hofstadter model. Note however that, in two dimensions, a particle can be carried around the Brillouin zone in an infinite number of different ways. In the following, we will make an educated guess and consider the Berry phase acquired when going around each plaquette of the Brillouin zone. We will show that the Hermiticity of the Hamiltonian constrains this sum in periodic systems.

\begin{figure}[t]
\centering
\includegraphics[width=\textwidth]{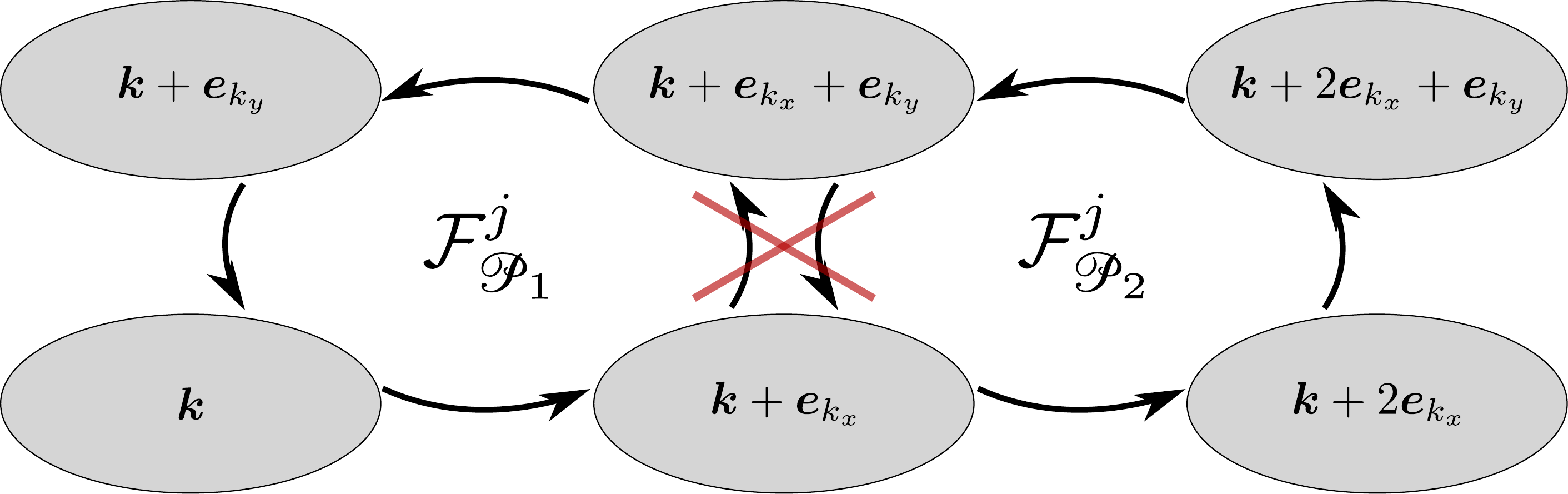}
\caption[Berry flux through two neighbouring plaquettes]{Illustration of the Berry fluxes, $\mathcal{F}_{\mathscr{P}_1}^j$ and $\mathcal{F}_{\mathscr{P}_2}^j$, acquired when going around the two neighbouring plaquettes in quasimomentum space $\mathscr{P}_1$ and $\mathscr{P}_2$, respectively. 
When summing these Berry fluxes, the phase acquired on all repeated links cancel out up to an integer multiple of $2\pi$.}
\label{fig:BerryFlux}
\end{figure}

Let $\ket{u_j(\kk)}$ be the $j^{\rm th}$ eigenstate of $\h_{\rm Hof}(\kk)$, Eq.~\eqref{eq:HofFullMom}. The Berry phase acquired when a particle is carried around the plaquette $\mathscr{P}$ in quasimomentum space, which has its lower-left corner at point $\kk$ in reciprocal space, is given by:
\begin{equation}
 \label{eq:BerryFlux}
 \begin{split}
  \mathcal{F}_\mathscr{P}^j = 
  -\arg \Big[ & \braket{u_j(\kk)}{u_j(\kk + \bm{e}_{k_x})}
  \braket{u_j(\kk + \bm{e}_{k_x})}{u_j(\kk + \bm{e}_{k_x} + \bm{e}_{k_y})}\\
  & \braket{u_j(\kk + \bm{e}_{k_x} + \bm{e}_{k_y})}{u_j(\kk + \bm{e}_{k_y})}
  \braket{u_j(\kk + \bm{e}_{k_y})}{u_j(\kk)} \Big]
 \end{split}
\end{equation}
This is the Berry flux. Due to the argument function, this quantity has a branch cut, such that:
\begin{equation}
 \label{eq:BranchCut}
 -\pi < \mathcal{F}_\mathscr{P}^j \leq \pi
\end{equation}
We sketch the Berry flux in the figure \ref{fig:BerryFlux}. As is illustrated in this figure, when summing the Berry fluxes of two neighbouring plaquettes, the link joining the plaquettes appears twice in the sum,
such that the phase acquired on this link cancels out, but only up to an integer multiple of $2\pi$ due to the branch cut of the argument, Eq.~\eqref{eq:BranchCut}.

The total phase acquired when going around each plaquette of the Brillouin zone is then simply:
\begin{equation}
\label{eq:FHSChern}
\gamma_j = \sum_{\mathscr{P} \in BZ} \mathcal{F}_\mathscr{P}^j = 2 \pi \mathcal{C}_j, \mathcal{C}_j \in \mathbb{Z}
\end{equation}
The last equality in Eq.~\eqref{eq:FHSChern} is due to the periodicity of the Brillouin zone. Indeed, it implies that all links appear twice in the sum over plaquettes, such that all acquired phases cancel out up to an integer multiple of $2\pi$
\parencite{Asboth2015}.

Thus, we see that the total Berry phase is quantised in two dimensions, as it was in the SSH models. This leads to an integer quantity appearing, $\mathcal{C}_j$, which we will refer to as the Chern number. This quantity is similar to the winding number in that it is a property of the Hamiltonian, and it is a global property because it is obtained by integrating over the entire Brillouin zone. What is more, the Chern number is obtained by adiabatically carrying a particle around the Brillouin zone, such that it can only be defined for a gapped system. These similarities are by no means coincidental: the Chern number is a topological invariant which characterises the phase of the Hofstadter model in the same way as the winding number characterises the phase of the SSH model. Note however that, in one dimension, the Berry phase was quantised due to symmetry constraints. In two dimensions, we did not need to invoke any such symmetry, and this is a crucial difference between the Chern number and the winding number.

While we chose to consider a discrete system, the definition of the Chern number, Eq.~\eqref{eq:FHSChern}, can be extended to continuous systems \parencite{Asboth2015}:
\begin{equation}
\label{eq:ChernFormula}
 \mathcal{C}_j = \frac{1}{2\pi} \int_{BZ} \bm{\mathcal{F}}_j(\kk) \cdot {\rm d}\bm{S}.
\end{equation}
For consistency with the literature, we have expressed the Berry curvature $\bm{\mathcal{F}}_j(\kk)$, which is the continuum analogue of the Berry flux, as a vector which is perpendicular to the surface of the Brillouin zone. This quantity is obtained by taking the continuum limit of Eq.~\eqref{eq:BerryFlux}:
\begin{equation}
 \label{eq:BerryCurvature}
 \bm{\mathcal{F}}_j = i \langle \partial_{\kk} u_j(\kk) | \times | \partial_{\kk} u_j(\kk) \rangle.
\end{equation}
The Berry curvature can be understood in an intuitive way by noticing that Eq.~\eqref{eq:BerryFlux} is mathematically identical to the Peierls phase acquired when moving around a plaquette, except that this plaquette exists in quasimomentum space rather than real space.
This analogy allows us to think of the Berry curvature as the magnetic flux traversing the Brillouin zone.
By analogy to Gauss's law, the Chern number, or equivalently, the total flux through a closed surface (the Brillouin zone) is then the system's charge in magnetic monopoles.

\subsubsection{Particle dynamics in a topological insulator}
\label{sec:anomalousVelocity}

We will now describe one important way in which the Chern number manifests itself in topological invariants. Specifically, we will show that the Berry curvature determines the particle motion in the Hofstadter model. It is actually through this mechanism that the first evidence of topological phenomena was identified \parencite{Klitzing1980}, though it was only later explained in Refs.\ \parencite{Laughlin1981,Thouless1983}. We are further motivated to study this effect because it is a clear manifestation of a system's topological properties, and can therefore be used to experimentally measure its topological invariant.

We begin by considering the Hofstadter model subject to a constant force $\bm{F} = (F_x, F_y)$:
\begin{equation}
\h = \h_{\rm Hof} - \bm{F}\cdot\hat{\bm{r}},
\end{equation}
where $\hat{\bm{r}}$ is the position operator. Note that the addition of $\bm{F}$ breaks translational invariance. If $\h_{\rm Hof}$ is gapped, however, and when $|\bm{F}| d$ is small compared to the band gap, we can neglect inter-band transitions. This allows us to restore translational invariance by performing the gauge transformation $\UU=\exp(-i \bm{F}\cdot\hat{\bm{r}} t /\hbar )$. The Hamiltonian then transforms as:
\begin{equation}
\label{eq:ForceGaugeTrans}
 \h \xrightarrow{} \UU \h \UU^\dagger + i \hbar (\partial_t \UU) \UU^\dagger = \UU \h_{\rm Hof} \UU^\dagger.
\end{equation}
In this gauge the quasimomentum $\kk(t)$ is a good quantum number and takes the form \parencite{Xiao}:
\begin{equation}
\label{eq:Meank}
\kk(t) = \kk(t=0) + \frac{1}{\hbar}\bm{F} t.
\end{equation}
Thus, we see that a small, constant force induces Bloch oscillations. This is known as the acceleration theorem.

The Hamiltonian $\h$ is block-diagonal in the quasimomentum basis, and its elements, $\h(\kk, t)$, are given by Eq.~\eqref{eq:HofFullMom}.
We define $\ket{u_j(\kk, t)}$, the $j^{\rm th}$ instantaneous eigenstate of $\h(\kk, t)$. To simplify the notation in the following, we will omit the $\kk$ and $t$ dependence when it is possible.

We wish to study the motion of atoms in this system. To this end, we define $\bm{\hat{J}}$, the current operator, which is given by \parencite{Asboth2015}:
\begin{equation}
 \label{eq:CurrentOperator}
 \bm{\hat{J}} = \frac{1}{N_P} \partial_{\kk} \h(\kk, t),
\end{equation}
with $N_P$ the number of particles.

Let us consider an atom initially in the instantaneous eigenstate $\ket{\kk_0}\otimes\ket{u_j(\kk_0, t=0)}$.
This state's quasimomentum subsequently evolves according to Eq.~\eqref{eq:Meank}. Its internal degree of freedom at time $t$ is given by Eq.~\eqref{eq:StateTimeT}. For a small force $\bm{F}$, we can expand this state to first order in $\dot{\kk}$, giving \parencite{Xiao}:
\begin{equation}
\label{eq:StateToFirstOrder}
 \ket{\Psi(t)} = \ket{u_j(t)} - i \hbar \sum_{j'\neq j} \ket{u_{j'}(t)}
 \frac{
 \braket{u_{j'}(t)}{\partial_t u_j(t)}
 }
 {E_j-E_{j'}},
\end{equation}
where we have neglected an overall phase. We can use Eq.~\eqref{eq:CurrentOperator} to calculate the atomic group velocity to first order in $\dot{\kk}$:
\begin{equation}
\label{eq:SemiClassicalGroupVel2}
 \bm{v}_j(\kk) = \bra{\Psi} \bm{\hat{J}} \ket{\Psi}
 = \frac{1}{\hbar} \partial_{\kk} E_j - i \left[ \braket{\partial_{\kk} u_j}{\partial_t u_j} - \text{H.c.} \right].
\end{equation}
This equation can be put into a simpler form by using $\partial_t = \dot{\kk} \partial_{\kk}$:
\begin{equation}
\label{eq:SemiClassicalGroupVel}
\bm{v}_j(\bm{k}) = \frac{1}{\hbar}\partial_{\bm{k}} E_j(\bm{k}) - \dot{\bm{k}} \times \bm{\mathcal{F}}_j(\bm{k}).
\end{equation}
The first term of the right hand side of Eq.~\eqref{eq:SemiClassicalGroupVel} is the familiar band velocity. The second term is known as the anomalous velocity, which is determined by the Berry curvature $\bm{\mathcal{F}}_j(\bm{k})$, Eq.~\eqref{eq:BerryCurvature}. For a filled energy band $j$, i.e., when all its Bloch states are uniformly occupied, the band velocity vanishes, 
such that the mean velocity of the centre of mass simply:
\begin{equation}
\label{eq:TKNN}
\mean{\bm{v}(t)}= -\frac{2\pi}{A_{BZ}} \left( \dot{\kk} \times \bm{e}_z \right) \CC_j,
\end{equation}
where $\CC_j$ is the Chern number, given by Eq.~\eqref{eq:ChernFormula} and $A_{BZ}$ is the area of the Brillouin zone. For the Hofstadter model, this is:
\begin{equation}
 \label{eq:AreaBZ}
 A_{BZ} = \frac{(2\pi)^2}{d^2 q}
\end{equation}
Thus, we have reached the surprising result that an insulating system with a filled band can present a non-zero atomic current when subject to a linear potential gradient. What is more, the current is quantised in units of $2\pi$ and perpendicular to the applied potential. The Eq.~\eqref{eq:TKNN} is in fact the celebrated TKNN formula \parencite{Thouless1982}, which explains the anomalous conductance of the integer quantum Hall effect.


To conclude this section, we see that the Berry curvature appears naturally in the centre of mass dynamics through Eq.~\eqref{eq:SemiClassicalGroupVel}. We anticipate that we can use this result to measure the system's Chern number. What is more, we can also gain a more intuitive understanding of the Chern number through the derivation presented in this section. Indeed, we have seen that, when we perform a single band approximation, the Berry curvature appears by expanding the atom's state to first order in perturbation theory.
The Chern number can therefore be understood as a residual effect from the system's empty bands.

\subsubsection{The FHS algorithm}
\label{sec:FHSalgo}

While the Chern number can, in some cases, be calculated analytically from Eq.~\eqref{eq:ChernFormula} \parencite{Sticlet2012}, this is in general not possible. We can, however, evaluate it numerically, e.g: by discretising Eqs.~\eqref{eq:ChernFormula} and \eqref{eq:BerryCurvature}. This approach is problematic because these equations involves gauge dependent quantities, such as the eigenstate's derivative, and 
it is difficult to choose a gauge in which these are smooth from one point in the Brillouin zone to the next \parencite{Asboth2015}.

It is in this situation that Eqs.~\eqref{eq:BerryFlux} and \eqref{eq:FHSChern} become useful. Indeed, in these equations, the Chern number is expressed in terms of gauge invariant quantities, such that the phase continuity of the wavefunction is not required when applying this formula. This method for numerically computing the Chern number was first suggested in Ref.\ \parencite{Fukui2005}. We will refer to it as the FHS algorithm.

Let us now consider running the FHS algorithm on the Hofstadter Hamiltonian.
The FHS algorithm is said to accurately predict the Chern number if Eq.~\eqref{eq:FHSChern} returns the same value as Eq.~\eqref{eq:ChernFormula}, computed for an infinite system. This is guaranteed to be the case if the continuous Berry curvature, given by Eq.~\eqref{eq:BerryCurvature}, integrated over any plaquette $\mathscr{P}$ of the Brillouin zone, is equal to the discrete Berry flux, Eq.~\eqref{eq:BerryFlux}, of the same plaquette. Due to the branch cut of the Berry flux, Eq.~\eqref{eq:BranchCut}, this is only true when:
\begin{equation}
\label{eq:FHSadmissibility}
 \left| \int_{\mathscr{P}} \bm{\mathcal{F}}_j(\kk) \cdot {\rm d} \bm{S} \right| \leq \pi, \forall \mathscr{P}.
\end{equation}
This is the admissibility condition of the FHS algorithm \parencite{Fukui2005}.


\subsection{Hofstadter model on a cylinder}
\label{sec:HofCylinder}

We saw in Sec.~\ref{sec:TopoBS} that a key feature of the SSH model with open boundary conditions is the presence of topological bound states at its boundaries. We anticipate that the topologically non-trivial Hofstadter model might present a similar behaviour at its boundaries. Motivated by this intuition, we will compare in this subsection the Hofstadter model on a torus to the Hofstadter model on a cylinder. In the latter case, we will set open boundary conditions in the $y$ direction and observe how the system's properties are affected.

\subsubsection{Dispersion of the Hofstadter Hamiltonian on a cylinder}

We begin by considering the specific example of the Hofstadter model which is periodic in the $x$ direction and has $N_y=27$ sites in the $y$ direction. The Hamiltonian of this model is given by Eq.~\eqref{eq:H0}, and we set the external magnetic flux to $\Phi=2\pi/3$ and the hopping amplitudes to $J_x=J_y$. For visualisation purposes, we will find it practical to rotate the unit cell such that it lies in the $y$ direction. This can be done through the gauge transformation:
\begin{equation}
\label{eq:GaugeTranform}
\hat{c}_{n,m} \xrightarrow{} e^{i \Phi m n} \hat{c}_{n,m}. 
\end{equation}
In the new gauge, the magnetic unit cell has dimensions $(1,q)$.

\begin{figure}[t]
\centering
\includegraphics[width=0.8\textwidth]{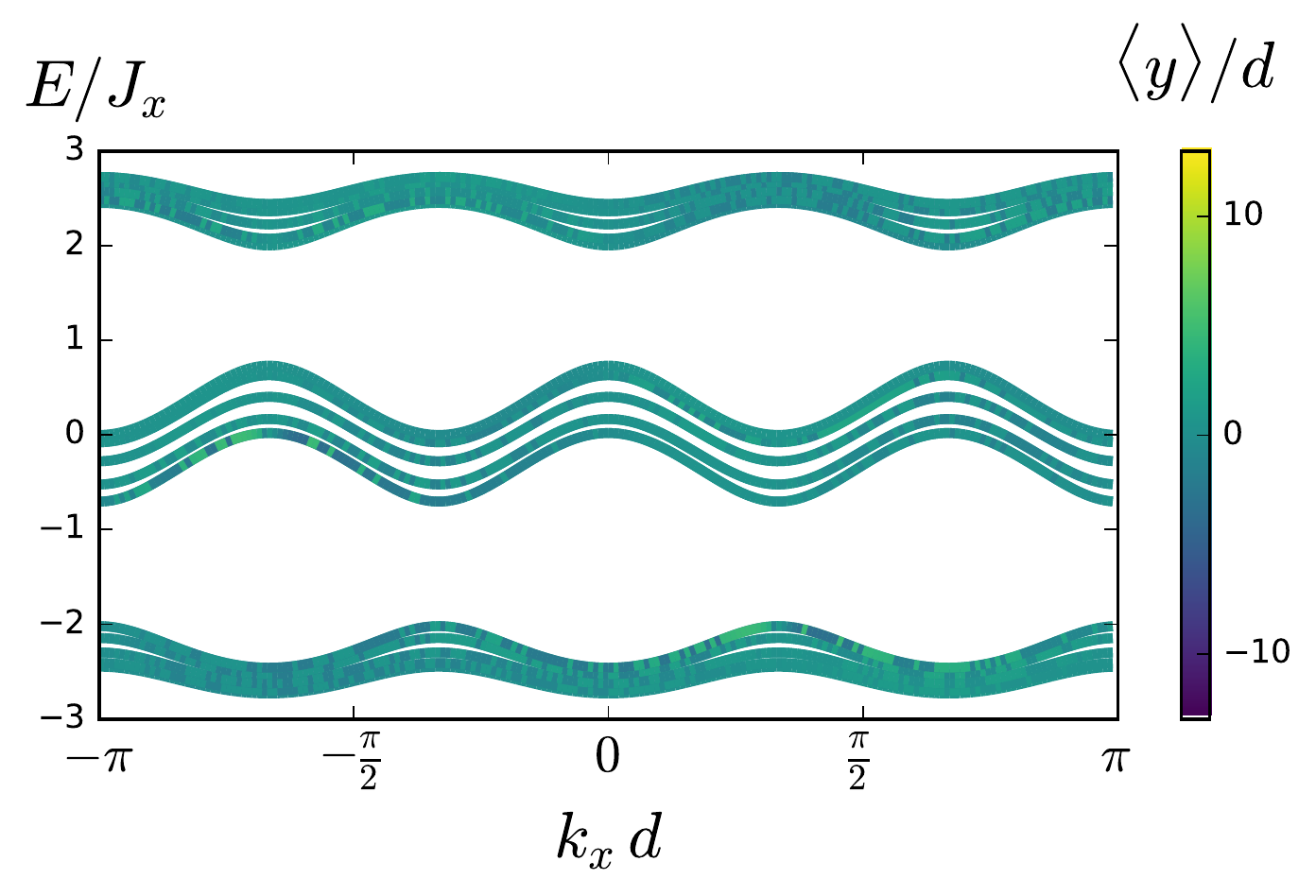}
\caption[Dispersion of the Hofstadter model on a torus]{\label{fig:cylinder_periodic}
Dispersion of the Hofstadter Hamiltonian on a torus, obtained by diagonalising Eq.~\eqref{eq:H0Mom} with $N_y=27$, $J_x = J_y = 1$, and $\Phi = 2\pi/3$. The colour coding indicates the states' mean position in the $y$ direction. The system presents two energy gaps.}
\end{figure}

By projecting onto the single particle subspace, and thanks to the Hamiltonian's translational invariance in the $x$ direction, we can partially Fourier transform $\h_{\rm Hof}$ along this direction. In the basis of states with well defined quasimomentum in the $x$ direction, Eq.~\eqref{eq:H0} is block-diagonal, and its elements are given by:
\begin{equation}
\label{eq:H0Mom}
\begin{split}
\h_{\rm Hof}(k_x) = \sum_n -2 J_x \cos(k_x d - n \Phi) \ket{n} \bra{n} - \left( J_y \ket{n+1} \bra{n} + \text{H.c.} \right),
\end{split}
\end{equation}
where $\ket{n}$ is the state which is well localised in the $y$ direction and belongs to the $n^{\rm th}$ unit cell. In this gauge, there is only one site per unit cell in the $x$ direction such that the quasimomentum values lie in the interval $k_x \in [-\pi/d, \pi/d]$. We plot the dispersion of this system with periodic boundary conditions in both the $x$ and $y$ directions (Hofstadter model on a torus) in Fig.~\ref{fig:cylinder_periodic}. For each value of $k_x$, $\h_{\rm Hof}(k_x)$ has $N_y$ discrete energy eigenvalues. As we already saw in Fig.~\ref{fig:dispersion}, these gather into $q$ bands which are separated by band gaps. Due to the system's periodicity, the eigenstates of this Hamiltonian are delocalised in the $y$ direction, such that they all have mean position $n\approx 0$ (as can be seen from the colour coding).

\begin{figure}[t]
\centering
\includegraphics[width=\textwidth]{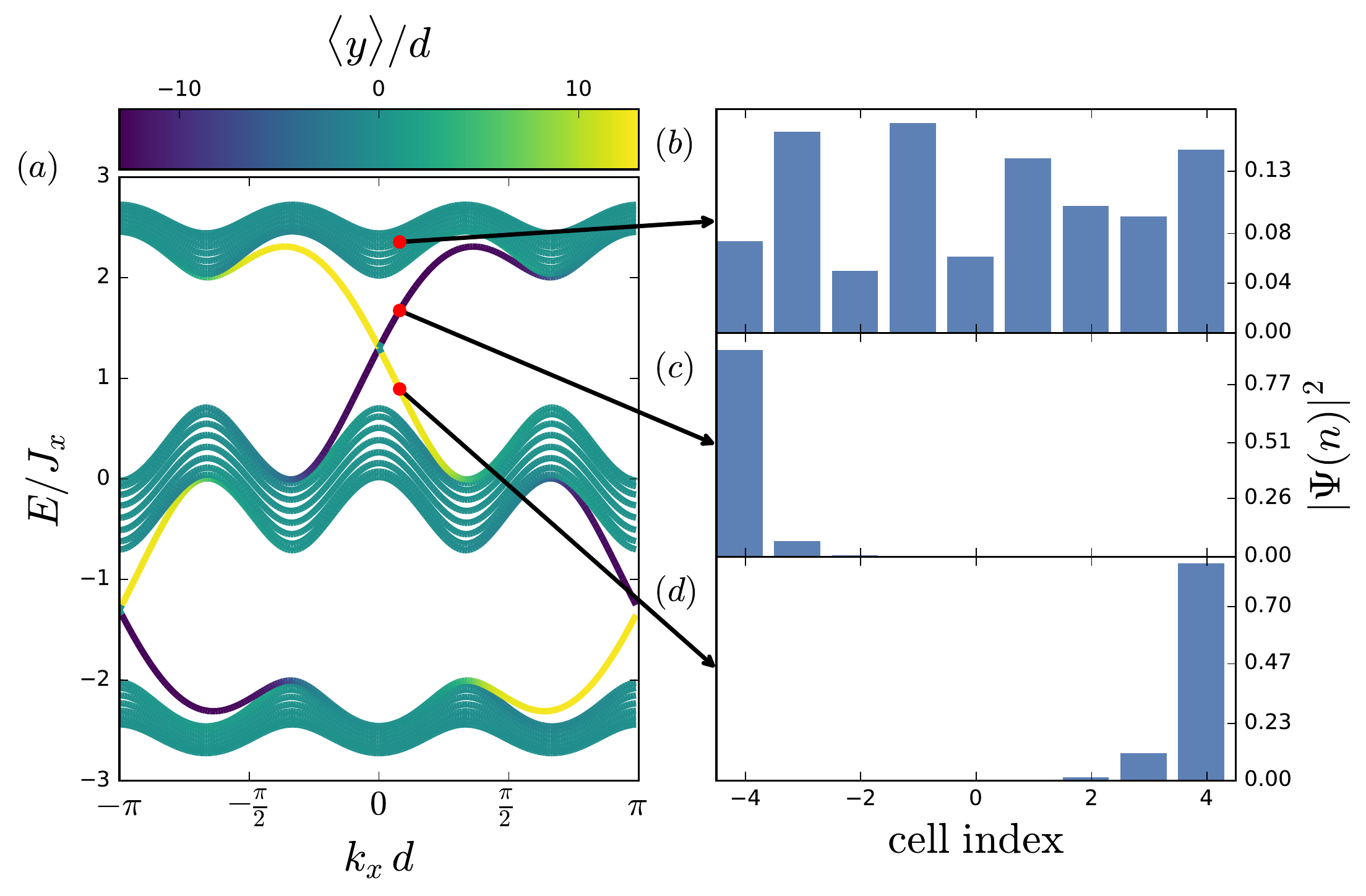}
\caption[Dispersion of the Hofstadter model on a cylinder]{\label{fig:cylinder_open}
Fig.~\ref{fig:cylinder_open}(a): dispersion of the Hofstadter Hamiltonian on a cylinder, obtained by diagonalising Eq.~\eqref{eq:H0Mom} with $N_y=27$, $J_x = J_y = 1$, and $\Phi = 2\pi/3$. The colour coding indicates the states' mean position in the $y$ direction. The energy distribution of bulk states is approximately the same as the dispersion shown in Fig.~\ref{fig:cylinder_periodic}. Bulk states present an energy gap in which we find states located at the system's edges. We plot the probability density distribution as a function of $y$ of a bulk state in Fig.~\ref{fig:cylinder_open}(b), and of states located at the left and right edges in Figs.~\ref{fig:cylinder_open}(c) and (d), respectively.}
\end{figure}

We plot the dispersion for the system with open boundary conditions in the $y$ direction in Fig.~\ref{fig:cylinder_open}(a). As is apparent from this figure, this system presents eigenstates which are delocalised in $y$ and have approximately the same dispersion as periodic system. These states are only perturbatively affected by the presence of the edges; for this reason, we will refer to them as bulk states. Note however that the cylindrical system is not gapped. Indeed, the bulk band gap is sparsely populated by states which are confined to the system's edges. These states form branches which bridge the band gap. We plot a sample bulk state in the Fig.~\ref{fig:cylinder_open}(b), and sample edge states in the Figs.~\ref{fig:cylinder_open}(c) and (d).
Because the edge states are energetically forbidden from existing in the system's bulk, their probability density function decays exponentially away from the edges.

\subsubsection{The number of edge branches is a topological invariant}

In this paragraph, we will briefly show that the number of edge states in a given bulk band gap is a topological invariant, and relate this quantity to the Chern number.

\begin{figure}
    \centering
    \includegraphics[width=\textwidth]{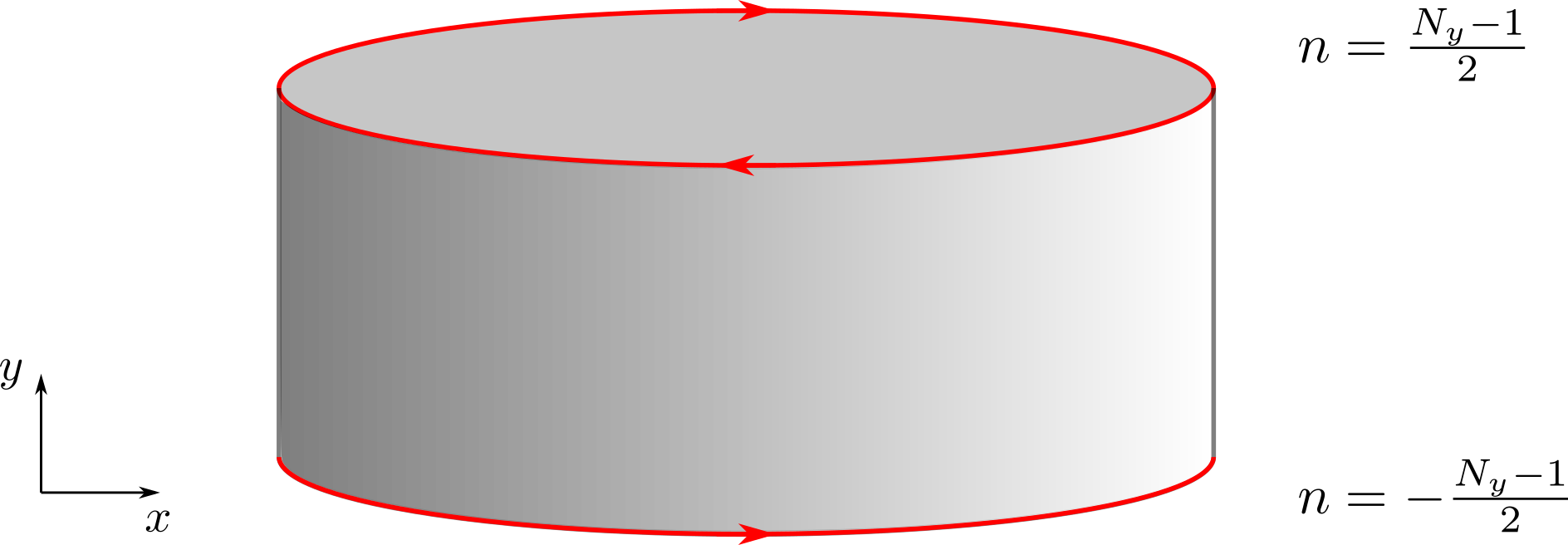}
    \caption[Sketch of edge states on a Hofstadter cylinder]{\label{fig:ChernToEdgeStates} Sketch of a Hofstadter cylinder, with dispersion plotted in Fig.~\ref{fig:cylinder_open}(a). The edge states belonging to the $j=2$ bulk band gap are coloured in red. At the $n=(N_y-1)/2$ edge, we find $N_+=0$ ($N_-=1$) edge states travelling in the positive (negative) $x$ direction. We deduce that $\mathcal{N}_2=-1$. This allows us to evaluate the third band's Chern number: $\mathcal{C}_3=1$.}
\end{figure}

As can be seen from Fig.~\ref{fig:cylinder_open}(a), the edge states are delocalised along the cylinder's edge, and their group velocity along the $x$ direction is determined by the edge branch's slope. In the second bulk band gap of Fig.~\ref{fig:cylinder_open}(a), we therefore have a branch of states which are bound to the $n=-(N_y-1)/2$ edge and move in the positive $x$ direction. One of these states is shown in Fig.~\ref{fig:cylinder_open}(c). Likewise, this band gap has a branch of edge states bound to $n=(N_y-1)/2$ moving in the negative $x$ direction, of which one is shown in Fig.~\ref{fig:cylinder_open}(d).
These states are sketched in Fig.~\ref{fig:ChernToEdgeStates}.

Consider the crossing of two edge branches. We assume these live in the same bulk band gap and are localised at the same edge [for instance, the $n=(N_y-1)/2$ edge]. If the two branches have slopes with opposite signs, they can hybridise and be lifted out of the band gap. If their slope has the same sign, however, they cannot turn into an avoided crossing as this would cause the edge branches to become multivalued functions of $k_x$ [this is derived in detail in Ref.\ \parencite{Asboth2015}]. The only way to destroy these edge branches is therefore to close the band gap. We conclude that, for a given edge and bulk band gap, two edge states travelling in the same direction are topologically protected, while edge states travelling in opposite directions are not. Note that this result relies on the assumption that edge states living at different edges have vanishing overlap. This is usually the case in the limit $N_y \gg 1$.

Let us define $N_+$ ($N_-$), the number of edge states in the $j^{\rm th}$ band gap, bound to the $n=(N_y-1)/2$ edge, which travel in the positive (negative) $x$ direction. From the reasoning above, we can see that $\mathcal{N}_j=N_+-N_-$ is a topological invariant.
In the example presented in Fig.~\ref{fig:ChernToEdgeStates}, we can see by inspection that $\mathcal{N}_2=-1$. As show in \parencite{Asboth2015}, this topological invariant can be related to the Chern number through: $\CC_j = \mathcal{N}_j - \mathcal{N}_{j-1}$. 

Thus, at a boundary between two topological insulators, there can exist robust edge states which are delocalised along the boundary. The number of these edge states is determined by the difference in Chern numbers at either side of the boundary. This behaviour is very similar to the one we observed in the SSH model, where the difference in winding numbers at an edge determines the number of topological bound states which exist at the boundary.



\section{Beyond the topological classification: Floquet systems}
\label{sec:TwoStepQW}

In Sec.~\ref{sec:1dTopoIntro}, we introduced the SSH model in detail. Through this model we understood how the Berry phase, when integrated over the entire system, could become quantised as a result of chiral symmetry. This lead to the appearance of an integer topological invariant which characterised the system's phase. Without demonstrating this, we indicate that the Hamiltonian, Eq.~\eqref{eq:SSH} is real, which leads to it being both time-reversal symmetric and particle-hole symmetric, in addition to chiral symmetric \parencite{Kitagawa2010a}. As a result, the SSH Hamiltonian belongs to the BDI class of Table \ref{tab:topo}, which can have a non-zero winding number in one spatial dimension \parencite{Schnyder2008b}.


In Sec.~\ref{sec:2dTopoIntro}, we studied the example of the Hofstadter model, through which we showed that topological properties can appear in two-dimensional systems through a mechanism which is similar to the SSH model. We showed that chiral symmetry was not necessary for the Berry phase to become quantised. In fact, the Hofstadter Hamiltonian can be shown to break chiral, time-reversal and particle-hole symmetries, such that it belongs to the A class of Table \ref{tab:topo}. Hamiltonians in this symmetry class can have non-zero Chern numbers in two dimensions \parencite{Schnyder2008b}.

As we will see in this section, however, the topological classification of Hamiltonians is not the whole story. We will show this by introducing a simple model which does not belong to any of the classes of Table \ref{tab:topo}. We will consider  the two-step quantum walk \parencite{Kitagawa2010a}, a system which escapes the classification due to its time dependence \parencite{Asboth2012}. In particular, we will focus on the particular properties that this system displays to understand how this new type of topological phase can be experimentally identified.

\subsection{The two-step quantum walk}

\begin{figure}[t]
\centering
  \includegraphics[width=0.5\textwidth]{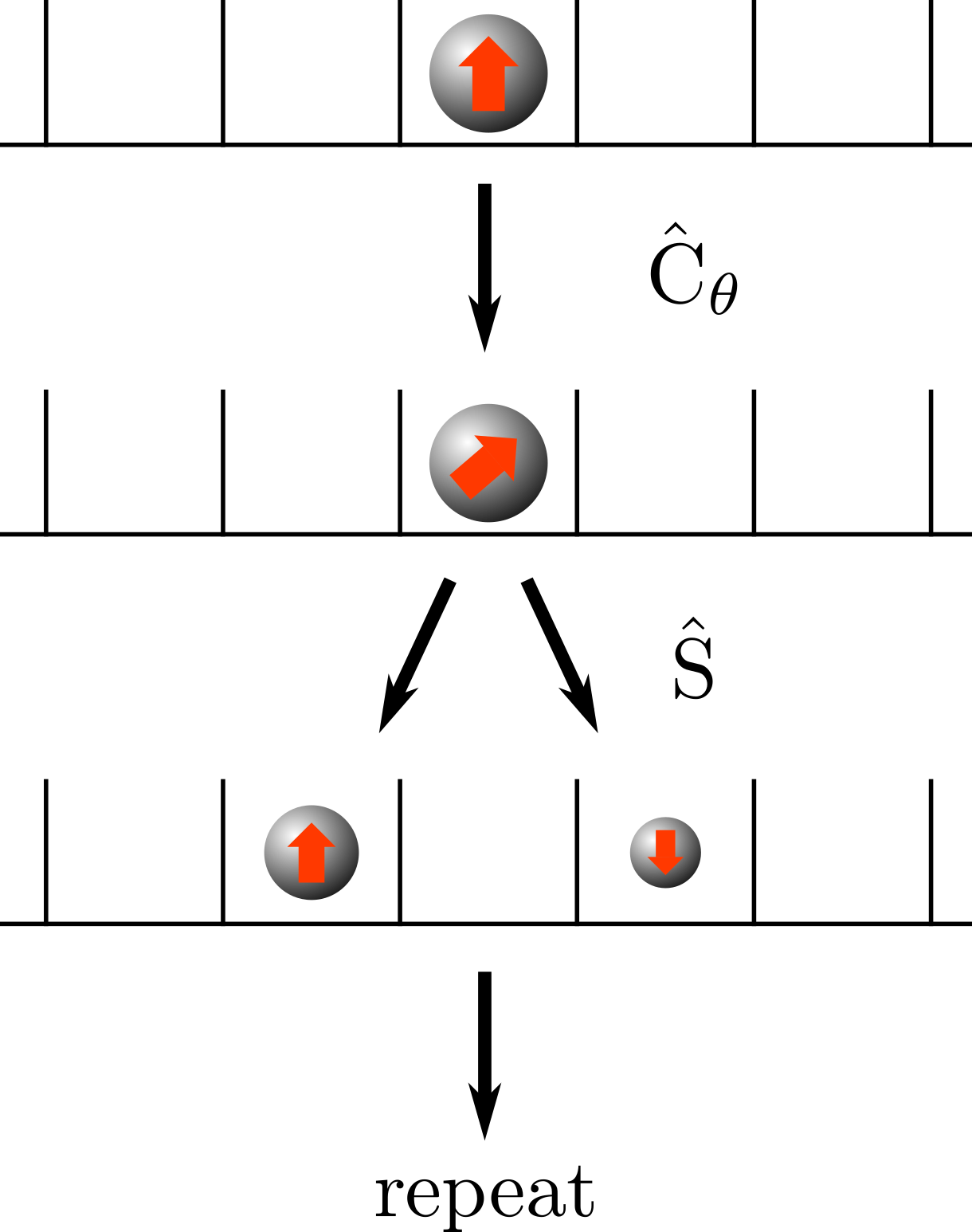}
  \caption[Quantum walk protocol]{\label{fig:QW_detail} Protocol for building a two-step quantum walk.}
\end{figure}

In the one dimensional, discrete classical random walk, a coin toss is performed at each time step. The outcome determines the direction in which the particle is translated. A quantum version of this system can be built, by bringing the particle into a superposition of going to the left and right at each time step. As a result, the system's time-evolution becomes deterministic. One way to build a quantum walk is to introduce an internal degree of freedom, for instance the particle's pseudo-spin. We will assume this spin has two available states, up or down, written $\ket{\uparrow}$ and $\ket{\downarrow}$ respectively. We then act on the particle, alternating between an on-site spin mixing operation \^C$_\theta$ and a spin dependent translation \^S, the effect of which are sketched in the figure \ref{fig:QW_detail}.

In a single particle system, the spin dependent translation sketched in Fig.~\ref{fig:QW_detail} can be put into the form:
\begin{equation}
\label{eq:TSQW_T}
\text{\^S} = \sum_{n} \ket{n+1} \bra{n} \otimes \frac{\sigma_0+\sigma_3}{2}
+ \ket{n} \bra{n+1} \otimes \frac{\sigma_0-\sigma_3}{2},
\end{equation}
where $\ket{n}$ is the state which is well localised at site $n\in [-N/2,~N/2]$, and $\sigma_i$, with $i=\{0, 1, 2, 3\}$, are the identity and the three Pauli matrices which act in spin space.

We will assume that the spin rotation, or coin matrix, takes the form:
\begin{equation}
\label{eq:TSQW_C}
\hat{\text{C}}_{\theta} = \sum_{n} \ket{n} \bra{n} \otimes   e^{-i \theta  \sigma_2}.
\end{equation}
We can block diagonalise Eqs.~\eqref{eq:TSQW_T} and \eqref{eq:TSQW_C} by going into the basis of quasimomentum states:
\begin{equation}
\label{eq:QW_TC_Mom}
 \begin{split}
 & \text{\^S}(k) = e^{i k d \sigma_3}, \\
 & \text{\^C}_\theta(k) = e^{-i \theta \sigma_2}.
 \end{split}
\end{equation}
The quasimomentum variables $k$ belong to the interval $[-\pi/d, \pi/d]$, where $d$ is the lattice spacing.

\subsection{Introduction to Floquet theory}
\label{sec:IntroFloquet}


We saw that the two-step quantum walk is generated by acting stroboscopically on the particle, alternating between operators $\text{\^S}$ and $\hat{\text{C}}_{\theta}$. We define the time evolution matrix as:
\begin{equation}
\label{eq:TSQW_U}
\hat{\text{U}} = \text{\^S}  \hat{\text{C}}_{\theta}.
\end{equation}
This operator captures the system's evolution over a period of the Hamiltonian.
The operator $\hat{\text{U}}$, being a series of discrete operations, is generated by a time dependent, periodic Hamiltonian \^H$(t)$.
We will find it useful in this context to resort to Floquet theory, which
allows us to approximate a periodic Hamiltonian \^H$(t)$, with period $T$, by a static effective Hamiltonian $\hat{\text{H}}_F$. The effective, or Floquet Hamiltonian is defined from the time evolution through the formula:
\begin{equation}
\label{eq:TSQW_Heff}
\hat{\text{U}}=e^{-i \hat{\text{H}}_F T/\hbar}
\end{equation}
The spectrum of the effective Hamiltonian can be found using the definition of the Floquet Hamiltonian Eq.~(\ref{eq:TSQW_Heff}). We first find the set of eigenvalues $\lambda_j$ of the time evolution \^U Eq.~(\ref{eq:TSQW_U}), then relate them to the effective Hamiltonian's eigenvalues $\varepsilon_j$ through:
\begin{equation}
\label{eq:QuasiEnergies}
\varepsilon_j=\frac{i \hbar}{T} \log\left(\lambda_j\right)
\end{equation}
In the following, we set $T/\hbar=1$. Note that for a unitary operator such as \^U, the eigenvalues $\lambda_j$ are complex with unit norm. This implies that the eigenvalues of \^H$_F$ defined using Eq.~(\ref{eq:QuasiEnergies}) can only be defined up to $2 \pi$. This periodicity of eigenstates is a property which is common to all Floquet Hamiltonians. Visibly the Floquet formalism, which is used to describe Hamiltonians which are periodic in time, is analogous to Bloch's theorem, which is used to describe Hamiltonians that are periodic in space. For this reason, we refer to the eigenstates of \^H$_F$ as quasienergies. Of particular interest to us is the fact that the system's topological properties can be seen by studying the Floquet Hamiltonian given by Eq.~(\ref{eq:TSQW_Heff}).

In this particular case, it is possible to analytically calculate the Floquet Hamiltonian. Substituting Eqs.~\eqref{eq:QW_TC_Mom} into Eq.~\eqref{eq:TSQW_U}, we can find \^U$(k)$:
\begin{equation}
\label{Uk_rotation}
\hat{\text{U}}(k) = e^{i  k d \sigma_3}   e^{-i  \theta  \sigma_2}
=e^{-i  \varepsilon(k)  \bm{\hat{n}}(k)\cdot\bm{\sigma}},
\end{equation}
where $\bm{\sigma}$ is the vector of Pauli matrices. In the last equality, we used the fact that
the Pauli matrices form a basis for the vector space of $2\times 2$ Hermitian matrices. Comparing with Eq.~\eqref{eq:TSQW_Heff}, we find:
\begin{equation}
\label{eq:H_TSQW}
 \h_F(k) = \varepsilon(k) \bm{\hat{n}}(k)\cdot \bm{\sigma}.
\end{equation}
Pauli matrices anti-commute, such that, for any vector $\bm{v}$:
\begin{equation}
\label{eq:SigmaAntiCommut}
e^{i \bm{v}\cdot\bm{\sigma}}=\sigma_0 \cos(|\bm{v}|)+i \frac{\bm{v}\cdot \bm{\sigma}}{|\bm{v}|} \sin(|\bm{v}|).
\end{equation}
Using this property, the precise form of the quasienergy $\varepsilon(k)$ and the unit vector $\bm{\hat{n}}(k)$ can be derived:
\begin{equation}
 \begin{split}
  & \cos(\varepsilon(k)) = \cos(\theta)\cos(k d), \\
  & \bm{\hat{n}}(k) = \frac{1}{\sin(\varepsilon(k))}\left[
  \sin(\theta)\sin(k d), \sin(\theta)\cos(k d), -\cos(\theta)\sin(k d)\right].
 \end{split}
\end{equation}

\subsection{Chiral symmetry in the two-step quantum walk}
\label{sec:QWChiralSym}

In this subsection, we will show that the Floquet Hamiltonian of the two-step quantum walk is chiral symmetric. We will do this from a purely geometric point of view, by showing that all the $\bm{\hat{n}}(k)$ vectors are constrained to a plane which contains the origin when plotted on the Bloch sphere. Indeed, we saw in Sec.~\ref{sec:ChiralSymmetry} that, when this is the case, $\h_F$ is chiral symmetric. This is an important observation as it will will allow us to show, building on the study presented in Sec.~\ref{sec:1dTopoIntro}, that the system has well defined winding number.

The operations constituting the time evolution operator in Eq.~\eqref{Uk_rotation} correspond to rotations on the Bloch sphere. Specifically, when time evolved for one time step, a state of well defined quasimomentum has its spin state on the Bloch sphere rotated by $2 \theta$ around the unit vector $\bm{e}_y$, then by $2 k d$ around $\bm{e}_z$. Let $R_y(2 \theta)$ and $R_z(2 k d)$ be the corresponding $3\times 3$ rotation matrices. As can be seen from the second equality of Eq.~\eqref{Uk_rotation}, this results in an effective rotation by $2  \varepsilon(k)$ around the unit vector $\bm{\hat{n}}(k)$, written $R_n(2  \varepsilon(k))$. To summarise:
\begin{equation}
R_z(2 k d)  R_y(2 \theta)=R_n(2  \varepsilon(k)).
\end{equation}
Let us now search for the eigenvectors of the time evolution, which is simply the axis of the effective rotation, $\pm \bm{\hat{n}}(k)$. These have the property:
\begin{equation}
\label{eq_eival_n}
R_z(2 k d)  R_y(2 \theta) \bm{\hat{n}}(k)=\bm{\hat{n}}(k),  ~\forall  k\in \left[-\frac{\pi}{d}, \frac{\pi}{d}\right].
\end{equation}
Note that the eigenstates of \^U$(k)$ are also eigenstates of $\h(k)$, such that we can deduce the system's winding number directly from the time evolution operator.

Rotations around the $\bm{e}_z$ axis do not change the $z$ component of a vector (loosely speaking, their latitude on the sphere). This implies a minimal requirement for Eq.~(\ref{eq_eival_n}) to hold: $\bm{\hat{n}}(k)$ must be such that the rotation $R_y(2 \theta)$ returns it to the same latitude. The set of vectors which satisfy this condition will be referred to as ``candidate eigenvectors''.
\begin{figure}[t]
\centering
\includegraphics[width = 0.7\textwidth]{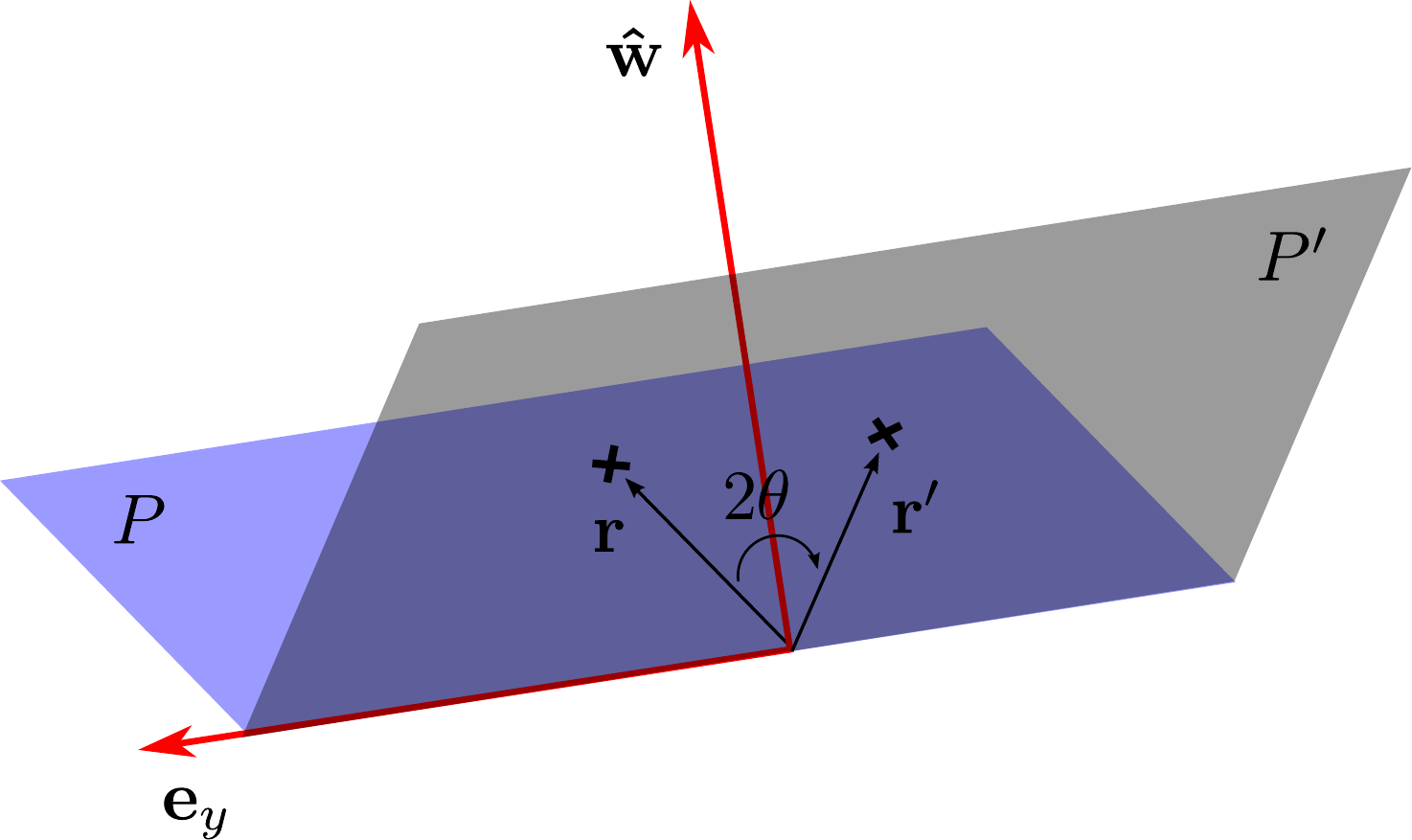}
\caption[Chiral symmetry illustrated as a geometric constraint]{The plane $P$ is rotated by $2 \theta$ around $\bm{e}_y$ to $P'$ (thereby sending each point $\bm{r}$ to $\bm{r}'$). Note that the projection of each vector $\bm{r}$ along $\bm{\hat{w}}$ is conserved.}
\label{composite_rotation_latitute}
\end{figure}

When a rotation $R_y(2 \theta)$ is applied to a set of points in a plane $P$, a new plane $P'$ is generated. If the vector $\bm{e}_y$ lies in $P$, it is always possible to find a vector $\bm{\hat{w}}$ such that the latitude along $\bm{\hat{w}}$ of each point of $P$ is conserved by the operation $R_y(2 \theta)$ (see figure \ref{composite_rotation_latitute} for visual support). Specifically, $\bm{r}\cdot \bm{\hat{w}} = \bm{r}'\cdot \bm{\hat{w}}$ if $\bm{\hat{w}}$ lies in the plane that makes an angle of $\theta$ with $P$. For a fixed angle $2 \theta$, there is one plane $P$ which contains $\bm{e}_y$ and makes an angle $\theta$ with the $\bm{e}_z$ axis; the points of this plane are all the points which have their latitude along $\bm{e}_z$ conserved under $R_y(2 \theta)$.

We are only interested in unit vectors (as the spin states we are considering lie on the Bloch sphere); the set of candidate eigenvectors is therefore given by the intersection of the plane $P$ and the Bloch sphere. Additionally, the latitude of points $\bm{r}=(0, 1, 0)$ and $\bm{r}=(0, -1, 0)$ are trivially conserved (these are eigenvectors of $R_y(2 \theta)$). Thus $P$ must contain the origin. These geometric requirements, together, constrain all vectors which satisfy Eq.~\eqref{eq_eival_n} to lie on a great circle of the Bloch sphere.

We conclude that the $\h_F$ has chiral symmetry (CS), and that its CS operator is given by:
\begin{equation}
\label{eq:TSQW_ChiralOperator}
\Gamma=\bm{a}(\theta)\cdot\bm{\sigma},
\end{equation}
where $\bm{a}(\theta)$ is the vector normal to $P$. Seeing as the $\bm{\hat{n}}(k)$ have the same quasimomentum periodicity as the Hamiltonian, they form a closed trajectory on the Bloch sphere, such that this Hamiltonian has a well defined winding number. As we will see in the next subsection, however, the system's topological phase cannot be characterised by this single topological invariant.

\subsection{Two topological invariants}
\label{sec:TwoInvariants}

If we naively calculate the system's winding number from Eq.~\eqref{eq:evaluate_winding}, we find that this system presents  a single topological phase, which winding number $\nu=1$ regardless of the system's parameters. Ref.\ \parencite{Kitagawa2012a}, however, found experimental evidence of a topological bound state when spatially varying $\theta$.

To understand this, we must go back to our study of chiral symmetric systems, in Sec.~\ref{sec:ChiralSymmetry}. There, we saw that states with energy $E=-E$ could exist without a chiral partner, which resulted in their symmetry protection.

In Floquet systems, however, the quasienergy is $2\pi$-periodic as a result of Eq.~\eqref{eq:QuasiEnergies}. There are therefore two quasienergies, $ \varepsilon=0$ and $ \varepsilon=\pi$, which satisfy the condition $\varepsilon=-\varepsilon$. We conclude from the arguments presented in Sec.~\ref{sec:TopoBS} that there can exist in this system bound states which are topologically protected with quasienergy $\varepsilon=0$ or $\varepsilon=\pi$ \parencite{Kitagawa2010a,Kitagawa2010}. It should also be clear from the arguments presented in this section that the $\varepsilon=0$ ($\varepsilon=\pi$) bound state cannot be removed from the system unless the band gap is closed in $\varepsilon=0$ ($\varepsilon=\pi$), or that CS is broken \parencite{Asboth2013}. We conclude that the two-step quantum walk with an $\varepsilon=0$ bound state cannot be adiabatically deformed into a quantum walk with an $\varepsilon=\pi$ bound state. There must therefore be two independent topological invariants, $\nu_0$ and $\nu_\pi$, which count the number of $\varepsilon=0$ and $\varepsilon=\pi$ bound states, respectively.

\subsection{Winding numbers of the two chiral time frames}
\label{sec:ChiralTimeFrames}

Let us consider a quantum walk on a finite lattice, such that the system presents topological boundaries. The aim of this subsection is to develop a method to relate the number of edge states at the boundaries to the system's topological invariants. This will allow us to understand why the Floquet nature of this system leads it to having two topological invariants.

Because we are counting the number of edge states at one of the system's boundaries, we need to define an invariant in a way which is robust to breaking translational invariance. The problem is that the Hamiltonian's chiral symmetry operator, given by Eq.~\eqref{eq:TSQW_ChiralOperator}, is in general $\theta$ dependent. This means that we cannot define a global chiral symmetry operator when varying $\theta$ spatially. When this is the case, the winding number is not well defined, such that it is not the right quantity to measure the number of edge states at the system's boundaries \parencite{Asboth2012}.


\begin{figure}[t]
\centering
\includegraphics[width=\textwidth]{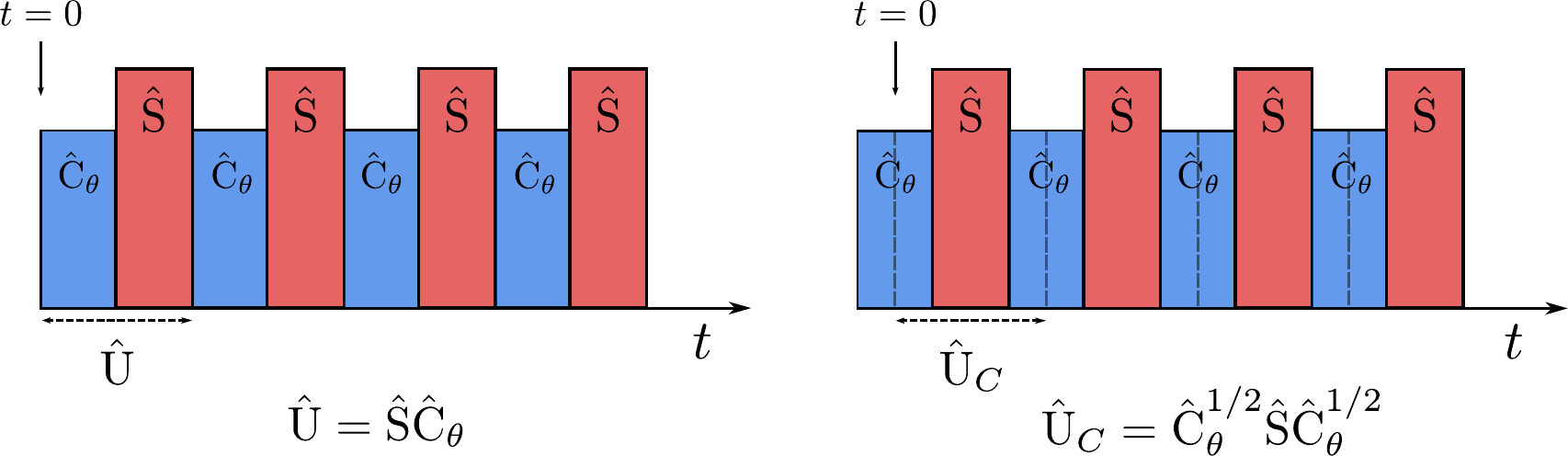}
\caption[Origin of time of a Floquet system]{\label{fig:TimeShift}
illustration of the shift in the origin of time which we have performed to bring \^U in the symmetric form of Eq.~\eqref{eq:TSQW_UC}. Note that this is essentially a redefinition of the initial state: $|\psi_0\rangle\rightarrow\text{\^C}_\theta^{1/2}|\psi_0\rangle$.}
\end{figure}

Note however that there is some ambiguity in the definition of the Floquet Hamiltonian. Indeed, it is possible to change the form of the time evolution (and therefore, the Floquet Hamiltonian) simply by redefining the initial state. This is sketched in the Fig.~\ref{fig:TimeShift}.
We can find two time frames in which the chiral symmetry operator, given by Eq.~\eqref{eq:TSQW_ChiralOperator}, is independent of $\theta$ \parencite{Asboth2013}:
\begin{eqnarray}
\label{eq:TSQW_UC}
\hat{\text{U}}_C & = & \hat{\text{C}}_\theta^{1/2} \text{\^S}  \hat{\text{C}}_\theta^{1/2},\\
\label{eq:TSQW_UT}
\hat{\text{U}}_S & = & \text{\^S}^{1/2} \hat{\text{C}}_\theta \text{\^S}^{1/2}.
\end{eqnarray}
In this time frame, the chiral symmetry operator takes on the global ($\theta$ independent) form $\hat{\Gamma}=\sigma_1$. Consequently, the winding number is guaranteed to be well defined (even if we break translational invariance).

Note that in Eqs.~\eqref{eq:TSQW_UC} and \eqref{eq:TSQW_UT}, the time evolution has a form that is symmetric in time. Specifically, when \^U is symmetric in time about an inversion point, it was shown in Ref.\ \parencite{Asboth2013} that it can always be put in the form:
\begin{equation}
\label{eq:ChiralForm}
\text{\^U}=\hat{\Gamma} \text{\^G}^\dagger \hat{\Gamma} \text{\^G},
\end{equation}
where $\hat{\Gamma}$ is the chiral symmetry operator. This is said to be in the chiral time-frame. 


Thus, we have identified the two ``proper'', or chiral time frames in which to compute the winding numbers, which we will refer to as $\nu_C$ and $\nu_S$. Inversion points of the time evolution come in pairs, which explains why we have two well defined topological invariants in this case. The winding number $\nu_C$ and $\nu_S$ count the number of bound states at the system's edges in the time frames of \^U$_C$ and \^U$_S$, respectively. 
They can be calculated by applying Eq.~\eqref{eq:evaluate_winding} directly in both of these times frames. Note that in general, these do not count the number of $\varepsilon=0$ and $\varepsilon=\pi$ bound states at the ends of the lattice. It is the subject of the next and final subsection to show how $\nu_C$ and $\nu_S$ are related to $\nu_0$ and $\nu_\pi$.


\subsection{Spin structure of the topological bound states}
\label{sec:nu0_nuPi}


In this subsection, we will study a quantum walk of a finite, $N$ site lattice, focusing on the dynamics of the $\varepsilon=0$ and $\varepsilon=\pi$ bound states which appear at the lattice edges. Our main goal is to develop a method to compute the $\nu_0$ and $\nu_\pi$ topological invariants. We are, however, further motivated by the notion that a detailed understanding of the physics governing the system's bound states could hold the key to identifying them experimentally.


As we saw in Sec.~\ref{sec:TopoBS}, a topological bound state is simultaneously an eigenstate of the Hamiltonian and the chiral symmetry operator.
We conclude that the spin part of the topological bound states, in a chiral time-frame of the two-step quantum walk, must be either $\ket{\uparrow}_x$ or $\ket{\downarrow}_x$ where $\ket{\uparrow}_x$ and $\ket{\downarrow}_x$ are the eigenstates of $\hat{\Gamma}=\sigma_1$, with eigenvalues $+1$ and $-1$ respectively.

Conversely, all other eigenstates of the Hamiltonian transform into their (orthogonal) chiral partner under CS; they must therefore have equal support over the $\ket{\uparrow}_x$ and $\ket{\downarrow}_x$ states. Consider now two overlapping eigenstates of CS with the same quasienergy $\varepsilon_0$. If these states are in the same spin state, their energy is symmetry protected. If, however, they have orthogonal spins, they can hybridise by forming a pair of chiral partners and move away from $\varepsilon_0$. We conclude that we can measure the bulk winding number by measuring the mean atomic spin at a boundary:
\begin{equation}
\label{eq:WindingIsPolarisation}
 \nu = N_{\uparrow, x} - N_{\downarrow, x},
\end{equation}
where $N_{\uparrow, x}$ ($N_{\downarrow, x}$) is the number of $\ket{\uparrow}_x$ ($\ket{\downarrow}_x$) states at a given boundary.

Let us now consider the dynamics of an atom in a topological bound state of the system. In the time-frame of \^U$_C$, the atom is in the state $\ket{\Psi}$, which is an eigenstate of CS:
\begin{equation}
 \hat{\Gamma}\ket{\Psi}=(-1)^f\ket{\Psi}, f\in\{0,1\}.
\end{equation}
After a period of the Hamiltonian, the atom is in the state:
\begin{equation}
\label{eq:BS_evolution1}
 \text{\^U}_C\ket{\Psi}= \hat{\Gamma}\text{\^G}^\dagger\hat{\Gamma}\text{\^G}\ket{\Psi}.
\end{equation}
We define $\ket{\Phi}=\text{\^G}\ket{\Psi}$, the state of the atom after half a period of the potential. This state is also a topological bound state of the system, such that:
\begin{equation}
 \hat{\Gamma}\ket{\Phi}=(-1)^g\ket{\Phi}, g\in\{0,1\}.
\end{equation}
Substituting into Eq.~\eqref{eq:BS_evolution1}:
\begin{equation}
\label{eq:BS_evolution2}
 \text{\^U}_C\ket{\Psi}= \hat{\Gamma}\text{\^G}^\dagger\hat{\Gamma}\ket{\Phi}
 = (-1)^g\hat{\Gamma}\text{\^G}^\dagger\ket{\Phi}
 = (-1)^{f+g}\ket{\Psi}.
\end{equation}
Finally, we know from Eq.~\eqref{eq:TSQW_Heff} that, if $\varepsilon=0$, \^U$_C=\sigma_0$, such that we must have $f+g$ even. Similarly, if $\varepsilon=\pi$, \^U$_C=-\sigma_0$ and $f+g$ is odd. We conclude that $\varepsilon=0$ bound states keep the same spin state in both chiral time-frames, while $\varepsilon=\pi$ bound states switch spin states at every half period of the Hamiltonian. We can therefore measure the $\nu_0$ ($\nu_\pi$) topological invariant by counting the number of bound states at a given boundary that keep the same spin (change spin) at every half period of the Hamiltonian. Using Eq.~\eqref{eq:WindingIsPolarisation}, we conclude that:
\begin{align}
\label{eq:nuC}
& \nu_0=\frac{1}{2}(\nu_C+\nu_S),\\
\label{eq:nuT}
& \nu_\pi=\frac{1}{2}(\nu_C-\nu_S).
\end{align}
We refer the reader to Ref.\ \parencite{Asboth2013} for an explicit demonstration.

To conclude this section, we have seen that the two-step quantum walk presents two inequivalent winding numbers due to its Floquet nature. Due to the presence of these two topological invariants, it does not belong to any of the topological classes of the standard classification, shown in Table \ref{tab:topo}. We have further shown that the bound states associated to the two winding numbers are spin polarised along the $x$ direction, and we can infer their quasienergy from their dynamics at half time-steps. 


\chapter{Review of selected experimental techniques in cold atoms} 
\label{Chapter3}

As we discussed in Chapter \ref{Chapter1}, a common problem in material science is to understand the collective behaviour of particles. If the system's Hamiltonian is known, but is too complex to be approached analytically, it can be engineered experimentally, such that the system's behaviour can simply be observed. This is the core principle of quantum simulators \parencite{Lewenstein2012}.

There exist many systems which can be turned into quantum simulators, each of which have their advantages in terms of experimental control. Amongst these, we find photonic systems, ionic systems, and superconducting circuits \parencite{Georgescu2014}. Our main interest in this thesis is the simulation of materials using gases of neutral atoms in optical lattices. These systems are particularly attractive due to their scalability, the ability to tune contact interactions, and the sheer amount of control which can be exerted on the system \parencite{Inguscio2013}. Thanks to this vast experimental toolbox, it has recently become possible to engineer extremely delicate states of matter, leading to an explosion in our understanding of interacting gases \parencite{Pitaevskii2003,Pethick2008}.

In this section, we will briefly review some of the experimental techniques which are relevant to our work. Specifically, we will discuss the way in which an atom interacts with a laser field, which will allow us to understand how optical potentials can be engineered \parencite{Grimm2000}. We will then discuss stimulated Raman scattering, which is the inelastic scattering of two photons by an atom \parencite{Grynberg}. These types of processes have an incredibly vast number of applications, which range from atomic cooling \parencite{Kasevich1992,Davidson1994} to laser stimulated tunnelling \parencite{Jaksch2003}. We will then study how optical traps can be used to build optical lattices, which are attractive because of their tunable tunnelling amplitudes and dimensionality, and that they are defect free systems \parencite{Bloch2005}. Finally, we will discuss how artificial gauge fields can be generated for gases of neutral atoms in optical lattices. We will consider doing this, either effectively by averaging over fast dynamics \parencite{Goldman2014}, or by imprinting a phase on the atoms using an optical field \parencite{Goldman2014a}. For a more detailed review on experimental tools for cold atoms in optical lattices, we refer the reader to Ref.\ \parencite{Bloch2012}.

\begin{framed}
In this chapter:
\begin{enumerate}
 \item We study the interaction of a single atom with a laser field, which will give us a basis to understand optical trapping and Raman coupling.
 \item We expose the core principles and advantages of optical lattices.
 \item We define gauge fields, and describe two methods to generate them for neutral atoms, namely lattice shaking and optical phase imprinting.
\end{enumerate}
\end{framed} 

\section{The atom-light interaction}

In this section, we will consider the basic interaction between an atom and a classical laser field. This type of interactions are central to the functioning of atomic clocks \parencite{Ludlow2015}, they enable us to generate versatile optical potentials \parencite{Neuman2004}, and form the basis for powerful measurement techniques such as radio-frequency spectroscopy \parencite{Chen2009}. The goal of this section is to understand two of the core tools for quantum simulation with cold atoms, namely optical trapping and Raman coupling.

\subsection{Off-resonant optical trapping}
\label{sec:AtomLightInteraction}

In this section, we will describe the interaction of an atom with an off-resonant laser. These types of interactions are widely used for optical trapping purposes, and consequently highly relevant to the field of ultracold atoms. In this section, we will study the simple example of a two level atom interacting with a laser field, and see that we can already understand some of the basic features of optical trapping. For a more complete study, we refer the reader to Refs.\ \parencite{Grimm2000, Goldman2014a}.

Let us consider an atom with two internal states, $\ket{a}$ and $\ket{b}$, with bare energies $E_a=0$ and $E_b=\hbar \omega_0$, respectively. The bare Hamiltonian therefore has the form:
\begin{equation}
\label{eq:BareHTwoLevel}
 \h_0 = \begin{pmatrix}
         0 & 0\\
         0 & \hbar \omega_0
        \end{pmatrix}.
\end{equation}
We will consider the interaction of this atom with the laser field:
\begin{equation}
\label{eq:ElectricField}
 \bm{\E}(\bm{r},t)=\bm{\E}(\bm{r})\cos(\omega t+\varphi(\bm{r})).
\end{equation}
The effect of this field is to make the atom instantaneously electrically polarised. In particular, if $\bm{\hat{D}}$ is the atomic electric dipole operator, then the interaction potential of the form: $\h_{\rm AL}=-\bm{\hat{D}}\cdot \bm{\E}(\bm{r}_0,t)$, where $\bm{r}_0$ is the atom's position.

Let us define the detuning $\delta = \omega - \omega_0$. In the limit where the laser is relatively close to resonance, such that $|\delta| \ll \omega_0$, we can vastly simplify the form of this interaction by neglecting fast oscillating terms using the rotating-wave approximation. In the co-rotating basis:
\begin{equation}
 \ket{a'}=e^{-i \delta t/2} \ket{a}, \ket{b'}=e^{i \delta t/2} \ket{b},
\end{equation}
the full Hamiltonian takes the simple form \parencite{Grynberg}:
\begin{equation}
\label{eq:HAtomLight}
 \h = \h_0 + \h_{\rm AL}
 = \frac{\hbar}{2}
 \begin{pmatrix}
 \delta & \Omega e^{i\varphi}\\
 \Omega e^{-i\varphi} & -\delta
 \end{pmatrix}.
\end{equation}
Note that we have introduced the Rabi frequency $\Omega$, which is proportional to the matrix element between $\ket{a}$ and $\ket{b}$ induced by the electric field.

Let us now assume that the system is in the strongly off resonant regime, such that $|\Omega/\delta|\ll 1$. In this limit, the eigenvalues of Eq.~\eqref{eq:HAtomLight} are given by:
\begin{equation}
 E_{\pm} = \pm \frac{\hbar}{2}\sqrt{\delta^2 + \Omega^2}
 \approx \pm \hbar \left[ \frac{\delta}{2}
  + \left( \frac{\Omega^2}{4 \delta} \right)
  + \mathcal{O} \left( \frac{\Omega}{\delta} \right)^2 \right].
\end{equation}
Thus we see that, to first order in $\Omega/\delta$, interaction with the electric field shifts the energies of the two states by an equal and opposite amount. What is more, this amount is inversely proportional to the detuning, such that, if the detuning is positive (blue detuning) the atom-light interaction is repulsive, such that the atom is pushed away from locations where the laser intensity is high. Conversely, for negative detuning (red detuning), the atom is attractively bound to locations with high laser intensity. It is through this mechanism that off-resonant optical traps can be designed.

As a final note, the reader might worry that level structure of atoms is in general much more complicated than the simple two level model presented here. Nonetheless, this model provides an excellent description of alkali atoms in an off-resonant laser field, provided that the laser detuning is much larger than the fine-structure splitting \parencite{Grimm2000}. The results derived here are therefore applicable to real world systems.

\subsection{Stimulated Raman scattering}
\label{sec:Raman}

As we saw in Sec.~\ref{sec:AtomLightInteraction}, two atomic levels can be coupled through a laser field. In some situations, however, the conservation of angular momentum forbids two atomic states from being coupled through the absorption of a single photon. When this is the case, an inelastic scattering event between the atom and two photons can be used to cause a transition between these states. These processes, which are known as Raman scattering, are immensely useful in ultra-cold gases, as they can 
be used for cooling purposes \parencite{Kasevich1992,Davidson1994}, imprint a phase onto the atom \parencite{Goldman2014a}, or, in discrete systems, engineer controlled tunnelling between sites \parencite{Jaksch2003}.

\begin{figure}[t]
\centering
\includegraphics[width=0.4\textwidth]{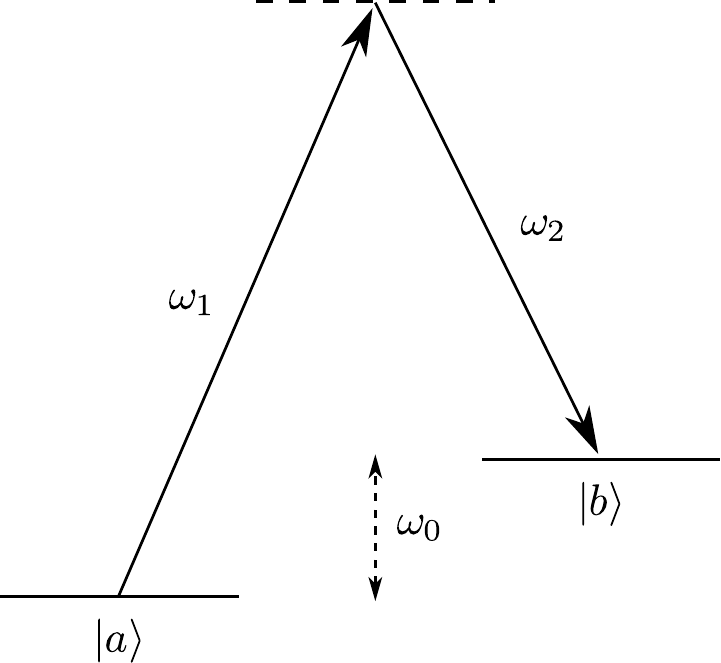}
\caption[Stimulated Raman scattering process]{\label{fig:Raman}
The states $\ket{a}$ and $\ket{b}$, which have energy separation $\hbar \omega_0$, are resonantly coupled through the two photon process with respective frequencies $\omega_1$, $\omega_2$ such that $\omega_0=\omega_1 - \omega_2$.}
\end{figure}

An example of a Raman process is sketched in the Fig.~\ref{fig:Raman}. As can be seen in this figure, two atomic states $\ket{a}$ and $\ket{b}$ can be coupled through the absorption of a photon with frequency $\omega_1$ and the stimulated emission of photon $\omega_2$. The process is near-resonant if $\omega_1-\omega_2\approx \omega_0$, where $\hbar\omega_0$ is the energy difference between states $\ket{a}$ and $\ket{b}$. When the laser frequencies $\omega_1$, $\omega_2$ do not correspond to any atomic internal transition, the amplitude with which levels are coupled by a single photon is extremely small. 
This allows us to define a reduced Hamiltonian, $\h_{\rm AL}^{\rm red}$, which describes the atom-light interaction, projected on the levels $\ket{a}$ and $\ket{b}$. Within the rotating-wave approximation, $\h_{\rm AL}^{\rm red}$ takes the form:
\begin{equation}
\label{eq:RamanHamiltonian}
 \h_{\rm AL}^{\rm red} 
 \approx \begin{pmatrix}
 h_{aa} & \hbar \Omega_R/2 \\
 \hbar \Omega_R^{\star}/2 & h_{bb}
 \end{pmatrix},
\end{equation}
where $\Omega_R$ is the effective Rabi frequency, and the diagonal elements $h_{aa}$, $h_{bb}$ are known as Stark shifts. These parameters can be determined with a perturbation theory expansion, with the single photon coupling amplitudes as the perturbative parameter. We will work exclusively in the resonant limit, $\omega_0=\omega_1 - \omega_2$, in which limit the Stark shifts are identically zero (to second order in the single photon coupling amplitudes) and we can choose $\Omega_R$ to be real. When this is the case, Eq.~\eqref{eq:RamanHamiltonian} takes the form of the interaction potential describing a two level atom in a resonant light field \parencite{Grynberg}. 
We can find the form of the full Hamiltonian, $\h_R$, by following a reasoning which is analogous to the one applied in Sec.~\ref{sec:AtomLightInteraction}:
\begin{equation}
\label{eq:RamanColinear}
\h_{\rm R} = \h_0 + \h_{\rm AL}^{\rm red} \approx \frac{\hbar\Omega_R}{2}\sigma_1,
\end{equation}
where $\h_0$ is the bare, two level Hamiltonian, Eq.~\eqref{eq:BareHTwoLevel} and $\sigma_1$ is the first Pauli matrix.

To conclude, we have seen that a pair of Raman lasers can induce a transition between different energy levels. In the following chapters, we will study some of the uses that these processes can be put to.
\section{Optical lattices}

To understand a many body system, it is essential to understand if its constituent particles display collective behaviour at equilibrium. In a number of extremely interesting cases, the existence of these phases can be attributed to the periodicity of the potential. These include, for instance, the Mott insulator, BCS superconductors and the Haldane model. Optical lattices provide an ideal framework to investigate these phases, as they provide defect free periodic potentials. Additionally, many of the lattice properties are experimentally accessible, such  that they can be dynamically tuned. This allows scientists to investigate experimentally the time scale of a phase transition simply by changing the parameters of the lattice in the course of an experiment. We will describe here the basic principles which optical lattices rely on. For a more complete description of the physics accessible using optical lattice, we refer the reader to Ref.\ \parencite{Bloch2005}.

Let us consider superimposing two counter propagating laser fields with equal wavelength $\lambda_L$, such that they form an optical standing wave. We will now consider the dynamics of a gas of neutral atoms in this laser field. We choose the lasers' wavelength such that it is off-resonant with the atomic species. Following the study performed in Sec.~\ref{sec:AtomLightInteraction}, locations with high optical intensity are either attractive or repulsive (depending on the lasers' detuning), such that the atoms see a potential of the form:
\begin{equation}
 V_{\rm lat}(x) = V_0 \sin^2(k_L x),
\end{equation}
where $V_0$ is the lattice depth and $k_L$ is the laser's wavenumber, with $k_L=2\pi/\lambda_L$. The depth of this one dimensional potential is simply controlled by the intensity of the applied lasers. This gives us an experimental handle to change the hopping amplitude of atoms in this lattice. Note that this optical lattice has a period of $\lambda_L/2$.

\begin{figure}[t]
\centering
\includegraphics[width=0.75\textwidth]{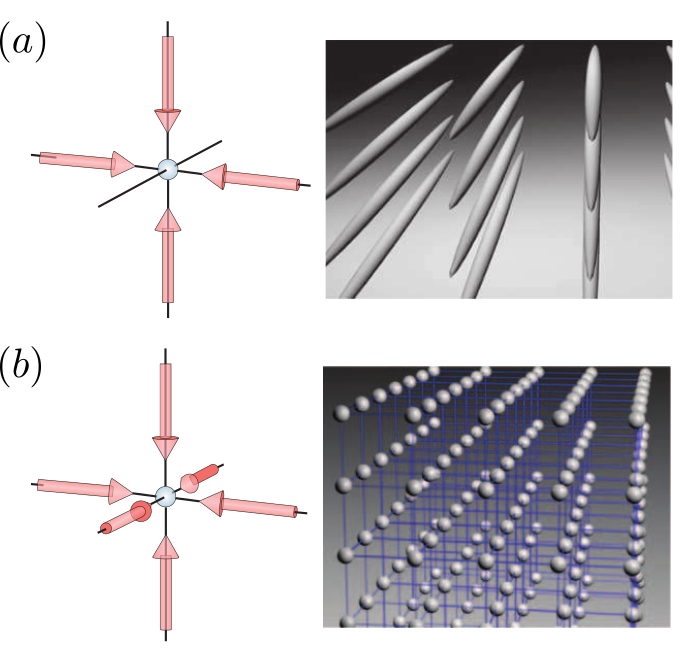}
\caption[2D and 3D optical lattices]{\label{fig:OpticalLattice}
(a): Two dimensional and (b): three dimensional optical lattices formed by interfering two and three pairs of mutually perpendicular laser, respectively. Adapted from \parencite{Bloch2005}.}
\end{figure}

Similarly, by using two pairs of lasers, a two dimensional periodic potential can be designed. If these pairs of lasers have the same wavelengths and are mutually perpendicular, they form a simple square lattice. A simple cubic lattice can also be designed, using three pairs of mutually perpendicular lasers with the same wavelength. This is sketched in the Fig.~\ref{fig:OpticalLattice}. More complicated lattice geometries can also be designed multiplying the lasers \parencite{Jotzu2014} or changing the angle between applied lasers \parencite{Petsas1994}. The Kagom\'e lattice, for instance, can be realised in this way \parencite{Santos2004}.

To conclude, we have seen that optical lattices form a defect free periodic potential for neutral atoms. These are extremely attractive experimentally due to the amount of control that can be exerted on these systems. Indeed, the lattice depth, geometry and even dimensionality can be readily modified.

\section{Artificial gauge fields}

A general property of gauge fields is that they cannot be measured directly. These fields can be transformed under a continuous group of transformations, called gauge transformations, which leave their associated (and measurable) physical quantities invariant. This property is known as gauge invariance. An example of such a field is the vector potential, which is not uniquely defined.
All known forces are mediated through the exchange of gauge bosons, such that the study of gauge theories are relevant at all length scales \parencite{Goldman2014a}. In this section, we will describe some of the means through which neutral atoms can be subjected to gauge fields. This will allow us to vastly extend the quantum simulator toolbox.


\subsection{Periodically modulated lattices}

As we saw in Sec.~\ref{sec:IntroFloquet}, a periodically modulated system can be well approximated by a static, or Floquet Hamiltonian. This realisation has enabled researchers to build new Hamiltonians, which were difficult to realise experimentally until now.
In particular, the ability of lattice shaking to generate artificial gauge fields was demonstrated in the early papers, Refs.\
\parencite{Hauke2012,Struck2012}, which has since become a standard method for engineering synthetic magnetic fields in gases of neutral atoms. In more recent years, these have been considered as a means for engineering spin-orbit coupling, as for instance in the experimental proposal by Anderson et al. \parencite{Anderson2013}, or the very general studies, Refs.\ \parencite{Goldman2014,Goldman2015b}.

Floquet systems naturally found an application for engineering topologically non-trivial materials
due to their ability to emulate spin-orbit coupling.
The famous topological insulator for instance, first suggested in \parencite{Kane2005}, was realised experimentally with photons in an array of waveguides \parencite{Cayssol2013}. The Floquet nature of this system becomes clear once the direction of propagation is treated as a time dimension. Indeed, the waveguides considered have a finite helicity which is averaged over to break time reversal symmetry, in much the same way as, in the Floquet formalism, an effective Hamiltonian is obtained by averaging over a rapidly oscillating one. Another experimental realisation of a topological insulator using a Floquet system was performed using a two dimensional atomic gas trapped in an optical lattice \parencite{Jotzu2014}. In this system, time reversal symmetry is broken by lattice shaking, thereby effectively inducing a complex tunnelling term, which is mathematically identical to a staggered external magnetic field.


\subsection{Raman coupling and the Hofstadter strip}
\label{sec:SyntheticDimensions}

Artificial gauge fields can also be generated by imprinting a phase on the atoms through optical radiation. Raman coupling is particularly interesting for generating uniform magnetic fields, a task which is tricky to do with lattice shaking. There are a vast number of ways to do this, which are described in detail in the reviews \parencite{Juzeliunas2012,Goldman2014a}. In this subsection, we will focus on a specific method for generating a synthetic gauge field, which relies on the principle of interpreting the spin degree of freedom as an extra dimension. This protocol was initially suggested in Ref.\ \parencite{Celi2013}, and implemented experimentally in Refs.\ \parencite{Mancini,Stuhl2015}.

Consider a non-interacting atomic gas in a one dimensional lattice, with magnetic sublevels labelled by the index $n$. For a total angular momentum of $F$, the magnetic sublevels lie in the interval $n \in[-F,F]$. The Hamiltonian describing these atoms is simply:
\begin{equation}
 \h_0 = \sum_{m,n} -J_x \hat{c}_{m+1,n}^\dagger \hat{c}_{m,n} + \text{H.c.}
\end{equation}
We have introduced $J_x$, the spatial hopping amplitude, and $\hat{c}_{m,n}^\dagger$ ($\hat{c}_{m,n}$) the operator which creates (destroys) an atom at site $m$ in the magnetic sublevel $n$.

\begin{figure}[t]
\centering
\includegraphics[width=\textwidth]{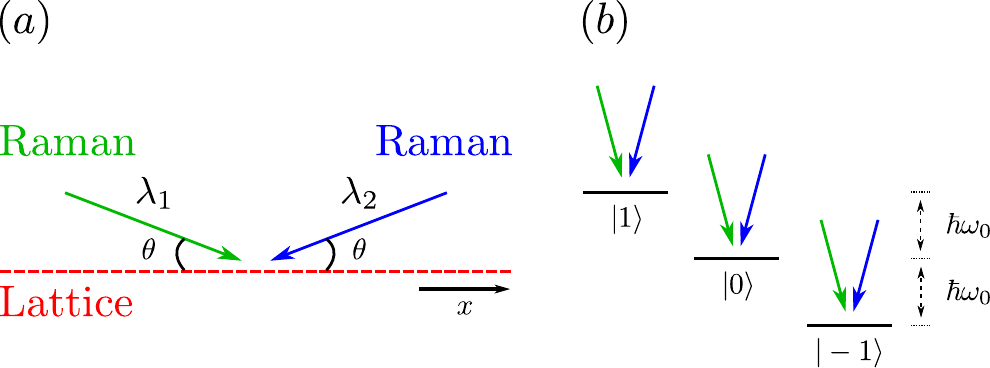}
\caption[Raman lasers at an angle]{\label{fig:RamanAngle}
(a): two Raman lasers, which are non-colinear but lie in the same plane, are at an angle $\theta$ with the $x$ axis. For every spin-flip event, the atoms acquire a recoil momentum $\Delta k$ given by Eq.~\eqref{eq:RamanMomentumKick2}. (b): the Zeeman field induces an energy splitting between the magnetic sublevels. These are coupled by the two Raman lasers, pictured in (a).}
\end{figure}

Let us now apply an external Zeeman field $\bm{B}=B_0 \bm{e}_z$. Under the effect of this field, the $n^{\rm th}$ magnetic sublevel acquires an energy $n \hbar \omega_0 = n g_F \mu_B B_0$, where $g_F$ is the Land\'e $g$-factor and $\mu_B$ is the Bohr magneton. We now illuminate the lattice with a pair of Raman lasers, and set their frequencies on resonance such that $\omega_0=\omega_1-\omega_2$.
We set the Raman lasers at an angle $\theta$ relative to $\bm{e}_x$, the axis of the system, as sketched in Fig.~\ref{fig:RamanAngle}. As a result, atoms which go from magnetic sublevel $\ket{n}$ to $\ket{n+1}$ acquire a recoil crystal momentum in the direction of the lattice, with amplitude:
\begin{equation}
\label{eq:RamanMomentumKick2}
\Delta k = 2\pi \cos\theta \left( \frac{1}{\lambda_1} + \frac{1}{\lambda_2} \right),
\end{equation}
with $\theta$ the angle of the Raman lasers with $\bm{e}_x$.
Including the recoil momentum in the expression of the atom-light Hamiltonian, given by Eq.~\eqref{eq:RamanColinear}, we find \parencite{Juzeliunas2012}:
\begin{equation}
\label{eq:RamanMomentumKick}
 \h_R = \frac{\hbar \Omega_R}{2} \sum_{m,n}\left( g_{F,n} e^{i \Delta k m d} \hat{c}_{m,n+1}^\dagger \hat{c}_{m,n} +\text{H.c.} \right),
\end{equation}
where $d$ the lattice spacing and the $g_{F,n}$ are the Clebsch-Gordan coefficients:
\begin{equation}
\label{eq:Clebsch-GordanCoeff}
 \hat{c}_{m,n+1}^\dagger \hat{c}_{m,n}\ket{m,n}=g_{F,n} \ket{m,n+1},
 g_{F,n}=\sqrt{F(F+1)-n(n+1)}.
\end{equation}
The full Hamiltonian is then:
\begin{equation}
 \h= \h_0 + \h_R=  \sum_{m,n} -J_x \hat{c}_{m+1,n}^\dagger \hat{c}_{m,n} +\frac{\hbar \Omega_R g_{F,n}}{2} e^{i \Phi m} \hat{c}_{m,n+1}^\dagger \hat{c}_{m,n} +\text{H.c.},
\end{equation}
where we have set $\Phi=\Delta k d$. Note that, in the particular case of $F=1$, we have $g_{F,n}=\sqrt{2}$ for all $n$.
In this case, we can interpret the magnetic sublevel index as a second spatial dimension, with hopping amplitude along this direction: $J_y=-\hbar \Omega_R / \sqrt{2}$. By doing this, we recover the Hofstadter Hamiltonian Eq.~\eqref{eq:H0}. Note that, when $F>1$, the mapping still holds, but the hopping amplitude in the new $y$ direction become site dependent.

Of particular interest, this system inherits all the topological properties of the Hofstadter model. Extremely thin samples with infinitely sharp edges can be realised this way, making this system, hereinafter referred to as the Hofstadter strip, ideal for studying the properties of edge states. As a result of its geometry, the Hofstadter strip may present novel bulk properties.
This is one of the subjects we discuss in detail in Chapter \ref{Chapter5}.


\chapter{Topologically non-trivial quantum walk with cold atoms}
\label{Chapter4}

Until recently, it was thought that any phase transition could be fully characterised by specifying the system’s local symmetries \parencite{Landau1936}. This is not the case, however, for topologically phase transitions \parencite{Wen2004QFT}, such that they became known as a new paradigm. While many efforts have gone toward building a general classification of topological phases, a general description of these systems remains elusive \parencite{Chiu2016a}. Amongst the systems which lie beyond the standard classification, established in Refs.\ \parencite{Schnyder2008,Kitaev2009a}, we find, for instance, interacting systems \parencite{Laughlin1983,Yoshioka2002,Wen2004QFT}, Floquet systems \parencite{Rudner2013}, and topological crystalline insulators \parencite{Fu2011}. The above observations underline the need to meticulously investigate the properties of systems which lie beyond the standard classification, such that we may develop a general theory of topological systems.

In this context, quantum simulators are a powerful tool to investigate phenomena which we do not have the resources to approach theoretically. Cold atomic gases, in particular, appear as a natural candidate for this type of application, as it is possible to exert extremely fine control over them \parencite{Bloch2012,Inguscio2013}. Additionally, cold atoms suffer from few losses relative to photons, and optical lattices have scalable size, allowing for a very long evolution with many time-steps. We further note that 1D atomic gases have proved to be a powerful tool for generating topologically non-trivial systems. Indeed, they can realise bound states in static systems \parencite{Ruostekoski2008, Jiang2011, Leder2016a}, topologically protected edge states by using a synthetic dimension \parencite{Celi2013, Mancini, Stuhl2015}, and topologically non-trivial atomic pumps \parencite{Lohse2015, Nakajima2016}.

In the following, we will develop a new topologically non-trivial system which is analogous to the two-step quantum walk, which we encountered in Sec.~\ref{sec:TwoStepQW}. Quantum walks, despite their simplicity, present rich topological phenomena \parencite{Rudner2009, Kitagawa2010a, Rapedius2012}, as they can realise all known topological classes in one and two dimensions. Discrete time quantum walks, being periodically driven systems, can present topological invariants which are not found in the topological classification of Hamiltonians, as was shown in \parencite{Kitagawa2010, Asboth2013}.

The problem which we will face is that topological properties in one dimensional systems are very difficult to observe. We plan to demonstrate that our system is topologically non-trivial by generating a symmetry protected bound state at its boundaries, which is the smoking gun of topologically non-trivial materials.
We further develop a general method to identify chiral bound states which relies on their heavily constrained spin distribution. This method can in principle be used to identify any bound state which is protected by chiral symmetry. It is therefore applicable to all experimental implementations of the two-step quantum walk, both photonic \parencite{Kitagawa2012a, Rechtsman2013, Cardano2015, Cardano2015a} and with cold atoms \parencite{Robens2015, Robens2015a}. This type of measure could be used, for instance, in combination with a method to directly measure the topological invariants of the two-step quantum walk \parencite{Cardano2016}, which was suggested after the publication of this work.


\begin{framed}
In this chapter:
\begin{enumerate}
 \item In Sec.~\ref{sec:Experiment}, we suggest an experimental protocol to realise a quantum walk with a single, two-state atom in a 1D optical lattice.
 \item In Sec.~\ref{sec:Model}, the equations governing the time evolution are presented. We numerical simulations of the system, and show that the atoms have the dynamics of a quantum walk.
 \item In Sec.~ \ref{sec:QW_topo_properties}, we show that this system is topologically non-trivial and derive its phase diagram. We find that the topological phase can be changed spatially, allowing us to generate a topological boundary.
 \item In Sec.~\ref{sec:BoundState}, we populate the bound state that appears at this interface, and suggest a method for measuring its presence.
 \item In Sec.~\ref{sec:JR_pair}, we consider an interesting limit of this Hamiltonian, which also presents bound states, despite being topologically trivial.
\end{enumerate}
\end{framed}  

\section{The atomic quantum walk}
\label{sec:Atomic_QW}

\subsection{Experimental proposal}
\label{sec:Experiment}

Our idea to implement a quantum walk is as follows: particles are allowed to evolve in a medium which accommodates right movers and left movers. By periodically applying a pulsed operation which has amplitude to interconvert right movers and left movers, we obtain path interference phenomena which are consistent with a quantum walk.

\begin{figure}[t]
\centering
\includegraphics[width=0.5\textwidth]{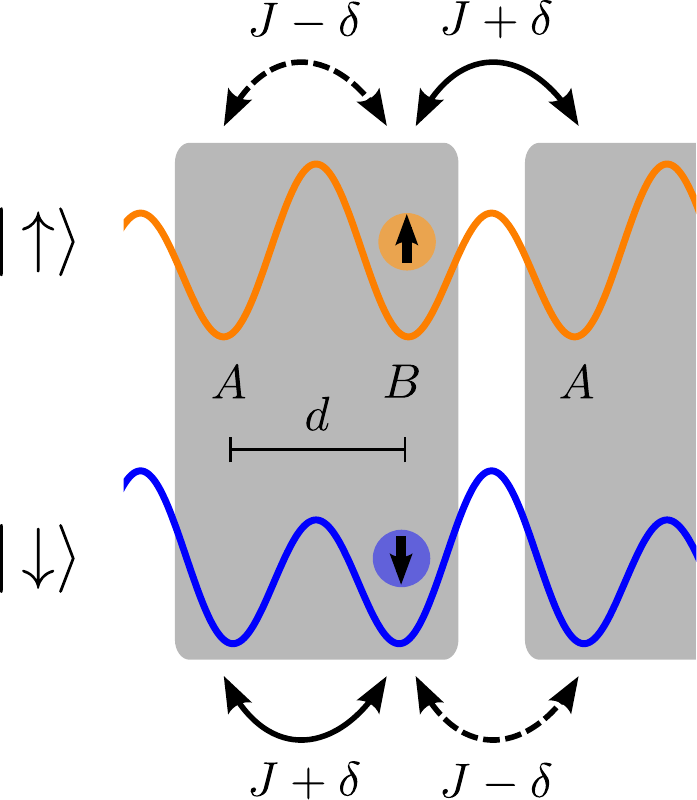}
\caption[Lattice used to build the spin dependent translation]{1D superlattice used to generate the spatial translation operation of the atomic quantum walk. The grey shaded boxes represent unit cells, in which there exists $A$ and $B$ sublattice sites. This lattice geometry is obtained by superimposing two lasers with wavelengths $\lambda_1$ and $\lambda_2$ such that $\lambda_1=2 \lambda_2$; the resulting intersite distance is $d=\lambda_2/2$.
Spin up ($\uparrow$) particles see the orange lattice (top), which has intra-cell hopping $J-\delta$ (dashed arrow) and inter-cell hopping $J+\delta$ (solid arrow). Spin down ($\downarrow$) particles see the blue lattice (bottom), which is identical to the orange lattice, but shifted by $d$. In the figure, $J$ and $\delta$ are depicted as positive parameters. Particles in this lattice are subject to the Hamiltonian \^H$_S$ Eq.~(\ref{eq:Ht}).}
\label{fig:spin_dependent_lattice}
\end{figure}

The specific background which is needed to obtain the topological properties we desire is the 1D lattice represented in Fig.~\ref{fig:spin_dependent_lattice}. We will consider the dynamics of an atom with two internal degrees of freedom, which we can refer to as the particle's spin. Spin up ($\uparrow$) particles see a superlattice with two sites per unit cell, with intra-cell hopping amplitude $J-\delta$ and inter-cell hopping $J+\delta$. This type of lattice is obtained by superimposing two standing waves $l_1$ and $l_2$, with wavelengths $\lambda_1=2 \lambda_2$ (as was done in Ref.\ \parencite{Lohse2015, Nakajima2016}). The easiest way to obtain $l_2$ is to frequency double $l_1$. The distance between neighbouring sites is $d=\lambda_2/2$, and the size of a full unit cell is $2 d$.

Spin down ($\downarrow$) particles see the same lattice as $\uparrow$ particles, but shifted by $d$. This can be done by making $l_1$ attractive for $\uparrow$ particles but repulsive for $\downarrow$ particles, effectively shifting it by a phase of $\pi$. To do this, set $\lambda_1$ to be the so called anti-magic wavelength of the atom, such that the lattice has an equal and opposite  detuning for $\uparrow$ and $\downarrow$ states.

Interesting candidates to play the role of our $\uparrow$ and $\downarrow$ states are the clock states of either Ytterbium or Strontium atoms. These electronic states have narrow transitions, making them long lived, and are well separated in energy, such that their anti-magic wavelength $\lambda_1$ is readily accessible, while $l_2$ remains at approximately the same amplitude for both species. Additionally, these states can be coupled using Raman beams without significantly heating the system.

We drive this system by periodically applying two laser pulses, which induce Raman transitions between $\uparrow$ and $\downarrow$ states. This is the coin operation \^C$_\theta$ of the atomic quantum walk. The amplitude of the coupling, controlled by the angle $\theta$, is proportional to the intensity of the lasers. Additionally, when the angle between the two lasers is non-zero, \^C$_\theta$ applies a momentum kick.

We will show in Sec.~ \ref{sec:TopoProp} that the topological properties of this system can be modified by changing the value of $\theta$ or $\delta$. We suggest to vary $\theta$ spatially to create a boundary between two regions with different topological properties. One way to do this is to create a gradient in the intensity of the lasers which generate the Raman operation. Because these beams can have beam waists of order the length of the system itself, this can be done simply by focusing the laser away from the centre of the lattice. For a typical optical lattice containing $\sim200$ sites, as we consider in our simulations, a Raman beam with waist of order $100 \mu$m should be used, a value which may easily be realised in current experiments. The amplitude of the spin mixing \^C$_\theta$ is proportional to the intensity of the Raman pulses; thus if the Raman lasers' intensity varies over the length of the system, the spin mixing angle $\theta$ will vary accordingly.

Finally, we will introduce a method to identify a topological bound state by performing a spin sensitive measure of the system's probability distribution. This information can be retrieved by averaging over measurements of individual atom positions at a time $t$. This can be done in a single measurement, by performing the experiment multiple times in an array of 1D tubes, then observing the atoms' position using e.g: a quantum gas microscope. It would also become possible to study the Floquet bound states' robustness to interactions and to faults in the periodic driving, thereby providing useful information on systems which are still poorly understood.

\subsection{Shift operation on wavepackets}
\label{sec:Model}

In this section, we present the model realising the atomic quantum
walk, and the operator controlling its time evolution. As shown in Fig.~\ref{fig:spin_dependent_lattice}, we divide the system into unit cells (indexed by $n$), and assume that the atoms only ever populate four quantum states per unit cell: within the $n^{\rm th}$ unit cell, the atom can have spin $\uparrow$ or $\downarrow$, and can reside in the motional ground state of the left/right potential well -- we will
refer to this as sublattice $A/B$. We gather these internal degrees of
freedom into a formal vector, and define vector creation operators as
\begin{equation}
\ccc_n^\dagger = (c^\dagger_{n\uparrow A},c^\dagger_{n\downarrow A}, 
c^\dagger_{n\uparrow B},c^\dagger_{n\downarrow B}).
\end{equation}
We use $\sigma_j$ and $\tau_j$ to denote the Pauli matrices acting in
spin space ($\uparrow$ and $\downarrow$) and sublattice space ($A$ and
$B$) respectively, with $j=\{1, 2, 3\}$, and $\tau_0$ and $\sigma_0$ are $2\times 2$ identity matrices. We will also use $\tau_\pm$ to represent
the sublattice index raising and lowering operators,
\begin{equation}
\label{eq:tau_pm}
\tau_\pm=\frac{1}{2}(\tau_1\pm i \tau_2).
\end{equation}

Consider first the Hamiltonian of the 1D bichromatic
optical lattice without the Raman pulses.  This reads
\begin{equation}
\label{eq:Ht}
\text{\^H}_S =\sum_{n} \ccc_n^\dagger \tau_1 \otimes (J \sigma_0 + \delta \sigma_3) \ccc_n
+ \left [ \ccc_{n+1}^\dagger \tau_+ \otimes (J \sigma_0-\delta \sigma_3)  \ccc_n + 
\text{H.c}\right].
\end{equation}
The parameters $J$ and $\delta$ control the particles' hopping amplitudes, as shown in Fig.~\ref{fig:spin_dependent_lattice}. In general, the full Hamiltonian will also contain a term proportional to the energy splitting between spin states $\Delta E$.

Since the Hamiltonian of Eq.~\eqref{eq:Ht} is translation invariant,
its eigenstates are plane waves with well defined quasimomentum $k\in [-\pi/(2  d), \pi/(2  d)]$. These states can be chosen to be fully polarised, either $\uparrow$ or $\downarrow$. In this way, each energy eigenvalue is doubly degenerate, since the system presents two identical lattices (one for each spin) with no tunnelling between them (see Fig.~\ref{fig:spin_dependent_lattice}).

\begin{figure}[t]
\centering
\includegraphics[width=0.75\textwidth]{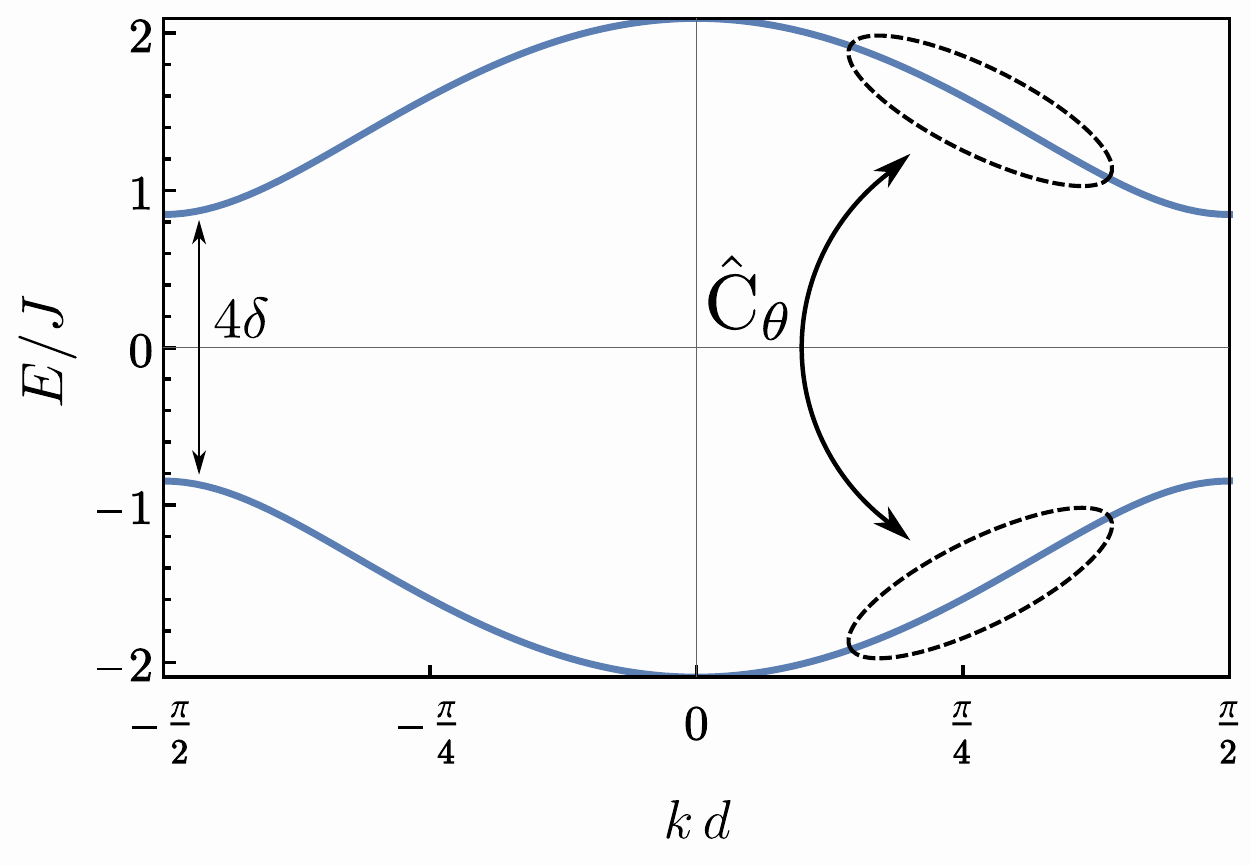}
\caption[Dispersion of \^H$_S$]{Dispersion of \^H$_S$ Eq.~(\ref{eq:Ht}) for $J=\pi/3$ and $\delta=0.42$. The eigenstates of \^H$_S$ can be chosen to be fully polarised (either $\uparrow$ or $\downarrow$), and the $\uparrow$ and $\downarrow$ bands overlap exactly; the value of the band gap is $4\delta$. Note that the dispersion is symmetric about $E=0$. This means that the slope of the two circled regions of the dispersion are equal and opposite. Thus the wavepackets existing in these regions, which are the states centred around $k=\pi/(4d)$, with energies $E$ and $-E$ respectively, move on average at equal and opposite velocities. The spin mixing \^C$_\theta$ couples states in these regions (illustrated by an arrow).}
\label{fig:Ht_dispersion}
\end{figure}

The object that undergoes a quantum walk is a wavepacket that is broad
in position space but restricted in momentum space to the vicinity of
the wavenumber $k=\pi/(4 d)$.  For this quasimomentum the Hamiltonian
is almost dispersionless, as shown in Fig.~\ref{fig:Ht_dispersion}.
Thus a wavepacket constructed with states from the lower branches of
the dispersion relation, with $k\approx \pi/(4 d)$, is translated
with a uniform velocity to the right, and its real-space width grows
only very slowly. It is therefore a right-mover. A wavepacket
similarly constructed, but belonging to the upper branches of the
dispersion relation, is a left-mover.

Including the energy splitting spin states, $\Delta E$, and the contribution from the Raman laser $\h_R$, we obtain the full, time dependent Hamiltonian:
\begin{equation}
\label{eq:Htimedependent}
\h(t) = \Hs + \frac{\Delta E}{2} \sum_n \ccc_n^\dagger \tau_0\otimes\sigma_3  \ccc_n +\h_R(t).
\end{equation}
We will analyse the Raman term in Sec.~\ref{sec:RamanPulse}, and thereby show how the energy difference $\Delta E$ can be eliminated.

\subsection{Rotation operation using the Raman pulses}
\label{sec:RamanPulse}

\subsubsection{Quasimomentum kick}

Consider next the effect of the two Raman lasers on the system. These induce a two photon tunnelling process between spin states with Rabi frequency $\Omega_R(n,t)$, which can in general vary spatially and be a function of time. Let $\omega$ be the frequency difference between the two Raman lasers, which is resonant with the energy separation between spin states: $\Delta E=\hbar \omega$. The coupling term induced by this Raman transition is of the form $\Omega_R(n,t) \cos(\omega t) \sigma_2$ as shown in Ref.\ \citep{Grynberg}. We have chosen the phase of $\Omega_R(n,t)$ such that the matrix which interchanges $\uparrow$ and $\downarrow$ densities appears in this expression as $\sigma_2$.

By setting these lasers at an angle, the Raman transition can impart momentum onto the atom. In the following, we will consider the specific case when the particle acquires a quasimomentum $\pi/d$ through this process. We will see in Sec.~\ref{sec:SimpleBasis} that this allows us to make a transformation which greatly simplifies the theoretical treatment of the system. Additionally, in Sec.~\ref{sec:DoubleQW}, we will see that this momentum kick introduces new physics which we could not achieve otherwise.

To understand the effect of the momentum kick, consider the state $\ket{k, \chi, s}$ with well defined quasimomentum $k\in [-\pi/2d, \pi/2d]$, with sublattice state $\ket{\chi}$ and spin state $\ket{s}$. We can express this state in terms of its $\ket{j, A, s}$ and $\ket{j, B, s}$, its components at site $j$ belonging to sublattices $A$ and $B$ respectively:
\begin{equation}
\ket{k, \chi, s} = \sum_j \left( \chi_A e^{i 2 j k d} \ket{j, A, s} + \chi_B e^{i (2 j +1) k d} \ket{j, B, s} \right).
\end{equation}
Note that we have introduced the amplitudes: $\chi_{A,B}= \braket{A,B}{\chi}$. Consider now the same atomic state, shifted by $\Delta k = \pi/d$:
\begin{equation}
\begin{split}
\ket{\left( k +\frac{\pi}{d} \right), \chi, s} & = \sum_j \left( \chi_A e^{i 2 j k d} \ket{j, A, s} - \chi_B e^{i (2 j +1) k d} \ket{j, B, s} \right) \\
& = \ket{k, \tau_3 \chi, s}.
\end{split}
\end{equation}
The effect of this specific momentum kick is therefore simply to change the sign of the wavefunction at every site belonging to sublattice $B$. From the above, we can express the coupling induced by the two Raman lasers as:
\begin{equation}
\label{eq:HRaman}
\h_R = \sum_n \Omega_R(n,t) \cos(\omega t) \ccc_{n}^\dagger \tau_3\otimes\sigma_2 \ccc_{n}.
\end{equation}

\subsubsection{Energy difference between spin states}
\label{sec:SpinStatesEnDiff}

In the following, we assume that the energy splitting between spin states $\Delta E$ is very large compared to the energy scales in the problem, and show that this term plays a  negligible role. We pay special attention to this term in this section, because it is proportional to $\sigma_3$, such that it could potentially break chiral symmetry and destroy the system's topological properties. Should $\Delta E$ be comparable to the other energy scales in the system, the energy shift should be eliminated, e.g., by applying an external Zeeman field.

\begin{figure}[t]
\centering
\includegraphics[width=0.8\textwidth]{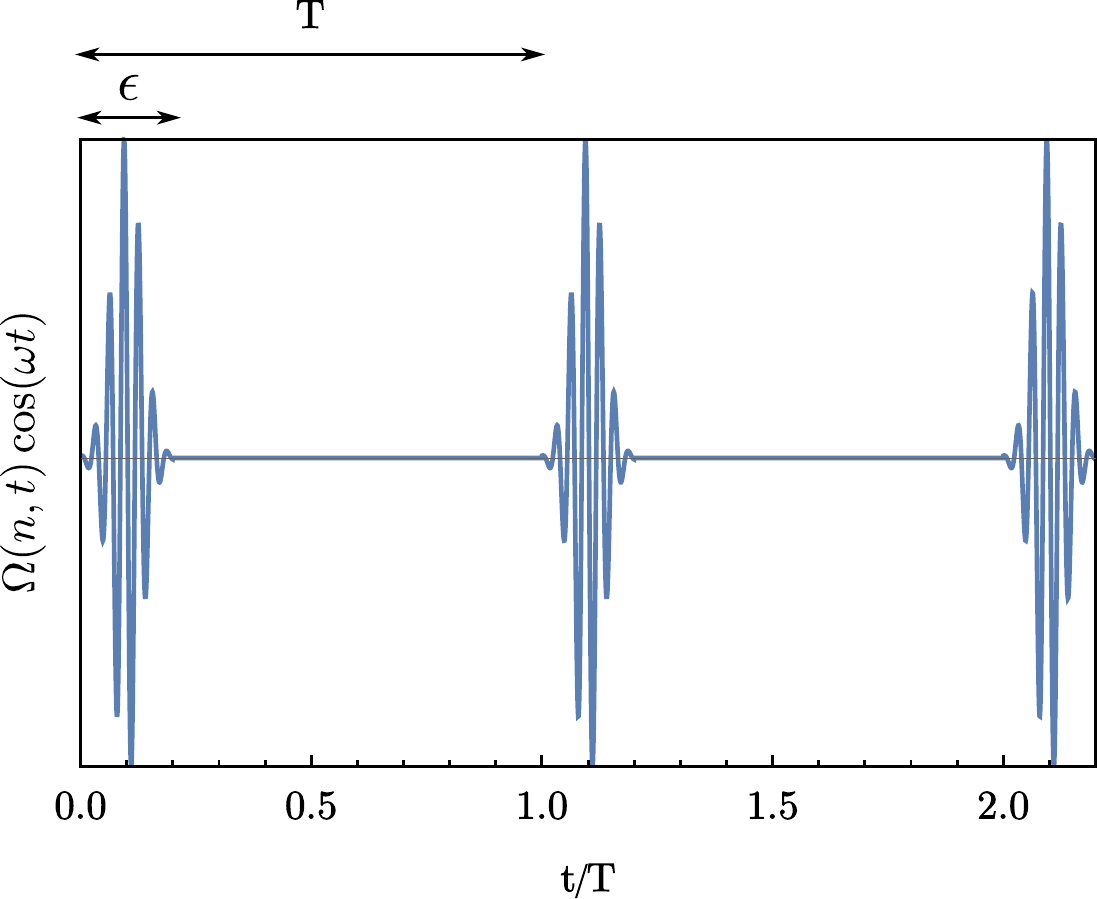}
\caption[Raman laser pulse sequence]{Pulse sequence of the two Raman lasers, with period $T$. Each pulse has duration $\epsilon$. The laser frequency is $\omega$, which is determined by the energy difference between spin states $\Delta E=\hbar \omega$.}
\label{fig:Raman_evelope}
\end{figure}

The Raman lasers are periodically applied, with period $T$, and the pulses have duration $\epsilon \ll T$. The resulting spin coupling amplitude is represented in the Fig.~\ref{fig:Raman_evelope}. Without loss of generality, we can include the envelope function of the pulse sequence in the definition of the (time dependent) Rabi frequency $\Omega_R(n,t)$.

We can eliminate the term proportional to $\Delta E=\hbar \omega$ in Eq.~\eqref{eq:Htimedependent} with the gauge transformation $\hat{M}= \exp( i \omega \sigma_3 t /2 )$:
\begin{equation}
\h(t) \rightarrow \hat{M}  \h(t)  \hat{M}^\dagger = \Hs + \hat{M}  \h_R  \hat{M}^\dagger.
\end{equation}
Substituting in Eq.~\eqref{eq:HRaman}, the transformed Hamiltonian Eq.~\eqref{eq:Htimedependent} takes the form:
\begin{align}
\label{eq:Ht_effective}
\h(t) = \Hs + \frac{1}{2} \sum_n \Omega_R(n,t) \ccc_{n}^\dagger \tau_3\otimes \sigma_2 \ccc_{n} + \mathcal{O} \left (e^{\pm 2 i \omega t} \right ),
\end{align}
where we have used the rotating wave approximation to eliminate terms which oscillate with frequency $\omega$. This step relies on the assumption that $\omega$ is much larger than any energy scale in the system, and in particular larger than any of the frequencies in $\Omega_R(n,t)$. This is the case in the limit $\omega \gg 1/\epsilon \gg 1/T$. To conclude, we have obtained a result analogous to Eq.~\eqref{eq:RamanMomentumKick} and shown that the energy difference between spin states is negligible in this limit.


\subsubsection{Effective coin operation}
\label{sec:CoinEff}

In the following section, we will derive an effective form for the Raman pulses. Assuming the laser pulses are intense enough, $\Hs$ can be neglected during the time they are switched on, and we have 
\begin{equation}
\label{eq:Htheta}
\h(t)=
\begin{cases}
\frac{1}{2}\sum_n \Omega_R(n,t) \ccc_{n}^\dagger \tau_3\otimes\sigma_2 \ccc_{n}
& \text{for } t \pmod{T} < \epsilon \\
\Hs & \text{for } t \pmod{T} \geq \epsilon. \\
\end{cases}
\end{equation}
We define the area of the pulse at unit cell $n$ by $\theta(n)$:
\begin{equation}
\theta(n) = \frac{1}{2}\int_{0}^{\epsilon} \Omega_R(n,t) dt,
\end{equation}
such that the effect of the whole pulse is given by:
\begin{align}
\label{eq:C}
\text{\^C}_\theta = \sum_n \ccc_{n}^\dagger 
\exp[-i \theta(n) \tau_3\otimes\sigma_2] \ccc_n \equiv\exp(-i \Htheta).
\end{align}
The effect of $\CCC$ is to couple right-moving states from the bottom branch of the dispersion relation to left-moving states of the top branch, as indicated in Fig.~\ref{fig:Ht_dispersion}. Interestingly, $\CCC$ also couples states on the same branch of the dispersion relation. We will discuss the consequences of this in Sec.~\ref{sec:DoubleQW}.

\subsection{Quantum walk with cold atoms}

\subsubsection{Complete sequence}

\begin{figure}[t]
\centering
\includegraphics[width=0.7\textwidth]{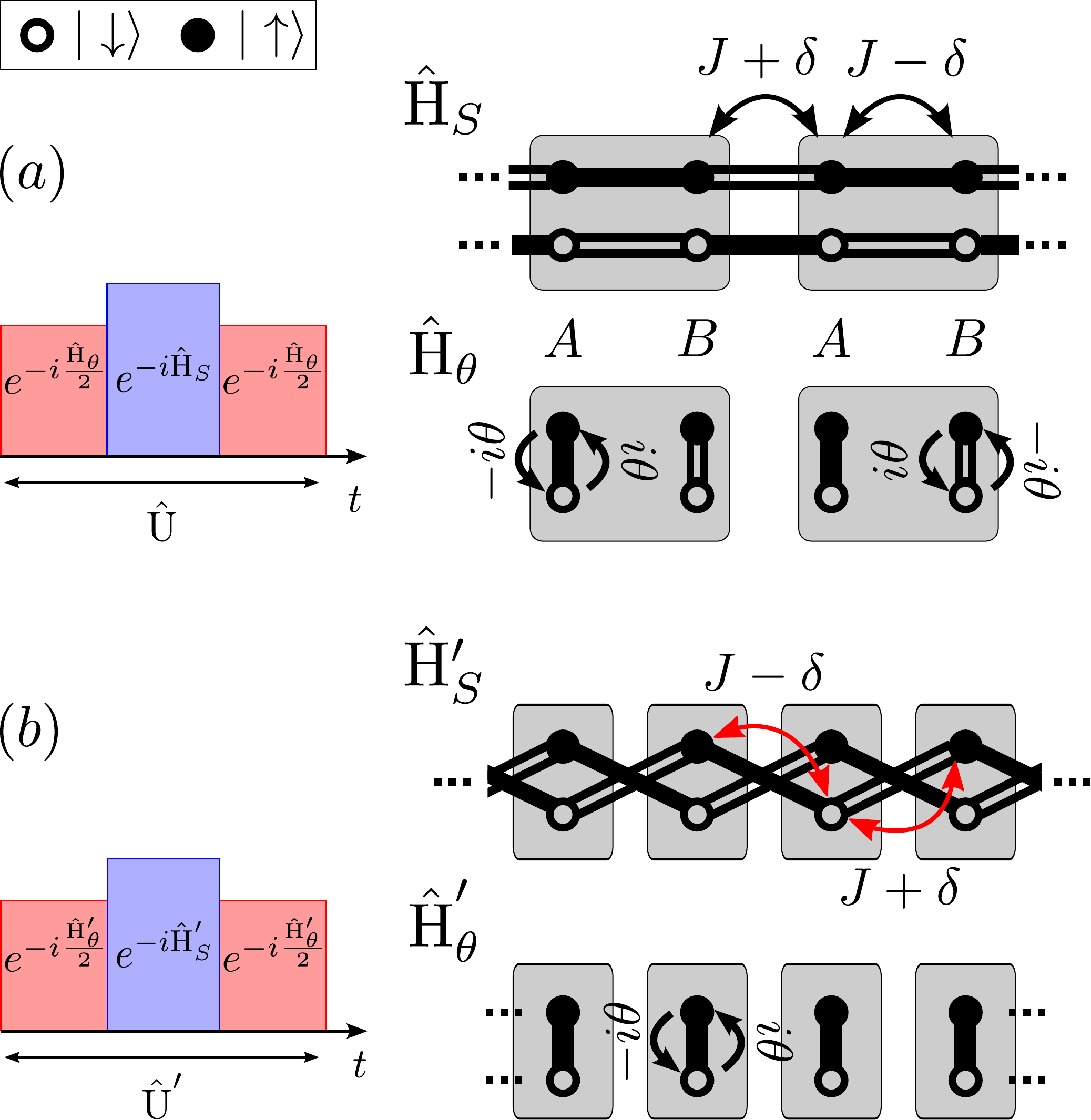}
\caption[Sketch of the tunnelling amplitudes in the atomic quantum walk]{Sketch of the tunnelling amplitudes in the atomic quantum walk. Tunnelling amplitudes are different along single and double lines. The tunnelling amplitudes are indicated by curved arrows, the colour of which is unimportant. Unit cells are represented by grey shaded boxes. (a) The time evolution \^U is controlled by \^H$_S$ and \^H$_\theta$, Eqs.~(\ref{eq:Ht}) and (\ref{eq:C}), respectively. The system has two sites per unit cell. (b) We flip the spins on every second site. the time evolution \^U$'$ is controlled by \^H$_S'$ and \^H$_\theta'$, given by Eqs.~(\ref{eq:Ht'}) and (\ref{eq:Htheta'}). The lattice has one site per unit cell. The system maps onto the Creutz ladder when $J=\delta$.}
\label{fig:tunnelling_sketches}
\end{figure}

As discussed in Sec.~\ref{sec:RamanPulse}, we obtain a quantum walk by switching on the Raman lasers for brief intense pulses of duration $\epsilon$ which follow each other periodically, with period $T \gg \epsilon$. As we saw in section \ref{sec:CoinEff}, in this limit, we can consider we are acting stroboscopically with $\Hs$ Eq.~\eqref{eq:Ht} and $\Htheta$ Eq.~\eqref{eq:C}. The unitary time evolution operator for one complete period, \^U, reads
\begin{equation}
\label{eq:U}
\text{\^U}=e^{-i\Htheta/2}e^{-i\Hs}e^{-i\Htheta/2}.
\end{equation}
Now and in the following, we use dimensionless units where $T/\hbar =1$. In writing down Eq.~(\ref{eq:U}), we have chosen the origin of time such that the sequence of operations defining the walk has the form of Eq.~\eqref{eq:TSQW_UC}. While this choice has little effect on the system's properties, the form of Eq.~\eqref{eq:U} makes it easier to find the symmetries of \^U. It is however important to always pick the same origin of time when averaging over multiple runs of the experiment, otherwise important details could be averaged out. The tunnelling amplitudes induced by \^H$_S$ and \^H$_\theta$ are sketched in the figure \ref{fig:tunnelling_sketches}(a).

We define the Floquet Hamiltonian \^H$_F$ as:
\begin{equation}
\label{eq:HF}
\text{\^U}=\exp(-i \text{\^H}_F).
\end{equation}
This static Hamiltonian describes the motion, integrated over a time step. This allows us to compute the system's spectrum. Because the eigenvalues of \^H$_F$ are defined from Eq.~\eqref{eq:HF}, the band structure of \^H$_F$ is $2\pi$ periodic. We will refer to the eigenvalues of \^H$_F$ as the system's quasienergies. To differentiate quasienergies  from energies, we will denote them by the letter $\varepsilon$.

At the end of this section we will find a change of basis which
simplifies \^H$_F$. Despite this, we will prefer to work in the basis
defined above when performing numerical simulations, so that our
results can be easily compared to experimental results from the
protocol described in Sec.~\ref{sec:Experiment}.

\subsubsection{Simulations}
\label{sec:Simulations}

\begin{figure}[t]
\centering
\includegraphics[width=0.9\textwidth]{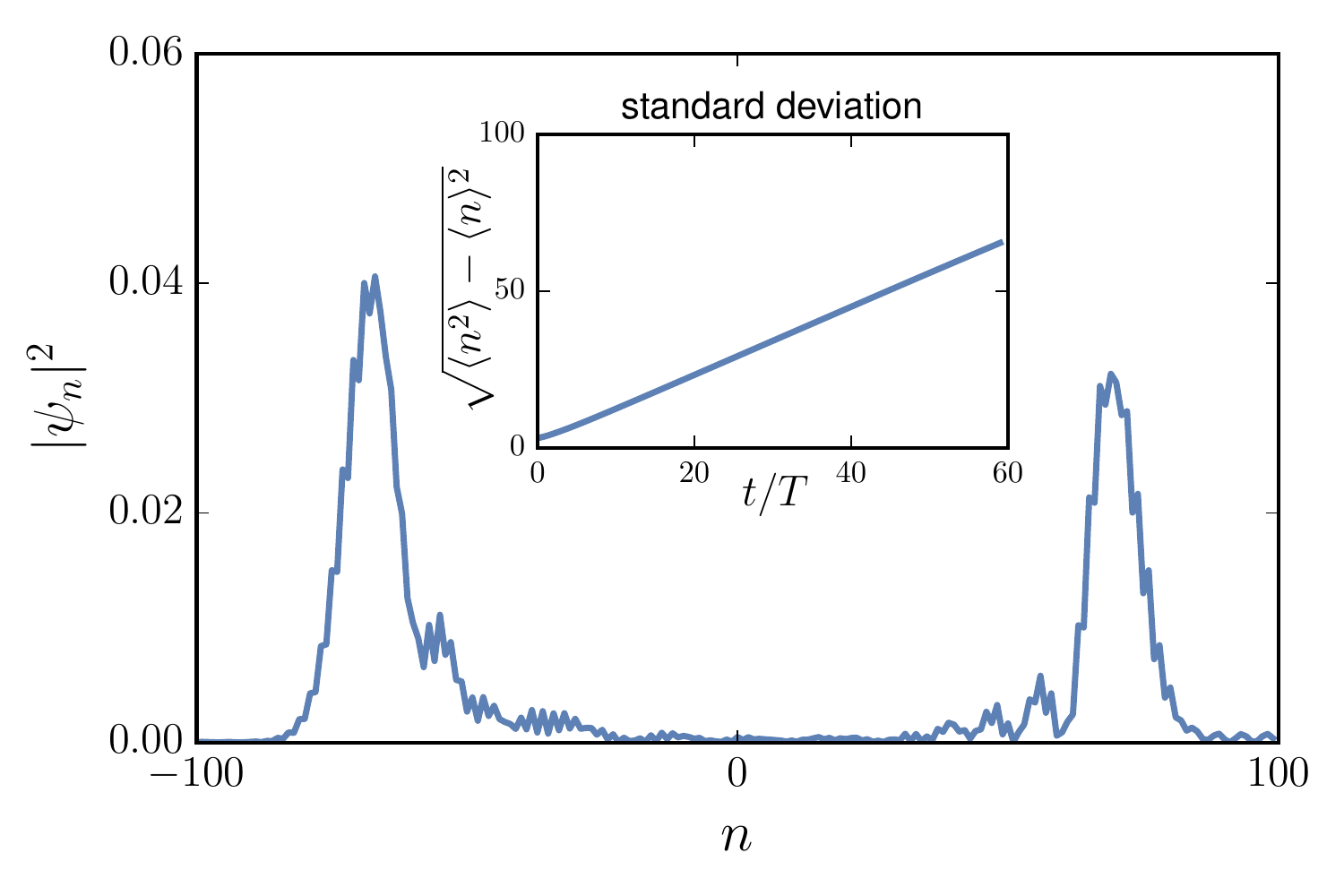}
\caption[Simulation of the atomic quantum walk]{Spatial probability distribution $|\psi_n|^2= \left|\braket{n}{\psi}\right|^2$ after $60$ time-steps. The interference pattern that the density distribution forms is typical of quantum walks. The walker's initial positions were sampled from the Gaussian wave function centred at the site $n=0$, with width $4  d$ and average quasimomentum $k=\pi/(4 d)$. We used the Hamiltonian's parameters: $J=\pi/3$, $\delta=0.42$ and $\theta=0.15$. Inset: the variance of the wavepacket scales linearly with time, which is a characteristic feature of quantum walks.}
\label{fig:AQW_simulation}
\end{figure}

Using the operators derived in Sec.~\ref{sec:Model}, we perform numerical simulations to determine the system's properties. As discussed above, repeated application of the time-step operator $\UU$ of Eq.~(\ref{eq:U}) on a wavepacket can be described as a quantum walk. We verified this using numerical simulations. As an example, we present the atom's final density distribution after $60$ time steps in Fig.~\ref{fig:AQW_simulation}. For this simulation, we initiated the system with a $\uparrow$ polarised Gaussian wavepacket $|\psi\rangle_{t=0}$ with width $4  d$ centred around $k=\pi/(4 d)$; we set $J=\pi/3$, $\delta=0.42$ and $\theta=0.15$. The final density distribution, given by $|\psi_n|^2= \left|\braket{n}{\psi}\right|^2$, shows sharp peaks where density is furthest from the origin. The probability to find the particle in any other region is inhibited due to back travelling waves, which interfere destructively with forward travelling ones. The inset shows the standard deviation of the position of the particle, which can be seen to increase linearly with time. Both the destructive interference and the ballistic expansion are well known feature of quantum walks (in contrast with classical random walks, which show diffusive expansion).

\subsubsection{A hidden unitary symmetry}

\begin{figure}[t]
 \centering
 \includegraphics[width=.75\textwidth]{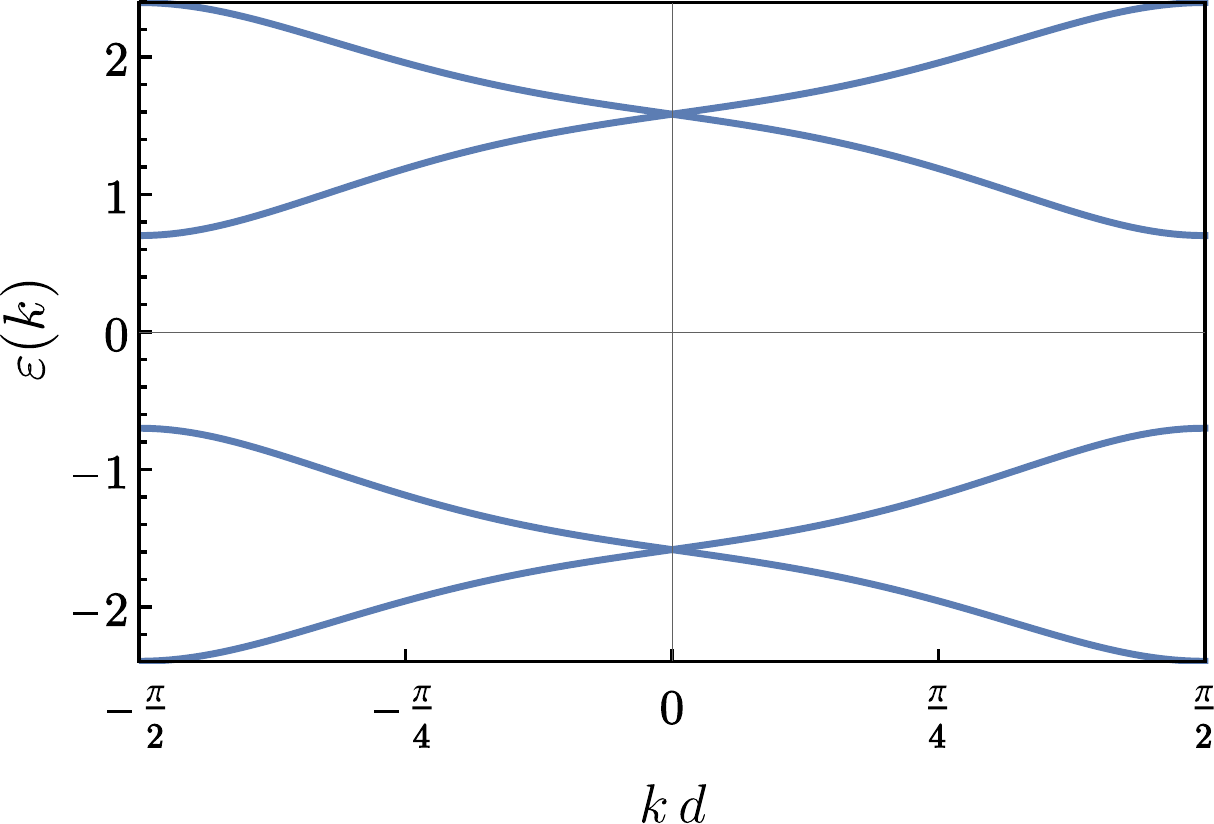}
 \caption[Dispersion of \^H$_F$]{Dispersion of \^H$_F$ Eq.~(\ref{eq:HF}) for $J=\pi/3$, $\delta=0.42$ and $\theta=1.5$. Introducing $\theta\neq 0$ lifts the degeneracy of the eigenvalues of \^H$_F$, but the two positive energy bands always have a band touching (as do the two negative energy bands). Thus, the system can be expressed as a two band model.}
 \label{fig:HF_dispersion}
\end{figure}

It this section, we will show that the atomic quantum walk can be reduced to a system with only two internal states. Consider the dispersion of $\h_F$ presented in Fig.~\ref{fig:HF_dispersion}. The two positive energy bands have a band touching for all values $\theta$, same as the two negative energy bands. This indicates that we can reduce the Hilbert space size simply by doubling the Brillouin zone.

We can show this formally by considering the unitary operator:
\begin{equation}
\label{eq:unitarySym}
 \text{\^M} = \sum_{n=1}^N \ccc_n^\dagger \tau_1\otimes\sigma_3 \ccc_{N-n-1}.
\end{equation}
Note that \^M satisfies:
\begin{align}
& \text{\^M}  \left(\sum_{n=1}^N \ccc_{n\pm1}^\dagger\ccc_{n}\right)  \text{\^M}^\dagger = \sum_{n=1}^N\ccc_{n}^\dagger\ccc_{n\pm 1},\\
& \text{\^M}  \ccc_{n}^\dagger\ccc_{n}  \text{\^M} = \ccc_{n}^\dagger\ccc_{n},
\end{align}
such that it commutes both with $\h_S$ and $\h_\theta$. As is visible from Eq.~\eqref{eq:U}:
\begin{equation}
 \text{\^M}  \text{\^U}   \text{\^M}^\dagger = \text{\^U},
\end{equation}
meaning that \^M is a unitary symmetry of the system. In the next section, we will perform a transformation which takes advantage of this symmetry to reduce the size of the Hilbert space.

\subsubsection{Gauge transformation for a smaller unit cell}
\label{sec:SimpleBasis}

In this section, we introduce a convenient unitary transformation
which simplifies the description of the system. By inspection of Fig.~\ref{fig:tunnelling_sketches}(a), we notice that it is possible to simplify the system's description by introducing the new vector creation operators:
\begin{align}
\cc_{2n}^\dagger &= (c^\dagger_{n\uparrow A},c^\dagger_{n\downarrow A});\\
\cc_{2n+1}^\dagger &= (c^\dagger_{n\downarrow B},c^\dagger_{n\uparrow B}).
\end{align}
Notice the inversion of the order of $\uparrow$ and $\downarrow$ on the odd sites.  In this basis, the tunnellings induced by the atomic quantum walk are represented in Fig.~\ref{fig:tunnelling_sketches}(b). Under this transformation, we see that the atomic quantum walk is reminiscent of the Creutz ladder \parencite{Creutz1999, Bermudez2009, Sticlet2014}, a 1D model which is known to support a non-zero
winding number. We will make this correspondence more obvious in
Sec.~\ref{sec:MapToCreutz}.

In the basis introduced above, the Hamiltonians \^H$_S$ and \^H$_\theta$ become:
\begin{eqnarray}
\label{eq:Ht'}
\text{\^H}_S & \rightarrow & \text{\^H}_S'=\sum_n \cc^\dagger_{n+1} (J
 \sigma_1 - i  \delta  \sigma_2) \cc_n +\text{H.c};\\
\label{eq:Htheta'}
\text{\^H}_\theta & \rightarrow & \text{\^H}_\theta'=\sum_n \cc^\dagger_n 
\theta(n)  \sigma_2 \cc_n.
\end{eqnarray}
While \^H$_\theta'$  Eq.~(\ref{eq:Htheta'}) is already in diagonal form, we can diagonalise \^H$_S'$ by Fourier transforming  Eq.~(\ref{eq:Ht'}):
\begin{equation}
\label{eq:Ht_Kspace}
\text{\^H}_S'(k) = 2  J \cos(k d) \sigma_1 + 2 \delta \sin(k d) \sigma_2,
\end{equation}
with $k\in [-\pi/d,\pi/d]$. Because of the smaller unit cell, the dispersion has now two non-degenerate branches. It is interesting to redefine our right- and left-walkers in this basis, such that the mapping of the system to a quantum walk can be made more obvious. This is done in Sec.~\ref{sec:DoubleQW}, where we show that the system has dynamics which is more complex than the two-step quantum walk (which we studied in Sec.~\ref{sec:TwoStepQW}) considered by Refs.\ \parencite{Karski2009, Kitagawa2010a, Robens2015}.

By analogy with Eqs.~\eqref{eq:U} and \eqref{eq:HF}, we can introduce the time evolution operator and the Floquet Hamiltonian in the new basis:
\begin{equation}
\label{eq:U'}
\text{\^U}' = e^{-i \text{\^H}_\theta'/2}  e^{-i \text{\^H}_S'}  e^{-i \text{\^H}_\theta'/2}=\exp(-i \text{\^H}_F').
\end{equation}
In the sections that follow, all analytical results will be obtained in the simplified basis of \^H$_F'$.

\subsection{Double atomic quantum walk}
\label{sec:DoubleQW}

As we explained in section \ref{sec:Model}, the system has the dynamics of a quantum walk. As we mentioned in this same section, however, our protocol is slightly more complex than the standard quantum walk considered in Refs.\ \parencite{Karski2009, Kitagawa2010a, Robens2015a}. Indeed, we saw that each eigenstate of $\Hs$ is doubly degenerate, implying that two wavepackets exist in each circled region of Fig.~\ref{fig:Ht_dispersion}. As a result of this, our quantum walk has four internal states, coupled by the coin operation $\CCC$. In the following, we will show that the system can be interpreted as two independent quantum walks using the simplified (two-state) basis. 

\begin{figure}[t]
\centering
\includegraphics[width=0.75\columnwidth]{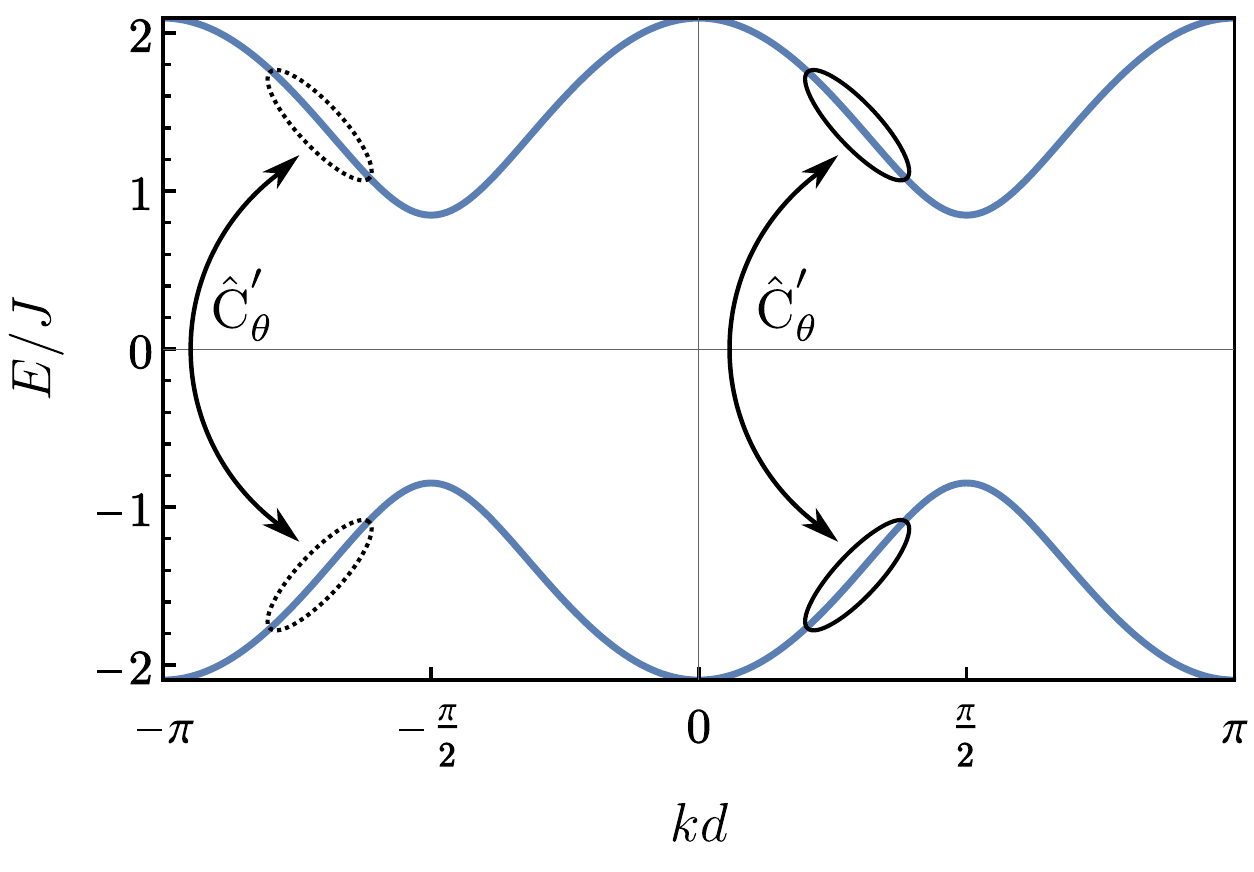}
\caption[Dispersion of \^H$_S'$]{Dispersion of \^H$_S'$ Eq.~(\ref{eq:Ht_Kspace}) for $J=\pi/3$ and $\delta=0.42$. The wavepackets centred around $k=\pi/(4d)$, where the slope is locally linear, are circled by a solid line. Due to the symmetry of the dispersion about $E=0$, these states move at equal and opposite average velocities. The regions circled in Fig.~\ref{fig:Ht_dispersion} correspond to a superposition of the wavepackets centred at $k=\pi/(4d)$ and the ones centred at $k=-3\pi/(4d)$ (circled by a dotted line). The couplings induced by \^C$_\theta'$ are illustrated by arrows. When both the states in the region of $k=\pi/(4d)$ and $k=-3\pi/(4d)$ are populated, the system performs two simultaneous, independent quantum walks.}
\label{fig:Hs'_dispersion}
\end{figure}

As a first step, consider the plane-wave eigenstates of the
Hamiltonian $\Hs'$ of Eq.~\eqref{eq:Ht_Kspace}. These can be
written as 
\begin{align}
& \ket{\pm',k}=2^{-1/2}(1,\pm e^{i\phi(k)}),\\
& \phi(k)=\arg [ J \cos(kd)+i\delta\sin(kd)],
\end{align}
where $\ket{+',k}$, $\ket{-',k}$ are the eigenstates belonging to the top
and bottom bands respectively, and ``arg'' denotes the phase of the
complex number. By the same arguments as in Sect.~\ref{sec:Model},
wavepackets centred narrowly around $k=\pi/(4d)$ can be used as right-
and left-movers for a quantum walk, if they are from the bottom or top
bands respectively, as represented on
Fig.~\ref{fig:Hs'_dispersion}. As previously, these are translated in
real space by $\Hs'$ at equal and opposite velocities.

To understand in what sense this system describes a quantum
walk, we express $\CCC'$ in the basis $\ket{+',k}$, $\ket{-',k}$. We find
\begin{align}
\CCC' &= e^{-i\Htheta'};\\
\label{eq:CthetaKspace}
\text{H}_\theta' &= \theta
\begin{pmatrix}
\sin[\phi(k)] & i \cos[\phi(k)] \\
-i \cos[\phi(k)] & -\sin[\phi(k)]
\end{pmatrix}.
\end{align}
This corresponds to a rotation by an angle $\theta$ about the axis $(0,-\cos[\phi(k)],$ $\sin[\phi(k)])$. 

To recapitulate, we have defined left- and right-moving states of a
walker, which are translated in real space at average equal and
opposite velocities. Our walkers are periodically coupled by the
$\CCC'$ operation, as indicated in Fig.~\ref{fig:Hs'_dispersion}. In
this sense the protocol fits exactly our definition of a quantum
walk. Note however that no choice of $\theta$ can in general fully
interchange right- and left-walkers; in this respect it is different
from the quantum walks considered in Refs.\ \parencite{Karski2009,
  Kitagawa2010a, Robens2015a}.

An additional subtlety comes from the choice of initial state. Let
$\ket{+,k,\uparrow}$, $\ket{+,k,\downarrow}$ be the eigenstates of $\Hs$
Eq.~\eqref{eq:Ht} in the superlattice basis, corresponding to the
state with quasimomentum $k$ belonging to the top band with spin
$\uparrow$ or $\downarrow$ respectively, and
$\ket{-,k,\uparrow}$, $\ket{-,k,\downarrow}$ their bottom band
counterparts. In general, these states are related to the eigenstates
of $\Hs'$ through:
\begin{align}
4\ket{\pm,k,\uparrow} & = \left(\tau_0+\tau_3\right)\otimes \left(\ket{\pm',k}+ \ket{\pm',k+\frac{\pi}{d}}\right)\\
& \left(\tau_0-\tau_3\right)\otimes\sigma_1   \left(\ket{\pm',k}- \ket{\pm',k+\frac{\pi}{d}}\right),
\end{align}
and
\begin{align}
4\ket{\pm,k,\downarrow} & = \left(\tau_0+\tau_3\right)\otimes \left(\ket{\pm',k}- \ket{\pm',k+\frac{\pi}{d}}\right)\\
& \left(\tau_0-\tau_3\right)\otimes\sigma_1   \left(\ket{\pm',k}+ \ket{\pm',k+\frac{\pi}{d}}\right),
\end{align}
where only the $\sigma_i$ matrices act on the states $\ket{\pm',k}$.

When simulating the atomic quantum walk, we chose as an initial state
the Gaussian wavepacket narrowly centred around $k=\pi/(4d)$ in the
superlattice basis, as circled on Fig.~\ref{fig:Ht_dispersion}. This
state can be expressed in the simpler basis of $\Hs'$ and $\CCC'$ as a
superposition of wavepackets centred about $k=\pi/(4d)$ and
$k=-3\pi/(4d)$, as represented on Fig.~\ref{fig:Hs'_dispersion} by
solid and dotted circles respectively. Thus with this initial state we
are performing two quantum walks at the same time in superposition
(the two right- and two left-moving states are shown in
Fig.~\ref{fig:Hs'_dispersion}). As states with different quasimomenta
are not coupled by $\CCC'$, these really are two independent quantum
walks. Since they correspond to orthogonal basis states, they cannot
interfere, and so the measured probability distribution after a number
of time-steps is a probabilistic mixture of these two walks. 

\section{Topological properties of the atomic quantum walk}
\label{sec:QW_topo_properties}

\subsection{Symmetries of the atomic quantum walk}
\label{sec:Symmetries}

In general in 1D, a system can display non-trivial topological behaviour only if it is constrained by certain symmetries. We are interested in whether or not the atomic quantum walk presents time reversal symmetry (TRS), particle hole symmetry (PHS) and chiral symmetry (CS), which determine the system's topological class. While we study the case specific to our system, a more general and complete study of symmetries and their relevance to topological phases can be found in Ref.\ \parencite{Haake2010}.
For simplicity, we will work in the basis described in Sec.~\ref{sec:SimpleBasis}, where \^U$'$ is given by Eq.~\eqref{eq:U'}.

The system has CS if there is a unitary operator $\hat{\Gamma}$ acting within a single unit cell which anti-commutes with the Hamiltonian:
\begin{equation}
\label{eq:AQW_ChiralSymm}
\hat{\Gamma}   \hat{\text{H}}_F'(k)   \hat{\Gamma}^\dagger = -\hat{\text{H}}_F'(k) ~ \Rightarrow ~ \hat{\Gamma}   \hat{\text{U}}'(k)   \hat{\Gamma}^\dagger = \hat{\text{U}}'(k)^\dagger.
\end{equation}
We further require that $\hat{\Gamma}$ acts only within a single unit cell, such that CS is robust to breaking translational invariance. Thanks to the symmetric form of Eq.~\eqref{eq:U'}, if an operator $\hat{\Gamma}$ simultaneously anti-commutes with \^H$_\theta'$ and \^H$_S'$, it automatically satisfies Eq.~(\ref{eq:AQW_ChiralSymm}):
\begin{equation}
\begin{split}
\text{\^U}'^\dagger & = e^{i \text{\^H}_\theta'/2}  e^{i \text{\^H}_S'}   e^{i \text{\^H}_\theta'/2}\\
\Rightarrow \hat{\Gamma} \text{\^U}'^\dagger \hat{\Gamma}^\dagger & = e^{i \hat{\Gamma} \text{\^H}_\theta' \hat{\Gamma}^\dagger /2}  e^{i \hat{\Gamma} \text{\^H}_S' \hat{\Gamma}^\dagger}   e^{i \hat{\Gamma} \text{\^H}_\theta' \hat{\Gamma}^\dagger /2}\\
& = \text{\^U}'.
\end{split}
\end{equation}
Thus, if $\hat{\Gamma}$ is a valid CS for \^H$_\theta'$ and \^H$_S'$, it is also a CS operator for \^U$'$. By inspection of Eqs.~(\ref{eq:Htheta'}) and (\ref{eq:Ht_Kspace}), we find that $\hat{\Gamma}=\sigma_3$ simultaneously anti-commutes with \^H$_\theta'$ and \^H$_S'$, and is therefore the CS operator in this basis.

But is it the only operator which satisfies Eq.~(\ref{eq:AQW_ChiralSymm})? Given a matrix \^U$'$ which has the form Eq.~(\ref{eq:U'}), it can happen that $\hat{\Gamma}$ satisfies Eq.~(\ref{eq:AQW_ChiralSymm}) without simultaneously anti-commuting with \^H$_S'$ and \^H$_\theta'$. Assuming this is true, however, leads to strong constraints on the form of \^H$_S'$ and \^H$_\theta'$. In particular, when \^H$_S'$ is a function of quasimomentum (as in the present case), there will in general exist no additional CS operator. This result must remain valid when $\theta$ varies spatially. Indeed,  breaking translational invariance cannot introduce new symmetries in the system.

We can now turn to the other symmetries of the system, starting with TRS. The system has TRS symmetry if there is an anti-unitary operator $\hat{\mathcal{T}}$ which commutes with the Hamiltonian, and acts only within a single unit cell. Without loss of generality, we can express $\hat{\mathcal{T}}$ as the product of a unitary operator $\hat{\tau}$ and $\hat{\mathcal{K}}$, the complex conjugation operator: $\hat{\mathcal{T}}=\hat{\tau}  \hat{\mathcal{K}}$. The complex conjugation operator is an anti-unitary operator which acts as:
\begin{equation}
\langle \hat{\mathcal{K}} n | \hat{\mathcal{K}} \psi\rangle = \psi_n^*.
\end{equation}
Using $|k\rangle=\sum_n \exp(-iknd)|n\rangle$, we find that:
\begin{equation}
\langle \hat{\mathcal{K}} k | \hat{\mathcal{K}} \psi\rangle =
\sum_n e^{-iknd} \psi_n^*
=\psi(-k)^*.
\end{equation}
Thus $\hat{\mathcal{K}}$ sends $k\rightarrow -k$. Searching for a TRS operator therefore amounts to finding $\hat{\tau}$ such that:
\begin{equation}
\label{eq:AQW_TRS}
\hat{\tau}   \hat{\text{H}}_F'(-k)^T   \hat{\tau}^\dagger = \hat{\text{H}}_F'(k) ~ \Rightarrow ~ \hat{\tau}   \hat{\text{U}}'(-k)^T   \hat{\tau}^\dagger = \hat{\text{U}}'(k).
\end{equation}
As previously, if an anti-unitary $\hat{\mathcal{T}}$ simultaneously commutes with \^H$_S'(k)$ and \^H$_\theta'$, it automatically satisfies Eq.~(\ref{eq:AQW_TRS}). In this case however, no such operator exists. This is not sufficient to say that the system does not have TRS. As was the case with CS however, the existence of $\hat{\mathcal{T}}$ which satisfies Eq.~(\ref{eq:AQW_TRS}) without simultaneously commuting with \^H$_S'(k)$ and \^H$_\theta'$ would imply strong constraints on these matrices. These are in general not satisfied when \^H$_S'(k)$ and \^H$_\theta'$ are functions of independent variables.

\begin{figure}[t]
 \centering
 \includegraphics[width=.75\textwidth]{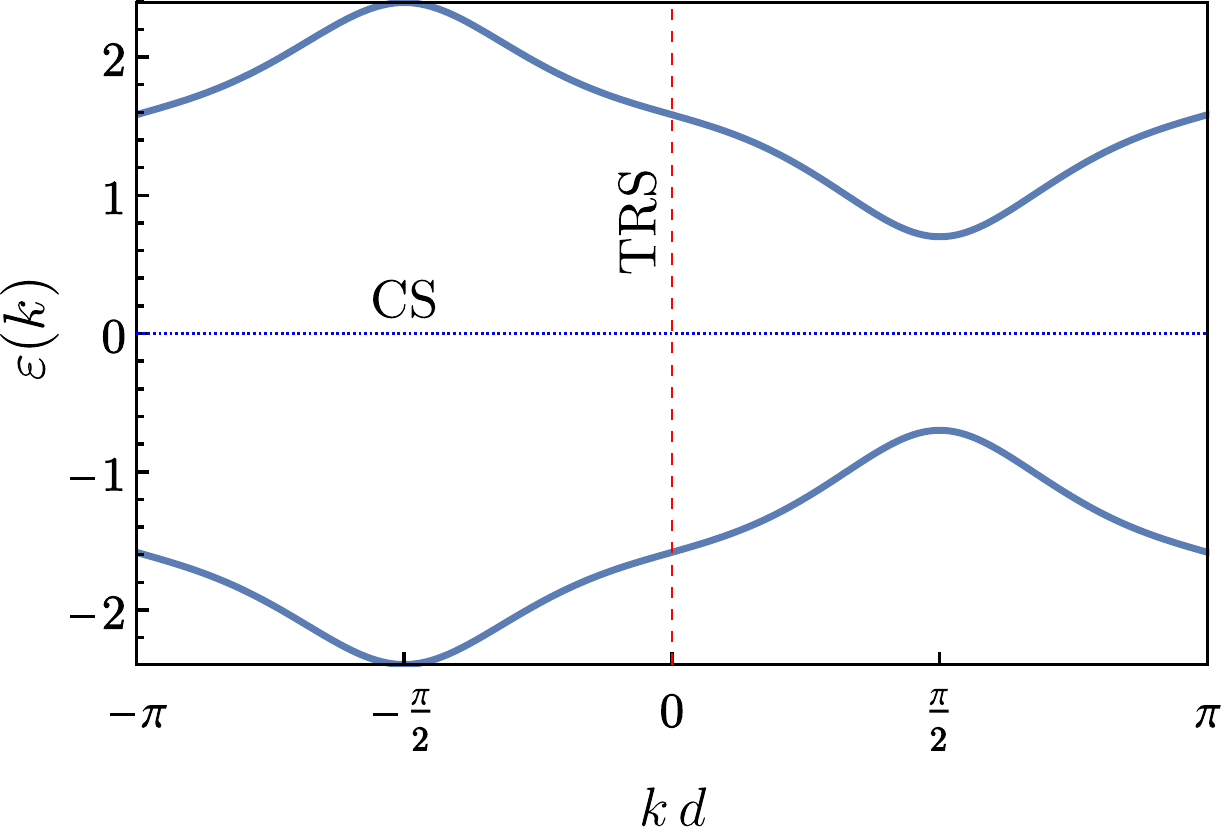}
 \caption[Dispersion in the two state basis]{Dispersion of $\h_F'$ for $J=\pi/3$, $\delta=0.42$ and $\theta=1.5$. A Hamiltonian possessing CS is symmetric around $\varepsilon=0$ (blue dotted line), while one with TRS is symmetric about $k=0$ (red dashed line). Clearly $\h_F'$ only has CS.}
 \label{fig:SimpleBasisDispersion}
\end{figure}

We can confirm numerically that the system does not have TRS by plotting the dispersion of \^H$_F'$. Indeed, TRS implies that any eigenstate of \^H$_F'$ has a partner eigenstate with equal energy and opposite quasimomentum. An example of the system's dispersion in this basis for $J=\pi/3$, $\delta=0.42$ and $\theta=1.5$ is shown in Fig.~\ref{fig:SimpleBasisDispersion}. Because the system's spectrum is manifestly not symmetric about $k=0$, we can conclude that the system does not present TRS.

Finally, the system has PHS symmetry if there is an anti-unitary operator $\hat{\mathcal{P}}$ which anti-commutes with the Hamiltonian. We define $\hat{\varrho}$, the unitary part of $\hat{\mathcal{P}}$, such that: $\hat{\mathcal{P}}=\hat{\varrho}  \hat{\mathcal{K}}$. This operator satisfies:
\begin{equation}
\label{eq:AQW_PHS}
\hat{\varrho}   \hat{\text{H}}_F(-k)^T   \hat{\varrho}^\dagger = -\hat{\text{H}}_F(k) ~ \Rightarrow ~ \hat{\varrho}   \hat{\text{U}}(-k)^T   \hat{\varrho}^\dagger = \hat{\text{U}}(k)^\dagger.
\end{equation}
We already know that there is no such operator due to the absence of TRS. Indeed, if \^U$'$ admitted both CS and PHS, their product would yield an anti-unitary matrix which commutes with the Hamiltonian, and this operator would satisfy Eq.~\eqref{eq:AQW_TRS}. As no such operator exists, we can conclude that PHS is also absent from this system. We conclude that the atomic quantum walk is constrained by CS and has broken TRS and PHS. Based on the study presented in Sec.~\ref{sec:TwoInvariants}, this system can have up to two non-zero winding numbers.

As a closing remark, we remind the reader that in Sec.~\ref{sec:CoinEff}, we approximated the full, time dependent Hamiltonian by Eq.~(\ref{eq:Htheta}). We were able to do this by saying that the spin mixing pulse is so short and intense that, when it is applied, \^H$_S$ is negligible. We now point out that, because the CS operator anti-commutes simultaneously with \^H$_S'$ and \^H$_\theta'$, even when our approximation breaks down, CS is not broken.

\subsection{Phase diagram}
\label{sec:TopoProp}

As we have seen in Sec.~\ref{sec:DoubleQW}, \^C$_\theta'$ brings the system into a superposition of right movers and left movers and in a superposition of the spin degree of freedom. As is visible from Eq.~(\ref{eq:CthetaKspace}), this results in spin-orbit coupling terms appearing in the Hamiltonian. This fact is essential for the non-trivial topological properties of the Floquet Hamiltonian to appear. We will see in this section that \^H$_F'$ can have non-zero winding numbers, which can be modified by changing the value of $\theta$ and $\delta$.

\subsubsection{Computation of the winding numbers}

We saw in Sec.~\ref{sec:Symmetries} that the atomic quantum walk presents chiral symmetry, enforced by the unitary chiral symmetry operator $\hat{\Gamma}$. Additionally, since the time evolution operator Eq.~(\ref{eq:U'}) has the form of Eq.~\eqref{eq:TSQW_UC}, it presents an inversion point. As we saw in Sec.~\ref{sec:nu0_nuPi}, this allows us to put it in the form:
\begin{equation}
\text{\^U}'=\hat{\Gamma}  \text{\^G}^\dagger  \hat{\Gamma}   \text{\^G},
\end{equation}
with
\begin{equation}
\label{eq:F}
\text{\^G}=e^{-i \text{\^H}_S'/2}  e^{-i \text{\^H}_\theta'/2}
=\begin{pmatrix}
a(k) & b(k)\\
c(k) & d(k)
\end{pmatrix},
\end{equation}
and where $a(k), ...,  d(k)$ are the entries of \^G, which are, in general, complex functions of $k$. We saw in Sec.~\ref{sec:nu0_nuPi} that a gapped Floquet system with one inversion point presents two topological invariants, the winding numbers $\nu_0$ and $\nu_\pi$. As was shown in Ref.\ \parencite{Asboth2014}, these can be calculated directly from \^G  using Eq.~\eqref{eq:evaluate_winding}. Specifically, in the basis where $\hat{\Gamma}=\sigma_3$, we have $\nu_0=\nu[b]$ and $\nu_\pi=\nu[d]$, with $\nu[b]$ and $\nu[d]$ the winding numbers of the $b(k)$ and $d(k)$ functions respectively.

Using Eq.~\eqref{eq:SigmaAntiCommut}, we can compute the exponential forms of Eqs.~(\ref{eq:Htheta'}) and (\ref{eq:Ht_Kspace}). Substituting into Eq.~(\ref{eq:F}), we find:
\begin{equation}
\label{eq:b(k)}
\begin{split}
b(k) &= \cos[\E(k)] \sin(\theta/2)\\
&-\sin[\E(k)] \cos(\theta/2) \frac{\delta \sin(k d)+i   J \cos(k d)}{\E(k)},
\end{split}
\end{equation}
and:
\begin{equation}
\label{eq:d(k)}
\begin{split}
d(k) &= \cos[\E(k)] \cos(\theta/2)\\
&+\sin[\E(k)] \sin(\theta/2) \frac{\delta \sin(k d)-i  J \cos(k d)}{\E(k)},
\end{split}
\end{equation}
where $\E(k)$ are the energy eigenvalues of \^H$_S'/2$:
\begin{equation}
\label{eq:EnHs}
\E(k)=\pm\sqrt{J^2 \cos^2(k d)+\delta^2 \sin^2(k d)}.
\end{equation}

\begin{figure}[t]
\centering
\includegraphics[width=0.7\textwidth]{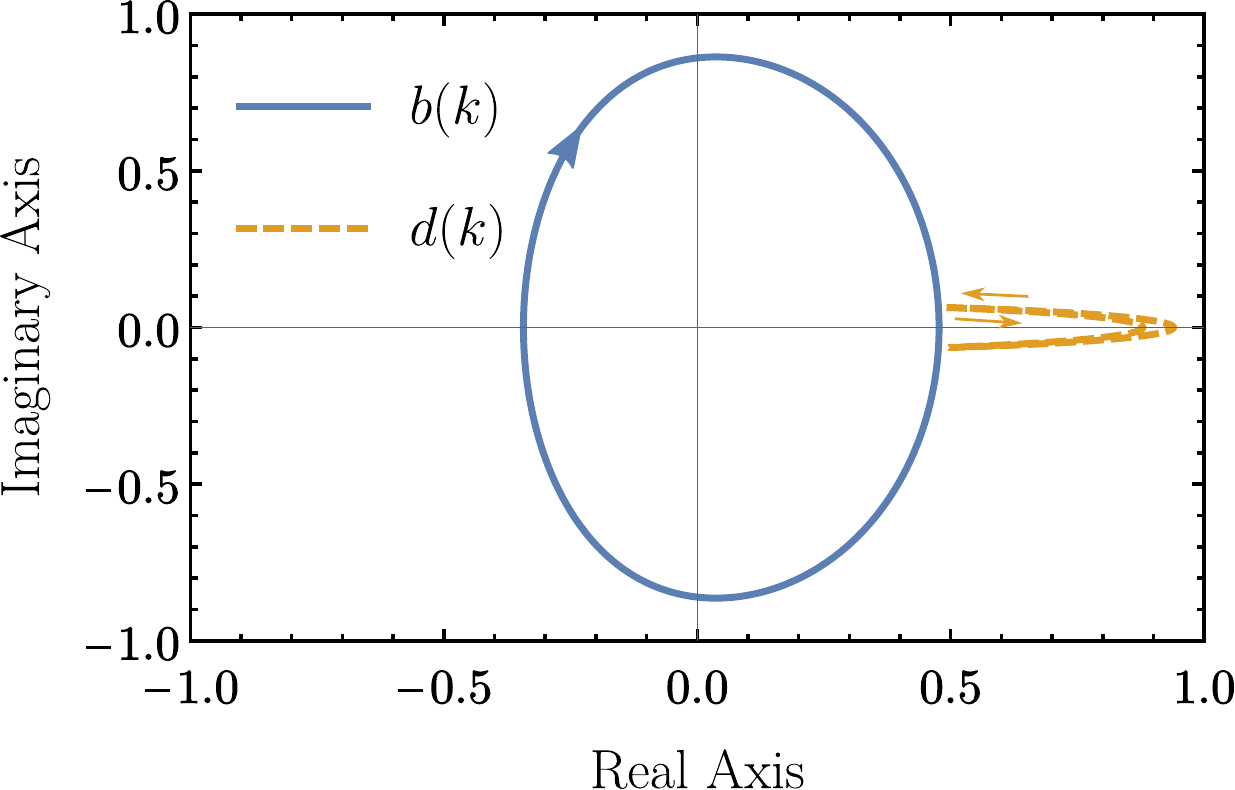}
\caption[Visualisation of the winding number]{Plots of functions $b(k)$ Eq.~(\ref{eq:b(k)}) (full blue) and $d(k)$ Eq.~(\ref{eq:d(k)}) (dashed orange) in the complex plane as $k$ goes from $-\pi/d$ to $\pi/d$ for $J=\pi/3$, $\delta=0.42$, and $\theta=0.15$. The number of times that $b(k)$ [$d(k)$] winds around the origin corresponds to the topological invariant $\nu_0$ ($\nu_\pi$). In this case we can read off $\nu_0=-1$ and $\nu_\pi=0$.}
\label{fig:winding_example}
\end{figure}

The simplest way to visualise the winding numbers $\nu_0$ and $\nu_\pi$ is to plot $b(k)$ and $d(k)$ in the complex plane as $k$ goes from $-\pi/d$ to $\pi/d$ (both bands have the same winding number). The system is topologically non-trivial if at least one of the curves winds around the origin. The curves that $b(k)$ and $d(k)$ form in the complex plane are presented for $J=\pi/3$, $\delta=0.42$, and $\theta=0.15$ in Fig.~\ref{fig:winding_example}. It is clear from this figure that the system can present non-zero winding numbers.

\subsubsection{Derivation of the phase boundaries}

The winding number $\nu_0$ cannot change unless both the real and imaginary parts of $b(k)$, Eq.~\eqref{eq:b(k)}, vanish simultaneously (see Fig.~\ref{fig:winding_example}). Thus the winding number can't change unless:
\begin{align}
\label{eq:Reb(k)}
\text{Re}[b(k)] & = \cos(\E(k)) \sin(\theta/2) -\delta \sin(k d) \cos(\theta/2) \frac{\sin[\E(k)]}{\E(k)}=0\\
\label{eq:Imb(k)}
\text{Im}[b(k)] & = -J \cos(k d) \cos(\theta/2) \frac{\sin[\E(k)]}{\E(k)}=0
\end{align}
Ignoring the trivial cases when $J=0$ or $\delta=0$, there are three special cases in which this can happen:

\begin{enumerate}

 \item When $\cos(kd)=0$, the imaginary part of $b(k)$ vanishes. From Eq.~\eqref{eq:EnHs}, this implies $\E=\pm \delta$. For Re$[b(k)]$, Eq.~\eqref{eq:Reb(k)}, to vanish simultaneously, we must have $\tan(\delta)=\pm\tan(\theta/2)$.
 
 \item When $\cos(\theta/2)=0$, the imaginary part of $b(k)$ vanishes. For Re$[b(k)]$, Eq.~\eqref{eq:Reb(k)}, to vanish simultaneously, we must have $\cos(\E(k))=0$. The band closing then occurs at $k=k_0$, when there exists a quasimomentum satisfying: 
 \begin{equation}
  \cos(k_0 d)^2 = \frac{(\pi/2)^2 - \delta^2}{J^2 - \delta^2}.
 \end{equation}
 As a consequence, this band touching only happens when $|\delta|\geq \pi/2 \geq |J|$ or when $|\delta|\leq \pi/2 \leq |J|$.
 
 \item When $\sin(\theta/2)=0$ and $\E(k)=\pm \pi$, then Eqs.~\eqref{eq:Reb(k)} and \eqref{eq:Imb(k)} vanish simultaneously. This happens at quasimomentum $k_0$ such that:
 \begin{equation}
  \cos(k_0 d)^2 = \frac{\pi^2 - \delta^2}{J^2 - \delta^2},
 \end{equation}
 which only exists when $|\delta|\leq \pi \leq |J|$ or when $|\delta|\geq \pi \geq |J|$.
 
\end{enumerate}

We can perform a similar analysis for the function $d(k)$ to understand under what conditions the winding number $\nu_\pi$ is allowed to change. The band touching events are summed up in the Table \ref{tab:BandTouching}. Note that we have excluded events where $J=0$ or $\delta=0$.

\begin{table}[t]
\centering
\caption[Band touching events of $\h_F'$]{Conditions for a band touching in the dispersion of $\h_F'$ to take place. For a particular band touching event, the first column shows which winding number can change, the second column shows the value of $\theta$ ($\theta$ is $2\pi$ periodic), the third column the value of $\E(k_0)$, Eq.~\eqref{eq:EnHs}, where $k_0$ is the quasimomentum value at which the band gap closing takes place, shown in the fourth column. The additional constraints on $J$ and $\delta$ for this event to be allowed are shown in the fifth column.}
\label{tab:BandTouching}
\begin{tabular}{ |c|c|c|c|c| }
\hline
$\Delta \nu$ & $\theta$ & $\E(k_0)$ & $\cos(k_0 d)$ & conditions \\
\hline
\hline
$\Delta \nu_0$ & $\theta=\pm 2 \delta$ & 
\multirow{2}{*}{$\pm \delta$} & 
\multirow{2}{*}{$0$} & 
\multirow{2}{*}{$\forall J, \delta$} \\
\cline{1-2}
$\Delta \nu_\pi$ & $\theta=\pm \left( 2 \delta \pm \pi/2 \right)$ & & & \\
\hline
$\Delta \nu_0$ & $\theta=\pm \pi$ &
\multirow{2}{*}{$\pm\pi/2$} &
\multirow{2}{*}{$\pm\sqrt{\frac{(\pi/2)^2 - \delta^2}{J^2 - \delta^2}}$} &
\multirow{2}{*}{\specialcell{$|J| \geq \pi/2 \geq |\delta|$ \\ or $|J| \leq \pi/2 \leq |\delta|$}}  \\
\cline{1-2}
$\Delta \nu_\pi$ & $\theta=0$ & & &\\
\hline
$\Delta \nu_0$ & $\theta=0$ &
\multirow{2}{*}{$\pm\pi$} &
\multirow{2}{*}{$\pm\sqrt{\frac{\pi^2 - \delta^2}{J^2 - \delta^2}}$} &
\multirow{2}{*}{\specialcell{$|J| \geq \pi \geq |\delta|$ \\ or $|J| \leq \pi \leq |\delta|$}}  \\
\cline{1-2}
$\Delta \nu_\pi$ & $\theta=\pm \pi$ & & &\\
\hline
\end{tabular}
\end{table}

Note that the band touching in the four lower rows of Table \ref{tab:BandTouching} satisfy $\theta=-\theta$. However, the trajectories of $b(k)$ and $d(k)$, represented, e.g., in Fig.~\ref{fig:winding_example}, are conserved under $\theta \rightarrow -\theta$. This can be explained by inspection of Eqs.~\eqref{eq:b(k)} and \eqref{eq:d(k)}:
\begin{equation}
\text{for } \theta \rightarrow -\theta \text{ and } k \rightarrow k +\pi :
\begin{cases}
b(k) \rightarrow -b(k)\\
d(k) \rightarrow d(k)
\end{cases}
\end{equation}
implying that the winding numbers must stay the same when $\theta \rightarrow -\theta$. Therefore, the topologically non-trivial band touchings only occur at $k= \pm \pi/2$ (first two rows of Table \ref{tab:BandTouching}).

\begin{figure}[t]
\centering
\includegraphics[width=0.6\textwidth]{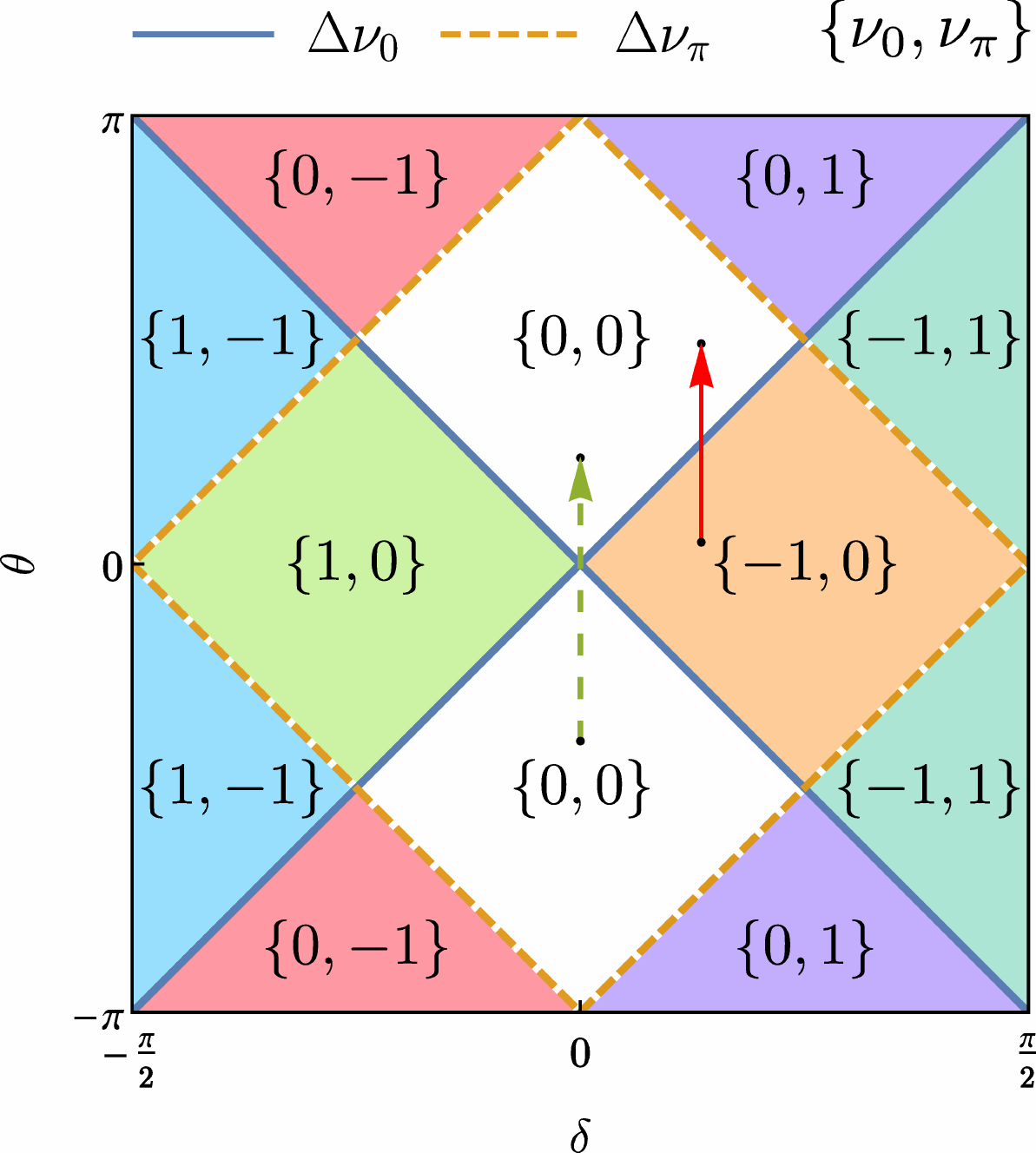}
\caption[Phase diagram]{Phase diagram of the atomic quantum walk. The function $b(k)$ [$d(k)$] vanishes along the full blue (dashed orange) lines, indicating a boundary along which the topological invariant $\nu_0$ ($\nu_\pi$) can change its value. The winding numbers for each region of the plot have been calculated using Eq.~(\ref{eq:evaluate_winding}), and are specified on the figure as $\{\nu_0,\nu_\pi\}$; similarly coloured regions have the same winding numbers. In Sec.~\ref{sec:BSsimulation}, we produce an $\varepsilon=0$ bound state by varying $\theta$ spatially along the path in parameter space indicated by a red arrow. The topological boundary $\theta=2 \delta$ is crossed in this process. In Sec.~\ref{sec:BSofHapprox}, we cross a trivial band gap closing which occurs for $\delta=0$. This path in parameter space is indicated by a dashed green arrow.}
\label{fig:phase_diagram}
\end{figure}

We can evaluate the winding numbers in each region of parameter space using the general formula Eq.~(\ref{eq:evaluate_winding}). This allows us to construct the topological phase diagram of the atomic quantum walk, which is presented in Fig.~\ref{fig:phase_diagram}. When $J=\pi/3$ (value used in the simulations presented in this paper), no topologically trivial band touchings occur in the interval considered in the Fig.~\ref{fig:phase_diagram}. As is visible from Fig.~\ref{fig:phase_diagram}, the winding number of the atomic quantum walk is a function of $\delta$ and $\theta$. In the following section, we will use this to create two regions with distinct topological properties.

\subsection{Mapping to the Creutz ladder}
\label{sec:MapToCreutz}

In this section, we will show that, under a simple change of basis, the system maps exactly onto a well known topologically non-trivial 1D system, the Creutz ladder \parencite{Creutz1999, Bermudez2009, Sticlet2014}. We will then use this knowledge to re-derive the phase diagram shown in Fig.~\ref{fig:phase_diagram}.

\begin{figure}[t]
\centering
\includegraphics[width=0.55\textwidth]{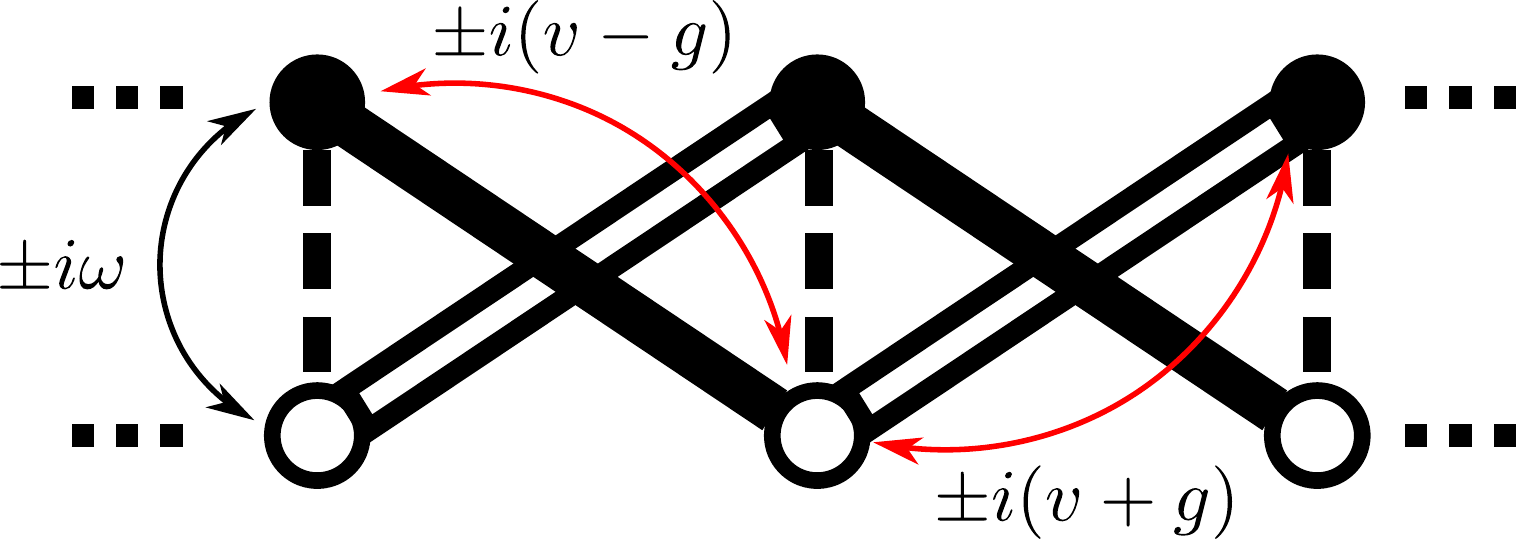}
\caption[Sketch of the Creutz ladder]{1D Creutz ladder. The particle has amplitude $\pm i w$ of tunnelling along the rungs of the ladder (dashed line), and diagonal tunnelling amplitudes $\pm i(v-g)$ (solid line) and $\pm i(v+g)$ (double line). The tunnelling amplitudes are indicated by curved arrows, the colour of which is unimportant. Sites belonging to the upper (lower) part of the ladder are indicated by full (empty) circles. This system is topologically non-trivial, with winding number given by Eq.~(\ref{eq:Creutz_winding}).}
\label{fig:Creutz_ladder}
\end{figure}

The Creutz ladder describes a spinless particle hopping in a 1D ladder, as sketched in Fig.~\ref{fig:Creutz_ladder}. The particularity of this model is that the particle has amplitude to hop in the diagonal directions.
The phase diagram of Fermions in the Creutz ladder was recently studied in Ref.~\parencite{Junemann}.

The Hamiltonian of the Creutz ladder has the form:
\begin{equation}
\label{eq:H_Creutz}
\text{\^H}_C = \frac{1}{2} \sum_n \left[ \cc_{n}^\dagger w  \sigma_2 \cc_n +\cc_{n+1}^\dagger(i v  \sigma_1-g \sigma_2)\cc_n + \text{H.c} \right].
\end{equation}
As previously, $\cc^\dagger_n$ ($\cc_n$) creates (annihilates) a particle with two internal states on site $n$. The $\sigma_i$ matrices denotes the Pauli matrices acting in the space of sites perpendicular to the axis of the ladder (represented in the vertical direction in Fig.~\ref{fig:Creutz_ladder}), with $i\in\{1,  2,  3\}$. We will assume that the hopping amplitudes $v$, $w$ and $g$ are all real parameters, and  block-diagonalise Eq.~(\ref{eq:H_Creutz}) by Fourier transformation:
\begin{equation}
\label{eq:H_Creutz_Kspace}
\text{\^H}_C(k) = -v \sin(k d)\sigma_1+[w-g \cos(k d)]  \sigma_2,
\end{equation}
where we have set the ladder's unit cell size to $d$. The Creutz ladder belongs to the BDI class of the topological classification of Hamiltonians \parencite{Sticlet2014}. As a result, this system admits a non-zero winding number  $\nu_C$ , the value of which depends on the system's parameters as:
\begin{equation}
\label{eq:Creutz_winding}
\nu_C=\frac{1}{2}[\text{sgn}(w+g)-\text{sgn}(w-g)].
\end{equation}

We will now show that we can map the atomic quantum walk to the Creutz ladder. We consider the translational invariant atomic quantum walk in the two-state basis presented in Sec.~\ref{sec:SimpleBasis}, and change the origin of momentum $k\rightarrow k+\pi/(2 d)$. In this basis, Eq.~(\ref{eq:Ht'}) becomes:
\begin{equation}
\label{eq:Ht''}
\text{\^H}_S'(k)\rightarrow \text{\~H}_S(k) = -2  J\sin(k d) \sigma_1 + 2 \delta\cos(k d) \sigma_2.
\end{equation}
The Hamiltonian \^H$_\theta'$ given by Eq.~(\ref{eq:Htheta'}) is not modified by this transformation. Note that this is a trivial gauge transformation that cannot change the topological properties of the system. Interestingly, in this basis, the system has CS, TRS and PHS. The operators implementing these symmetries are detailed in the Table \ref{tab:AQW_Sym}. The method used to determine the system's symmetries is detailed in Sec.~\ref{sec:Symmetries}.

\begin{table}[t]
\centering
\caption[Symmetries of \~H$_F$]{In the first column, the operators implementing various symmetries of $\tilde{\text{H}}_F$ are listed. We list the squares of these operators in the second column.}
\label{tab:AQW_Sym}
\begin{tabular}{ |c|c|c| }
\hline
symmetry & operator & square\\
\hline
Chiral symmetry & $\sigma_3$ & $\sigma_0$ \\
Time reversal symmetry & $\sigma_3  \hat{\mathcal{K}}$ & $\sigma_0$ \\
Particle hole symmetry & $\hat{\mathcal{K}}$ & $\sigma_0$ \\
\hline
\end{tabular}
\end{table}

From Table \ref{tab:AQW_Sym}, we find that the system presents CS, TRS and PHS, all of which square to the identity. This tells us immediately that we are in the BDI class of the topological classification of Hamiltonians, i.e the same symmetry class as the Creutz ladder.

\begin{figure}[t]
\centering
\includegraphics[width=0.6\textwidth]{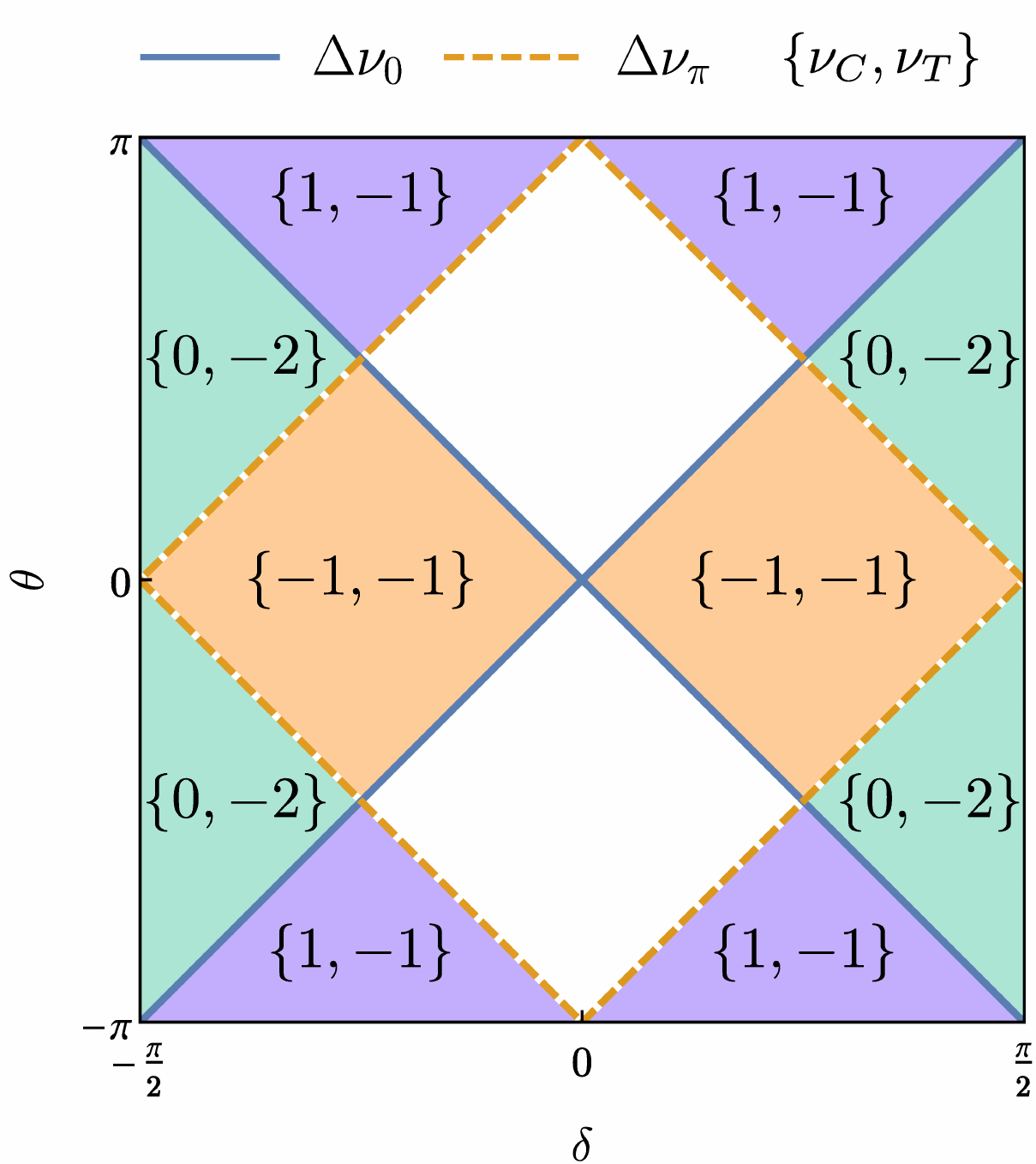}
\caption[Phase diagram (using the mapping to the Creutz ladder)]{Winding numbers $\{\nu_C,\nu_T\}$ of the atomic quantum walk when $J=\delta$. In the time-frame Eq.~\eqref{eq:U'}, the system maps onto the Creutz ladder and $\nu_C$ is accurately predicted by Eq.~(\ref{eq:Creutz_winding}). $\nu_T$ is the winding number in the other symmetric time frame Eq.~\eqref{eq:timeframe2}. These are related to the winding numbers $\nu_0,\nu_\pi$ through Eqs.~\eqref{eq:nuC} and \eqref{eq:nuT}.}
\label{fig:Creutz_phase_diagram}
\end{figure}

We will now proceed to show that \~H$_F$, the Floquet Hamiltonian, maps onto the Creutz ladder when $J=\delta$. We can find \~H$_F$ by substituting Eqs.~(\ref{eq:Htheta'}) and (\ref{eq:Ht''}) into Eq.~(\ref{eq:U'}). In the limit $J=\delta$, \~{H}$_F$ is:
\begin{equation}
\begin{split}
& \text{\~H}_F= 
\pm\frac{\tilde{\E}(k)}{\sin(\tilde{\E}(k))}\Big\{-\text{sgn}(\delta)\sin(2\delta)\sin(k d) \sigma_1\\
& +[-\cos(2\delta)\sin(\theta)+\text{sgn}(\delta) \cos(\theta)\sin(2\delta)\cos(k d)] \sigma_2\Big\}.
\end{split}
\end{equation}
Apart from the upfront $\tilde{\E}(k)/ \sin(\tilde{\E}(k))$, this is exactly the Creutz ladder Hamiltonian Eq.~(\ref{eq:H_Creutz_Kspace}), with:
\begin{eqnarray}
\label{eq:CreutzTunnelling1}
v & = & \text{sgn}(\delta)\sin(2\delta),\\
\label{eq:CreutzTunnelling2}
w & = & -\text{sgn}(\delta)\cos(2\delta)\sin(\theta),\\
\label{eq:CreutzTunnelling3}
g & = & -\cos(\theta)\sin(2\delta).
\end{eqnarray}
By substituting these values in Eq.~(\ref{eq:Creutz_winding}), we can calculate the winding number $\nu_C$ and deduce the phase diagram in this time frame, which is represented in Fig.~\ref{fig:Creutz_phase_diagram}. While we expect the upfront term $\tilde{\E}(k)/ \sin(\tilde{\E}(k))$ to deform the band structure, it does not change the symmetry properties of \~H$_F$, and therefore cannot change its topological properties as long as it does not close the band gap.

In expressing \^U$'$ in the symmetric form for Eq.~\eqref{eq:U'}, which has the form of Eq.~\eqref{eq:TSQW_UC}, we made a choice of time frame. There exists another time frame which has an inversion point in time and has the form of Eq.~\eqref{eq:TSQW_UT}:
\begin{equation}
\label{eq:timeframe2}
\text{\^U}_T'=e^{-i\Hs'/2}e^{-i\Htheta'}e^{-i\Hs'/2}.
\end{equation}
While the system does not map onto the Creutz ladder in this time frame, it does have a winding number $\nu_{T}$, which is the second topological invariant represented on Fig.~\ref{fig:Creutz_phase_diagram}. By comparing Figs.~\ref{fig:phase_diagram} and \ref{fig:Creutz_phase_diagram} when $\delta>0$, we notice that $\nu_C$ and $\nu_T$ are related to $\nu_0$ and $\nu_\pi$ by Eqs.~\eqref{eq:nuC} and \eqref{eq:nuT}, in agreement with Sec.~\ref{sec:nu0_nuPi}. What is more, we can construct the $\delta<0$ half of the phase diagram simply by setting $J=\delta<0$. When we do this, all winding numbers from Fig.~\ref{fig:phase_diagram} change sign, such that Eqs.~\eqref{eq:nuC} and \eqref{eq:nuT} also hold when $\delta<0$. 


\section{Detection of the topological bound state}
\label{sec:BoundState}

\subsection{Experiment suggested}
\label{sec:DetectionSuggestion}

As we saw in Sec.~\ref{sec:TopoProp}, the topological properties of the atomic quantum walk can be modified by changing the spin mixing angle $\theta$. By using spatially inhomogeneous Raman lasers, it is therefore possible to create two regions in the system with distinct topological properties. At the boundary between two regions which have different winding numbers, there lives a robust bound state which is protected by the system's symmetries and pinned either at $\varepsilon=0$ or $\varepsilon=\pi$. In the following, we suggest a method for experimentally generating this topological bound state, and identifying it using its characteristic spin distribution. The density distribution can either be retrieved in a single measurement, if the quantum walk is performed with a gas of non-interacting atoms, or by repeating the experiment many times and averaging if a single atom is used. We will be interested in the spin populations of sublattices $A$ and $B$, meaning that the position measurement must have single site resolution and be sensitive to the atom's spin. The location of $A$ and $B$ sites is fixed by the lasers generating the optical lattice (see Fig.~\ref{fig:spin_dependent_lattice}), so that their position will remain the same from one experiment to the next.

\begin{figure}[t]
\centering
\includegraphics[width=0.8\textwidth]{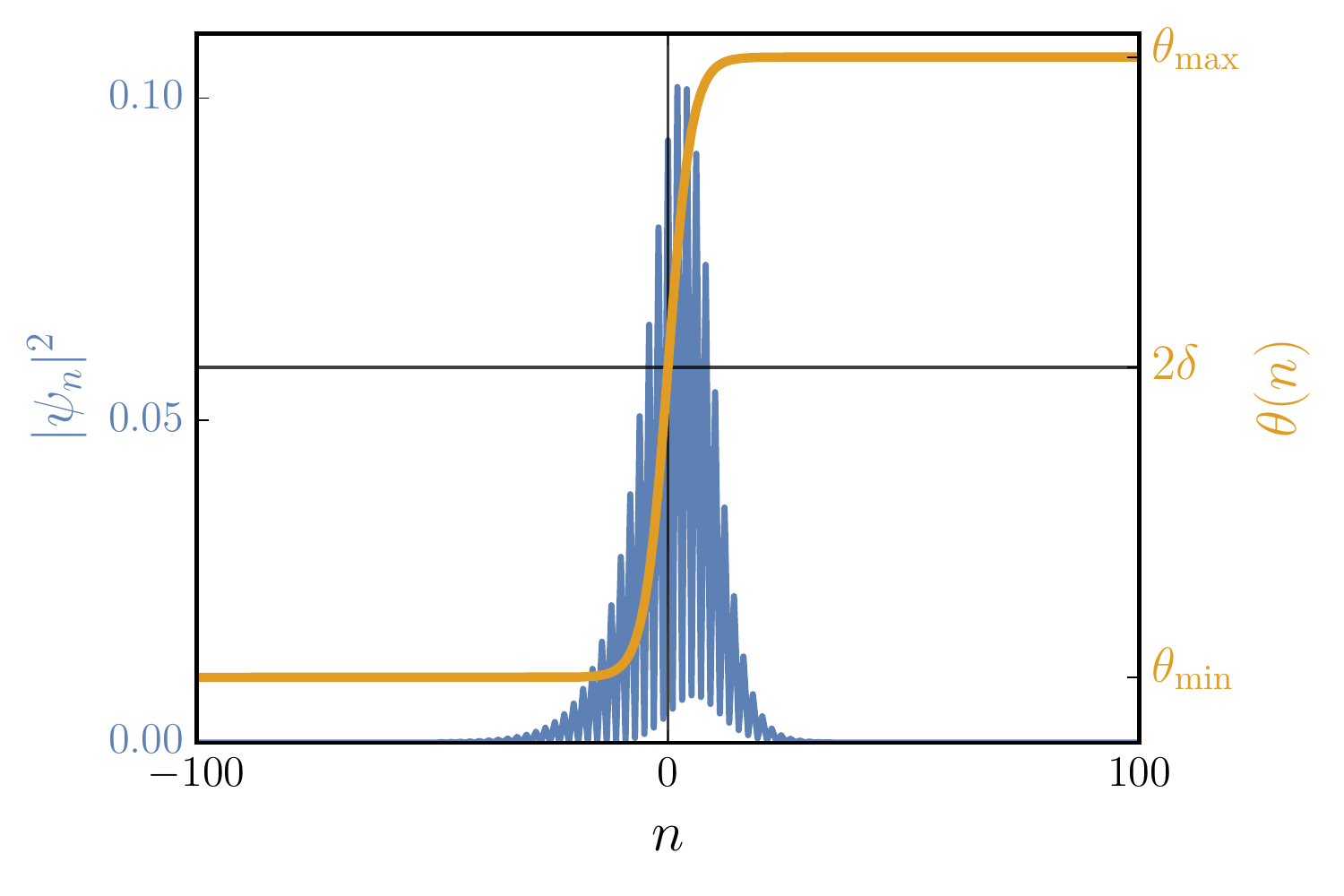}
\caption[Topological interface realised by varying $\theta$]{Topological interface created by varying $\theta$ spatially from $\theta_{\rm min}$ to $\theta_{\rm max}$. In orange: $\theta$ varies spatially according to Eq.~(\ref{eq:theta_function}). The topological boundary occurs at $n=0$, which is the point where $\theta=2 \delta$. In blue: topological bound state occurring at this boundary, obtained by diagonalising \^H$_F$ for $J=\pi/3$, $\delta=0.42$, $\theta_{\rm min}=0.15$, $\theta_{\rm max}=1.5$ and $\xi=10d$. This path in parameter space is indicated on Fig.~\ref{fig:phase_diagram} by a red arrow.}
\label{fig:theta_function}
\end{figure}

When performing this experiment, we expect a portion of the density to remain pinned to the topological boundary. This corresponds to the part of the initial state which overlaps with the bound state wave function. The rest of the density is translated ballistically away from the topological interface. We verify this numerically by performing simulations, where $\theta$ is varied according to:
\begin{equation}
\label{eq:theta_function}
\theta(n)=\frac{\theta_{\rm max}+\theta_{\rm min}}{2}+\frac{\theta_{\rm max}-\theta_{\rm min}}{2} \tanh\left(\frac{n d}{\xi} \right),
\end{equation}
where $\xi$ determines the width of the region where $\theta(n)$ changes value. This function is represented in Fig.~\ref{fig:theta_function}. Note that the existence of the bound state only requires that $\theta(n)$ crosses $2\delta$, and does not depend otherwise on the precise form of Eq.~\eqref{eq:theta_function}. Our simulations are performed in the basis where \^H$_S$ and \^H$_\theta$ have the form given by Eqs.~(\ref{eq:Ht}) and (\ref{eq:C}) respectively.

\subsection{Simulation}
\label{sec:BSsimulation}

\begin{figure}[t]
\centering
\includegraphics[width=\textwidth]{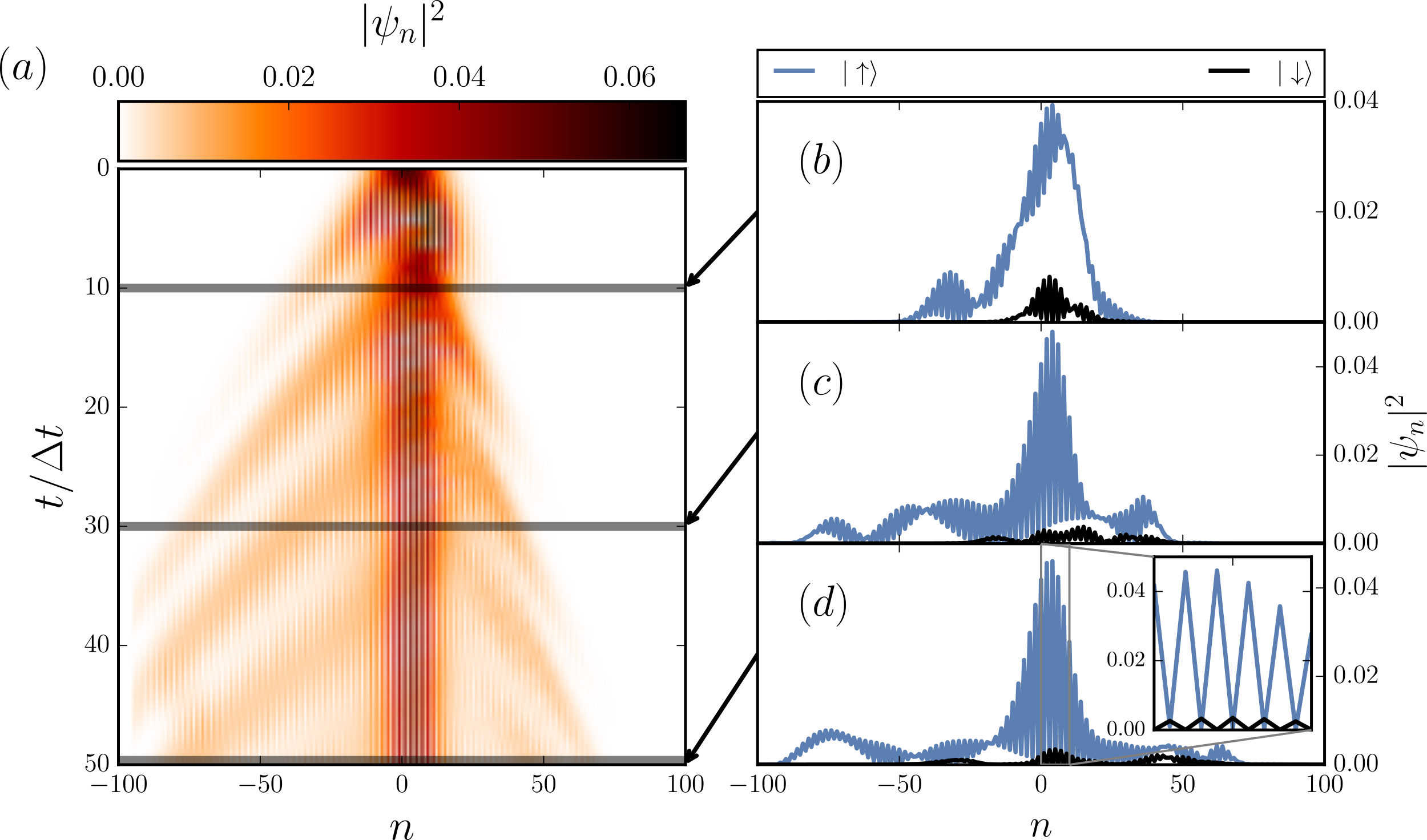}
\caption[Total density distribution when a bound state is populated]{(a) Atom density in position space versus time. After $t/T=50$ time-steps, some density has escaped ballistically to infinity but the probability density function remains sharply peaked around the origin. (b), (c) and (d) Probability density function at times $t_1=10$, $t_2=30$ and $t_3=50$ respectively, with $\uparrow$ represented in blue (grey) and $\downarrow$ in black. The inset of (d) shows the density in the interval $n\in [0,  10]$. We observe that $\uparrow$ ($\downarrow$) states have non-zero density only on even (odd) sites. This simulation was realised with $J=\pi/3$, $\delta=0.42$, and $\theta$ varying spatially from $\theta_{\rm min}=0.15$ to $\theta_{\rm max}=1.5$. The initial state was a Gaussian wavepacket centred around site $n=0$ with mean quasimomentum $k=\pi/(2d)$, in $\uparrow$ state with equal support on $A$ and $B$ sublattices.}
\label{fig:pdf_BS}
\end{figure}

Fig.~\ref{fig:pdf_BS}(a) shows an example of the evolution of the atomic density during 50 time-steps. As expected, we find that a portion of the density remains pinned to the region of $n=0$, which is the location of the topological boundary. Atoms which do not populate the bound state are transported ballistically away from $n=0$. The spreading is anisotropic due to the initial state we chose, which is fully $\uparrow$ polarised. The spreading density shows interference patterns between forward and backward travelling atoms, as is characteristic for quantum walks (indeed, we already observed this behaviour in Fig.~\ref{fig:AQW_simulation}).

The density at specific moments in time is represented in Fig.~\ref{fig:pdf_BS}(b), (c) and (d). We observe that at late times, the probability distribution is exponentially peaked at the location of the topological boundary. Importantly, the relative population of this state does not decrease at later times. We have verified this numerically by computing the overlap between the atom's wave function and the $\varepsilon=0$ eigenstate of \^H$_F$, and found that it is time independent.

\begin{figure}[t]
\centering
\includegraphics[width=0.7\textwidth]{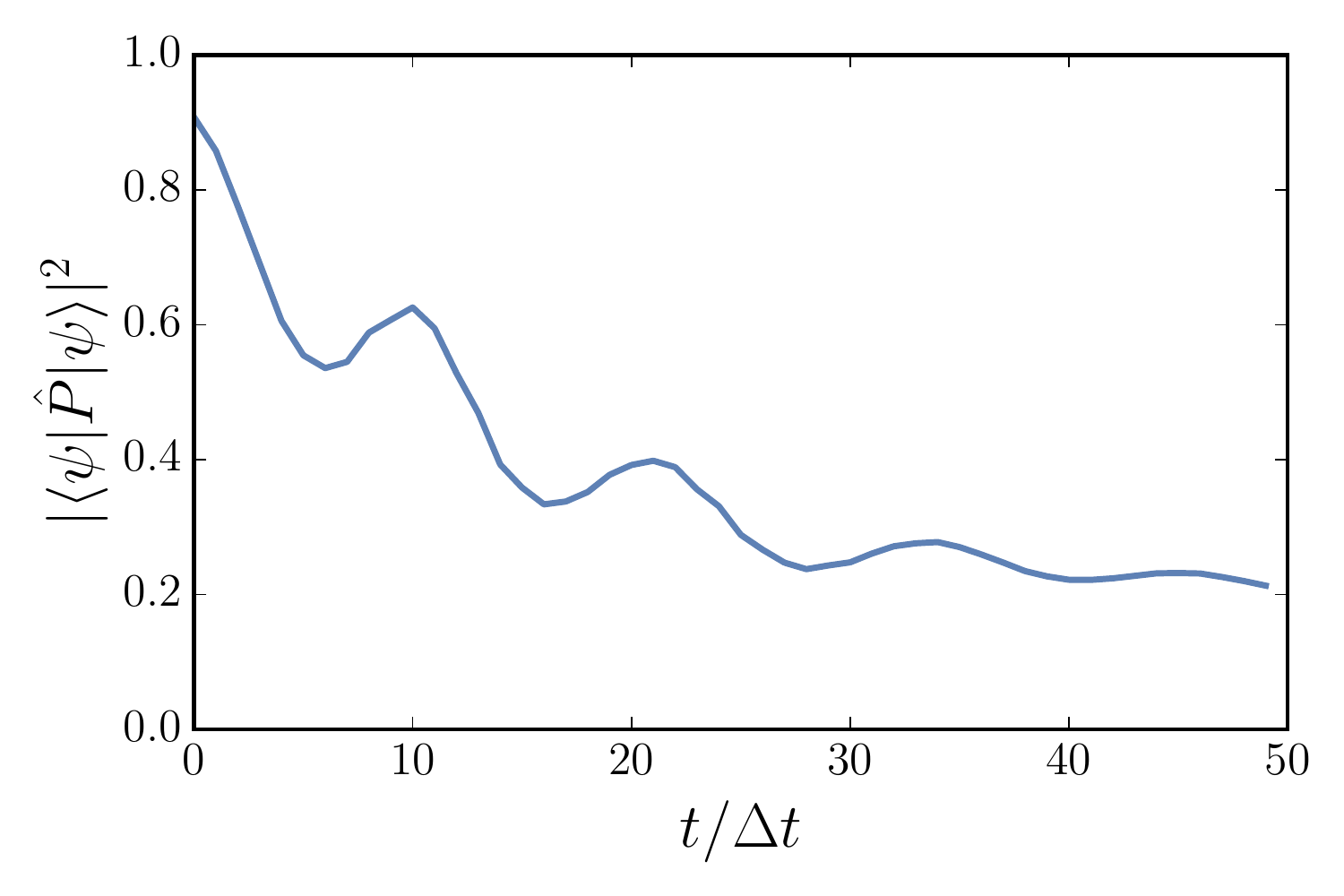}
\caption[Full atomic density in neighbourhood of the bound state]{Full atomic density, obtained from the simulation Fig.~\ref{fig:pdf_BS}, for $50$ time-steps in the region $n\in[-15, 15]$. This interval corresponds to the neighbourhood of the topological boundary, and $\hat{P}$ is the projector onto this region. At early times we observe oscillations as atoms which are not trapped in the bound state leave the region of the boundary. At late times, we see that the total density converges to a non-zero value.}
\label{fig:FullDensBS}
\end{figure}

Experimentally, this can be verified by plotting the total density in the neighbourhood of the topological boundary, which is displayed in Fig.~\ref{fig:FullDensBS}. We see that at early times, the total density near the origin decreases rapidly as atoms which do not populate the bound state are transported away from the topological boundary. During this period, we notice that the total density presents oscillations; these are due to the interference of ballistically transported atoms. At late times, the total density near the topological boundary converges to a non-zero value, which is a sign that atoms populating the bound state do not leak into other states of the system.

\subsection{Identification of the topological protection}
\label{sec:TopoIdentification}

A strong experimental signature of topological protection is the mapping of the phase boundaries. Indeed, by inspecting Fig.~\ref{fig:phase_diagram} we can deduce the parameter regimes in which we expect to find topological bound states. By knowing how the parameter $\theta$ varies, we can determine the position around which the bound states are centred. We can explore the system's parameter space by sampling different values for $\delta$ and $\theta$, and verify that exponentially bound states occur at all topological boundaries. This provides a straightforward way to verify that we are indeed observing topological bound states, and relies only on imaging the atoms' probability density function.

An alternative method for verifying that a state is a topological bound state is to verify that it is an eigenstate of the chiral symmetry (CS) operator $\hat{\Gamma}$. Indeed, we saw in Sec.~\ref{sec:1dTopoIntro} that if a state verifies the following conditions, it is a topological bound state:
\begin{enumerate}
\item The state is an eigenstate simultaneously of $\hat{\Gamma}$ and the Floquet Hamiltonian \^H$_F$.
\item The state has vanishing overlap with any other state of equal energy which satisfies condition 1.
\end{enumerate}
In the following, we describe a method to identify an eigenstate of $\hat{\Gamma}$, thereby providing a strong way of identifying a topological bound state.

We can verify that the bound state we are observing is a single eigenstate of \^H$_F$ directly from the time evolution of probability density distribution. Indeed, if this state was a superposition of states, we would observe Rabi oscillations, while it is clear from figure Fig.~\ref{fig:pdf_BS}(a) that the density distribution near the origin remains the same after a period of driving. Additionally, it can be seen from the inset of Fig.~\ref{fig:pdf_BS}(d) that the state which is found near the origin at late times has an extremely interesting structure. Indeed, we find that $\uparrow$ states only occur on even sites, while $\downarrow$ states only occur on odd sites. This is a direct result of being an eigenstate of CS, which has the form $\hat{\Gamma}=\tau_3\otimes\sigma_3$ in this basis, leading to a constrained spin distribution. Importantly, all other eigenstates of \^H$_F$ must transform into their chiral partner under the action of CS. The only states that can do this have equal $\uparrow$ and $\downarrow$ density on each site. Consequently, the state we are measuring near $n=0$ can only be a topological bound state.

Note that this type of spin structure is reminiscent of the one encountered in Sec.~\ref{sec:nu0_nuPi}. Indeed, all bound states protected by chiral symmetry show a spin distribution which is determined by the chiral symmetry operator. As a result, this method can be generalised to identify any bound state protected by chiral symmetry.

As can be read off from Fig.~\ref{fig:AQW_simulation}, an evolution time of $50$ time steps on an optical lattice of $200$ sites is sufficient to observe the topological bound state in this system. This observation should not be greatly affected by a small variation of these values.

\begin{figure}[t]
\centering
\includegraphics[width=0.85\textwidth]{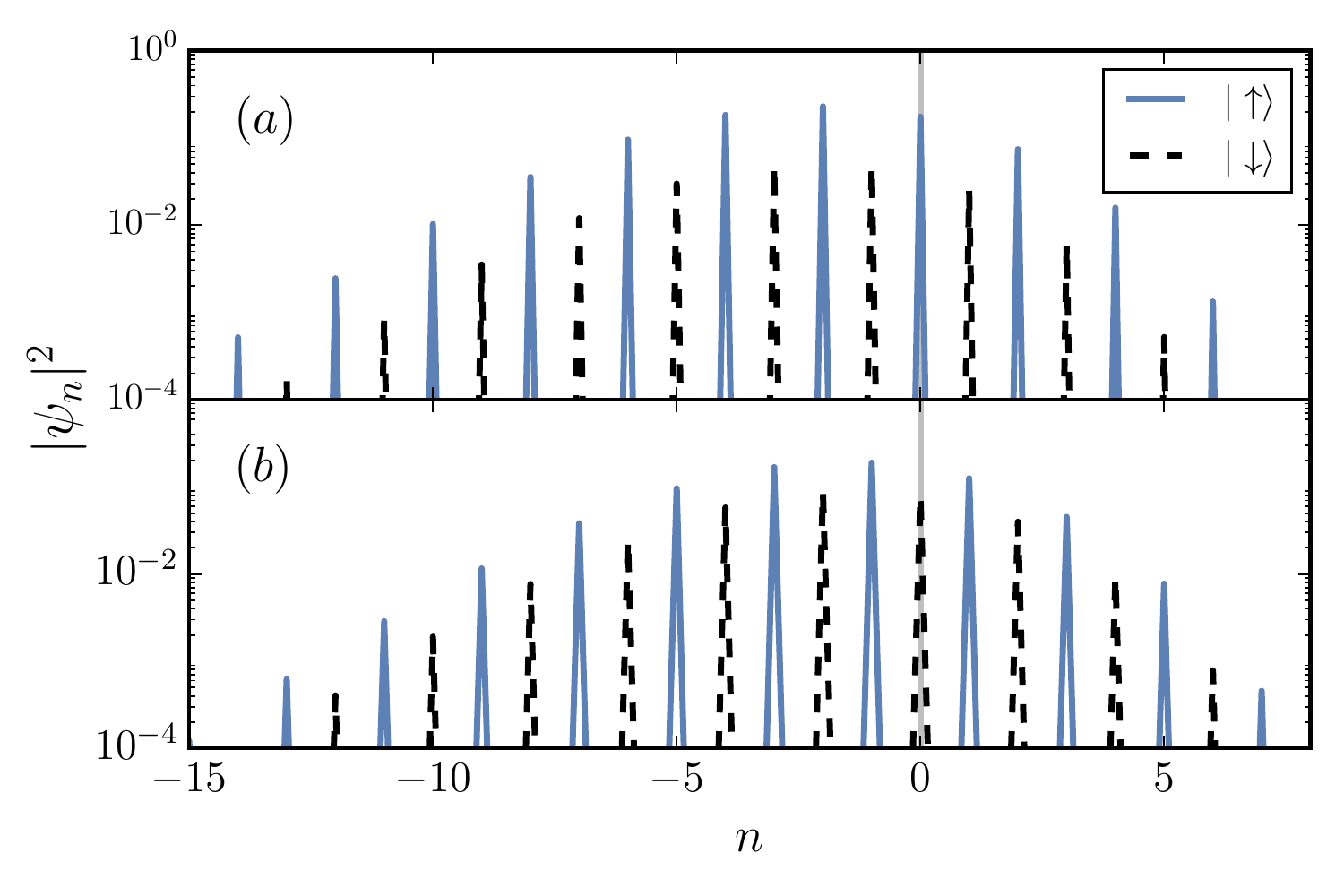}
\caption[Bound state at half time step]{(a) Density distribution of an $\varepsilon=\pi$ topological bound state (plotted on a log scale), obtained by diagonalising \^H$_F$ (not \^H$_F'$). $\uparrow$ ($\downarrow$) density is represented in blue (black). The state is exponentially localised in the neighbourhood of $n=0$. $\uparrow$ ($\downarrow$) states are only found on even (odd) sites. Only eigenstates of the chiral symmetry operator have this  complex spin distribution. (b) The same state, after evolution through half a time-step. $\uparrow$ ($\downarrow$) states are only found on odd (even) sites. For clarity, a grey vertical line has been drawn at $n=0$. States with $\varepsilon=\pi$ show an inversion in their spin density distribution at half time-steps.}
\label{fig:eigenstate_half_time_step}
\end{figure}

A spin sensitive density measurement therefore provides a direct
method to identify an $\varepsilon=0$ or an $\varepsilon=\pi$ bound
state protected by CS in a 1D system. Using the exact same measurement
at half time-steps, it is possible to discriminate between $\varepsilon=0$ and
$\varepsilon=\pi$ energy states. A half time-step is performed by applying a
spin rotation $\theta/2$, followed by time evolving with \^H$_S$ for a
time $T/2$. As we showed in Sec.~\ref{sec:nu0_nuPi}, the dynamics
of the bound state between two time-steps is sensitive to the state's
energy. As shown in Fig.~\ref{fig:eigenstate_half_time_step}(a), at
integer time-steps, the $\uparrow$ ($\downarrow$) density of $\varepsilon=\pi$
states only has support on even (odd) sites. At half time-steps, the
spin structure of $\varepsilon=\pi$ states is reversed, with spin $\uparrow$
($\downarrow$) states only on odd (even) sites. This is shown in
Fig.~\ref{fig:eigenstate_half_time_step}(b). This contrasts with the
behaviour of $\varepsilon=0$ states, which keep the same spin structure at
integer time-steps and half integer time-steps. Thus, performing a
position measurement with single site resolution which is sensitive to
the spin state does not only identify a topological
bound state of the system, but it also provides a method to
differentiate an $\varepsilon=0$ from an $\varepsilon=\pi$ state.

To recapitulate, we suggested a method to generate topologically protected bound states experimentally, and simulated this protocol numerically. We saw that it is possible to identify topological bound states by correlating their occurrence with Fig.~\ref{fig:phase_diagram} and mapping out the phase boundaries by changing $\theta, \delta$. Alternatively, we can recognise topological bound states thanks to their unique density distribution, which can also be used to differentiate $\varepsilon=0$ from $\varepsilon=\pi$ states. This means that we can experimentally construct the entire phase diagram of the system by identifying its topological bound states. 

\section{Pair of momentum separated Jackiw-Rebbi states}
\label{sec:JR_pair}

In this section, we will see that not all $\varepsilon= 0, \pi$ states which are exponentially localised are topologically protected. We will consider the atomic quantum walk in the limit $\delta=0$. As can be seen from Fig.~\ref{fig:phase_diagram}, when following the path in parameter space indicated by a dashed green arrow, a topologically trivial band gap closing occurs when $\theta$ change sign. Indeed, the winding numbers $\{\nu_0,\nu_\pi\}$ are the same in both regions. Surprisingly, we will show that the Hamiltonian in this limit also presents $\varepsilon\approx 0$ and $\varepsilon\approx \pi$ bound states. By finding an approximate expression for the Floquet Hamiltonian \^H$_F'$, we will understand that the $\varepsilon\approx 0$ bound states correspond to solutions of the Jackiw-Rebbi model in the continuum case. Because these states occur in pairs, they can hybridise and move symmetrically away from $\varepsilon=0$, thus destroying the states' topological protection. Importantly, these trivial bound states are not eigenstates of chiral symmetry, and therefore do not have the same spin distribution as the states presented in Fig.~\ref{fig:eigenstate_half_time_step}. In a recent publication, interacting Hamiltonians which host Majorana modes were investigated \parencite{Jevtic2017}. These bound states can overlap, causing a loss of topological protection and affecting their spin structure. In this sense, these behave similarly to the chiral bound states studied here.

In the limit of $\delta=0$, \^H$_S'$ from Eq.~\eqref{eq:Ht'} and \^H$_\theta'$ from Eq.~\eqref{eq:Htheta'} have the form:
\begin{eqnarray}
\label{eq:Ht_SQW}
\text{\^H}_S' &=& \sum_n J \cc^\dagger_{n+1}  \sigma_1  \cc_n +\text{H.c};\\
\label{eq:C_SQW}
\text{\^H}_\theta'&=&-\sum_n \cc^\dagger_{n} \theta(n)  \sigma_2\cc_n .
\end{eqnarray}

\subsection{Symmetries}

Using the method presented in Sec.~\ref{sec:Symmetries}, we find the symmetries of the system in this limit. These are presented in Table \ref{tab:SQW_Sym}. Comparing these results with Table \ref{tab:topo}, we find that \^H$_F'$ in the limit $\delta = 0$ belongs to the BdG symmetry class CI of the classification of single-particle Hamiltonians. This symmetry class is topologically trivial in one dimension, implying that, for all values of $J$ and $\theta$ in this limit, \^H$_F'$ can only explore one topological phase: the trivial one. As a result, this system cannot host topologically protected states when $\delta=0$, in agreement with Fig.~\ref{fig:phase_diagram}.

\begin{table}[t]
\centering
\caption[Symmetries of \^H$_F'$ for $\delta=0$]{In the first column, the operators implementing various symmetries of \^H$_F'$ are listed in the limit $\delta=0$. We list the squares of these operators in the second column.}
\label{tab:SQW_Sym}
\begin{tabular}{ |c|c|c| }
\hline
symmetry & operator & square\\
\hline
Chiral symmetry & $\sigma_3$ & $\sigma_0$ \\
Time reversal symmetry & $\sigma_1  \hat{\mathcal{K}}$ & $\sigma_0$ \\
Particle hole symmetry & $-i \sigma_2 \hat{\mathcal{K}}$ & $-\sigma_0$ \\
\hline
\end{tabular}
\end{table}

\subsection{Topologically trivial bound states}

\begin{figure}[t]
\centering
\includegraphics[width=0.8\textwidth]{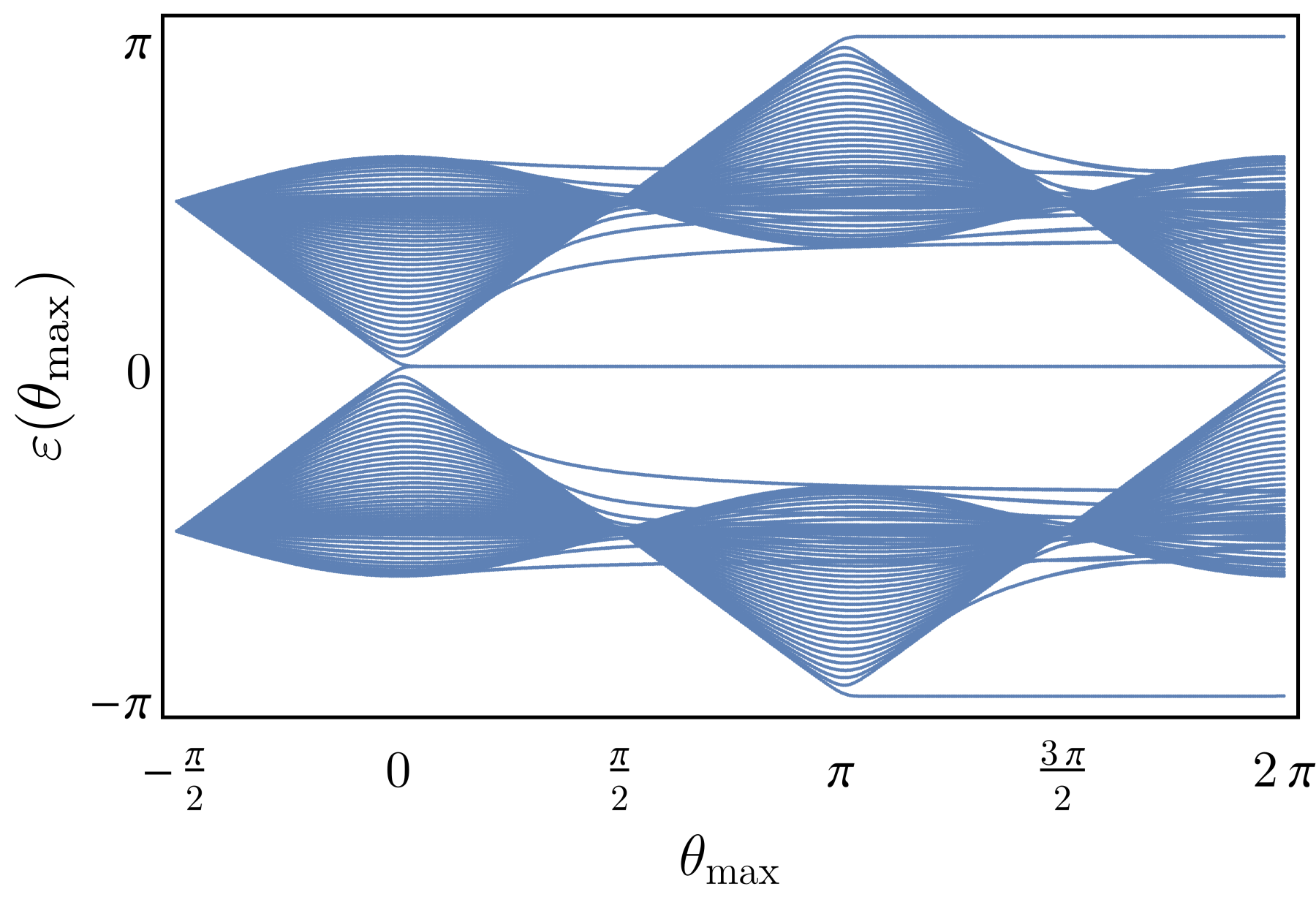}
\caption[Bound states in the topologically trivial limit]{Eigenvalues of $\hat H_F'$ for fixed $\theta_{\rm min}=-\pi/2$, as a function of $\theta_{\rm max}$, with open boundary conditions on an $N=300$ site lattice. Pairs of $\varepsilon\approx 0$ and $\varepsilon\approx \pi$ states appear in the band gaps and are exponentially localised around the lattice site where $\theta=0$ and $\theta=\pi$, respectively.}
\label{fig:SQW_InhomoSpectrum}
\end{figure}

Despite this, the system displays $\varepsilon\approx 0$ and $\varepsilon\approx \pi$ bound states when the spin mixing angle $\theta$ is varied spatially. We can verify this numerically by considering a chain of length $L$ and $\theta=\theta(n)$, a continuous function of position, going from $\theta_{\rm min}$ to $\theta_{\rm max}$ over a length scale $\xi$ according to Eq.~(\ref{eq:theta_function}). We assume $d \ll \xi \ll L$, where $d$ is the lattice spacing. The spectrum of \^H$_F'$ from Eq. \eqref{eq:U'} is shown in Fig.~\ref{fig:SQW_InhomoSpectrum}. We observe that a band gap is open around $\varepsilon=0$, except for $\theta_{\rm max}= 0 \pmod{2\pi}$. A pair of chiral partner zero-energy states appears in the spectral gap whenever $\theta_{\rm min}<0<\theta_{\rm max}$, exponentially localised around the site where $\theta(n)=0$. Similarly, $\pi$ energy states, centred at $\theta(n)=\pi$, appear if $\theta_{\rm min}<\pi<\theta_{\rm max}$.

\subsection{Approximate Floquet Hamiltonian}

We will now derive an approximate expression of \^H$_F'$ in the limit $d \ll \xi \ll L$, and use it to explain the origin of the $\varepsilon\approx 0$ bound states in this model. We will find it useful to go to the continuous limit, where we are able to find an exact eigenstate such that \^H$_F'|\psi\rangle=0$. We will then obtain a discrete ansatz from this solution and evaluate its energy, which we can compare to results from numerical simulations. 

For constant $\theta(n)=\theta$, the band gap closes when $\theta= 0 \pmod{2\pi}$ at $\pm k_{\rm bs}=\pm\pi/(2 d)$, where $d$ is the lattice spacing. When $\theta(n)$ is slowly varying near $\theta(n)=0$, we can apply the envelope function approximation (which we introduced in Sec.~\ref{sec:JRIntroduction}) to obtain an approximate expression for the $\varepsilon\approx 0$ eigenstates of \^H$_F'$:
\begin{equation}
\label{eq:SQW_FermiState}
|\psi_\pm\rangle=\sum_{n} e^{\pm i k_{\rm bs}  n  d} 
\vec{\varphi}_\pm(n) |n\rangle,
\end{equation}
where $|n\rangle$ is the state which is well localised at site $n$, and $\vec{\varphi}_\pm(n)$ is the envelope function spinor, which we assume to be slowly varying.

When \^H$_S'$ acts on $|\psi_\pm\rangle$, we obtain:
\begin{equation}
\label{eq:SQW_HPsiDiscrete}
\text{\^H}_S'|\psi_\pm\rangle =
\pm i J\sum_n  e^{\pm i k_{\rm bs}  n  d} \sigma_1 \left[ \vec{\varphi}_\pm(n+1)-\vec{\varphi}_\pm(n-1) \right]  |n\rangle.
\end{equation}
Assuming that $\vec{\varphi}_\pm(n)$ varies slowly compared to the lattice spacing $d$, we can take the continuous limit of Eq.~(\ref{eq:SQW_HPsiDiscrete}) by sending $d\rightarrow 0$, $n\rightarrow \infty$ such that $n  d = x$ is constant:
\begin{equation}
\label{eq:SQW_HPsiContinuous}
\text{\^H}_S'|\psi_\pm\rangle\rightarrow
\pm 2 i \tilde{v} \sigma_1 \int dx ~e^{\pm i k_{\rm bs}  x} 
\partial_x  \vec{\varphi}_\pm(x)  |x\rangle,
\end{equation}
where we have introduced the velocity parameter $\tilde{v}=Jd$.
Note that \^H$_S'$ appears within an exponential in Eq.~\eqref{eq:U'}; we therefore also need to consider higher powers of \^H$_S'$ when acting on $|\psi_\pm\rangle$. This can be done by repeating the above procedure, yielding:
\begin{equation}
\label{eq:SQW_HnPsiContinuous}
\left(\text{\^H}_S'\right)^m|\psi_\pm\rangle\rightarrow
(\pm 2 i \tilde{v} \sigma_1)^m \int dx ~e^{\pm i k_{\rm bs}  x} 
\partial_x^m  \vec{\varphi}_\pm(x)  |x\rangle,
\end{equation}
Thus, terms of the form of Eq.~(\ref{eq:SQW_HnPsiContinuous}) will be small for $m\ge 2$ when $\theta(x)$ varies slowly. Recalling that $|\psi_\pm\rangle$ is bound to the region $x\approx x_0$ where $\theta(x)$ vanishes, we find that:
\begin{equation}
\begin{split}
&\text{\^H}_F'|\psi_\pm\rangle =
i\log\left(e^{-i \text{\^H}_\theta'/2}  e^{-i \text{\^H}_S'}  e^{-i \text{\^H}_\theta'/2} \right) |\psi_\pm\rangle\\
& = \left[ \text{\^H}_S' +\text{\^H}_\theta' +\mathcal{O}(\text{\^H}_S'^2)+\mathcal{O}(\theta(x_0)^2)+\mathcal{O}(\theta(x_0)  \text{\^H}_S')\right] |\psi_\pm\rangle.
\end{split}
\end{equation}
Thus \^H$_F'$ is well approximated by:
\begin{equation}
\label{eq:HToyModel}
\text{\^H}_F'\approx \text{\^H}_{\rm approx}'= \text{\^H}_S' +\text{\^H}_\theta',
\end{equation}
when acting on states near $\varepsilon=0$ which are localised in the neighbourhood of $\theta(x)=0$. Thus, we have derived a static, approximate Hamiltonian, \^H$_{\rm approx}'$.

\subsection{Low energy eigenstates of \texorpdfstring{\^H$_{\rm approx}'$}{Happrox}}
\label{sec:BSofHapprox}

Using Eq.~(\ref{eq:SQW_HPsiContinuous}), we can apply this Hamiltonian to $|\psi_\pm\rangle$:
\begin{equation}
\label{eq:ContinuousHPsi}
\text{\^H}_{\rm approx}'|\psi_\pm\rangle=\int dx ~ e^{\pm i k_{\rm bs} x} \left[ \pm 2 i \tilde{v} \sigma_1 \partial_x \vec{\varphi}(x)
 -\theta(x) \sigma_2  \vec{\varphi}(x) \right] |x\rangle.
\end{equation}
The right hand side of this equation is reminiscent of the Jackiw-Rebbi model, which we encountered in Sec.~\ref{sec:JRIntroduction}. Building on the results presented in this section, there always exists a zero solution to the right hand side of Eq.~(\ref{eq:ContinuousHPsi}) when the sign of $\theta(x)$ is different for $x\rightarrow +\infty$ and $x\rightarrow -\infty$ \parencite{Jackiw1976}. When this is the case, \^H$_{\rm approx}'$ has an $E=0$ solution. If $\lim_{x\rightarrow\pm \infty}\text{sgn}[\theta(x)]=\pm 1$, then $\vec{\varphi}_\pm(x)$ takes the form:
\begin{equation}
\label{eq:JR_envelope}
\vec{\varphi}_\pm(x)=\psi_0 \exp \left[-\frac{1}{2\tilde{v}}\int_0^x dx'\theta(x')\right] |\mp\rangle,
\end{equation}
where we have defined the spin states: $|+\rangle=(1, 0)^T$ and $|-\rangle=(0, 1)^T$, and $\psi_0$ is the normalisation of the wave function. In the following, we will restrict our study to $\theta(x)$ varying as:
\begin{equation}
\label{eq:ContinuousBeta}
\theta(x)=\alpha+\beta \tanh\left(\frac{x}{\xi}\right),
\end{equation}
with $\beta>\alpha$ and $\beta>0$. This is the continuous version of Eq.~\eqref{eq:theta_function}, with $2\alpha=\theta_{\rm max}+\theta_{\rm min}$ and $2\beta=\theta_{\rm max}-\theta_{\rm min}$. From Eqs.~(\ref{eq:JR_envelope}), (\ref{eq:ContinuousBeta}), we see that $\vec{\varphi}_\pm(x)$ has the form:
\begin{align}
& \vec{\varphi}_\pm(x)=\varphi(x)|\mp\rangle, \\
\label{eq:ContinuousEnvelope}
& \varphi(x)= \psi_0\exp \left(-\frac{\alpha  x}{2\tilde{v}}\right)\cosh\left(\frac{x}{\xi}\right)^{-\beta \xi/(2\tilde{v})}.
\end{align}
By discretising the above result, sending $x\rightarrow n d$, we obtain an ansatz wave function for the two zero energy states that we are observing:
\begin{align}
\label{eq:DiscreteAnsatz}
& |\psi_\pm\rangle=
\sum_n e^{\pm i k_{\rm bs} n d}~ \varphi_n~|n,~ \mp\rangle,\\
\label{eq:DiscreteEnvelope}
& \varphi_n= \psi_0\exp \left(-\frac{\alpha  n}{2 J}\right)\cosh\left(\frac{n  d}{\xi}\right)^{-\beta \xi/(2 J  d)}.
\end{align}

Let's take a moment to recapitulate what we have done so far. We have defined two states, $|\psi_+\rangle$ and $|\psi_-\rangle$, which are centred around the momenta $k=\pm k_{\rm bs}$. We have shown that when $\theta(x)$ changes sign, these states are eigenstates of the Hamiltonian with eigenvalue $E=0$. In the Jackiw-Rebbi model, there is only one such state, which is pinned at $E=0$ by chiral symmetry. In our system, spatially varying $\theta(x)$ leads to terms which mix the states $|\psi_+\rangle$ and $|\psi_-\rangle$, causing them to hybridise and move symmetrically away from $E=0$.

\begin{figure}[t]
\centering
\includegraphics[width=0.7\textwidth]{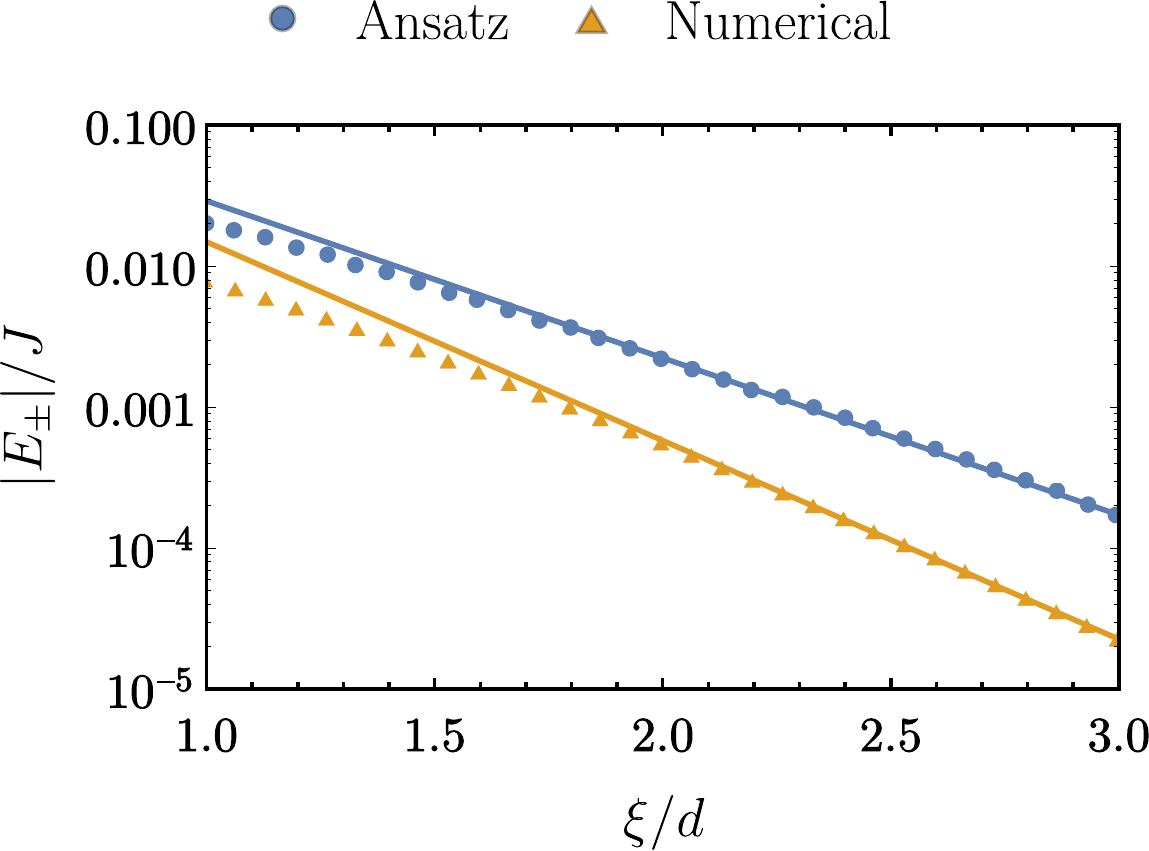}
\caption[Calculated bound state energy]{Energy of the Jackiw-Rebbi states $|E_\pm|$ versus $\xi$, the inverse rate of change of $\theta(n)$. Blue dots: value of $|E_\pm|$ from Eq.~(\ref{eq:HredEigenvals}). Yellow triangles: Eigenvalues of \^H$_{\rm approx}'$ closest to $E=0$ (obtained numerically). The lines are exponential fits to the data.
Both these curves approach each other exponentially (which is clear because the scale of the $y$ axis becomes exponentially small). In both cases, $\alpha=-1/4,~\beta=1$.}
\label{fig:JR_energies}
\end{figure}

We can find the energy of these hybrid states by studying the eigenvalues of \^H$_{\rm red}'$, the projected Hamiltonian on the basis of states $|\psi_+\rangle$ and $|\psi_-\rangle$.
\begin{equation}
\label{eq:Hreduced}
\text{\^H}_{\rm red}'=\begin{pmatrix}
\langle \psi_+|\text{\^H}_{\rm approx}'|\psi_+\rangle &&
\langle \psi_+|\text{\^H}_{\rm approx}'|\psi_-\rangle \\
\langle \psi_-|\text{\^H}_{\rm approx}'|\psi_+\rangle &&
\langle \psi_-|\text{\^H}_{\rm approx}'|\psi_-\rangle
\end{pmatrix}.
\end{equation}
This matrix has eigenvalues:
\begin{equation}
\label{eq:HredEigenvals}
E_\pm=\pm \sum_n  (-1)^{n} \varphi_n^*
\left[ J(\varphi_{n+1}-\varphi_{n-1})+\theta(n) \varphi_n \right].
\end{equation}
The energy $|E_\pm|$ from Eq.~(\ref{eq:HredEigenvals}) is plotted versus $\xi$ in Fig.~\ref{fig:JR_energies}. Alongside this estimate of the lowest eigenstate's energy, we have diagonalise \^H$_{\rm approx}'$ and plotted its eigenvalues. Visibly there is a good agreement between the estimate Eq.~(\ref{eq:HredEigenvals}) and the exact energies, suggesting that our hypothesis was indeed correct, and that the states we are observing are indeed two Jackiw-Rebbi states separated in momentum, the energies of which behave as:
\begin{equation}
\label{eq:Eapprox}
|E_\pm|\approx J \exp\left(-\frac{c \xi}{d}\right),
\end{equation}
where $c$ is a positive constant.

From the results presented here, we can therefore conclude that the system presents a pair of zero energy bound states when $\theta(n)$ changes sign, despite being in a trivial topological configuration. We find that when $\theta(n)$ changes slowly ($\xi \gg d$), we have two non-overlapping (in $k$ space) bound states, one associated to each Dirac cone. The overlap increases for faster changing $\theta(n)$ (smaller $\xi$), resulting in both states hybridising and moving away from $E=0$ without breaking chiral symmetry. For this reason, these bound states do not benefit from the robustness displayed by topologically bound states.

This type of behaviour is generally observed in systems which present two degenerate energy levels. When the two states are coupled, their energies split symmetrically, proportionally to the matrix element between them (the off diagonal term of Eq.~(\ref{eq:Hreduced})). In our system, we are able to directly control the tunnelling between these states by modifying the length scale over which the potential varies. Thus our model gives us access to a single parameter which controls the states' degree of hybridisation.

\section{Conclusions}

In this chapter, we designed a new topologically non-trivial system which is analogous to the two-step quantum walk. In our protocol, spin $1/2$ cold atoms are trapped in a 1D optically generated bipartite lattice, which is driven periodically by a pair of Raman lasers. We verified analytically that the atomic quantum walk has two distinct topological invariants, such that it can realise two flavours of symmetry protected bound states.

We verified, through numerical simulations, that topological bound states can be generated by spatially varying the amplitude of the spin coupling. What is more, we showed that density populating a bound state can be identified by searching for eigenstates of the chiral symmetry operator, which have a heavily constrained spin distribution. We further showed that this type of measure can discriminate between both flavours of bound states, allowing us to reconstruct the full phase diagram.

Finally, we mentioned that it is possible for bound states to appear at band gap closings where none of the topological invariants change value. These states can hybridise to move away from $\varepsilon=0,\pi$. Their degree of hybridisation is controlled by the rate of change of the spin coupling amplitude. Because these states are not eigenstates of the chiral symmetry operator, their spin structure is different from that of topological bound states.

\chapter{Measure of the Chern number in a Hofstadter strip}
\label{Chapter5}

The detection of topological properties is at the very centre of the study and application of topologically non-trivial materials \parencite{Hasan2010}. Indeed, in some cases, this is the only way to verify that interesting topological properties survive complex effects, such as strong interactions or random perturbations. The bulk-boundary correspondence tells us that the number of topologically protected states at the edge of a material is determined by the bulk topological invariant. As a result of this theorem, we can split methods for the detection of topological properties into two main categories:
\begin{enumerate}
 \item Measures performed on the system's bulk, to determine the topological invariant.
 \item Measures performed on the system's boundary, to identify the system's edge states.
\end{enumerate}
In solid state systems, experimental measurements of topological properties usually rely on transport measurements \parencite{Hasan2010,Qi}. Unfortunately, these types of experiments are difficult to engineer in cold atomic systems \parencite{Brantut2012}, such that alternative ways need to be found to measure the topological invariant. When simulating the integer quantum Hall effect with cold atoms, the Chern number can be measured using time of flight measurements \parencite{Alba2011,Hauke2014,Flaschner2016,Flaschner}. Alternatively, the Berry curvature can be related to the system's polarisation \parencite{King-Smith1993, Coh2009}, allowing us to measure the Chern number directly from the centre of mass dynamics of the atomic cloud \parencite{Price2012,Dauphin,Aidelsburger2014}.
Finally, Berry curvature effects can also be observed in one dimensional systems which are periodic in time. This was demonstrated recently using cold atoms by realising Thouless pumping \parencite{Lohse2015,Nakajima2016}, spin pumping \parencite{Schweizer2016}, and geometrical pumping \parencite{Lu2016}.

For very narrow systems, bulk measurements become problematic. Indeed, topological invariants are defined in the limit of infinite or periodic systems. Let us consider the example of the Hofstadter strip with open boundary conditions, which we encountered in Sec.~\ref{sec:SyntheticDimensions}. This system's bulk is only a few sites wide, such that it is not clear that its topological invariant is well defined. On the other hand, we showed in Sec.~\ref{sec:SyntheticDimensions} that it maps onto the Hofstadter model, which has well known topological properties, and the robustness of its  edge states has been shown numerically \parencite{Celi2013} and experimentally \parencite{Stuhl2015,Mancini}. We can therefore legitimately ask: does this system have a well defined Chern number?
If so, does this parameter measure the number of topologically protected edge states?



In the strip geometry, earlier works have suggested observing the system's topological properties by performing hybrid time-of-flight and in-situ measurements \parencite{Wang2013}, or by exploiting strong contact interactions between atoms \parencite{Zeng, Barbarino2015a, Taddia}.


The topological invariant is a difficult quantity to measure in non interacting systems with a small bulk because it is a global property. As such, it is necessary to integrate the dynamics over an entire band to observe it. This is usually done by considering a sample which presents a filled band. In thin samples, however, it is difficult to fill a band homogeneously without populating the edges due to the large edge to bulk surface ratio. To circumvent this problem, we determine the Chern number from a displacement measurement after a period of Bloch oscillations.
Surprisingly, we find that our method is successful in measuring the Chern number even in a single atom system.

\begin{framed}
In this chapter:
\begin{enumerate}
 \item In Sec.~\ref{sec:HofstadterStrip}, we briefly present the Hofstadter strip in the limits relevant to our study.
 \item In Sec.~\ref{sec:ChernMeasure}, we show that the Chern number can in general be measured from an atom's mean position, and thereby identify the assumptions necessary to reach this result.
 \item In Sec.~\ref{sec:ExperimentalSuggestion}, we use this result to develop an experimental protocol for measuring the Chern number. We describe in detail the system's preparation and the measurement method, and, in doing so, we identify the relevant parameter regime.
 \item In Sec.~\ref{sec:StripSimulations} and \ref{sec:EdgeEffects}, we demonstrate the success of our method through numerical simulations, and identify the main sources of error. We also show that our method can measure the Chern number of some higher bands through atomic pumping.
 \item In Sec.~\ref{sec:HigherChernMeasure}, we extend our study by considering a system in the limit where the FHS algorithm breaks down \parencite{Fukui2005}, and show that our method remains valid in this limit.
 \item In Sec.~\ref{sec:AtomicGas}, we show that our results can be generalised to gases of non-interacting particles.
 \item In Sec.~\ref{sec:broken_trans_inv}, we demonstrate the measure's robustness both to disorder and the presence of a harmonic trap.
\end{enumerate}
\end{framed}

\section{The Hofstadter strip}
\label{sec:HofstadterStrip}

The focus of our study will be the Hofstadter model \parencite{Hofstadter1976}, which is sketched in the Fig.~\ref{fig:Hofstadter}. As was discussed in detail in the Sec.~\ref{sec:Hofstadter}, it describes spinless, non-interacting atoms in a two dimensional lattice subject to a uniform external magnetic field. In this setting, a flux $\Phi=2\pi p/q$ traverses every plaquette (with $p$ and $q$ coprime integers). In the Landau gauge, the magnetic unit cell has dimensions $(q,1)$ and the Hofstadter Hamiltonian is given by Eq.~\eqref{eq:H0}.

We saw that the Hofstadter Hamiltonian is in the unitary class of the topological classification of Hamiltonians \parencite{Schnyder2008}. This means that in two dimensions, it presents a topological invariant, the Chern number, which can be non-zero. As we mentioned in Sec.~\ref{sec:HofCylinder}, this invariant counts the number of topologically protected states which occur at the system's boundary when the system presents open boundary conditions.

\subsection{The Hofstadter model with a synthetic dimension}

\begin{figure}
    \centering
    \includegraphics[width=0.9\textwidth]{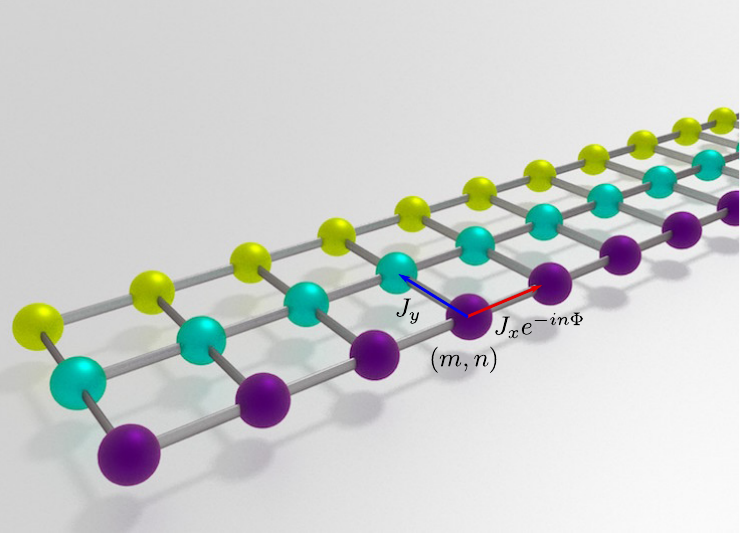}
    \caption[Hofstadter strip in the gauge of Eq.~\eqref{eq:H0Mom}]{\label{fig:HofstadterStrip} Sketch of the hoppings of the Hofstadter strip with $N_y=3$ sites in the $y$ direction. The site indices in the $x,y$ directions are $m,n$ respectively. The total flux through each plaquette is $\Phi=2\pi p/q$ with $p, q \in \mathbb{Z}$.} 
\end{figure}

Recently, a method for building the Hofstadter model was suggested which uses a one dimensional gas of spinful atoms, the so called Hofstadter strip \parencite{Celi2013,Stuhl2015,Mancini}. In this model, which we described in detail in Sec.~\ref{sec:SyntheticDimensions}, 
a single atom evolves in a one-dimensional optical lattice. Its magnetic sublevels, which we will refer to as spin degrees of freedom, are coupled with Raman lasers, which imprint a position dependent phase on the atom. By interpreting the atom's spin as an extra spatial dimension, the system's Hamiltonian is exactly equivalent to the Hofstadter Hamiltonian, Eq.~\eqref{eq:H0}. Topological insulators with extremely thin bulks can be realised in this way.

\subsection{The Hofstadter strip on a cylinder}
\label{sec:SyntheticDimensions_GaugeTransform}

In the following sections, we will study in detail the dynamics of a single atom evolving in the Hofstadter strip, controlled by the Hamiltonian Eq.~\eqref{eq:H0}. Without loss of generality, we will set $x$ to be the spatial dimension, composed of $N_x \gg 1$ sites, and set $y$  to be the synthetic, or spin dimension, composed of $N_y$ sites with $N_x \gg N_y$. 

\begin{figure}[t]
    \centering
    \includegraphics[width=0.75\textwidth]{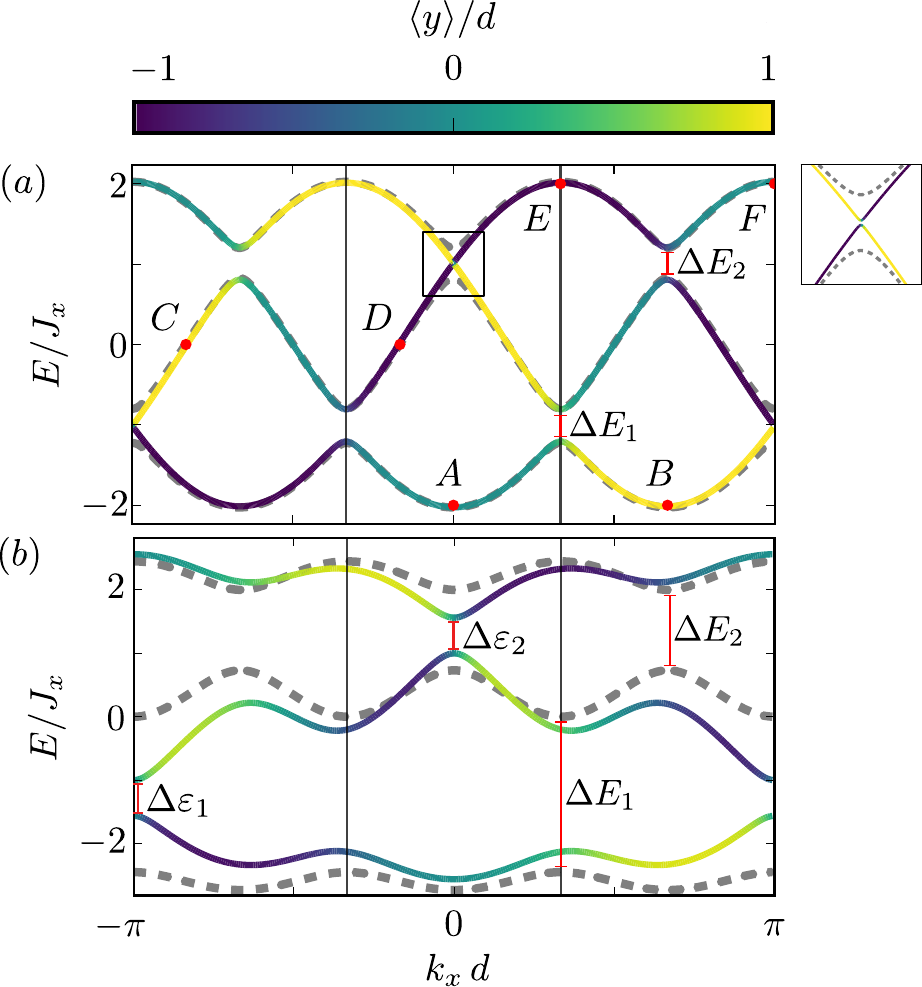}
    \caption[Dispersion of the Hofstadter strip with $\Phi=2\pi/3$]{\label{fig:dispersions} Dispersion of the Hofstadter model with $\Phi=2\pi/3$, $N_y=3$ for $J_y=J_x/5$ (a) and $J_y=J_x$ (b). Grey: dispersion with periodic boundary conditions. Coloured: system with open boundary conditions. The colour coding indicates the mean eigenstate's position along $y$. Well localised edge states exist in the gaps of the periodic system [inset, (a)]. Edge states overlap, giving rise to hybridisation gaps $\Delta \varepsilon_j < \Delta E_j$ , where $\Delta E_j$ denote the band gaps. In the following, we will induce Bloch oscillations to pump the atomic density from points $A-F$.
    }
\end{figure}

We will assume, as we did in Sec.~\ref{sec:HofCylinder}, that the model has open boundary conditions along the $y$ direction, but periodic boundary conditions in the $x$ direction. This allows us to partially Fourier transform the Hamiltonian along the spatial direction, to obtain Eq.~\eqref{eq:H0Mom}. We insist upon the fact that, to obtain this expression, we performed a gauge transformation, Eq.~\eqref{eq:GaugeTranform}. The magnetic unit cell in this gauge has dimensions $(1,q)$, such that the quasimomentum variable belongs to the interval $k_x\in[-\pi/d,\pi/d]$.

We present two limits of this Hamiltonian's dispersion for $\Phi = 2\pi/ 3$: in Fig.~\ref{fig:dispersions}(a) we set $J_y=J_x/5$, while in Fig.~\ref{fig:dispersions}(b) we choose instead symmetric couplings, $J_y=J_x$. In both cases, the spectrum is composed of $N_y$ discrete energy values for each value of $k_x$. When the system has periodic boundary conditions (plotted in grey), these are grouped in $q$ distinct bands, separated by band gaps $\Delta E_j$, $j \in [1,q-1]$. We observe that the magnitude of the band gaps increases for larger $J_y/J_x$, such that the $\Delta E_j$ are larger in Fig.~\ref{fig:dispersions}(b) than in Fig.~\ref{fig:dispersions}(a).

The coloured lines in Figs.~\ref{fig:dispersions} depict the spectrum of the system with open boundary along $y$. In both figures, we see the appearance of states which bridge the band gap. Because these are sharply localised at the edges of the system (as indicated by the line colouring), we will refer to them as edge states. We find four edge states: a pair in the first band gap which cross at $k_x=\pm \pi/d$, and a pair in the second band gap which cross at $k_x=0$. These states have amplitude in the bulk, such that they overlap and form an avoided crossing, with gap magnitude $\Delta \varepsilon_j$ in the $j^{\rm th}$ band gap. While this hybridisation gap is negligible in the limit $J_y \ll J_x$, we will see that this gap profoundly affects the dynamics for $J_y=J_x$.

\section{Measure of the Chern number}
\label{sec:ChernMeasure}

In this section, we will consider the dynamics of a two-dimensional insulating system controlled by a periodic Hamiltonian $\h_0$ in the presence of a constant force $\bm{F} = F_x \bm{e}_x$. We will show that, under certain conditions, the atomic density populating the system's bulk is spatially pumped, and that the total displacement in a well defined time $T$ is determined by the system's Chern number.
For generality, we will assume that $\h_0$ has $q_x$ ($q_y$) sites per unit cell in the $x$ ($y$) direction. We do this to later underline a subtle point which comes from the choice of gauge in the Hofstadter model.

\subsection{A linear potential gradient causes adiabatic transport}
\label{sec:adiabaticTransport}

We will assume that the force $\bm{F}$ is applied in the $x$ direction, indexed by the variable $m \in [1,N_x]$, and has the form:
\begin{equation}
\label{eq:H}
\h =  \h_0 - F_x d \sum_m m \hat{c}_{m,n}^\dagger \hat{c}_{m,n}.
\end{equation}
As discussed in Sec.~\ref{sec:anomalousVelocity}, when $|F_x| d$ is much smaller than the band gap, we can neglect inter-band transitions. In this case, translational invariance can be restored by performing a gauge transformation. In the new gauge, quasimomentum is a good quantum number and takes the form of Eq.~\eqref{eq:Meank}.
If $\h_0$ has $q_x$ sites per unit cell, then $k_x \in [-\pi/(q_x d), \pi/ (q_x d)]$ and it takes a time $T$ to pump states adiabatically around the Brillouin zone, with:
\begin{equation}
 \label{eq:periodGeneral}
 T=\frac{2\pi \hbar}{q_x d |F_x|}.
\end{equation}
We will refer to $T$ as the period of the potential.

\subsection{Semiclassical equations of motion}
\label{sec:Laughlin}

In this subsection, we show that it is possible to measure the Chern number of a Hamiltonian of the form Eq.~\eqref{eq:H} simply by monitoring the displacement of the atom's centre of mass. For a periodic system in this setup, the group velocity of a Bloch state in the $j^{\rm th}$ band is given by Eq.~\eqref{eq:SemiClassicalGroupVel}.
If the system presents a filled band, we saw in Sec.~\ref{sec:anomalousVelocity} that the mean group velocity is determined by the band's Chern number. Refs.\ \parencite{Dauphin,Aidelsburger2014} used this property to measure the Chern number by studying an insulator which presents a filled band.

Inspired by this result, we will now compute the mean displacement of a single atom in this system. We will show that, with a careful choice of the atom's initial state, we are able to integrate the group velocity, Eq.~\eqref{eq:SemiClassicalGroupVel}, over the entire Brillouin zone, such that the Chern number of the occupied band appears in the atom's mean displacement.

We denote the state of the atom at time $t$ by $\ket{\psi(\kk,t)}$. The translation velocity of the atom's centre of mass is given by:
\begin{equation}
\label{eq:MultiBandVel}
\mean{\bm{v}(t)} = \sum_j \int_{BZ} \bm{v}_j (\bm{k}) \rho_j (\kk, t) {\rm d}^2\kk,
\end{equation}
where
\begin{equation}
\label{eq:SingleBandDensity}
\rho_j(\kk,t) = |\braket{ u_j(\kk) }{ \psi(\kk,t) } |^2
\end{equation}
is the density of particles with quasimomentum $\kk$ belonging to the $j^{\rm th}$ band, and $\ket{u_j(\kk)}$ is the normalised wavefunction of the $j^{\rm th}$ Bloch band of the bare Hamiltonian $\h_0$.

Let us prepare the particle in a state strongly localised along $y$ and very extended along $x$, and with initial momentum $k_{x0}$. The corresponding momentum space wavefunction $\ket{\psi(\kk,t=0)}$ will  be sharply peaked in the $x$ direction around $k_{x0}$, while it will occupy uniformly the Brillouin zone in the $y$ direction, so that $\rho_j(\kk,t)\approx\rho_j(k_x,t)$. Let us further assume that the particle wavefunction at $t=0$ has support only on the $j^{\rm th}$ band. As discussed above, as long as the strength of the force is smaller than the band gaps, inter-gap transitions are strongly suppressed, so that we may drop the sum over $j$ in Eq.~\eqref{eq:MultiBandVel}. 

The mean displacement over one period of the potential is:
\begin{equation}
\label{eq:meanDisplacement}
\mean{\Delta \bm{r}}\equiv
\mean{\bm{r}(T) - \bm{r}(0)} = \int_0^T \mean{\bm{v}(t)} {\rm d}t =  \int_{BZ} \bm{v}_j(\bm{k}) \left( \int_0^T \rho_j\left(k_x(t)\right) {\rm d}t \right) {\rm d}^2\kk.
\end{equation}
Under the adiabatic approximation, the atomic density is rigidly transported around the Brillouin zone in a time $T$. The mean occupation, averaged over a period of the potential, is therefore the same for all $k_x$. This means we can replace $\rho_j(k_x)$ in the above expression by the homogeneous distribution:
\begin{equation}
\label{eq:mean_rho}
 \frac{1}{T}\int_0^T \rho_j\left(k_x(t)\right) {\rm d}t=\frac{1}{A_{BZ}},
\end{equation}
where the area of the Brillouin Zone is $A_{BZ}= (2\pi)^2 / (d^2 q_x q_y)$. Substituting Eq.~\eqref{eq:mean_rho} into Eq.~\eqref{eq:meanDisplacement}:
\begin{equation}
\label{eq:ChernMeasureGeneral}
\mean{\Delta \bm{r}}
= \frac{T}{A_{BZ}} \int_{BZ} \bm{v}_j(\bm{k}) {\rm d}^2\kk 
= {\rm sgn}(F_x) d q_y \CC_j \bm{e}_y.
\end{equation}
Note that, in the second equality of Eq.~\eqref{eq:ChernMeasureGeneral}, we have used the fact that the first term of Eq.~\eqref{eq:SemiClassicalGroupVel} integrates to zero due to the periodicity of $E(\kk)$. 
Thus, a constant force in the $x$ direction pumps particles in the $y$ direction. Furthermore, the number of magnetic unit cells that the atomic centre of mass is pumped after a time $T$ is exactly the Chern number of the occupied band \parencite{King-Smith1993}. This means that we can measure the Chern number with a single measurement of the centre of mass position along $y$. It was shown in Ref.\ \parencite{Wang2013} that a similar mechanism could be used to measure the $\mathbb{Z}_2$ topological invariant of a two dimensional topological insulator.

As a final note, we saw in Sec.~\ref{sec:SyntheticDimensions_GaugeTransform} that the size of the magnetic unit cell in the Hofstadter model is gauge dependent. This leads Eq.~\eqref{eq:ChernMeasureGeneral}
to be more constraining in the gauge where the magnetic unit cell has dimensions $(q_x,q_y)=(q,1)$. In the rest of this chapter, whenever we apply Eq.~\eqref{eq:ChernMeasureGeneral}, it will be in this gauge.

\section{Charge pumping in the Hofstadter strip}
\label{sec:ExperimentalSuggestion}

While the study performed in Sec.~\ref{sec:Laughlin} is only strictly applicable to periodic systems, we are motivated by these results to measure the Chern number from the bulk dynamics.  In particular, we suggest using Eq.~\eqref{eq:ChernMeasureGeneral} as a definition of the Chern number in narrow systems with open boundary conditions. Further in this chapter, we will verify through numerical simulations that this definition is consistent and reproduces our expectations.

\begin{figure}[t]
\centering
\includegraphics[width=\textwidth]{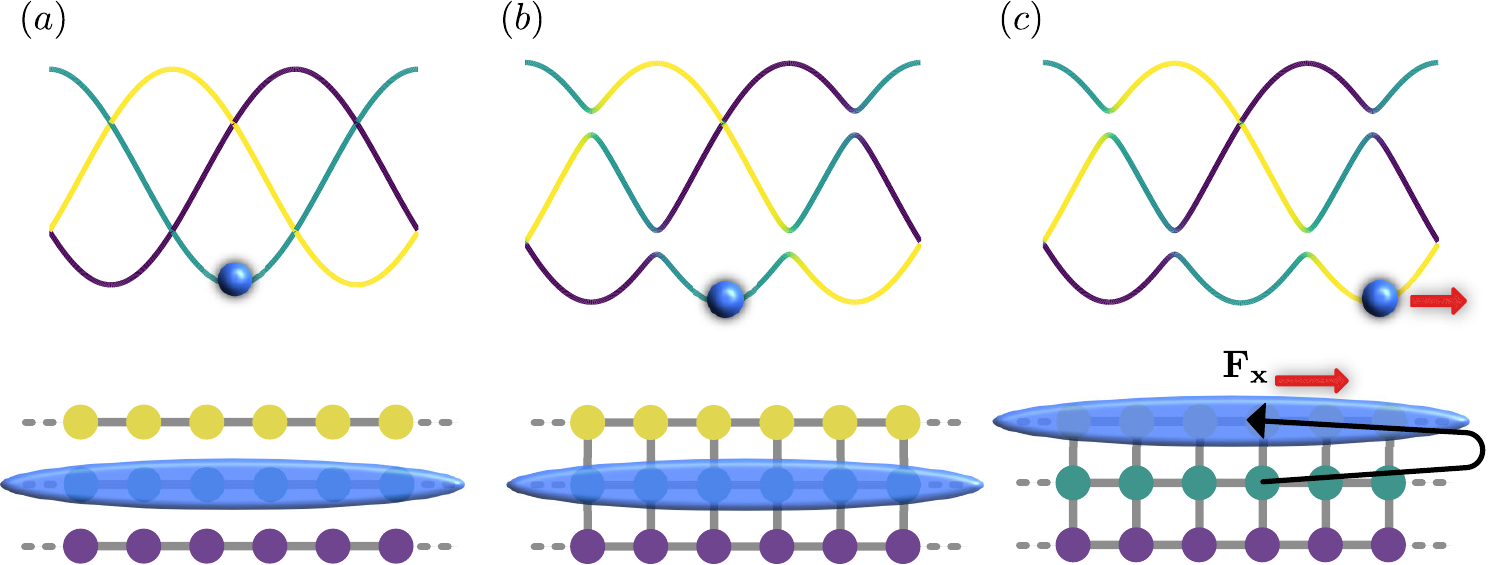}
\caption[Sketch of the pumping dynamics in the Hofstadter strip]{\label{fig:sketch_method}(a) Initially, the Hamiltonian has no tunnelling in the $y$ direction. The state is prepared to be localised at $y=0$. (b) By adiabatically turning on the tunnelling in the $y$ direction, the atom is loaded into the ground state of the lattice (c) A weak force is applied in the $x$ direction. After a Bloch oscillation, the state has moved of a number of sites proportional to the Chern number of the lowest energy band (as indicated by the black arrow).}
\end{figure}

The assumption we are making here is that the bulk dynamics of the system with open boundary conditions is no different from that of the periodic system, as long as the edge states are not populated. Under this assumption, for the results of the previous section to be applicable, we must consider 
an atom which is both localised in the bulk and belongs to the lowest energy band. The initial state preparation and subsequent experimental sequence is represented in the Fig.~\ref{fig:sketch_method} for a generic Hofstadter strip subject to an external magnetic flux of $\Phi$ .

The atom is initially spin polarised with $y=0$ and is in the lowest eigenstate of the Hamiltonian with no tunnelling in the $y$ direction [Fig.~\ref{fig:sketch_method}(a)]. We load the atom adiabatically into the bottom of the lattice by linearly switching on the Raman lasers, such that the spin tunnelling reaches a value of $J_y$ in a time $\tau$ [Fig.~\ref{fig:sketch_method}(b)]. To minimise the transfer to higher bands, we 
require that $\hbar/\tau$ is much smaller than the local energy gap $E_2(k_{x0})-E_1(k_{x0})$. 
Because the atom is in a potential minimum throughout the loading sequence, the atomic centre of mass is not displaced during the loading procedure.

At the end of the loading sequence, we induce Bloch oscillations by applying a weak, constant force with magnitude $F_x$ along the $x$ direction. After one Bloch oscillation ($k_x\xrightarrow{}k_x+\pi/(qd)$), the state is adiabatically translated from points $A$ to $B$, such that its centre of the mass is displaced by $\CC_1$ sites in the $y$ direction [Fig.~\ref{fig:sketch_method}(c)]. For the single band approximation to hold, we must require that the magnitude of the force is small compared to the first energy gap, such that $|F_x| d \ll \Delta E_1$. What is more, in a finite system, Eq.~\eqref{eq:Meank} is only valid far away from the system's edges in the $x$ direction. Consequently, we demand that the edges in the $x$ direction are not populated during the timespan of the experiment.





The different steps of this loading sequence can be realised with state of the art experimental techniques. Refs.\ \parencite{Stuhl2015,Mancini}, which studied the dynamics of bulk and edge states in the Hofstadter strip, prepared their initial states in a similar way.
Our protocol requires as additional ingredient a constant force, which should be identical for all the spin states. It can be implemented either by employing a moving optical lattice \cite{BenDahan1996,Morsch2001}, a linear potential realised optically, or simply the projection of gravity along the lattice direction.


The centre of mass displacement can be measured with single site precision along $y$ by applying a Stern-Gerlach pulse to separate the spin species before imaging. We will show in Sec \ref{sec:AtomicGas} that the Chern number can be measured with the same experimental protocol if a non-interacting Bose-Einstein condensate is loaded into the lattice instead of a single atom. This would allow us to perform partial-transfer absorption imaging to repeatedly measure the cloud's mean spin without destroying the system. In Ref.\ \parencite{Lu2016} for instance, a small number of atoms were transferred using a microwave pulse from internal state $|F, n\rangle=|1,-1\rangle$ to $|2,0\rangle$, where $F$ is the total angular momentum and $n$ is the magnetic orbital angular momentum. Atoms in the state $|2,0\rangle$ were then absorption imaged. Applying a similar technique to our system would allow experimenters to monitor the displacement of the centre of mass at all times in a single experiment, thus recovering useful information about the density distribution in quasi-momentum space.


\section{Pumping Dynamics}
\label{sec:StripSimulations}

In this section, we demonstrate numerically the success of our experimental protocol, described in Sec.~\ref{sec:ExperimentalSuggestion} by measuring the Chern number of the Hofstadter strip. To do this, we simulate Eq.~\eqref{eq:H0} for an atom with three degrees of freedom, corresponding to $N_y=3$, and open boundary conditions in the synthetic dimension. This case was realised experimentally in Ref.\ \parencite{Stuhl2015}. We will focus particularly on the edge states of the Hofstadter strip, and how these affect the dynamics.

\subsection{Weak spin tunnelling amplitude limit}
\label{sec:decoupled_pumping}

We consider the three-legged Hofstadter strip, subject to a magnetic flux of $\Phi=2\pi/3$. The initial state is constrained to $w_x=30$ sites along the $x$ direction, and is spin polarised such that $n=0$. We load the lattice by adiabatically turning on the tunnelling amplitude in the synthetic dimension to $J_y=J_x/5$. The dispersion of this system is presented in Fig.~\ref{fig:dispersions}(a). After the loading sequence, we apply a constant force in the $x$ direction, with magnitude $F_x=0.03 \Delta E_1/d$, such that $|F_x| d$ is much smaller than the first band gap $\Delta E_1 = 0.42 J_x$. As we saw in Sec.~\ref{sec:Laughlin}, under these conditions, the force induces Bloch oscillations, the period of which is given by Eq.~\eqref{eq:periodGeneral}.

\begin{figure}
    \centering
    \includegraphics[width=\textwidth]{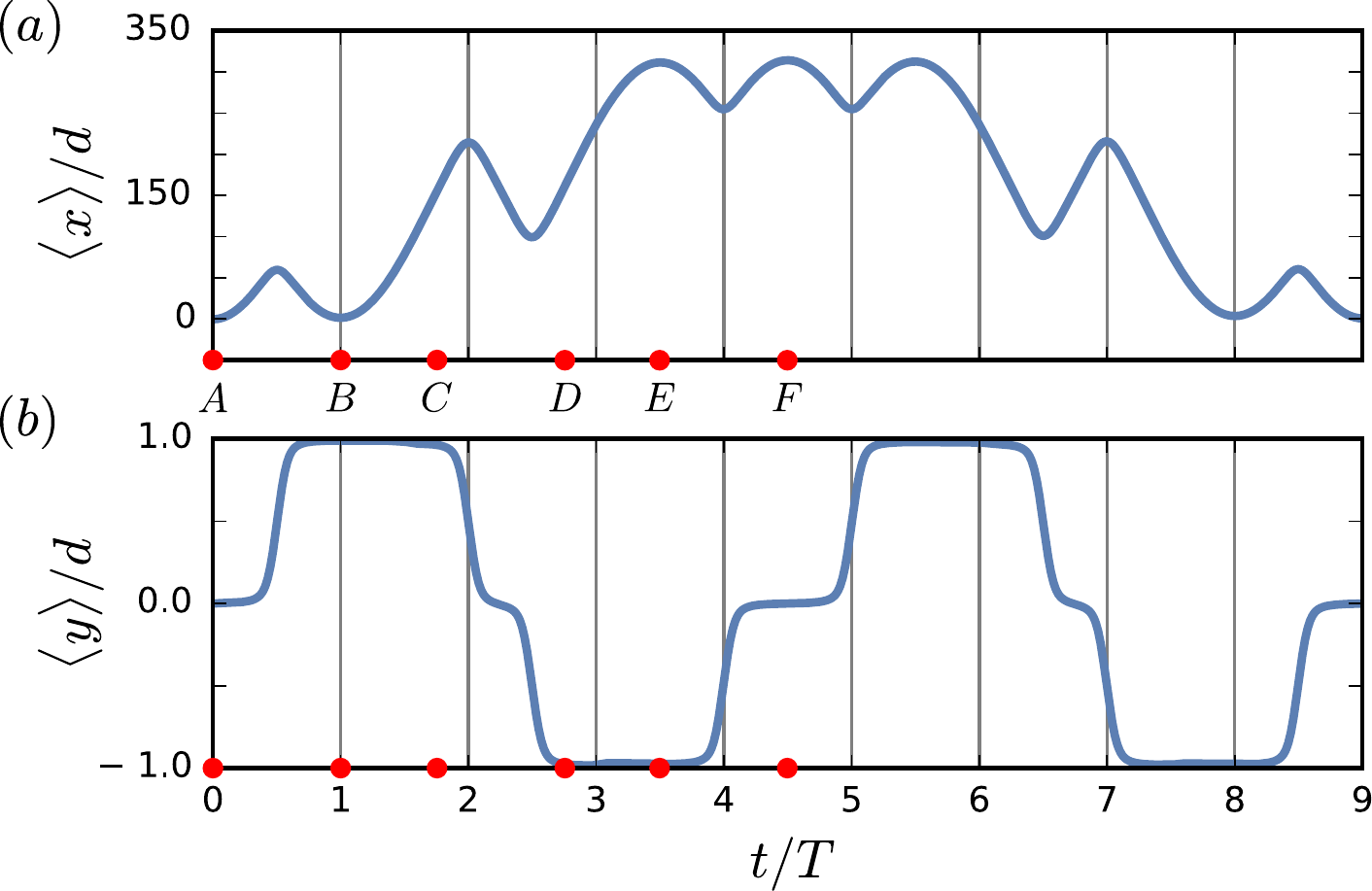}
    \caption[Centre of mass displacement versus time for $J_y \ll J_x$]{\label{fig:COM_decoupled} Mean position of the atom as a function of time, for $J_y=J_x/5$. Time is measured in units of $T$, the period associated to the force. We measure the atom's mean displacement in the $y$ direction between point $A-F$. We estimate the Chern number of each band this way; these are specified in Table \ref{tab:Cherns}.} 
\end{figure}
\begin{figure}
    \centering
    \includegraphics[width=0.8\textwidth]{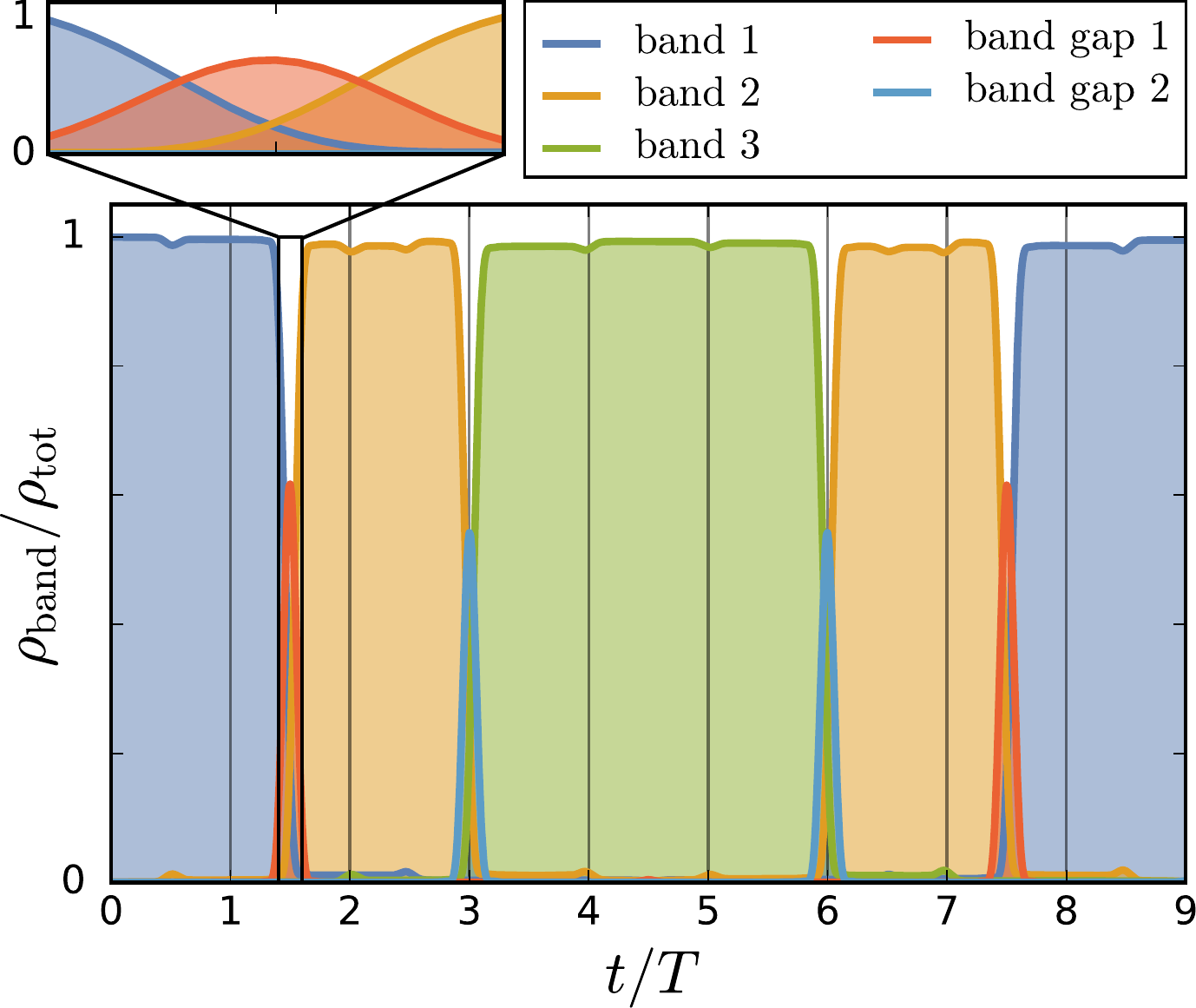}
    \caption[Band and edge state populations versus time for $J_y=0.2 J_x$]{\label{fig:density_decoupled} Band and edge state populations as a function of time, for $J_y=J_x/5$, classified according to the energies of the periodic system. Time is measured in units of $T$, the period of the potential. Inset: the edge states of the first band gap are populated for $t/T\in [1.4, 1.6]$.} 
\end{figure}

In Fig.~\ref{fig:COM_decoupled}, we plot the centre of mass positions recorded at every time of the system's evolution. As we will see shortly, these can be understood when studied jointly with the different bands' populations, which is shown in Fig.~\ref{fig:density_decoupled}. We constructed this figure by numerically diagonalising $\h_{\rm Hof}(k_x)$, given in Eq.~\eqref{eq:H0Mom} and expressing these eigenstates in real space using Bloch's theorem. The atom's wavefunction was then projected at all times on each eigenstate, giving us the population of states in the system. We then defined the energetic range of the bands and band gaps, allowing us to associate each state to either a band or a band gap. We found these energy ranges from the extremum energy values of the periodic system, which we plotted in grey in Fig.~\ref{fig:dispersions}(a).

In the first period of the motion, we can see from Fig.~\ref{fig:density_decoupled} that almost all of the atomic density is in the lowest energy band. Following the study of Sec.~\ref{sec:Laughlin}, the Chern number of the ground band can be extracted by measuring the displacement transverse to the force over the first period. As we can see by studying Fig.~\ref{fig:COM_decoupled}(b), the centre of mass is pumped from $\langle y \rangle=0$ to $\langle y \rangle=0.99$ in the first period, yielding the estimate $\CC_1=0.99$. We can compare this to the value calculated using the Fukui-Hatsugai-Suzuki (FHS) algorithm \parencite{Fukui2005} (see Sec.~\ref{sec:FHSalgo}) which yields $\CC^{\rm FHS}_1 = 1$. Thus, we see that, in this case, our measure agrees extremely well with the expected value.

As we can see from Eq.~\eqref{eq:Meank}, the density in quasimomentum space is rigidly displaced as: $k_x \xrightarrow{} k_x + 2\pi/(3d)$. In the dispersion in Fig.~\ref{fig:dispersions}(a), this corresponds to being pumped from the $y=0$ well (annotated by the point $A$) to the $y=1$ well (point $B$). In this case, we can read off the wavepacket's mean transverse displacement over a period of the potential simply from the colour coding of Fig.~\ref{fig:dispersions}(a). This was alerady noted by \parencite{Taddia}, and is generally true in quasi one-dimensional systems with open boundary conditions.

In the derivation presented in Sec.~\ref{sec:Laughlin}, we were able to ignore edge effects by considering a system with periodic boundary conditions. In the present case, however, after one period of the potential all of the atomic density is located at the system's edge. As a result, the atom cannot be pumped any further in the $y$ direction. What is more, we saw that the atom's centre of mass does not have amplitude to travel backwards in the bulk. Consequently, it must be pumped to another energy band. Indeed, once the atom reaches the edge it is pumped along the edge state to the second band, as can be seen from the inset of Fig.~\ref{fig:density_decoupled}. During this period, the displacement saturates at the edge value $\mean{y} \approx 1$. Thanks to this, our measure of the Chern number is robust to small errors in the time of measure.

While the edge state is populated, we observe a rapid displacement of the atom's centre of mass in the $x$ direction. We can calculate the mean edge velocity by assuming that there is vanishing overlap between the edge states. In this case, in the neighbourhood of $k_x=\pi/d$, the state $\ket{k_x, n=\pm 1}$ is an approximate eigenstate of the Hamiltonian and Eq.~\eqref{eq:H0Mom} reduces to:
\begin{equation}
\begin{split}
 & \h_{\rm Hof} \ket{k_x, n=\pm 1} \approx E_n(k_x) \ket{k_x, n=\pm 1}, \text{ for } k_x \approx \pi/d, J_y \ll J_x \\
 & E_n(k_x) = -2 J_x \cos(k_x d - n \Phi).
\end{split}
\end{equation}
This allows us to compute the mean velocity along the edge in the $x$ direction:
\begin{equation}
 \mean{v_x}=\frac{1}{\hbar} \frac{\partial E_n(k_x)}{\partial k_x}\bigg|_{k_x=\frac{\pi}{d}, n=\pm 1}
 \approx \frac{\sqrt{3} J_x d}{\hbar}.
\end{equation}
We can compare this value to the slope of $\mean{x}$ between $t=1.5 T$ and $t=2T$, which we measure from Fig.~\ref{fig:COM_decoupled}(a). This value is found to be within $4\%$ of our estimate, such that we are able to estimate the edge state velocity to a high accuracy.

For $t=1.75 T$, the wavepacket is centred at point $C$ in Fig.~\ref{fig:dispersions}(a); it is subsequently pumped to point $D$, which it reaches at $t=2.75 T$. We can deduce the Chern number of the second band by measuring the centre of mass positions at these times. This provides the estimate $\CC_2 \approx -1.94$, which is once again in excellent agreement with the value obtained using the FHS algorithm of $\FHS_2=-2$.
Subsequently, almost all the density is promoted to the third band along the $n=-1$ edge state. Between times $t=3.5 T$ and $t=4.5 T$, the atomic density is pumped from point $E$ to $F$ in Fig.~\ref{fig:dispersions}(a). As previously, we estimate the Chern number of the third band, yielding: $\CC_3 \approx 0.97$, while the FHS algorithm gave: $\FHS_3=1$. A total $q N_y = 9$ periods are necessary to return the system to its initial state.

\begin{table}[t]
\centering
\caption[Measured Chern numbers for $J_y\ll J_x$]{Chern numbers measured from Fig.~\ref{fig:COM_decoupled}.}
\label{tab:Cherns}
\begin{tabular}{c|c|c}
\hline
Band & Formula & Value \\
$n=1$ & $\mean{y}_{t=T} - \mean{y}_{t=0}$ & $\CC_{1} \approx 0.99$ \\
$n=2$ & $\mean{y}_{t=2.75 T}  - \mean{y}_{t=1.75 T}$ & $\CC_{2} \approx -1.94$ \\
$n=3$ & $\mean{y}_{t=4.5 T} - \mean{y}_{t=3.5 T}$ & $\CC_{3} \approx 0.97$ \\
\hline
\end{tabular}
\end{table}

These measured Chern numbers are recorded in Table \ref{tab:Cherns}. We can compare these values to the number of edge states which exist at each boundary. Indeed, we saw in Sec.~\ref{sec:HofCylinder} that, in long cylindrical systems, the Chern number is a measure of the number of topologically protected edge states at any given boundary. By inspecting, for instance, the first band gap of the dispersion presented in Fig.~\ref{fig:dispersions}(a), we see that the system has a single edge branch localised at the $n=(N_y-1)/2$ edge with positive slope. Following the discussion of Sec.~\ref{sec:HofCylinder}, we deduce that the ground band's Chern number must be $\mathcal{C}_1=1$. By applying this reasoning to higher bands, we can see that, in this case, the Chern numbers from Table \ref{tab:Cherns} accurately measure the number of edge states. This means that, even in such a small system, the bulk transport properties are intimately linked to the properties at the edge of the system.

Thus, in this case, we were able to use the system's topological properties to pump the atomic density to any band. This allowed us to measure the Chern numbers of each band to an extreme accuracy. What is more, we did this starting from an initial state which is simple to prepare.
Strictly speaking, the derivation presented in Sec.~\ref{sec:Laughlin} is only valid for an infinite system, or for one which presents periodic boundary conditions. As the results presented in this section show, however, this method is extremely successful at measuring the Chern number even in a system as small as this one. Heuristically, we can say that, in a finite system, reducing the tunnelling amplitude a direction amounts to making that direction effectively longer. Thus, even though the Hofstadter strip had $N_y \ll N_x$, we effectively reduced this extreme anisotropy by setting $J_y \ll J_x$. This gives us an intuitive way of understanding why the measurement method we developed in Sec.~\ref{sec:ExperimentalSuggestion} was able to estimate the Chern number of the Hofstadter strip with such accuracy.

\subsection{Hofstadter strip with isotropic tunnelling amplitudes}
\label{sec:coupled_pumping}

We consider a Hofstadter strip subject to a magnetic field of $\Phi=2\pi/3$ with $N_y=3$ sites in the $y$ direction. The initial wavepacket occupies $w_x=30$ sites in the spatial direction, and is spin polarised with $n=0$. We adiabatically load the lattice following the method described in Sec.~\ref{sec:ExperimentalSuggestion}. In the light of the arguments presented at the end of Sec.~\ref{sec:decoupled_pumping}, we set $J_y=J_x$. The dispersion of this system is presented in Fig.~\ref{fig:dispersions}(b).
We anticipate that in this limit, edge effects which were negligible in the previous section profoundly modify the atom's dynamics.
After the loading sequence, we apply a constant force along the $x$ direction with magnitude $F_x d=\Delta E_1/100$, where $\Delta E_1 = 2.45 J_x$ is the first bulk energy gap.

\begin{figure}
\centering
\includegraphics[width=0.9\textwidth]{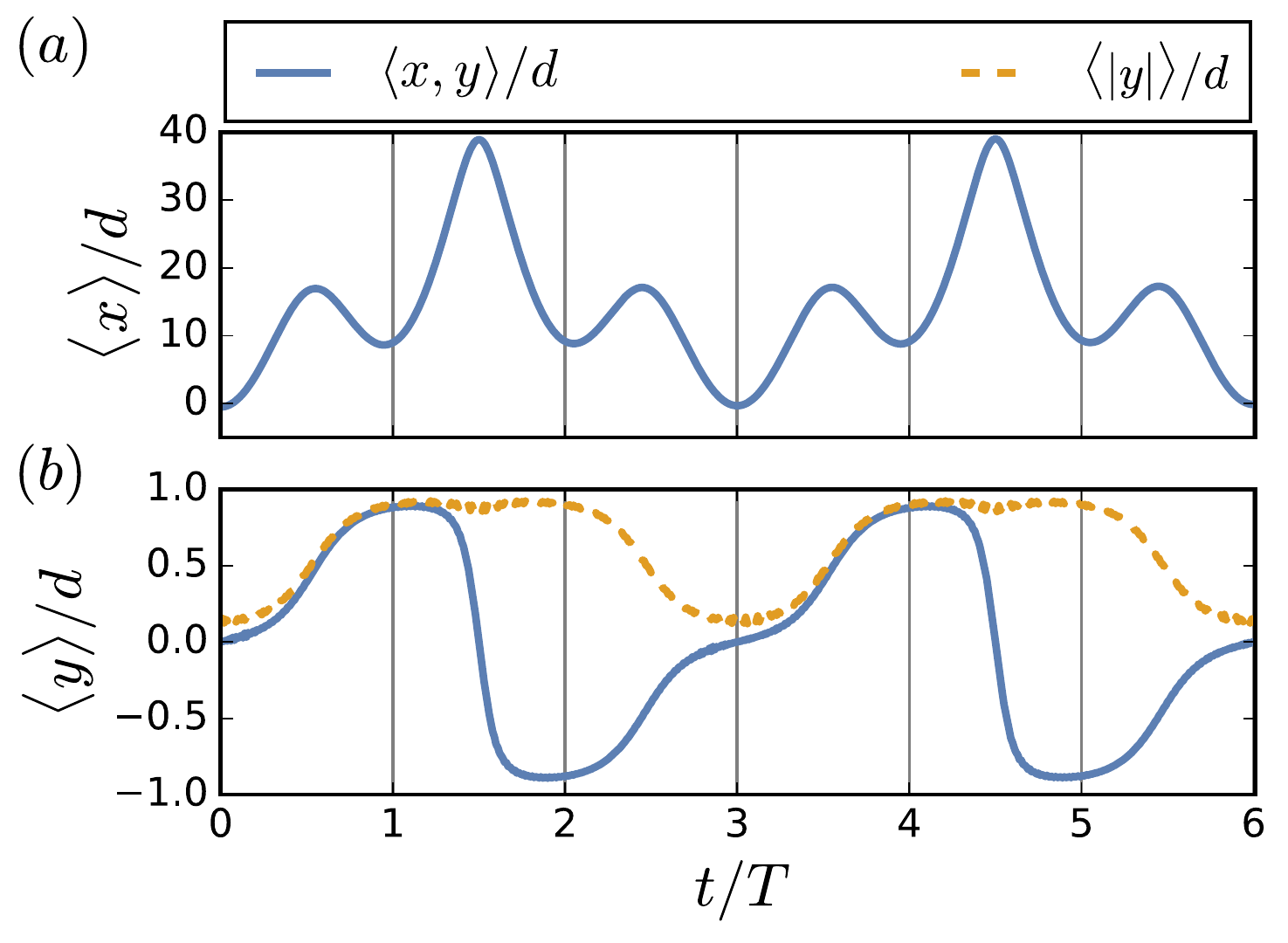}
\caption[Centre of mass in Hofstadter strip with hybridised edge states]{\label{fig:COM_coupled} Mean atomic position (solid blue) in the $x$ (a) and $y$ (b) directions for $J_y=J_x$. Vertical grey lines indicate integer periods of the vector potential. The mean distance from the origin (dashed orange) is presented in (b).}
\end{figure}
\begin{figure}
\centering
\includegraphics[width=0.8\textwidth]{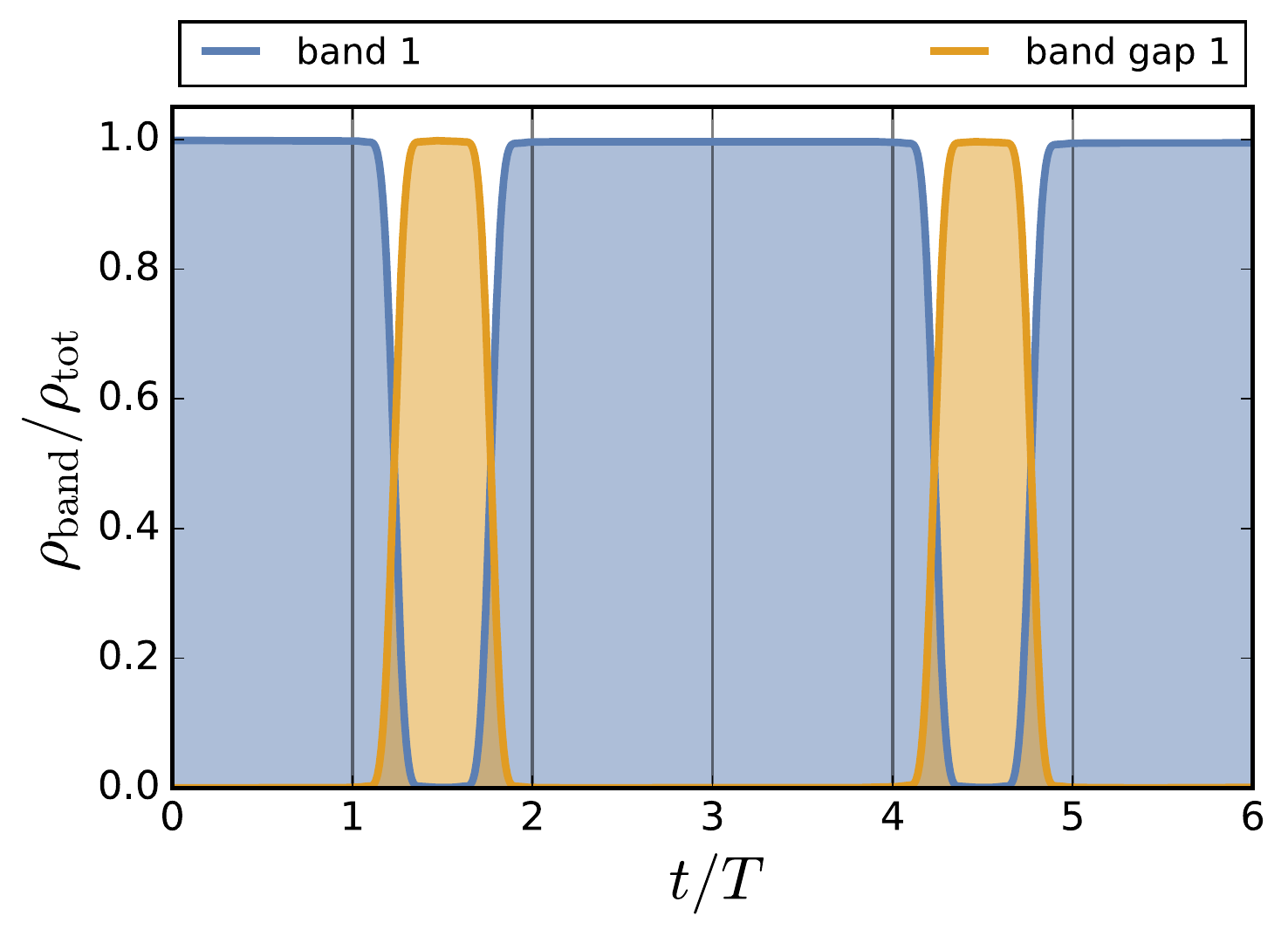}
\caption[Density fraction in Hofstadter strip with hybridised edge states]{\label{fig:density_coupled} Density fraction per band for $J_y=J_x$. Vertical grey lines indicate integer periods of the vector potential.}
\end{figure}

The atomic displacement, represented in Fig.~\ref{fig:COM_coupled}, can be understood when studied jointly with the energy band populations, which are plotted in Fig.~\ref{fig:density_coupled}. In the first period of the motion, the atom's centre of mass is pumped to the $n=1$ edge, resulting in the mean displacement $\mean{y(T)}-\mean{y(0)}=0.88$. This constitutes a measure of the Chern number $\CC_1$ which is vastly underestimated relative to the expected value of $\FHS_1=1$. In the following, we will show that this is due to the delocalisation of the edge states in spin space.


When the atom reached an edge in Sec.~\ref{sec:decoupled_pumping}, it had to be pumped to a different band. The reason for this can be understood from Eq.~\eqref{eq:ChernMeasureGeneral}, which tells us that the atom's probability density function in the bulk is pumped in one direction. In this case, however, the edge states are significantly hybridised, such that they form a band gap $\Delta \varepsilon_1 \gg |F_x| d$. This allows the atom to be adiabatically pumped from the $n=1$ edge directly to the $n=-1$ edge, without leaving the first energy band. What is more, we can see from Fig.~\ref{fig:COM_coupled}(b) that $\mean{|y|}$, the atom's mean distance from the origin, remains approximately constant for $t\in [T, 2T]$. The hybridisation of the system's edge states therefore introduces a new hopping term, coupling the $n=1$ and $n=-1$ edges. 
This feature is unique to systems that are short along one direction.


We notice that the process of pumping the centre of mass from one edge to the other along the edge states takes a time $T$. This is easiest understood if we consider the tunnelling term from one edge to the other as an effective periodic boundary condition. In this case, the atom can be indefinitely pumped in one direction as it never reaches the edge. Following this logic, the nearest neighbour of site $n=1$ in the positive $y$ direction is $n=-1$. From the derivation presented in Sec.~\ref{sec:Laughlin}, we know that the atom is adiabatically transported $\CC_1=1$ sites in the positive $y$ direction per period $T$, regardless of the tunnelling amplitude. As a result, it takes exactly a time $T$ for the centre of mass to be adiabatically pumped from the $n=1$ edge to the $n=-1$ edge, provided of course $|F_x| d$ is much smaller than the band gap, in this case, $\Delta \varepsilon_1$.


\section{Edge effects}
\label{sec:EdgeEffects}


In Sec.~\ref{sec:decoupled_pumping}, we studied the Hofstadter strip subject to a flux of $\Phi=2\pi/3$ in the limit of small spin tunnelling, $J_y \ll J_x$. Of particular interest, we were able to measure this system's Chern number with extreme accuracy. We did this by measuring the atom's mean transverse displacement, which we obtained from Fig.~\ref{fig:COM_decoupled}(b). In Sec.~\ref{sec:coupled_pumping}, we repeated this study on a Hofstadter strip with isotropic tunnelling, with $J_y=J_x$, and found that the estimated Chern number, which we obtained from Fig.~\ref{fig:COM_coupled}, was vastly underestimated.

The systems which we studied in Sec.~\ref{sec:StripSimulations} both had very few sites in the synthetic dimensions. It is natural that interaction of the atomic density with the systems edges would be very important in this limit, particularly if we are trying to measure a bulk property of the system such as the Chern number. In this section, we focus on the edge effects which can adversely affect our measurement protocol.

We hypothesise that it is the coupling of spin states which falsifies our measure, by causing the Hamiltonian's eigenstates to be delocalised in spin space. As a corollary, we expect our measure to converge to the correct answer as $J_y$ goes to zero. In the following, we calculate the mean atomic displacement over one period using perturbation theory, and verify that it predicts how our measure decays with $J_y$.

\subsection{Bloch wavefunctions to second order in perturbation theory}
\label{sec:JyPerturbation}

We will consider in the following the $J_x\gg J_y$ limit in the three-legged Hofstadter strip ($N_y=3$). By treating $(J_y/J_x)$ as a small parameter, we can use perturbation theory to find the approximate eigenstates of $\h_{\rm Hof}(k_x)$, given by Eq.~\eqref{eq:H0Mom}.
When $J_y=0$, the eigenstates of $\h_{\rm Hof}(k_x)$ are simply given by $\ket{n}$ for all $k_x$, i.e: the spin polarised state, with $n\in \{-1,0,1\}$. For arbitrary $J_y$, let $\ket{\varphi_n(k_x)}$ be the eigenstate of $\h_{\rm Hof}(k_x)$ which converges to $\ket{n}$ in the limit $J_y \xrightarrow{} 0$. To simplify the notation, we will omit the dependence of $\ket{\varphi_n}$ on $k_x$.

Let us initiate the system in the state with well defined momentum $\ket{\Psi_{t=0}} = \ket{k_x=0}\otimes \ket{\varphi_{0}}$. This state is an eigenstate of $\h_{\rm Hof}$; its corresponding eigenvalue is annotated in the Fig.~\ref{fig:dispersions}(a) by the point $A$. After a period of the potential, the atom is adiabatically pumped to $\ket{\Psi_{t=T}} = \ket{k_x=2\pi/3}\otimes \ket{\varphi_{1}}$ (annotated by the point $B$).

The total displacement in the $y$ direction over one period of the potential is therefore simply given by:
\begin{equation}
 \label{eq:total_y_displacement}
\mean{\Delta y} = \bra{\Psi_{t=T}}\hat{n}\ket{\Psi_{t=T}} - \bra{\Psi_{t=0}}\hat{n}\ket{\Psi_{t=0}},
\end{equation}
where $\hat{n}$ is the position operator in the spin direction. Note that, due to the symmetry of the Hofstadter strip around $n=0$, the mean spin of $\ket{\varphi_{0}}$ is always zero, such that $\bra{\Psi_{t=0}}\hat{n}\ket{\Psi_{t=0}} = 0$.

In this setting, it is therefore sufficient to find the approximate expression of the eigenstate $\ket{k_x=2\pi/3}\otimes\ket{\varphi_{1}}$ to find the mean transverse displacement over a period. Because this state doesn't occur at a degeneracy point, we can do this using non-degenerate perturbation theory. Let:
\begin{equation}
 E_{n,n'}(k_x)= \cos(k_x d - n \Phi) - \cos(k_x d - n' \Phi)
\end{equation}
and
\begin{equation}
 \lambda = \frac{J_y}{2 J_x}
\end{equation}
We find:
\begin{equation}
 \begin{split}
 \ket{\varphi_{1}} =
 &\left[ 1 - \frac{\lambda^2}{2 E_{1,0}(k_x)^2}\right] \ket{n=1}
 + \frac{\lambda}{E_{1,0}(k_x)} \ket{n=0}\\
 & + \frac{\lambda^2}{E_{1,-1}(k_x)E_{1,0}(k_x)}\ket{n=-1}
 + \mathcal{O}\left( \lambda^3 \right).
 \end{split}
\end{equation}
Note that this state is normalised (discarding terms of order $\lambda^3$ and higher). Substituting this into Eq.~\eqref{eq:total_y_displacement}, we find:
\begin{equation}
 \label{eq:total_y_displacement2}
 \mean{\Delta y} = \bra{\varphi_{1}}\hat{n}\ket{\varphi_{1}}\Big{|}_{k_x=\frac{2\pi}{3}} = 1 - \left( \frac{J_y}{3 J_x} \right)^2 + \mathcal{O}\left( \frac{J_y}{J_x} \right)^3. 
\end{equation}

\subsection{Comparison to the measured Chern number}

\begin{figure}[t]
\centering
\includegraphics[width=.85\linewidth]{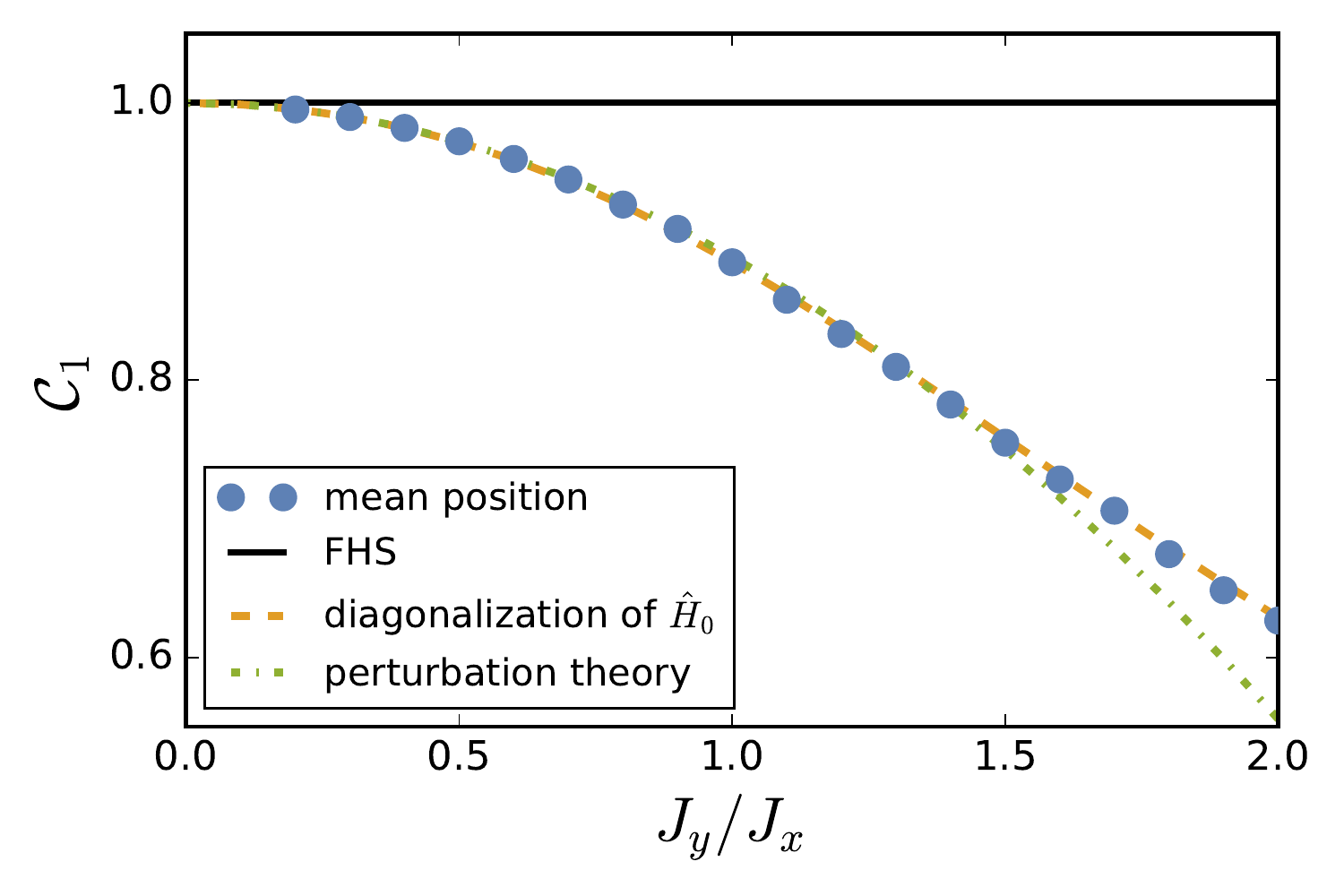}
\caption[Measured Chern number versus varying $J_y/J_x$]{Measured Chern number versus varying $J_y/J_x$ obtained from mean position measurements (blue dots), from Eq.~\eqref{eq:total_y_displacement2}, calculated using perturbation theory (dash-dotted green), calculated by diagonalising $\h_{\rm Hof}(k_x)$, Eq.~\eqref{eq:H0Mom} (dashed orange), and computed with the FHS algorithm (solid green).}
\label{fig:ChernVsJy}
\end{figure}

We now estimate the Chern numbers for a range of $J_y/J_x$ values, and plot these as blue dots in the Fig.~\ref{fig:ChernVsJy}. For these simulations, we used a three-legged Hofstadter strip subject to a magnetic flux $\Phi=2\pi/3$. When considering a single period of the force, this lattice size is sufficient for edge effects in the spatial direction to be negligible. The atom is initially in the $n=0$ state, and occupies $w_x=30$ sites. We load this atom into the lattice adiabatically by switching on the spin tunnelling amplitude. We then apply a force along the $x$ direction with amplitude to be $F_x d = 0.02 \Delta E_1$, such that the single band approximation is applicable.


In Sec.~\ref{sec:JyPerturbation}, we calculated analytically the mean spin of the lowest energy eigenstate of $\h_{\rm Hof}(k_x=2\pi/3)$ for $N_y=3$. This result, given by Eq.~\eqref{eq:total_y_displacement2}, is plotted in dashed dotted green alongside our measure. Additionally, we diagonalise the Hamiltonian, as given by Eq.~\eqref{eq:H0Mom}, and plot in dashed orange the mean spin of its lowest energy eigenstate at at $k_x=2\pi/3$. We find an excellent agreement between both these estimates
and the Chern number measurement obtained from simulations.
We conclude that our hypothesis is verified, such that it is the coupling between different spin states is responsible for the reduced $y$ displacement which we observe in Fig.~\ref{fig:COM_coupled}.
Importantly, we can see that the parabola converges to $\FHS_1=1$ as $J_y \xrightarrow{} 0$, indicating that our measure converges to the correct answer in the limit $J_y \ll J_x$.



It is important to understand that the corrections to the atom's displacement which we are observing are in fact an edge effect. Indeed, our measure of the Chern number relies on a correlation between the atom's quasimomentum $k_x$ and its mean spin \parencite{Wang2013}.
This correlation is not destroyed when the bulk eigenstates of $\h_{\rm Hof}$ are delocalised in the synthetic dimension. When the eigenstates have amplitude at the edges of the system, however, delocalisation in the spin direction is highly anisotropic. The consequence, as we observed in Sec.~\ref{sec:JyPerturbation}, is that their mean position becomes dependent on the tunnelling amplitude in the spin direction. It is this effect which falsifies our measure of the Chern number.

\section{Higher Chern number measurement}
\label{sec:HigherChernMeasure}

In this section, we study a Hofstadter strip which presents a ground band Chern number $|\CC_1| > 1$. The Hofstadter model can in general present large Chern numbers in the ground band, but this is always accompanied by a very reduced band gap, making this measurement problematic. The aim of this section is two fold. First, we would like to show that our measurement method is applicable even under such extreme conditions. Secondly, for this particular flux, the Chern number calculated using the FHS algorithm changes as the system size is reduced. We would like to see if these jumps in the Chern number are physical, i.e: if they are reflected in the pumping properties of the bulk, or if they are artificial.

In the following subsections, we will consider a system traversed by a magnetic flux of $\Phi=4\pi/5$. We prepare an atom restricted to $w_x=50$ sites in the spatial direction and belongs to the $n=0$ site in the spin dimension. This state is adiabatically loaded into the lattice by ramping up the spin tunnelling amplitude to $J_y = J_x /2$. At the end of the loading sequence, this system has an extremely small first energy gap with amplitude $\Delta E_1 = 0.11 J_x$. For the adiabatic approximation to be valid, we apply an extremely weak force, with amplitude $F_x=0.01\Delta E_1 /d$.

The ground band's Chern number, for a system with a large number of sites in the $y$ direction, is known to be $\CC_1=-2$. We will discuss in detail the system with $N_y=5$ sites in the spin direction in Sec.~\ref{sec:LargeChernNy5}, then with $N_y=4$ in Sec.~\ref{sec:LargeChernNy4}, and finally compare to the results of the FHS algorithm in Sec.~\ref{sec:LargeChernFHSComparison}.

\subsection{Hofstadter strip with \texorpdfstring{$\Phi=4\pi/5$}{Phi=4pi/5}}
\label{sec:LargeChernNy5}

\begin{figure}[t]
  \centering
  \includegraphics[width=0.75\textwidth]{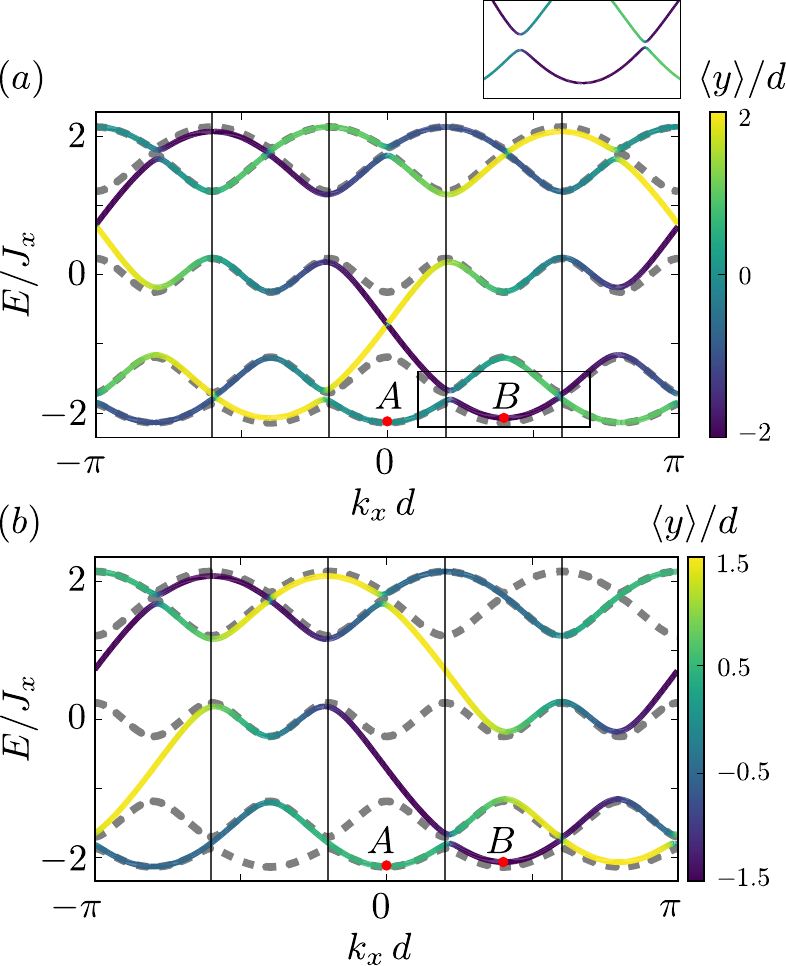}
  \caption[Hofstadter strip dispersion for $\Phi=4\pi/5$]{\label{fig:dispersions_qx5} Dispersion of the Hofstadter model with $\Phi=4\pi/5$, $J_y=J_x/2$, and (a): $N_y=5$, (b): $N_y=4$ sites in the $y$ direction. Dashed grey: dispersion with periodic boundary conditions (with $N_y=5$). Coloured: system with open boundary conditions. The colour coding indicates the mean eigenstate's position along $y$. In the first band gap, a pair of well localised edge states cross at $k_x=3\pi/5$ [(a), inset]. The vertical lines indicate the change in $k_x$ in a period of Bloch oscillations. In this way, an atom starting at point $A$ can be pumped to $B$.}
\end{figure}

In this subsection, we focus on the Hofstadter strip subject to a flux of $\Phi=4\pi/5$ with $N_y=5$ sites in the $y$ direction. 
The system's dispersion is presented in Fig.~\ref{fig:dispersions_qx5}(a). The periodic system, plotted in grey, presents five bands for $\Phi=4\pi/5$. Two pairs of edge states are clearly visible between the first and second bands, which cross at $k_x = \pm 3\pi/5$ (see inset).
Specifically, we find two forward travelling edge states in the first band gap, which are bound to the $n<0$ edge. These occur at $k_x = -3\pi/5$ ($k_x = 3\pi/5$) and are centred about $n=-1$ ($n=-2$). Likewise, we find two backwards travelling edge states, bound to the $n>0$ edge. These occur at $k_x = -3\pi/5$ ($k_x = 3\pi/5$), and are centred about $n=2$ ($n=1$). Following the discussion presented in Sec.~\ref{sec:decoupled_pumping}, this is what we would expect from a system with a Chern number of $\CC_1=-2$.

Let us now simulate this system and estimate the Chern number from the bulk dynamics. After the atom is loaded into the lattice, it is in a superposition of eigenstates narrowly centred around the point $A$ in Fig.~\ref{fig:dispersions_qx5}(a). The atom is subsequently pumped to $n=-2$ site of the bottom band during the first period of evolution, which is annotated by point $B$. From the location of the edge states, we can see that the pumping from points $A$ to $B$ takes place without the atom leaving the periodic system's bottom band. This allows us to measure this band's Chern number, following the method of Sec.~\ref{sec:Laughlin}. As previously, we measure the displacement of the atomic centre of mass in the $y$ direction, which yields $\CC_1\approx -1.98$. This estimate is extremely close to the value expected from the bulk-boundary correspondence.

\subsection{Generalisation to \texorpdfstring{$q>N_y$}{q>Ny}}
\label{sec:LargeChernNy4}



The Hofstadter strip with $N_y=4$ is extremely interesting because $\h_{\rm Hof}(k_x)$, Eq.~\eqref{eq:H0Mom}, only has $N_y=4$ eigenstates for each quasimomentum value $k_x$. For this value of $N_y$, the index $n$ can take values $n\in \{-\frac{3}{2},-\frac{1}{2},\frac{1}{2},\frac{3}{2}\}$. The dispersion of this system is plotted in the Fig.~\ref{fig:dispersions_qx5}(b). Notice that each eigenvalue of this dispersion can be associated either to one of the $q=5$ bands of the periodic system, or to one of the system's edge states. As previously, two pairs of edge states can be identified in the first energy gap. These can be found either by inspecting the dispersion or through analytical calculations \parencite{Hatsugai1993a}. Two forward travelling edge states can be found at $k_x = -3\pi/5$ ($k_x = 3\pi/5$). These are centred about $n=-0.5$ ($n=-1.5$). Similarly, two backward travelling edge states occur at $k_x = -\pi/5$ ($k_x = 3\pi/5$). These are centred about $n=0.5$ ($n=1.5$). Note that these are shifted in quasimomentum space relative to Fig.~\ref{fig:dispersions_qx5}(a) due to the suppression of a site. As in the previous subsection, we would expect a system with a Chern number of $\CC_1=-2$ to display this edge state structure.

Interestingly, it is still possible to measure the lowest band's Chern number, provided we ensure that all of the dynamics takes place in this band. As previously, the atom is initially in a superposition of states narrowly centred around the annotated point $A$ in the Fig.~\ref{fig:dispersions_qx5}(b). We drive the system with the same force as in the previous subsection. This pumps the atomic density adiabatically from the point $A$ in Fig.~\ref{fig:dispersions_qx5}(b) to point $B$. As can be seen by inspecting the dispersion,
all the eigenstates which the atom explores during this pumping sequence can be associated to the bottom band of the periodic system. This means that, in this setting, we are able to measure the first band's Chern number from the atom's displacement, yielding the value of $\CC_1\approx -1.96$. Note that this is an extremely surprising observation: we are able to measure the system's Chern number even though its bands are not well distinguished. The measured value is, here again in excellent agreement with the value expected from the bulk-boundary correspondence.

\subsection{Comparison to the FHS algorithm results}
\label{sec:LargeChernFHSComparison}

\begin{figure}[t]
\centering
\includegraphics[width=\textwidth]{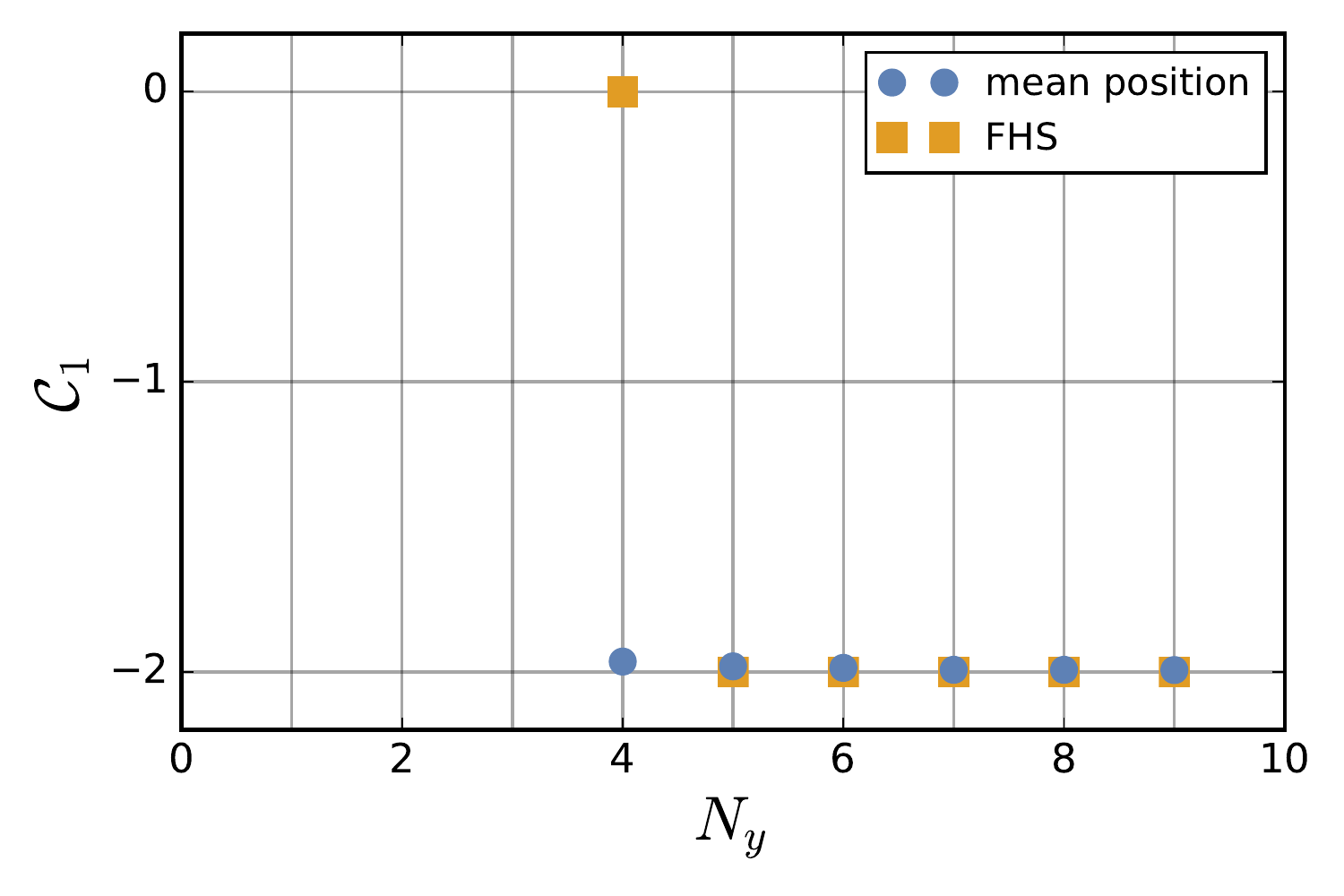}
\caption[Breakdown of the FHS algorithm]{\label{fig:higherChern} We estimate the lowest band Chern number for $N_y \in [4, 9]$, measured using atomic pumping (blue dots) and with the FHS algorithm (orange squares). For $N_y > 5$, both results agree, but differ for for $N_y < 5$. This is due to the breakdown of the FHS method.}
\end{figure}

Using the same initial state as in the previous subsections, we measure the Chern number in this system for $N_y \in [4,9]$. These are plotted as blue dots in the Fig.~\ref{fig:higherChern}. As $N_y$ is increased, we observe that the measured value tends asymptotically to $\CC_1=-2$.

Alongside these estimates, we plot as orange squares the Chern number calculated using the FHS algorithm. Note that, in the limit $N_y<5$, this value does not correspond to the Chern number of the infinite systems. As we discussed in Sec.~\ref{sec:FHSalgo}, this is simply because, for very small systems, the Berry flux through some plaquettes can exceed $\pm \pi$, such that the FHS algorithm cannot record them accurately \parencite{Fukui2005}. We could ask, however, if the Chern number in this system could change as the system size is reduced. From the results presented here, we can see that our measure does not agree with the FHS algorithm for $N_y=4$. We conclude that the Chern number calculated using the FHS algorithm is not related to the system's transverse conductance in this limit. What is more, we can say from the results of Sec.~\ref{sec:LargeChernNy4} that the FHS algorithm fails to accurately count the number of edge states which the system displays for $N_y=4$. In light of these arguments, it would seem like the FHS algorithm does not accurately predict the Chern number in this system for $N_y=4$.

\section{Generalisation to non-interacting atomic gases}
\label{sec:AtomicGas}

In this section, we consider the dynamics of a non-interacting gas of Fermions in the Hofstadter strip. We will concentrate on the case where a single one of the system's bands is homogeneously filled, and show that, when this is the case, we can measure the Chern number through pumping experiments, using the method from Sec.~\ref{sec:ExperimentalSuggestion}.

We explore this possibility because it is likely that an experimental implementation of the scheme described in Sec.~\ref{sec:ExperimentalSuggestion} is not genuinely single particle, but uses a dilute gas of atoms instead. For this reason, it is useful to analyse both the difference in dynamics with the single particle case, and the difficulties which can be expected to arise.

There are, however, two significant advantages to performing this experiment with a Fermionic gas. Indeed, as we will show in the Sec.~\ref{sec:FilledBandGroupVelocity}, studying a system with a filled band gives us access to a new way of measuring the Chern number. Unfortunately, the implementation of this method on the Hofstadter strip will prove problematic. Secondly, it was mentioned in Sec.~\ref{sec:ExperimentalSuggestion} that using an atomic cloud to perform this experiment could give us access to partial density transfer techniques. These would allow us to measure the cloud's mean position without destroying the experiment, such that the atoms' mean position can be monitored at all times in a single experiment.

\subsection{Cloud preparation and Brillouin zone population}
\label{sec:GasBZPopulation}

To underline the differences from the single particle case, we consider the same system which we studied in detail in Sec.~\ref{sec:decoupled_pumping}. As previously, we consider a system with open boundary conditions, presenting $N_y=3$ sites in the synthetic dimension and subject to a flux of $\Phi=2\pi/3$. Initially the atoms are in the $n=0$ spin state, and are restricted to $w_x=30$ sites in the $x$ direction. We set the Fermi energy to be $E_F=-1.3 J_x$, and adiabatically load the lattice by linearly increasing the spin tunnelling amplitude to $J_y=0.2 J_x$. This Fermi energy is slightly lower than the highest energy state of the first band for a system with periodic boundary conditions and $J_y/J_x=0.2$ [dashed grey line in the Fig.~\ref{fig:dispersions}(a)]. We fill every available state which are consistent with these requirements, such that the system presents $N_P=9$ atoms.

\begin{figure}[t]
    \centering
    \includegraphics[width=\textwidth]{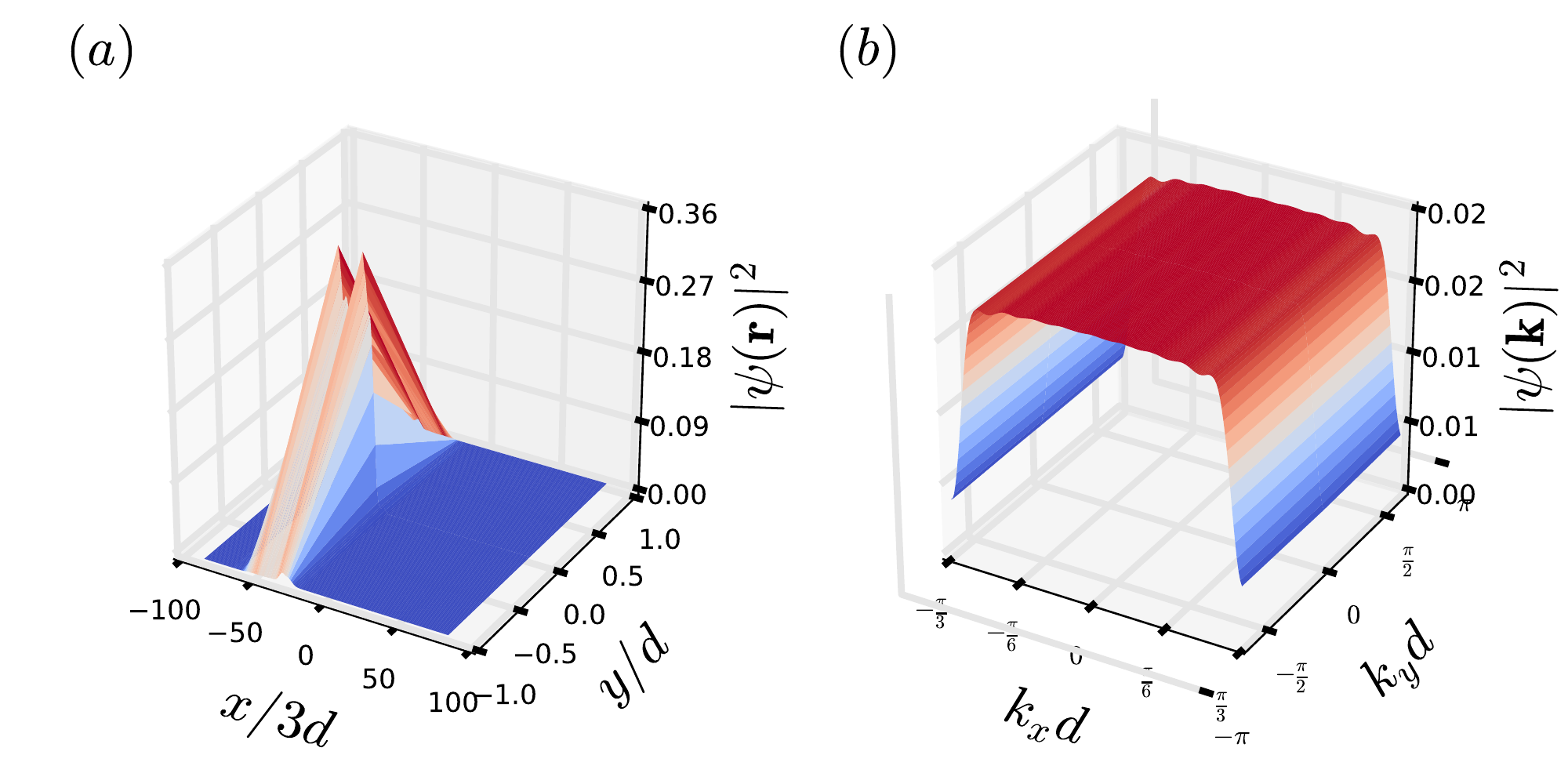}
    \caption[Initial atomic cloud]{\label{fig:MultiParticleInitialState} Density distribution of the initial atomic cloud in (a): real space and (b): quasimomentum space. These are presented in the gauge of Eq.~\eqref{eq:H0}. The atomic cloud is initially restricted to $w_x = 30$ sites in the $x$ direction, and spin polarised with $n=0$.}
\end{figure}

The probability density distribution at the end of the loading sequence is presented in Fig.~\ref{fig:MultiParticleInitialState}. The real space distribution, shown in Fig.~\ref{fig:MultiParticleInitialState}(a), is sharply peaked around $n=0$, resulting in a quasimomentum distribution, shown in Fig.~\ref{fig:MultiParticleInitialState}(b), which has equal amplitude for all $k_y$. The quasimomentum distribution in the $x$ direction presents a density depletion around $k_x=\pm \pi / (3d)$. This is because the system's dispersion, Fig.~\ref{fig:dispersions}(a), admits no state with energy smaller than $E_F$ for these quasimomenta. As a result, these eigenstates were discarded during the initial cloud preparation. Note that, aside from this density depletion, we have an almost homogeneous population of the Hofstadter strip's first band.

\subsection{Chern number, measured from the mean displacement} 
\label{sec:decoupled_pumping_gas}


\begin{figure}[t]
    \centering
    \includegraphics[width=0.9\textwidth]{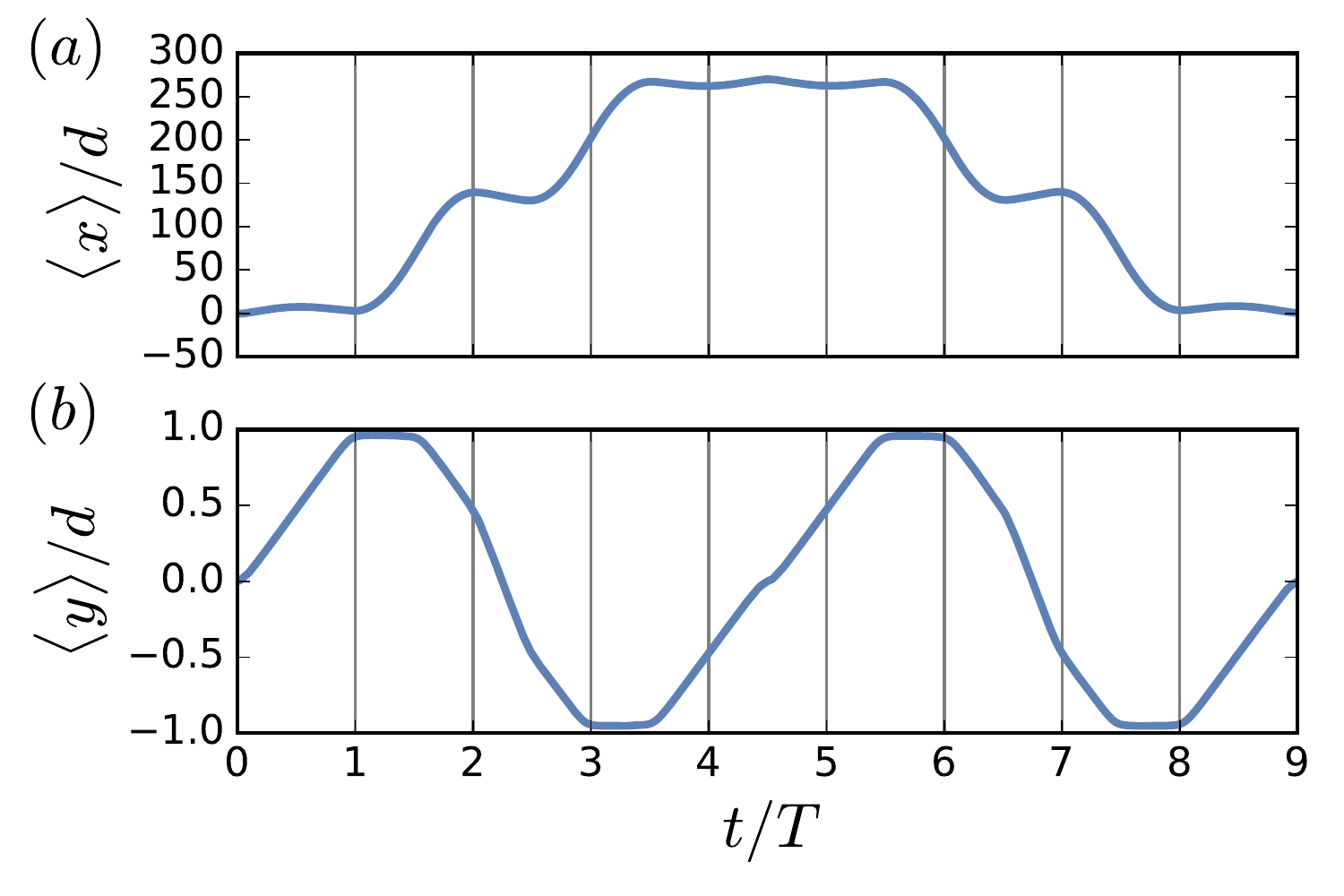}
    \caption[Mean atomic cloud position]{\label{fig:MultiParticleGroupVel} Mean position of the atomic cloud as a function of time in the (a) spatial and (b): synthetic directions. Time is measured in units of $T$, the period associated to the force.}
\end{figure}
\begin{figure}[t]
    \centering
    \includegraphics[width=0.9\textwidth]{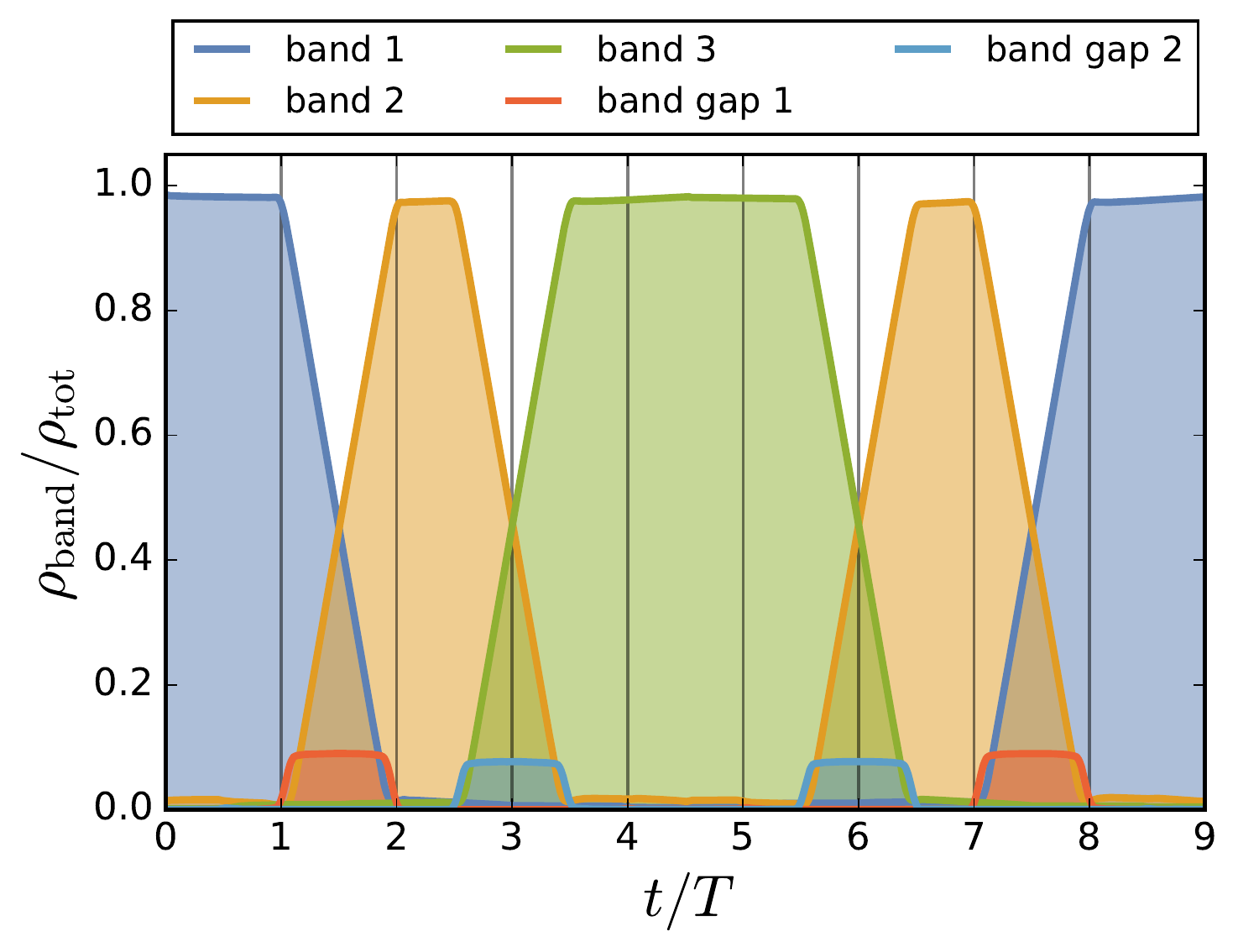}
    \caption[Band population of the atomic cloud]{\label{fig:MultiParticleDensityFraction} Band and edge state populations as a function of time, for $J_y=J_x/5$. Time is measured in units of $T$, the period associated to the force.}
\end{figure}

Following the protocol detailed in Sec.~\ref{sec:Laughlin}, we induce Bloch oscillations by applying a weak force in the $x$ direction. We choose a force magnitude $F_x d= 0.03 \Delta E_1$, such that the potential gradient is much smaller than the first bulk band gap, $\Delta E_1=0.42 J_x$. The atomic centre of mass position is plotted in Fig.~\ref{fig:MultiParticleGroupVel}. This figure can be understood when studied jointly with the different levels' populations, shown in Fig.~\ref{fig:MultiParticleDensityFraction}. Qualitatively, the system's behaviour is very similar to the dynamics observed in Sec.~\ref{sec:decoupled_pumping}. In the first period, the centre of mass is pumped from $n=0$ to the $n=1$ edge. We record the displacement transverse to the force, which provides an estimate $\CC_1=0.95$ of the first band's Chern number.

During the second period of motion, the atomic cloud is pumped along the edge state to the second band.
While density is being pumped to the second energy band, it remains located at the $n=1$ edge, but moves rapidly in the $x$ direction.

\begin{table}[t]
\centering
\caption[Measured Chern numbers for a Fermionic gas]{Chern numbers $\CC_j$ measured from the mean displacement and $\CC_j^{\rm grad}$ measured from the slope of Fig.~\ref{fig:MultiParticleGroupVel}(b), with $j\in\{1,2,3\}$ the band index.}
\label{tab:MultiParticleCherns}
\begin{tabular}{c|c|c|c}
\hline
Band & Time interval & Displacement & Slope \\
$n=1$ & $t/T \in [0,1]$ & $\CC_{1} \approx 0.95$ & $\CC_1^{\rm grad} \approx 1.03$ \\
$n=2$ & $t/T \in [2,2.5]$ & $\CC_{2} \approx -1.88$ & $\CC_2^{\rm grad} \approx -1.99$ \\
$n=3$ & $t/T \in [3.5,4.5] $ & $\CC_{3} \approx 0.94$ & $\CC_3^{\rm grad} \approx 1.01$ \\
\hline
\end{tabular}
\end{table}

As previously, we can measure the second band's Chern number by measuring the displacement of the centre of mass. Note however that the only times for which the second band is maximally populated is for $t\in [2T, 2.5T]$. By measuring the displacement over this half period, we deduce $\CC_2=-1.88$.
For $t\in [2.5T, 3.5T]$, the atomic density is promoted to the third band along the $n=-1$ edge state. We estimate the third band's Chern number to be $\CC_3=0.94$. These values are recorded in the Table \ref{tab:MultiParticleCherns}.


Note however that, while all these estimates
are excellent, they are not as close to the expected values as the single particle estimates recorded in Table \ref{tab:Cherns}. This is a result of a small portion of the atomic density being promoted to higher bands during the loading sequence.
This is due to the loading time considered in Sec.~\ref{sec:decoupled_pumping} is too short for an atomic gas which presents such a broad distribution of quasimomenta.

\subsection{Chern number, measured from the mean velocity}
\label{sec:FilledBandGroupVelocity}

A noteworthy difference between the single and many body cases is that the mean transverse displacement in Fig.~\ref{fig:MultiParticleGroupVel}(b) is linear with time.
This is expected from the TKNN formula, Eq.~\eqref{eq:TKNN}, which tells us that, in a system which presents a filled band, the mean transverse velocity is constant and proportional to the Chern number. It was suggested in Ref.\ \parencite{Dauphin} to measure the Chern number from the gradient of the displacement.

In the following, we will use Eq.~\eqref{eq:TKNN} to measure the Chern number in the Hofstadter strip. The main advantage of this measurement method is that it does not require time evolving the system for a full period of the dynamics. Note however that many measurements of the atoms' position is necessary. This either means that the experiment must be performed many times, or that non-destructive measurement techniques must be used.

The Chern number can be estimated by measuring the slope of the displacement in the Fig.~\ref{fig:MultiParticleGroupVel}(b) over the first period of the potential and substituting it into Eq.~\eqref{eq:TKNN}. This returns $\CC_1^{\rm grad} \approx 1.03$. We can similarly estimate the Chern numbers of the second and third band. By performing a linear fit of the displacement between $t/T \in[2,2.5]$, we obtain $\CC_2^{\rm grad} \approx -1.99$, and by fitting the displacement for $t/T \in[3.5,4.5]$, we find $\CC_3^{\rm grad} \approx 1.01$. These values are recorded in Table \ref{tab:MultiParticleCherns}.

While these measures are very close to the results we obtained by measuring the mean atomic displacement, they tend to be overestimated. This is because of the erroneous assumption that the Brillouin zone is homogeneously populated.
As we saw in Sec.~\ref{sec:GasBZPopulation}, the Brillouin zone population presents a narrow density depletion in the $k_x$ direction. The Berry curvature is largest in the region around $k_x = 2\pi/(qd)$, such that when the density depletion lies in this region, the instantaneous mean velocity is smaller than Eq.~\eqref{eq:TKNN}. It is for this reason, for instance, that the Fig.~\ref{fig:MultiParticleGroupVel}(b) presents a clear plateau around $t=4.5T$. 

Conversely, when the density depletion is away from $k_x = 2\pi/(qd)$, fast moving states, i.e: ones which exist in a region of the Brillouin zone which presents a large Berry curvature, have a more important relative population than they would in a system with a filled band. They therefore contribute more importantly to the mean velocity. Here, the density depletion is very narrow in $k_x$ space, such that the mean velocity extracted from fitting Fig.~\ref{fig:MultiParticleGroupVel}(b) is larger than Eq.~\eqref{eq:TKNN}.

To conclude, we have shown that a system's Chern number can also be measured from the mean displacement of a non-interacting Fermionic gas. The main drawback is that loading the lattice adiabatically is problematic due to the increased population of the band. Additionally, preparing a system with a homogeneously filled band provides an alternative way to measure the Chern number.

\section{Perturbations which break translational invariance}
\label{sec:broken_trans_inv}

A fundamental property of topologically non-trivial systems is their robustness to local perturbations.
In this section, we study the robustness of our measure to two types of perturbations which are extremely relevant from an experimental point of view. Specifically, in Sec.~\ref{sec:StaticDisorder}, we will look at the effect of local disorder, which is present to some extent in all real world systems, and in Sec.~\ref{sec:HarmonicTrap}, we will study the system in the presence of a harmonic trapping potential, which is used in most ultracold atom experiments to keep the atomic cloud localised.

\subsection{Static disorder}
\label{sec:StaticDisorder}

\begin{figure}[t]
\centering
\includegraphics[width=0.85\linewidth]{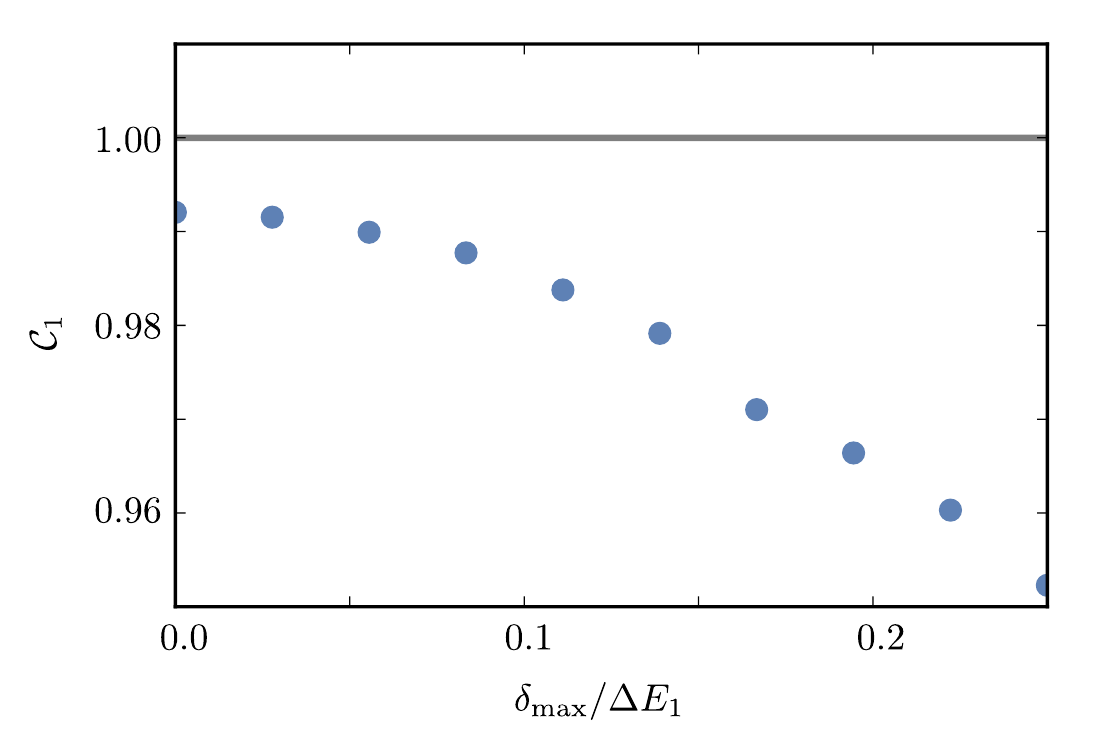}
\caption[Chern number versus disorder magnitude]{\label{fig:disorder}
Chern number of the ground band measured by Laughlin pumping, in presence of static onsite magnetic disorder of amplitude  $\delta_{\rm max}$. Each data point is an average of $20$ realisations, and $N_y=3$.
}
\end{figure}

A fundamental property of topologically non-trivial systems is their robustness to local perturbations. In particular, the interplay of disorder and topologically phases is an extremely rich area of research \parencite{Niu1985,Prodan2011}.
In this subsection, we study to what extent our measure is affected by the presence of disorder.

To this end, we consider a three-legged ladder ($N_y=3$), traversed by an external magnetic flux of $\Phi=2\pi/3$, in presence of static onsite disorder:
\begin{equation}
\h_\text{dis}=\h+\hat{V}_\text{dis}=\h+\sum_{m,n}\delta_{m,n}\hat{c}^\dagger_{m,n} \hat{c}_{m,n},
\end{equation}
where $\delta_{m,n}$ is uniformly distributed in the interval $[0,\delta_\text{max}]$ and corresponds to uncorrelated disorder in both the spatial and synthetic dimensions. 
We consider an atom which is initially spin polarised in the $n=0$ state, and constrained to $w_x=60$ sites in the $x$ direction. This state is adiabatically loaded   into the lattice by switching on the spin hopping amplitude to $J_y=0.2 J_x$. We then subject the atom to a constant force with amplitude $F_x d=0.03 \Delta E_1$, where the first bulk band gap has amplitude $\Delta E_1= 1.65 J_x$.

In Fig.~\ref{fig:disorder}, we plot the measured Chern number of the lowest band, averaged over $20$ realisations, as a function of $\delta_{\rm max}$.  For increasing amplitudes of the random potential, the measurement of the Chern number deviates steadily from the expected result, and we identify the broadening of the wavepacket as the main source of error. Remarkably, even for disorder magnitudes as high as $\delta_{\rm max} = \Delta E_1/4$, we still measure a Chern number which is within $94\%$ of its actual value. For larger magnitudes of the random potential, tunnelling events between bands (Landau-Zener transitions) become the main source of error in our measure. We conclude that, as is generally the case in topological systems, disorder does not significantly affect topological measurements in Hofstadter strips up to disorder amplitudes comparable to the band gap.


\subsection{Harmonic trap}
\label{sec:HarmonicTrap}

\begin{figure}[t]
\centering
\includegraphics[width=0.7\linewidth]{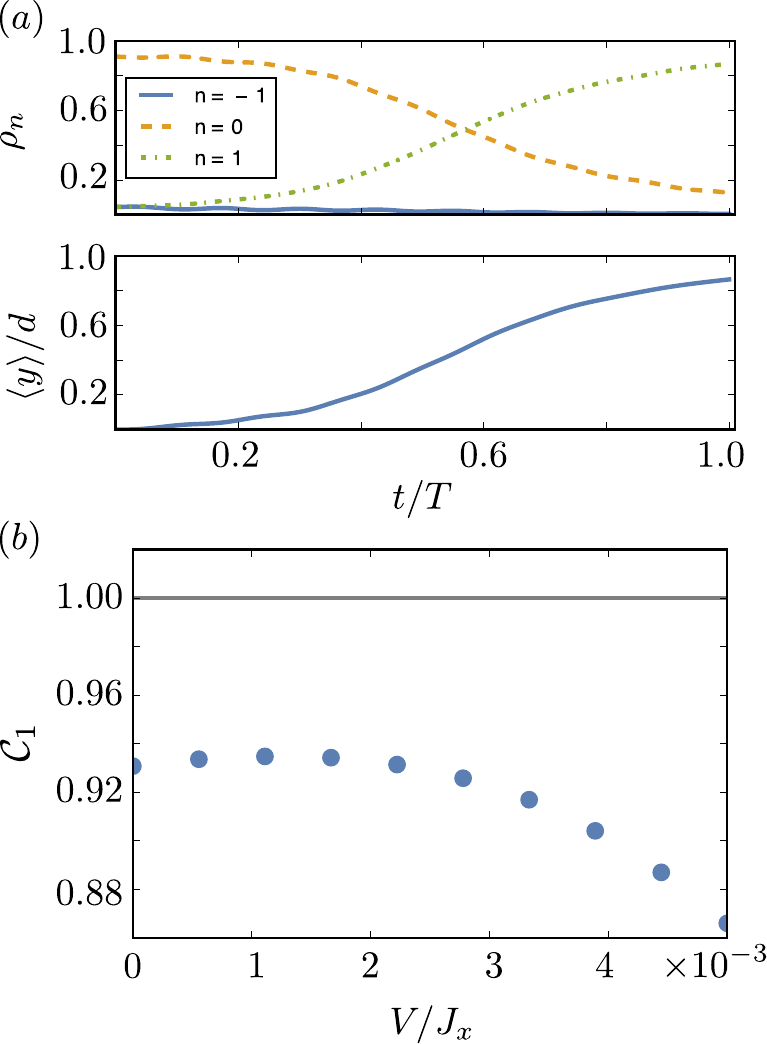}
\caption[Pumping in a harmonic trap]{\label{fig:SpinPopTrap}
Pumping dynamics in a trap. (a): Spin populations $\rho_n$ (top) and mean displacement along $y$ (bottom), plotted as a function of time, in a trap of strength $V=5\times 10^{-3} J_x$. (b): Measured ground band Chern number versus typical trap depths $V$.
}
\end{figure}

In ultracold atom experiments, the atomic gas is usually spatially constrained using a harmonic trap. To ensure that this potential is not disruptive to our measure, we must verify that the dynamics it induces are negligible on the time scale of a Bloch oscillation. Specifically, a harmonic trap with frequency $\omega/2\pi$ induces dipolar oscillations which can spoil Bloch oscillations, which are a crucial element of our protocol (see Sec.~\ref{sec:Laughlin}). To limit their effect,  we must ensure that $\omega\ll2\pi/T$.

The Hamiltonian for an atom of mass $M$ in a harmonic trap reads:
\begin{equation}
\label{eq:Htrap}
\h_{\rm ho} =  \h + \hat{V} = \h + V \sum_{m,n} (m-N_x/2)^2 \cc_{m,n}^\dagger \cc_{m,n},
\end{equation}
with $V=M \omega^2 d^2/2$ the trap depth.

For concreteness, let us consider $^{39}$K atoms in an optical lattice with lattice spacing $d=532\, $nm. Typical spatial hopping amplitudes and trap frequencies are, respectively, $J_x/h \approx 100\, $Hz and $\omega/2\pi \approx 30 \, $Hz,  leading to characteristic trapping energies on the order of $10^{-3} J_x$.

Let us consider $V=5\times 10^{-3} J_x$, and estimate the Chern number for a three-legged Hofstadter strip subject to an external magnetic flux of $\Phi=2\pi/3$.
We set the hopping amplitudes' ratio to $J_y/J_x=0.7$, such that the band gap is $\Delta E_1 = 1.65 J_x$, and apply a constant force with magnitude $F_x d=\Delta E_1/10$. Under these conditions, we have $T^{-1} \approx 50\,\text{Hz} \gg \omega / 2\pi$.

As in previous sections, we adiabatically load the lattice with an atom which is spin polarised, with $n=0$. Due to the presence of the harmonic trap, we do not need to spatially constrain the initial state to obtain a well localised state. In Fig.~\ref{fig:SpinPopTrap}(a), typical spin populations $\rho_n, n\in\{-1,0,1\}$ (top) and mean atomic spin (bottom) are plotted against time for $V=5\times 10^{-3} J_x$. The total mean $y$ displacement in a period provides an estimate of the Chern number, in this case $\CC_1=0.87$. Once again this agrees well with the expected value of 1.

As shown in Fig.~\ref{fig:SpinPopTrap}(b), the extracted Chern number only weakly depends on the trap depth $V$, proving that the proposed measure of the Chern number is robust in a wide range of trap amplitudes. This study shows that our protocol is experimentally realistic, and can provide an accurate measure of the Chern number with present day technology.

\section{Conclusions}
\label{sec:HofConclusion}

In this chapter, we have investigated the topological properties of narrow two dimensional topologically non-trivial systems, and their differences relative to the corresponding extended model. In order to address this issue quantitatively, we have presented a method to measure the Chern number in integer quantum Hall systems with open boundary conditions which are extremely short in one direction. This situation is of direct experimental relevance as it was recently shown that, under certain conditions, spinful atoms in one dimension could be mapped to spinless atoms on a narrow strip subject to the two dimensional Hofstadter model. We have presented a detailed experimental protocol to measure the Chern number on a Hofstadter strip, which relies on adiabatically loading a spin polarised atom into the ground state of the lattice, and measuring the mean displacement after a period of Bloch oscillations. We have shown through simulations that this simple experimental protocol can measure the Chern number to an extreme accuracy, despite the reduced transverse extension of the strip. In particular, we observed that the Chern number counts the number of edge states accurately (when these can be easily identified), which suggests that the bulk-boundary correspondence remains valid even for such narrow systems.

In order to probe the experimental feasibility of our protocol, we have tested its robustness against disorder and the presence of a harmonic confinement.
We verified that the measurement of the Chern number remains accurate for disorder magnitudes smaller than the band gap, and for realistic harmonic traps.

\chapter{Conclusion and outlook}
\label{Chapter6}


The central theme of this thesis was the use of cold atoms as a tool to explore topology in quantum mechanics. We used these both to create a system which could detect topological bound states and to motivate new questions.


We approached this theme from three different angles, the first of which was \emph{the design of new topologically non-trivial systems}, using the vast experimental toolbox of cold atoms.
By designing our own specific example, the atomic quantum walk, we detailed the appearance of new topological phases as a result of the system’s Floquet nature. We further showed that this system could host topological bound states. The key to doing this was creating a topological boundary by spatially modulating the Raman coupling. We believe that this study underlines some of the important tools and considerations to be taken into account when designing a new topological system, and as such can be used as a road map for the design of these new phases.


The second topic we discussed was the \emph{detection of topological properties}. Because our understanding of topological phases in open or interacting systems is still incomplete, this subject is of extreme importance to experimentally construct the phase diagram of these complex systems. In the atomic quantum walk, we showed that topological bound states had a unique spin structure as a result of their symmetry protection. By identifying this structure through a spin-dependent position measurement, we were able to identify a topological bound state of the Hamiltonian. This method is general to systems which have bound states protected by chiral symmetry. What is more, we showed that it is possible to differentiate different flavours of bound states in Floquet systems by performing measurements at half time-steps. This means that we can reconstruct experimentally the entire phase diagram using this method.

In the penultimate chapter of this thesis, we developed a method for measuring directly the topological invariant in the Hofstadter strip. While this is a bulk property, and as such only strictly defined for periodic systems, we saw that we could measure it to an extreme accuracy from the atomic dynamics, as long as these took place away from the system's edges. Our method relies on adiabatically loading a spin polarised atom into the system's lowest band, inducing a period of Bloch oscillations, and performing a single measurement of the atom's mean position. This protocol should, in principle, not pose any experimental difficulty in modern setups. We further found that, when the atoms had amplitude at the system's edges, they were pumped to higher bands through the system's edge states. This observation allowed us, in some cases, to measure the topological invariant of higher bands.


The last topic we discussed in this thesis was the \emph{topological characterisation of small systems with open boundary conditions}. We were concerned that for a system with an extremely thin bulk, such as the Hofstadter strip, it would not be possible to define a topological invariant. To our surprise, we found that this topological invariant could be measured through methods inspired from the integer quantum Hall effect. We further found that this integer invariant accurately counted the number of edge states at the system's boundary (when these could be clearly identified). This observation suggests that, even in systems which present such a high edge to bulk surface ratio, the bulk-boundary correspondence continues to apply.


There are several interesting directions in which this work could be extended. The road map which we developed could, for instance, be used to develop new methods to generate long-lived Floquet Majorana bound states, which are highly relevant to the field of quantum computations.

An equally interesting perspective would be to extend our bound state identification method to symmetries other than chiral symmetry. One application of this would be to recognise Majorana Fermions, which are extremely difficult to identify unambiguously.


Finally, our work on the Hofstadter strip underlines the need to construct topological invariants which do not rely on periodic boundary conditions. This would allow us to identify when the transition takes place from topologically non-trivial to topologically trivial, i.e: when the bulk topological invariant ceases to be well defined, and the edge states can no longer be identified.





\printbibliography[heading=bibintoc]


\end{document}